\documentclass[11pt,urlcolor=blue, linkcolor=blue]{article} 
\usepackage{cite}

\usepackage{amsmath, amsthm, amssymb,slashed}
\usepackage{ifpdf}
\ifpdf
  \usepackage[pdftex]{graphicx}
  \usepackage{epstopdf}
\else
  \usepackage[dvips]{graphicx}
\fi
\usepackage[letterpaper]{geometry}
\parskip=\baselineskip

\usepackage{tablefootnote}

\usepackage[usenames, dvipsnames]{color}
\usepackage[svgnames]{xcolor}
\usepackage[colorlinks,citecolor=RoyalBlue, urlcolor=RoyalBlue, linkcolor=RoyalBlue ]{hyperref} 

\allowdisplaybreaks[1]
\numberwithin{equation}{section}

\newcommand{\cblue}[1]{\textcolor{blue}{#1}}
\newcommand{\cred}[1]{\textcolor{black}{#1}}
\newcommand{\ccred}[1]{\textcolor{red}{#1}}
\newcommand{\ccblue}[1]{\textcolor{blue}{#1}}

\usepackage{upgreek}

\usepackage{tabularx}

\DeclareMathAlphabet{\mathpzc}{OT1}{pzc}{m}{it}

\newcommand\hcup[1]{\underset{{\scriptscriptstyle #1}}{\cup}}

\newcommand{\bea}{\begin{eqnarray}}
\newcommand{\eea}{\end{eqnarray}}
\def\be{\begin{equation}}
\def\ee{\end{equation}}

\def\RP{{\mathbb{RP}}}
\def\CP{{\mathbb{CP}}}

\definecolor{red}{rgb}{1,0,0}
\definecolor{blue}{rgb}{0,0,1}
\definecolor{dblue}{rgb}{0,0,0.4}
\definecolor{green}{rgb}{0,1,0}
\definecolor{black}{rgb}{0,0,0}
\definecolor{white}{rgb}{1,1,1}

\definecolor{brn}{rgb}{.8,.4,.0}
\definecolor{redo}{rgb}{1,.5,.0}
\definecolor{ddgrn}{rgb}{0,0.4,0}
\definecolor{dgrn}{rgb}{0,0.55,0}
\definecolor{dbl}{rgb}{0,0,0.5}

\usepackage[bbgreekl]{mathbbol}
\usepackage{amscd}

\newcommand{\Z}{\mathbb{Z}}
\newcommand{\C}{\mathbb{C}}
\newcommand{\R}{\mathbb{R}}

\newcommand{\ii}{\hspace{1pt}\mathrm{i}\hspace{1pt}}
\newcommand{\dd}{\hspace{1pt}\mathrm{d}}
\newcommand{\<}{\langle} 
\renewcommand{\>}{\rangle} 

\newcommand{\Refe}[1]{Ref.~\cite{#1}}
\newcommand{\Eq}[1]{Eq.~(\ref{#1})} 
\newcommand{\eq}[1]{(\ref{#1})} 
\newcommand{\eqn}[1]{Eq.~(\ref{#1})}

\newcommand{\Tr}{{\rm Tr}}

\newcommand{\bpm}{\begin{pmatrix}}
\newcommand{\epm}{\end{pmatrix}}
\newcommand{\bmm}{\begin{matrix}}
\newcommand{\emm}{\end{matrix}}

\newcommand{\cB}{ {\cal B} }
\newcommand{\cC}{ {\cal C} } 
\newcommand{\cD}{ {\cal D} }

\newcommand{\cP}{ {\cal P} }

\newcommand{\cT}{ {\cal T} }

\newcommand{\al}{\alpha}

\def\CB{{\cal B}}

\def\CD{{\cal D}}

\def\Z{{\mathbb{Z}}}

\def\R{{\mathbb{R}}}
\def\C{{\mathbb{C}}}

\def\Tr{{\mathrm{Tr}}}

\def\bZ{{\mathbf{Z}}}

\usepackage{slashed}
\usepackage[makeroom]{cancel}
\usepackage{soul}

\DeclareRobustCommand\clefG{\includegraphics[height=3.95ex]{clefG}}
\DeclareRobustCommand\clefF{\includegraphics[height=3ex]{clefF}}
\DeclareRobustCommand\clefC{\includegraphics[height=2.6ex]{clefC}}

\newcommand{\Gfootnote}[1]{%
\let\oldthefootnote=\thefootnote%
\stepcounter{mpfootnote}%
\addtocounter{footnote}{-1}%
\renewcommand{\thefootnote}{\clefG}
\footnote{#1}%
\let\thefootnote=\oldthefootnote%
}

\newcommand{\Ffootnote}[1]{%
\let\oldthefootnote=\thefootnote%
\stepcounter{mpfootnote}%
\addtocounter{footnote}{-1}%
\renewcommand{\thefootnote}{\clefF}
\footnote{#1}%
\let\thefootnote=\oldthefootnote%
}

\newcommand{\Cfootnote}[1]{%
\let\oldthefootnote=\thefootnote%
\stepcounter{mpfootnote}%
\addtocounter{footnote}{-1}%
\renewcommand{\thefootnote}{\clefC}
\footnote{#1}%
\let\thefootnote=\oldthefootnote%
}

\DeclareRobustCommand\sWan{\includegraphics[height=4.85ex]{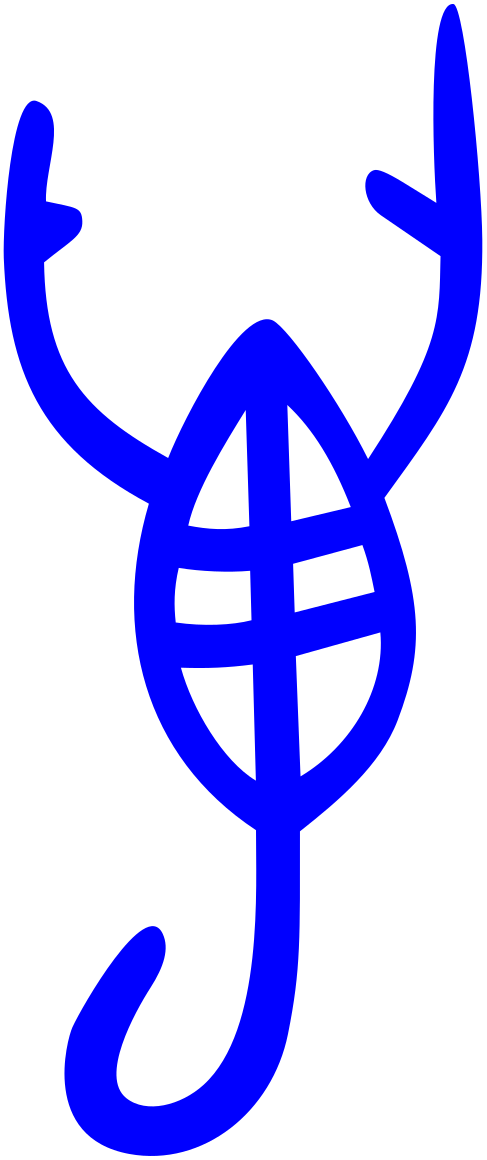}}
\DeclareRobustCommand\sWang
{\cblue{\includegraphics[height=4.3ex]{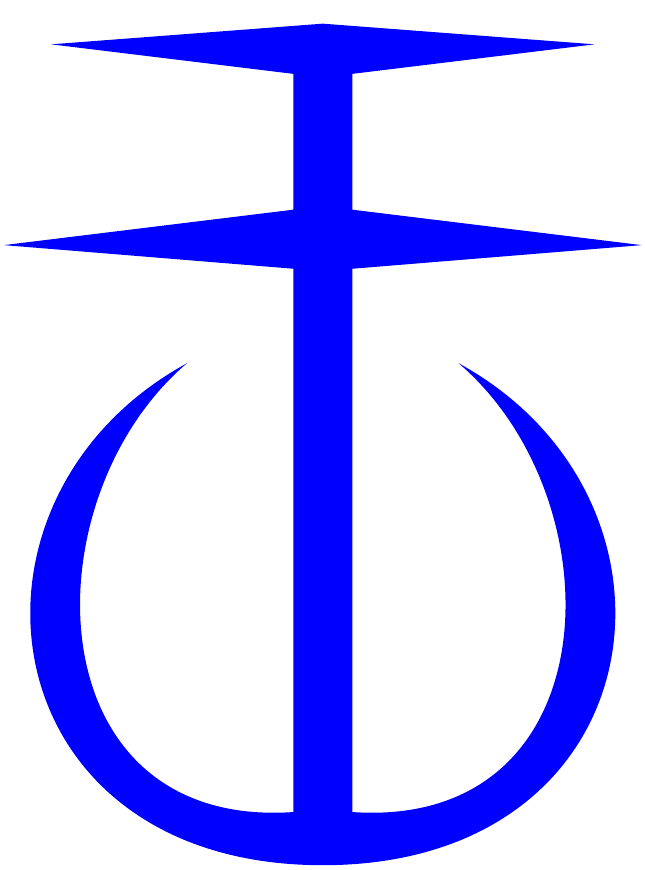}}}
\DeclareRobustCommand\sZheng{\includegraphics[height=4.55ex]{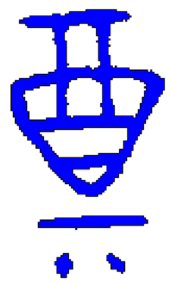}}

\newcommand{\Wanfootnote}[1]{%
\let\oldthefootnote=\thefootnote%
\stepcounter{mpfootnote}%
\addtocounter{footnote}{-1}%
\renewcommand{\thefootnote}{\sWan}
\footnote{#1}%
\let\thefootnote=\oldthefootnote%
}

\newcommand{\Wangfootnote}[1]{%
\let\oldthefootnote=\thefootnote%
\stepcounter{mpfootnote}%
\addtocounter{footnote}{-1}%
\renewcommand{\thefootnote}{\sWang}
\footnote{#1}%
\let\thefootnote=\oldthefootnote%
}

\newcommand{\Zhengfootnote}[1]{%
\let\oldthefootnote=\thefootnote%
\stepcounter{mpfootnote}%
\addtocounter{footnote}{-1}%
\renewcommand{\thefootnote}{\sZheng}
\footnote{#1}%
\let\thefootnote=\oldthefootnote%
}

\usepackage{enumitem,cleveref}
\newcommand{\nn}{\nonumber}

\usepackage{comment}

\def \- {\!\smallsetminus\!}

\def\TP{\mathrm{TP}}
\def\Sq{\mathrm{Sq}}
\def\B{\mathrm{B}}
\newcommand{\tO}{{\rm O}}
\newcommand{\rO}{{\rm O}}
\newcommand{\rE}{{\rm E}}

\newcommand{\SO}{{\rm SO}}
\newcommand{\Spin}{{\rm Spin}}
\newcommand{\U}{{\rm U}}

\newcommand{\SU}{{\rm SU}}
\newcommand{\PSU}{{\rm PSU}}
\newcommand{\Pin}{{\rm Pin}}

\newcommand{\tE}{{\rm E}}

\newcommand{\W}{{\rm W}}
\newcommand{\rN}{{\rm N}}

\def \Hom{\operatorname{Hom}}

\def \H{\operatorname{H}}

\def \im{\mathrm{i}}

\def \Z{\mathbb{Z}}
\def \A{\mathcal{A}}
\def \RP{\mathbb{RP}}
\def \CP{\mathbb{CP}}

\def\Ext{\operatorname{Ext}}

\newcommand{\Sec}[1]{Sec.~\ref{#1}} 

\newcommand{\Fig}[1]{Fig.~\ref{#1}} 

\newcommand{\fig}[1]{Fig.~\ref{#1}}

\usepackage{tikz}
\usetikzlibrary{calc}

\usepackage[all,cmtip]{xy}
\usepackage{tikz-cd}

\usetikzlibrary{matrix}
\usetikzlibrary{knots}

\usetikzlibrary{calc}

\usepackage[all,cmtip]{xy}
\usepackage{tikz-cd}

\usepackage{datetime}

\newcommand\finline[3][]{\begin{myfont}[#1]{#2}#3\end{myfont}}%
\newenvironment{myfont}[2][]{\csname#2\endcsname[#1]}{}

\usepackage{makecell} 

\begin{document}
\begin{titlepage}
\vskip 1.mm

\begin{center}

{\bf\LARGE{Quantum 4d Yang-Mills Theory and\\[5.5mm] 
Time-Reversal Symmetric \\[5.5mm]  
 5d Higher-Gauge Topological Field Theory:
 }\\[5.5mm]} 
{\Large{Anyonic-String/Brane Braiding Statistics to Link Invariants
}\\[7.5mm]}


\vskip-0.mm 
\quad\quad\quad
\Large{
{Zheyan Wan$^{1,2}$,{\Wanfootnote{e-mail: {\tt wanzheyan@mail.tsinghua.edu.cn} 
\quad\quad \quad\quad\quad\quad \quad\quad \quad\quad\quad\quad \quad\quad \quad \quad\quad \quad\quad\quad\quad\quad\quad\quad\quad\quad\quad
 March 2019}} \quad
Juven Wang$^{3,4}$,{\Wangfootnote{e-mail: {\tt jw@cmsa.fas.harvard.edu} 
(Corresponding Author)
}} and
Yunqin Zheng$^{5}${\Zhengfootnote{e-mail: {\tt {yunqinz@princeton.edu}} 
\flushleft
 \quad\quad \emph{Dedicated to Professor Shing-Tung Yau {\includegraphics[height=3.25ex]{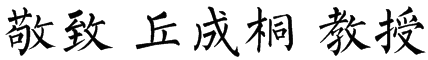}}
and his 70th Birthday celebration on April 4th, 2019.} \quad\quad\quad\quad \quad\quad  }} 
} 
\vskip.5cm
{\small{\textit{$^1${Yau Mathematical Sciences Center, Tsinghua University, Beijing 100084, China}\\}}
}
\vskip.2cm
{\small{\textit{$^2${School of Mathematical Sciences, USTC, Hefei 230026, China}\\}}
}
 \vskip.2cm
  {\small{\textit{$^3$School of Natural Sciences, Institute for Advanced Study,  Einstein Drive, Princeton, NJ 08540, USA}\\}}
 \vskip.2cm
 {\small{\textit{$^4${Center of Mathematical Sciences and Applications, Harvard University,  Cambridge, MA 02138, USA} \\}}
}
 \vskip.2cm
 {\small{\textit{$^5$Physics Department, Princeton University, Princeton, New Jersey 08544, USA \\}}
}}

\end{center}
\begin{abstract}

We explore various 4d Yang-Mills gauge theories (YM) living  
as boundary conditions of 5d gapped short/long-range entangled (SRE/LRE) topological states.
Specifically, we explore 4d time-reversal symmetric pure YM of an SU(2) gauge group with a second-Chern-class topological term at $\theta=\pi$ (SU(2)$_{\theta=\pi}$ YM), by turning on the background fields for both the time-reversal (i.e., on unorientable and non-spin manifolds) and  the 1-form center global symmetry.
We find Four Siblings of {time-reversal and Lorentz symmetry-enriched} SU(2)$_{\theta=\pi}$ YM with distinct couplings to background fields, labeled by $(K_1, K_2)$: 
$K_1=0,1$ specifies Kramers singlet/doublet Wilson line and new mixed higher 't Hooft anomalies; 
$K_2=0,1$  specifies boson/fermionic Wilson line and a new 
Wess-Zumino-Witten-like counterterm. 
Higher anomalies indicate that 
 in order to realize all higher $n$-global symmetries locally on $n$-simplices, 
the 4d theory becomes a boundary of a 5d higher-symmetry-protected topological state (SPTs, as
{an invertible topological quantum field theory (iTQFT)
or a cobordism invariant} in math, 
or as a 5d higher-symmetric interacting \emph{topological superconductor} in condensed matter).
Via Weyl's gauge principle, by dynamically gauging the 1-form symmetry, 
we transform a 5d bulk SRE SPTs into an LRE symmetry-enriched topologically ordered state (SETs); 
thus we obtain the 4d SO(3)$_{\theta=\pi}$ YM-5d LRE-higher-SETs coupled system with dynamical higher-form gauge fields. 
We further derive new exotic anyonic statistics of extended objects such as 2d worldsheet of strings and 3d worldvolume of branes, physically characterizing the 5d SETs. 
We discover triple and quadruple link invariants potentially associated with the underlying 5d higher-gauge TQFTs, hinting at a new intrinsic relation between 
non-supersymmetric 4d pure YM and topological links in 5d.
We provide 4d-5d lattice simplicial complex regularizations and bridge to 4d SU(2) and SO(3)-gauged quantum spin liquids as 
3+1 dimensional 
realizations. 
We constrain YM low-energy gauge dynamics by higher anomalies and a higher symmetry-extension method.

\end{abstract}
\end{titlepage}


 






\renewcommand{\eqref}{\eqn}

  \pagenumbering{arabic}
    \setcounter{page}{2}
    
\tableofcontents   

\newpage
\section{Introduction and Summary}

\label{sec:intro}

The world where we reside, to our best present understanding, can be described by quantum theory and the underlying long-range entanglement.
Quantum field theory (QFT) and in particular
quantum gauge field theory, under the spell of \emph{Gauge Principle} following the insights since Maxwell, Hilbert, Weyl, Pauli, and others (see a historical review \cite{RevModPhys13.203Pauli}), 
embodies the quantum, special relativity and long-range entanglement into a systematic framework.
Yang-Mills (YM) gauge theory \cite{PhysRev.96.191-YM}, generalizing the U(1) gauge group of quantum electrodynamics to a non-abelian Lie group,   has been proven to be powerful to describe the Standard Model physics.

The Euclidean partition function of a pure YM theory with an SU(N) gauge group  on a 4-dimensional  (i.e., 4d)\footnote{
We denote $n$d for an $n$-dimensional spacetime.
We denote $m+1$D for an $m$-dimensional space and 1-dimensional time.
We denote $m$D for an $m$-dimensional spatial object.
} 
spacetime $M^4$
and a second-Chern-class topological term $S_{\theta}$ labeled by $\theta$, i.e., SU(N)$_{\theta}$-YM, 
is
\begin{multline}
 \label{eq:YM-pi}
\bZ^{\text{$4$d}}_{{\text{YM}}}
\equiv\int [{\cal D} {a}] \exp\big( - S_{\text{SU(N)}_{\theta}}[a] \big)\equiv\\
\int [{\cal D} {a}] \exp\big( - S_{\text{SU(N)}_{\theta=0}}[a] \big)\exp\big( - S_{\theta}[a] \big)
\equiv
\int [{\cal D} {a}] \exp   \Bigg(- \int\limits_{M^4} \frac{1}{g^2}\text{Tr}\,F_a\wedge \star F_a
+  \int\limits_{M^4} \frac{\ii  \theta}{8 \pi^2}  \text{Tr}\,F_a\wedge F_a \Bigg),
\end{multline}
where $a$ is the $\SU(\rN)$ gauge field, $F_a=\dd a-\ii a \wedge a$ is the SU(N) field strength, and $g$ is the gauge coupling constant. 
See the footnote\footnote{
\label{ft1}
$a$ is locally a 1-form SU(N) connection obtained from parallel transporting the principal-SU(N)  bundle over the spacetime manifold $M^4$.
Locally $a = a_\mu \dd x^\mu= a_\mu^\alpha T^\alpha \dd x^\mu$ with
 $T^\alpha$ is the hermitian generator of su(N) Lie algebra, satisfying the commutator $[T^\alpha,T^\beta]=\ii f^{\alpha \beta \gamma} T^\gamma$ where $f^{\alpha \beta \gamma}$ is a fully anti-symmetric structure constant.
Locally $\dd x^\mu$ is a differential 1-form. $a_\mu=a_\mu^\alpha T^\alpha$ is the Lie algebra valued gauge field.
The path integral $\int [{\cal D} {a}]$ is meant to integrate over all the configurations of  $a(t,x)$ modding out the SU(N)  gauge transformation $a\to 
\mathfrak{g}^{-1}a \mathfrak{g} -\ii \mathfrak{g}^{-1}\dd \mathfrak{g}$.  
The $\frac{1}{g^2}\text{Tr}\,(F_a\wedge \star F_a)$ is the Yang-Mills Lagrangian \cite{PhysRev.96.191-YM} (a non-abelian generalization of Maxwell U(1) gauge theory) where $\star F_a$ is the Hodge dual of $F_a$.
Tr stands for the trace as an invariant quadratic form of the Lie algebra su(N).
The term   $S_\theta[a]=\frac{ -\ii  \theta}{8 \pi^2}  \text{Tr}\,F_a\wedge F_a$ is related to the second-Chern-class $c_2(V_{\SU(\rN)})$ of the SU(N) gauge bundle via $S_{\theta}[a]=-\ii \theta c_2(V_{\SU(\rN)})$. 
This path integral is sensible for physicists, but may not be mathematically well-defined.
We will also point out how to grasp the meaning of YM path integral on unorientable manifolds in \Sec{sec:SPT}. 
}
for further explanations of the notations. When $\theta=0$, the SU(N)$_{\theta=0}$  YM theory is believed to be in the confined phase \cite{YMMP-Jaffe-Witten}, have an energy gap, and have a single ground state on any manifold. In particular, there is no 't Hooft anomaly \cite{tHooft:1980xss}. 
Recently,  \Refe{Gaiotto2017yupZoharTTT1703.00501} discovered that  for
SU(N)$_{\theta=\pi}$-~YM with even N,
there is a subtle 't Hooft anomaly involving the time-reversal symmetry $\Z_2^T$ and 1-form center global symmetry $\Z_{\rN,[1]}^e$.\footnote{We use the subscript $[1]$ to indicate that the symmetry is a 1-form generalized global symmetry \cite{Gaiotto2014kfa1412.5148}, 
and the superscript $e$ to indicate the symmetry is electric as opposed to magnetic (i.e., the charged line operators are the Wilson lines rather than the 't Hooft lines).  When we say  {\emph{symmetry}} in this article, we always mean \emph{global symmetry} unless we state otherwise.} 
The  't Hooft anomaly of a 4d theory is captured by a 5d topological term through the anomaly inflow mechanism \cite{Callan:1984sa}. 
Schematically, \Refe{Gaiotto2017yupZoharTTT1703.00501} suggested a 5d  topological term linear in the time-reversal background field $\cT$ and quadratic in the $\Z_{\rN,[1]}^e$ background field $B$, 
\bea
\sim \cT B B. \label{eq:TBB}
\eea
The 5d topological term characterizes the 5d \emph{short-rangle-entangled} (SRE) phase. See \Sec{sec:SPT} for more rigorous definitions of the background fields and the 5d topological term. 

Further recently, \Refe{Wan2018zql1812.11968} suggested that
there are additional new higher 't Hooft anomalies for some 4d SU(N)$_{\theta=\pi}$ theories at even N:
From one perspective,  \Refe{Wan2018zql1812.11968} suggested that when $\rN = 2$,
there is a mixed anomaly captured by a 5d topological term which is  \emph{cubic} in $\cT$ and linear in   $B$, which is schematically denoted as 
\bea
\sim \cT \cT \cT  B. \label{eq:TTTB} 
\eea
From another perspective, \Refe{Wan2018zql1812.11968} suggested that the SU(N)$_{\theta=\pi}$ YM at an even integer $\rN \geq 4$ 
contains new mixed anomalies involving
 $\Z_2^T$, $\Z_{\rN,[1]}^e$ and a 0-form charge conjugation (i.e., a $\Z_2$ outer-automorphism) symmetry. 
For example, at N $=4$, another anomaly can be  captured by the 5d topological terms schematically as 
${\sim A_C \cT \cT B.}${\footnote{\cred{Note that
so far $\SU(\rN)$ YM for N $=4$, \Refe{Wan2018zql1812.11968} finds a new 4d anomaly expressible by a term 
$$ A_C w_1(TM)^2 B \sim  A_C \cT \cT B.$$ 
The $w_j(TM)$ is the $j$-th Stiefel-Whitney (SW) class of spacetime $M$'s tangent bundle $TM$.
Although there is still a possibility that another 4d anomaly may exist 
$$
A_C^2 \beta_{(2,4)} B \sim  A_C A_C \cT B.
$$
More precisely, these two 4d anomalies are captured by 5d \emph{invertible} topological quantum field theories (iTQFTs) 
$ A_C w_1(TM)^2 B$ and $A_C^2 \beta_{(2,4)} B$
on a 5d closed manifold respectively.
Here $\beta_{(n,m)}:\H^*(-,\Z_{m})\to\H^{*+1}(-,\Z_{n})$ is the Bockstein homomorphism associated with the extension $\Z_n\stackrel{\cdot m}{\to}\Z_{nm}\to\Z_m$ where $\cdot m$ is the group homomorphism given by multiplication by $m$. In particular, $\beta_{(2,2^n)}=\frac{1}{2^n}\delta\mod2$.
The detailed derivation of $\SU(\rN)$ YM for $\rN >2$ will be left for the future work \cite{WWZ2019-2}.}} 
}
%
%
Here $A_C$ is  a 1-form background field 
for the 0-form $\Z_2^C$ charge conjugation symmetry.
In the following, we will make the above schematic 5d topological terms  
\eqn{eq:TBB} and \eqn{eq:TTTB} 
mathematically precise,  following the setup in
\Refe{Wan2018zql1812.11968} and \Refe{Wan2018bns1812.11967}.

The above 5d topological terms can be regarded as semi-classical partition functions (definable on closed 5-manifolds with appropriate structures)
whose functional values depend on the \cred{external values} to global symmetry-background probe fields. 
These 5d topological terms are also {known as} \emph{invertible} topological quantum field theories (iTQFTs) in the literatures.\footnote{
By iTQFT, physically it means that the absolute value of partition function 
$|\bZ|=1$ on any closed manifold. Thus this $\bZ$ can only be a complex phase $\bZ = \rm{e}^{\ii \theta}$, which can thus be inverted
and cancelled by $ \rm{e}^{- \ii \theta}$ as another iTQFT.
}
In the present work, we will further dynamically gauge
the  1-form symmetry $\Z_{\rN,[1]}^e$ associated to the coupled systems of 4d YM and 5d topological terms above. After gauging $\Z_{\rN,[1]}^e$, 
the 5d SRE topological terms become  5d  \emph{long-range entangled}  (LRE) topological quantum field theories (TQFT). 
We further apply the methods developed in 
Refs.~\cite{1602.05951, Putrov2016qdo1612.09298, Wang2018edf1801.05416} 
to analytically compute the physical observables of the higher-gauge 5d TQFTs. 
The physical observables of 5d TQFTs include, for example, (i) the partition functions $\bZ[M^5]$ on closed manifolds $M^5$,
(ii) braiding statistics of anyonic strings and anyonic branes whose spacetime trajectories forming 2d worldsheets and 3d worldvolumes respectively, 
and link invariants of these 2d worldsheets and 3d worldvolumes in $M^5$.
We uncover new spacetime braiding processes and link invariants, 
including triple and quadruple linkings
analogous to previous works \cite{Putrov2016qdo1612.09298, JWangthesis, 1602.05951, Guo2018vij1812.11959, Wang1901.11537}.
\footnote{
Here we comment on the physical and mathematical meanings of fractional statistics and non-abelian statistics associated with the spacetime braiding processes involving
0D anyonic particles, 1D anyonic strings,  2D anyonic branes and other extended objects. 
In the discussions below, we take a generalized 
definition of \emph{anyonic}.  \\
$\bullet$ In a more \emph{restricted definition},  anyonic means the self-exchange statistics can go beyond 
bosons or fermions \cite{Wilczek:1990ik}.  \\
$\bullet$ In our \emph{generalized definition}, anyonic means that either self-exchange statistics (of identical objects) 
or the mutual statistics (of multiple distinguishable objects, may involving more than two objects)
can go beyond bosonic or fermionic statistics. 
\\
--- In 3d (2+1D) spacetime $M^3$,
braiding statistics of particles can be fractional (such as the exchange statistics of two identical particles,
or mutual statistics of two different particles) which are called anyonic particles (see an excellent historical overview \cite{Wilczek:1990ik}).
As an example, this can be understood from a 3d Chern-Simons action with  1-form gauge field $a$ integrated over $M^3$
$$\sim \int_{M^3} a_I \dd a_J$$
which modifies the quantum statistics of particle worldline whose open ends host the  anyonic particles.
\\
--- In 4d (3+1D) spacetime $M^4$,
braiding statistics of particles \emph{cannot} be fractional as the two 1-worldlines cannot be linked in 4d.
Thus there is no anyonic particle and no fractional particle statistics beyond bosons or fermions in 4d. %
However, 
the braiding statistics of strings \emph{can} be fractional which we dub anyonic strings.
As an example, the fractional statistics of strings can be understood from a 4d TQFT with a 1-form gauge field $a$ and a 2-form gauge field $b$, as
$$\sim \int_{M^4} b \dd a$$
which modifies the mutual quantum statistics of a 0D particle from 1-worldline $W_1= e^{i \int_{\gamma} a}$ linked with a 1D string from 2d worldsheet $U_2= e^{i \int_{S^2}b}$ in $M^4$.\\
Since a particle cannot carry a fractional charge in 4d,
we can interpret the above theory as the anyonic string carrying a fractional flux in 4d. 
Another way to interpret the fractional statistics of anyonic strings is through the dimensional reduction from 4d to 3d.  Let $M^4=S^1\times M^3$ with the size of $S^1$ much smaller than the size of $M^3$ and let  the closed anyonic strings wrap around $S^1$  \cite{Wang1403.7437, Jiang:2014ksa, Wang1404.7854} , then the anyonic strings in $M^4$  reduce to anyonic particles in $M^3$.  \\
From the field theory side, the 4d TQFTs 
$$\sim \int_{M^4} a_I a_J \dd a_K, \quad \sim \int_{M^4} a_I a_J  a_K a_L$$ 
can modify the braiding statistics of strings 
\cite{Putrov2016qdo1612.09298, 1602.05951,1405.7689, Gu2015lfaGWW1503.01768, Ye1508.05689, RyuTiwariChen1603.08429, He1608.05393, Wang2018iwz1810.13428}. See the relations between Dijkgraaf-Witten's group cohomology theory \cite{Wittencohomology} and these TQFTs discussed in  \cite{Putrov2016qdo1612.09298, 1602.05951,1405.7689}. Furthermore, there are  4d gauge invariant topological terms with 2-form gauge field $b$ \cite{Putrov2016qdo1612.09298, Ye1410.2594, Gaiotto2014kfa1412.5148}
$$\sim \int_{M^4} b_I b_J.$$
\\
--- In 5d (4+1D) spacetime $M^5$, for example, there exist self and mutual coupling type of 5d topological terms with 2-form gauge fields $b$, $b_I$, $b_J$, etc., 
$$
\sim \int_{M^5} b \dd b, \quad \sim \int_{M^5} b_I \dd b_J.
$$
The self coupling term $\int_{M^5}  b \dd b$ leads to anyonic strings within the \emph{restricted definition}, where the self-exchange statistics goes beyond bosonic and fermionic\cite{Wilczek:1990ik}. 
The mutual coupling term $\int_{M^5} b_I \dd b_J$ leads to anyonic strings within the \emph{generalized definition}, where anyonic means that mutual statistics of distinguishable 1D strings
can go beyond bosonic or fermionic statistics. 
Both terms modify the 
quantum statistics of string worldsheet whose open ends host the 1D anyonic string.\\
We can have
another Aharonov-Bohm like topological term with local 1-form gauge field $a$ and 3-form gauge field $c$.
$$\sim \int_{M^5} c \dd a,$$
When the above term appears together with other  topological terms like $\int_{M^5} a_I a_J a_K \dd a_L$, 
the statistics of 2D brane (attached to the end of 3d worldvolume) can have fractional statistics within the general definition, while 
the statistics of 0D particle (attached to the end of 1-worldline $e^{i \int a}$) remains  non-anyonic. 
Again the anyonic brane in 5d can reduce to anyonic particles in 3d by compactifying along $T^2$ in $M^5$ \cite{Wang1403.7437, Jiang:2014ksa, Wang1404.7854}. 
There are many other terms allowed in 5d\cite{1405.7689}. For a general dimension $d$, there exists the topological term 
$$ \int c_{m} \dd c_{n}  \sim \int c_{n} \dd c_{m}, ~~~ n+m=d-1, ~~~ n\leq m $$ 
where $c_n$ is a $n$-form gauge field. we always take the higher-dimensional object from the $c_{m}$-field to have fractional statistics (the analogs of fractional flux),
while we take the lower-dimensional object from the $c_{n}$-field to have a regular statistics (the analogs of integrally quantized charge).
}



{Now let us take a step back to digest the physical meanings of these 5d topological terms \eqn{eq:TBB}-\eqn{eq:TTTB}.}
The $d$ dimensional  't Hooft anomaly of ordinary 0-form 
global symmetries is known to be captured by a $(d+1)$ dimensional iTQFT. 
In the condensed matter literatures, these  $(d+1)$d iTQFTs describe Symmetry-Protected Topological states (SPTs)\footnote{We abbreviate both Symmetry-Protected Topological \emph{state}(s) 
as SPTs.
We abbreviate Symmetry-Enriched Topologically ordered \emph{state}(s) 
as SETs.} \cite{Chen2011pg1106.4772, Senthil1405.4015, Wen1610.03911, Weneaal3099}.  
The relations between the SPTs and the response probe field-theoretic partition functions have been \emph{systematically} studied, selectively, in 
 \cite{QiHughesZhang, 1405.7689, Wen1410.8477, Metlitski:2015yqa, Witten:2015aba, Witten2016cio1605.02391, 2017arXiv171111587GPW} (and References therein), and climaxed 
 to the hint of cobordism classification of SPTs\cite{Kapustin2014tfa1403.1467, Kapustin1406.7329}.
Recently the iTQFTs and SPTs are found to be systematically classified
by a powerful cobordism theory
framework of Freed-Hopkins \cite{Freed2016.1604.06527}, following the earlier work of Thom-Madsen-Tillmann spectra \cite{thom1954quelques,MadsenTillmann4}.

Further recently, \Refe{Wan2018bns1812.11967} generalized the
Thom-Madsen-Tillmann-Freed-Hopkins cobordism theory  \cite{thom1954quelques,MadsenTillmann4,Freed2016.1604.06527}
to the cases with generalized higher global symmetries \cite{Gaiotto2014kfa1412.5148}.
The generalized cobordism group computation \cite{Wan2018bns1812.11967}, which involves the bordism group of higher classifying spaces and their fibrations, 
e.g. $\B\mathbb{G}$,
can capture 
the $d$ dimensional higher 't Hooft anomaly of generalized global symmetries $\mathbb{G}$
by $(d+1)$ dimensional bordism invariants
(i.e., generalized symmetric iTQFTs or higher symmetric iTQFTs). 
{In the following, we also call the generalized symmetric iTQFTs as \emph{higher-SPTs}. 
The $d$d boundary of $(d+1)$d \emph{higher-SPTs} has $d$d \emph{higher-anomalies}. 
Gauging the higher-symmetry of \emph{higher-SPTs} gives rise to \emph{higher-gauge theories}.}
Earlier pursuits on a systematic framework of the generalized symmetric iTQFTs through cobordism theories and cohomology theories
include, but not limited to,  \Refe{2013arXiv1309.4721K, Sharpe2015mja1508.04770,
Kapustin2017jrc1701.08264, Tanizaki2017qhf1710.08923, Cordova2018cvg2group1802.04790, Delcamp2018wlb1802.10104, Benini2018reh1803.09336, Yonekura2018arXiv180310796Y, Zhu2018kzd1808.09394, Delcamp2019fdp1901.02249} and references therein.
In this work, we identify the
4d anomalies of SU(N)$_{\theta=\pi}$ YM \eqn{eq:TBB}-\eqn{eq:TTTB} with the  mathematically precise 5d bordism invariants,\footnote{
For the mathematical terminology, we
call: \\
$\bullet$ the \emph{bordism group generators} as the manifolds or manifold generators, which generate finite Abelian groups, e.g., $\Z_n$.\\
$\bullet$ the \emph{cobordism group generators} as the topological terms or iTQFTs, which generate Abelian groups, e.g., $\Z_n$ or $\Z$, etc.\\
$\bullet$ the \emph{co/bordism invariants} (people call bordism invariants as cobordism invariants with the same meaning)
mean that they are invariants
under the bordism class of manifolds, 
thus co/bordism invariants mean the topological terms or iTQFTs, which again generate Abelian groups, $\Z_n$ or $\Z$, etc.
}  obtained in Refs.~\cite{Wan2018zql1812.11968} and \cite{Wan2018bns1812.11967}. 

{Let us rephrase the higher anomalies into a condensed matter language on the lattice: 
Higher anomaly indicates that in order to realize all higher $n$-global symmetries locally on $n$-simplices, 
the theory needs to become a boundary of a one-higher dimensional higher-symmetry-protected topological state (higher-SPTs). 
If a theory has a higher anomaly, then the theory in its own dimension has some higher $n$-global symmetries acting non-locally (on $n$-simplices). 
Then there is an obstruction to gauge such a non-local symmetry, hence the name of higher ('t Hooft) anomaly \cite{tHooft:1980xss}.
More in \Sec{sec:lattice}.}

\subsection{The Outline}

Here are the outlines of the present work.

\noindent
$\bullet$ \Sec{sec:SPT} --- We identify the 5d bordism invariants with the 't Hooft anomalies of 
4d SU(N)$_{\theta=\pi}$ YM theory (where we focus on $\rN=2$), as a version of higher-anomaly matching.  

\noindent
$\bullet$ \Sec{sec:SU2} ---
We clarify and enumerate possible distinct classes of 4d SU(2)$_{\theta=\pi}$ YM theories. We take the  condensed matter viewpoint, where we regard the SU(2)$_{\theta=\pi}$ YM theories as  infrared theories emerging from 
{high-energy ultraviolet (UV) bosonic systems with a lattice cutoff}, as opposed to fermionic systems. We thus call the UV system as \emph{bosonic YM theories}. 
These bosonic YM theories still allow Wilson line operators as worldlines of particles 
being (1) either bosonic or fermionic in quantum statistics, (2) either Kramers doublet or Kramers singlet under the time-reversal symmetry.
This supplements as a \emph{partial classification} of 4d SU(2)$_{\theta=\pi}$ bosonic YM theories.
We apply the tools in \Refe{2017arXiv171111587GPW}
to understand the relation between gauge bundle constraint and the properties of line/surface operators. 

From \Sec{sec:SPT} and \Sec{sec:SU2}, we will see that there are at least four closely related
 4d SU(2)$_{\theta=\pi}$ non-supersymmetric pure YM theories 
 (which we nickname them as Four Siblings of 4d SU(2)$_{\theta=\pi}$ YM theories) with bosonic UV completions.
 Each of them carries \emph{either} a distinct 4d 't Hooft anomaly associated with a 5d higher-SPTs \emph{or} 
 a distinct 4d counterterm.
 The distinct 5d higher-SPTs labeled by distinct the 5d bordism invariants
 are actually the physical 
 analogs of the 5d (4+1D) one-form-center-symmetry-protected interacting topological superconductors in the condensed matter language.
%

\noindent
$\bullet$ \Sec{sec:5dTRTQFT} and  \Sec{sec:link-inv} ---  We dynamically gauge the 1-form center symmetry $\Z_{2,[1]}^e$. This turns the
4d SU(2)$_{\theta=\pi}$ YM/5d-higher-SPTs coupled systems in \cite{Wan2018zql1812.11968}
into
4d SO(3)$_{\theta=\pi}$ YM/5d-higher-SETs coupled systems, where SETs stand for 
the \emph{symmetry-enriched topologically ordered states}.\footnote{
Symmetry-Protected Topological state (SPTs) is a short-ranged entangled (SRE) gapped quantum state that can be defined on a lattice. 
(UV complete such as defining the quantum state on a triangulable manifold or a simplicial complex.) 
If we preserve  the global symmetry,
SPTs cannot be deformed to a trivial tensor product state under finite steps of local unitary transformations, assuming without closing any energy gap and without level crossing.  
Once we 
break the global symmetry,
SPTs can be deformed to a trivial tensor product state under finite steps of local unitary transformations, even without closing any energy gap and without energy level crossing.  
Symmetry-enriched topologically ordered state (SETs) 
is a long-ranged entangled (LRE) gapped quantum state that can be defined on a lattice. (Here we only discuss the SETs that are anomaly free.)
Even if the global symmetry is completely broken,
SETs cannot be deformed to a trivial tensor product state under finite steps of local unitary transformations, assuming without closing any energy gap and without level crossing.
The SETs have the same LRE nature as topologically ordered states.
See recent reviews \cite{Chen2011pg1106.4772, Senthil1405.4015, Wen1610.03911, Weneaal3099}. 
}
We then explore the detailed properties of various 5d higher-SETs.
The 5d higher-SETs are described by 5d time-reversal symmetric higher-TQFTs with emergent 2-form dynamical gauge fields.
We mainly focus on the Four Siblings of 5d higher-SETs, while  also consider other highly relevant exotic 5d higher-SETs.
To characterize these 5d higher-SETs, we study the following aspects: 
\begin{enumerate}
\item
Partition function $\bZ[M^5]$ without extended operator (1-line, 2-surface, 3-submanifold) insertions on 5-manifold $M^5$. We compute 
$\bZ[M^5]$ following the techniques and tools built from
 \cite{Putrov2016qdo1612.09298} and \cite{Wang2018edf1801.05416}. In particular, when $M^5=M^4\times S^1$, the partition function $\bZ[M^4\times S^1]$ is the topological ground state degeneracy (topological GSD) on a spatial $M^4$. 
This issue is addressed in \Sec{sec:5dTRTQFT}.


\item 
Braiding statistics involving anyonic 1D string/2D branes, and the associated link invariants of the spacetime 2d worldsheet/3d worldvolume. 
Here we compute the path integral $\bZ[M^5; W, U, \dots]$ with extended-operator insertions ($W, U, \dots$), 
 following the techniques and tools built from
\cite{1602.05951, Putrov2016qdo1612.09298, Guo2018vij1812.11959, Wang1901.11537}.
This issue is addressed in \Sec{sec:link-inv}.

\end{enumerate}

\noindent
$\bullet$ \Sec{sec:link-conf} --- We provide the exemplary 
spacetime braiding processes of anyonic string/brane in 5d, and the link configurations of extended operators,
which can be detected by the link invariants that we derived in \Sec{sec:link-inv}.

\noindent
$\bullet$  
\Sec{sec:SO3-SET} --- 
 We re-examine the coupled system of
 4d SO(3)$_{\theta=\pi}$ YM theories and 5d-higher-SETs appeared in \Sec{sec:5dTRTQFT}.

\noindent
$\bullet$ 
\Sec{sec:lattice}  --- We construct the lattice regularization and UV completion of some of our systems.
This includes 
a lattice realization of 5d higher-SPTs and higher-gauge SETs by implementing on 5d simplicial complex spacetime path integral,
and a 4+1D ``condensed matter'' realization on the spatial Hamiltonian operator.
We also provide a
lattice regularization of (1) higher-symmetry-extended and (2) higher-symmetry-preserving anomalous 3+1D topologically ordered gapped boundaries 
by generalizing the method of \cite{Wang2017locWWW1705.06728}. 
The higher-symmetry-extension method was also developed in \cite{Wan2018djlW2.1812.11955}.

\noindent
$\bullet$ 
\Sec{sec:conclude} --- We conclude and make connections to physics and mathematics in other perspectives.

Before we proceed to the detailed discussions in the main text, we first give a quick overview on
more colloquial and pedestrian summaries in terms of schematic descriptions and Table \ref{table:TQFTlink}, 
in \Sec{sec:table}.
Readers who are not familiar with certain mathematical information or physical motivations
may seek for additional helps from Refs.~\cite{2017arXiv171111587GPW} (and its Appendices), \cite{Wan2018zql1812.11968} and \cite{Wan2018bns1812.11967}.

\subsection{Summaries and Tables}
\label{sec:table}

As we mentioned, in \Sec{sec:SPT} and \Sec{sec:SU2}, we will see that there are at least four closely related
 4d SU(2)$_{\theta=\pi}$ non-supersymmetric pure YM theories 
 (nicknamed the Four Siblings of 4d SU(2)$_{\theta=\pi}$ YM theories are  
 labeled by $(K_1, K_2)  \in (\Z_2,\Z_2)$) with a bosonic UV completion.
{They carry either distinct 4d higher 't Hooft anomalies\footnote{
Distinct 4d higher 't Hooft anomalies correspond to distinct 5d higher-SPTs/counterterms labeled by distinct 5d bordism invariants:
 physical analogs of 5d (4+1D) one-form-center-symmetry-protected interacting ``topological superconductors'' in a condensed matter language.
 In condensed matter, topological superconductors refer to electronic systems with time-reversal symmetry but 
 without U(1) electron charge conservation symmetry (see an overview \cite{Senthil1405.4015,  Wen1610.03911}), for example due to the Cooper pairing breaking 
 U(1) charge symmetry down to a discrete subgroup or down to nothing.
 } or distinct 4d counterterms.}
All these anomalies that we will discuss below are the mod 2 non-perturbative global anomalies, similar to the old and the new 
SU(2) anomalies \cite{Witten:1982fp, Wang:2018qoyWWW}; except that instead of an ordinary global symmetry,
now we require a higher 1-form symmetry $\Z_{2,[1]}^e$ to probe higher anomalies.
Here we advertise these results in a colloquial and pedestrian manner.
\begin{enumerate}
\item
\underline{$(K_1, K_2)=(0,0)$.}
The 1st Sibling of 4d SU(2)$_{\theta=\pi}$ with Kramers singlet ($T^2=+1$) bosonic Wilson line has the  4d anomaly/5d bordism invariant schematically as: 
\bea
\sim w_1(TM) B B, \label{eq:w1BB}
\eea
with the $j$-th Stiefel-Whitney (SW) class $w_j(TM)$ of spacetime manifold $M$'s tangent bundle $TM$.
Here $B \in \H^2(M,\Z_2)$ is a degree-2 cohomology class obtained from restricting the 2-form $\CB$ field via $\CB \sim \pi B$
and $\oint_\Sigma \CB=\pi \Z$ for any closed surface $\Sigma$.
More rigorously, $w_1(TM) B B$ stands for $\frac{1}{2} \tilde w_1(TM)\cup \mathcal P( B)$, explained in \Sec{sec:SPT}, with a twisted SW class $\tilde w_1$,
a cup product $\cup$,
and a Pontryagin square $P( B)$.
\item
\underline{$(K_1, K_2)=(1,0)$.}
The 2nd Sibling  of 4d SU(2)$_{\theta=\pi}$ with Kramers doublet ($T^2=-1$) bosonic Wilson line 
has the  4d anomaly/5d bordism invariant schematically as: 
\bea
\sim w_1(TM) B B + w_1(TM) ^3  B. \label{eq:w13B}
\eea
We note that the the 4d anomaly associated with the 5d $w_1(TM) ^3  B$ term is highly related to the 
2d charge conjugation anomaly associated to the 3d cubic $A^3$ term for a $\Z_2$-valued 1-cohomology class $A$.
See the relevant studies of 2d anomaly from the 3d cubic $A^3$ term in 
\cite{WangSantosWen1403.5256, Gu2015lfaGWW1503.01768, Komargodski2017dmc1705.04786, Metlitski2017fmd1707.07686, Wan2018zql1812.11968} and References therein.
\item
\underline{$(K_1, K_2)=(0,1)$.}
The 3rd Sibling  of 4d SU(2)$_{\theta=\pi}$  with Kramers doublet ($T^2=-1$) fermionic Wilson line
has the  4d anomaly/5d bordism invariant schematically as: 
\bea
\sim w_1(TM) B B + \frac{1}{2} \delta (w_2(TM)   B). \label{eq:w3B}
\eea
{Here $\delta$ is a coboundary operator, sending a $j$-cochain in the cochain group $C^j(M,\Z_n)$
to a $(j+1)$-coboundary in the coboundary group $B^{j+1}(M,\Z_n)$. 
Note that there are maps $M \to \B \tO$ and $M \to \B^2 \Z_2$, 
so $w_2(TM)B$ in the cohomology group $\H^4(\B \tO\times \B^2 \Z_2, \Z_2)$ 
can be pulled back to another cohomology group $\H^4(M,\Z_2)$, with O the orthogonal group O$(d)$ for $d$-manifold.
In this case, the $w_2(TM)  B$ is a cohomology class in
$\H^4(M,\Z_2)$. 
 Meanwhile
$\frac{1}{2} \delta(w_2(TM)  B)$ sends $w_2(TM)  B$ to a cohomology class in $\H^5(M,\Z_2)$. 
The $\frac{1}{2} \delta$ is mathematically precisely a Steenrod square $\Sq^1$ \cite{Steenrod1947}.}

\item 
\underline{$(K_1, K_2)=(1,1)$.}
The 4th Sibling  of 4d SU(2)$_{\theta=\pi}$  with Kramers singlet ($T^2=+1$) fermionic Wilson line
has the  4d anomaly/5d bordism invariant schematically as: 
\bea 
\sim w_1(TM) B B + w_1(TM) ^3  B + \frac{1}{2} \delta (w_2(TM)   B). \label{eq:w13B}
\eea
\end{enumerate}
We remark that our investigations on Kramers time-reversal properties and bosonic/fermionic statistics of line operators (for non-abelian gauge theories here)
give rise to a further refined
classification of gauge theories somehow beyond the previous framework of \Refe{AharonyASY2013hda1305.0318} and \cite{Gaiotto2014kfa1412.5148}.
See  \Refe{Wang2016cto1505.03520, Zou2017ppq1710.00743} for the case of abelian U(1) gauge theories.
See also  \cite{2017arXiv171111587GPW}, \cite{SWW2018}
and \cite{GPW2018} for other examples of non-abelian gauge theories.

\begin{figure}[!h]
	\centering
	\includegraphics[width=16.cm]{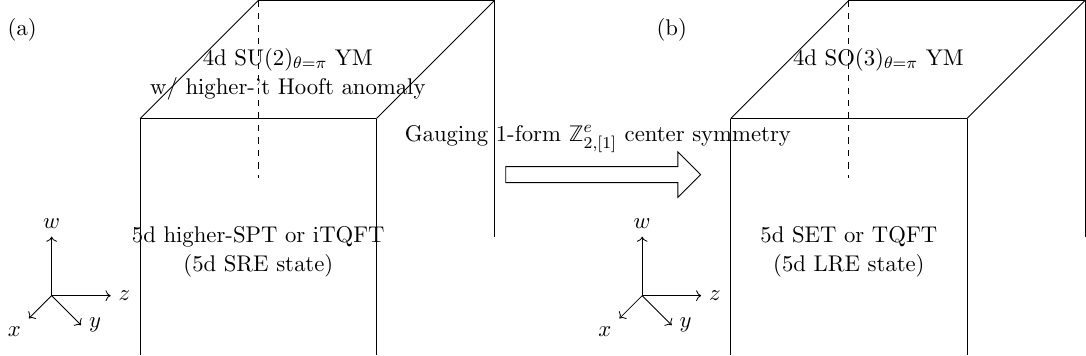}
	\caption[caption]{\\ \hspace{\textwidth}
	{(a) Schematic illustration of 
	4d-5d coupled system: 
	4d SU(2)$_{\theta=\pi}$ YM and 5d SRE higher-SPTs coupled systems. 
	There are  Four Siblings of such systems with bosonic UV completion, summarized in Table \ref{table:TQFTlink}.
	We use $x,y,z$ to label the spatial coordinates of 4d (3+1D) YM,
	and we introduce an extra coordinate $w$ to label the additional dimension of 5d higher-SPTs.}\\ 
	{(b) Schematic illustration of 
	4d-5d coupled system:
	4d SO(3)$_{\theta=\pi}$ YM-5d LRE higher-SETs
	coupled systems via gauging 1-form  $\mathbb{Z}_{2,[1]}^e$ center symmetry in \Fig{Fig-SU(2)-SPT} (a).
	There are Four Siblings of such 5d SET systems with bosonic UV completion, summarized in Table \ref{table:TQFTlink}.
	We use $x,y,z$ to label the spatial coordinates of 4d (3+1D) YM,
	and we introduce an extra coordinate $w$ to label  the additional dimension of 5d higher-SETs.
	See also \Fig{Fig:curve-SU(2)-SO(3)}.
	}
	}
	\label{Fig-SU(2)-SPT}
\end{figure}

{The schematic $\int_{M^5} \frac{1}{2} \delta (w_2(TM)   B)$ term in \eqn{eq:w3B} and \eqn{eq:w13B} is written as 
mathematically precisely
$\int_{M^5} \Sq^1 (w_2(TM)   B)$ 
on a 5-manifold ${M^5}$ in \Sec{sec:SPT}.
We will see that
such a term $\int_{M^5} \Sq^1 (w_2(TM)   B)$  vanishes (as the 0 mod 2), when ${M^5}$ is a closed 5-manifold.
However, \Sec{sec:SPT} shows that when ${M^5}$ has a boundary $M^4 = \partial M^5$,
$\int_{M^5} \Sq^1 (w_2(TM)   B)$ transforms nontrivially 
under  $B \to B+ \delta \lambda$ where $\lambda$ is  a 1-cochain. 
This nontrivial transformation is essential to cancel the noninvariance of  the 4d YM theory. 
 %
This observation 
indicates a subtle fact that $\int_{M^5} \Sq^1 (w_2(TM)   B)$ \emph{cannot} be dropped and should be kept as a 
certain physical term, since
we are studying the physics on a 5d manifold with 4d boundary. To summarize:
\begin{itemize}
\item $\int_{M^5} \Sq^1 (w_2(TM)   B)$ vanishes as 0 (mod 2) on a closed 5-manifold ${M^5}$. This can be interpreted in many distinct but related ways. It describes a trivial gapped vacuum with no SPT order, or a trivial gapped insulator in condensed matter language, or a trivial iTQFT on $M^5$. 
\item However, $\int_{M^5} \Sq^1 (w_2(TM)   B)$ has \emph{essential} physical effects on a 5-manifold $M^5$ with a nontrivial boundary
 $M^4 = \partial M^5$. Under the background gauge transformation $B \to B+ \delta \lambda$,
the gauge variant is non-zero.
\item $\int_{M^5} \Sq^1 (w_2(TM)   B)=\int_{M^5} \frac{1}{2} \delta (w_2(TM)   B)$ on an $M^5$ with boundary $M^4= \partial M^5$
may \emph{behave} like $\int_{M^4} \frac{1}{2} (w_2(TM)   B)$ --- which is half of a 4d bordism invariant $w_2(TM) B$.  
%
Twice of this fractional term $\sim 2 \int_{M^4} \frac{1}{2} (w_2(TM)   B)$ $\sim \int_{M^4} (w_2(TM)   B)$ is a 4d bordism invariant,
and quadruple of this fractional term $4 \int_{M^4} \frac{1}{2} (w_2(TM)   B) \sim 2\int_{M^4} (w_2(TM)   B) =0 $ mod 2 
is a trivial 4d bordism invariant. Thus $\int_{M^5} \Sq^1 (w_2(TM)   B)$ cannot be interpreted as a 4d local counter term.
Instead, we interpret it as a \emph{non-local} counter term, a \emph{fractional} counter term, or a \emph{fractional} SPTs on $M^4$. 
%
This is analogous to a certain
\emph{Wess-Zumino-Witten (WZW)-like} term\footnote{We thank Ho Tat Lam for an inspiring conversation on this issue.}
with the following new features: \\ 
(i) The standard WZW term \cite{Wess1971yuWZ, Witten1983twGlobalACA} is labeled by an integer,
but here $\int_{M^5} \Sq^1 (w_2(TM)   B)$  is labeled by a $\Z_2$ number. 
\\
(ii) The standard WZW term is written in terms of dynamical fields, 
but the WZW-like term here is written in terms of the background fields of the time-reversal symmetry $\Z_2^T$ and a higher symmetry $\Z_{2,[1]}^e$. \\
\item Similar to the standard WZW term, our WZW-like term affects the symmetry quantum numbers of physical observables, i.e., the statistics and Kramers degeneracy (i.e. singlet or doublet) of the Wilson lines. 
\end{itemize}
}
A schematic illustration of 4d SU(2)$_{\theta=\pi}$ YM-5d SRE higher-SPTs coupled system
is shown in \Fig{Fig-SU(2)-SPT}.
See Table \ref{table:TQFTlink} for a short summary for the Four Siblings of 4d SU(2)$_{\theta=\pi}$ YM theories and their coupling to the
5d systems, as well as their physical properties.
See Table \ref{table:link-cong} for a summary of the link invariants and link configurations of 5d TQFTs.


\newpage
\newgeometry{left=0.5cm, right=0.5cm, top=0.5cm, bottom=1.5cm}
\begin{table}[!h]
\centering
\finline[\fontsize{10}{10}]{Alpine}{
\noindent
\makebox[\textwidth][c] 
{
\begin{tabular}{lcc} 
\hline\\[.25mm]
 \multicolumn{3}{c}{
 $\begin{matrix}
 \text{\bf Four Siblings of } 
 \text{\bf 5d SRE-higher-SPTs-4d SU(2)$_{\theta=\pi}$ YM coupled systems}\\[2mm]
   \text{\bf and their gauged analogous }\\[2mm]
  \text{\bf   Four Siblings of } 
 \text{\bf 5d LRE-higher-SETs-4d SO(3)$_{\theta=\pi}$ YM coupled systems}\\[2mm]
 \end{matrix}$
 }\\[.25mm]
\hline
\hline\\[.25mm]
$\begin{matrix}
\text{(i). 5d higher-anomaly polynomial}\\[1mm]
\text{(5d bordism invariants of $\Omega_5^{\tO}(\B^2\Z_2)$)}\\[1mm]
\text{involving 1-form center $\Z_{2,[1]}^e$}\\[1mm]
\text{time-reversal $\Z_2^T$-symmetries}\\[1mm]
\hline
\hline\\[-2mm]
\text{5d iTQFT / SPT partition function}:\\
\text{$\bZ_{\text{SPT}_{(K_1,K_2)}}^{5\text{d}}[M^5]$.}
\\[1mm]  
\hline
\hline\\[-2mm]
\text{5d TQFT / SET path integral}:\\
\text{$\bZ_{\text{SET}_{(K_1,K_2)}}^{5\text{d}}[M^5]$.}
\\[2mm]
\end{matrix}$  & 
$\begin{matrix}\text{(ii). 4d SU(2)$_{\theta=\pi}$ YM obtained via}\\ 
\text{dynamical gauging 4d SPTs}\\
\text{(4d bordism invariants of $\Omega^{G'}_4$)}\\[1mm] 
\hline
\hline\\[-2mm]
\text{$G'$ for a group extension:}\\[1mm]
\text{$1 \to \SU(2) \to  G' \to \tO(d) \to 1.$}\\[1mm]
\hline
\hline\\[-2.mm]
\text{Gauge and spacetime}\\ 
\text{bundle/connection constraints}\\[1mm]
\hline
\hline\\[-2mm]
\text{Wilson line operator $W$ properties}
\end{matrix}$
&
$\begin{matrix}
\text{(iii). 5d-spacetime-braiding process}\\[1mm] 
\text{of anyonic-1D-strings/2D-branes}\\[1mm] 
\text{from 2d worldsheet and 3d worldvolume}\\[1mm] 
\text{of 5d Higher-Gauge TQFTs/SETs:}\\[1mm] 
\text{Path-integral $\bZ[M^5,\text{Link}]$/$\bZ[M^5]$}\\[1mm]
\text{ $\equiv \langle\text{Link}\rangle$}\\[1mm]
\hline
\hline\\[0mm]
\text{New 5d Topological Link Invariants }
\end{matrix}$ \\[2mm]
\Xhline{2\arrayrulewidth}
\Xhline{2\arrayrulewidth}
\hline\\[-1mm]
%
$\begin{matrix}
{\text{\bf 1st system } (K_1=0,K_2=0):} \quad\quad\quad\quad\quad\\[1mm]
\frac{1}{2} \tilde w_1(TM) \mathcal P( B){=B\Sq^1B+\Sq^2\Sq^1B}\\[1mm]
{\sim w_1  B B}
\\[1mm]
\hline
\hline\\[-2mm]
\text{iTQFT: $\bZ_{\text{SPT}_{(0,0)}}^{5\text{d}}[M^5]$ of \eqn{eq:5dSPT-all}}\\[1mm]  
\hline
\hline\\[-2mm]
\text{TQFT: 
$\bZ_{\text{SET}_{(0,0)}}^{5\text{d}}[M^5]$ of \eqn{eq:5dSET-all}
}\\[1mm]
\end{matrix}$  & 
$\begin{matrix}
{\text{\eqn{Eq.refinedGBC-YM-Z}}}
\\[1.5mm]
\hline
\hline\\[-2mm]
G'=\tO(d) \times \SU(2) \text{ in \eqn{eq:bordism4O}}\\[1mm]  
\hline
\hline\\[0mm]
{w_2(V_{\PSU(2)})=B
}
\\[1mm]
\hline
\hline\\[-2mm]
\text{Kramers singlet ($T^2=+1$) bosonic $W$}
\end{matrix}$ 
&
$\begin{matrix}
{\text{\eqn{w1PB-Link}}}
\\[1.5mm]
\hline
\hline\\[0mm]
\text{ \small{\(  \#(V^4_X\cap V^3_{U_{\bf (i)}}\cap V^3_{U_{\bf (ii)}}) 
\)}}  \\
{\equiv\text{Tlk}^{(5)}_{{w_1 B B} }(\Sigma^3_X,\Sigma^2_{U_{\bf (i)}},\Sigma^2_{U_{\bf (ii)}})}
\\[1mm]
\end{matrix}$ \\
%
%
\hline
\hline
\hline\\[-1mm]
%
$\begin{matrix}
{\text{\bf 2nd system } (K_1=1,K_2=0):}\quad\quad\\[1mm]
{\frac{1}{2} \tilde w_1(TM) \mathcal P( B)+w_1(TM)^3 B}\\[1mm] 
{=B\Sq^1B+w_2(TM) \Sq^1 B}\\[1mm]
{\sim w_1  B B + (w_1)^{3} B}
\\[1mm]
\hline
\hline\\[-2mm]
\text{iTQFT: $\bZ_{\text{SPT}_{(1,0)}}^{5\text{d}}[M^5]$ of \eqn{eq:5dSPT-all}}\\[1mm]  
\hline
\hline\\[-2mm]
\text{TQFT: 
$\bZ_{\text{SET}_{(1,0)}}^{5\text{d}}[M^5]$ of \eqn{eq:5dSET-all}
}\\[1mm]
\end{matrix}$  & 
$\begin{matrix}
{\text{\eqn{Eq.refinedGBC-YM-Z}}}
\\[1.5mm]
\hline
\hline\\[-2mm]
G'=\tE(d) \times_{\Z_2} \SU(2)  \text{ in \eqn{eq:bordism4E}}\\[1mm]  
\hline
\hline\\[0mm]
{w_2(V_{\PSU(2)})=B+w_1(TM)^2
}
\\[1mm]
\hline
\hline\\[-2mm]
\text{Kramers doublet ($T^2=-1$) bosonic $W$}
\end{matrix}$ 
&
$\begin{matrix}
{\text{\eqn{eq:link-all-K1K2}}}
\\[1.5mm]
\hline
\hline\\[0mm]
{ \frac{1}{2}\#(V^3_{U_h}\cap \Sigma^2_{U_b})}\\[1mm]
{\equiv \frac{1}{2}\text{Lk}^{(5)}_{w_2\dd B}(\Sigma^2_{U_h},\Sigma^2_{U_b})},
\\[3mm]
{ \#(V^3_{U_b}\cap \Sigma^2_{U_b})}\\[1mm]
{\equiv \text{Lk}^{(5)}_{B\dd B}(\Sigma^2_{U_b},\Sigma^2_{U_b})}
\\[1mm]
\end{matrix}$ \\
%
%
\hline
\hline
\hline\\[-1mm]
%
$\begin{matrix}
{\text{\bf 3rd system } (K_1=0,K_2=1):}\quad\quad\\[1mm]
{\frac{1}{2} \tilde w_1(TM) \mathcal P( B)+\Sq^1(w_2(TM) B)}\\[1mm] 
{\sim w_1  B B +  \frac{1}{2} \delta( w_2 B) }
\\[1mm]
\hline
\hline\\[-2mm]
\text{iTQFT: $\bZ_{\text{SPT}_{(0,1)}}^{5\text{d}}[M^5]$ of \eqn{eq:5dSPT-all}}\\[1mm]  
\hline
\hline\\[-2mm]
\text{TQFT: 
$\bZ_{\text{SET}_{(0,1)}}^{5\text{d}}[M^5]$ of \eqn{eq:5dSET-all}
}\\[1mm]
\end{matrix}$  & 
$\begin{matrix}
{\text{\eqn{Eq.refinedGBC-YM-Z}}}
\\[1.5mm]
\hline
\hline\\[-2mm]
G'={\rm Pin}^+(d) \times_{\Z_2} \SU(2)  \text{ in \eqn{eq:bordism4Pin+}}\\[1mm]  
\hline
\hline\\[0mm]
{w_2(V_{\PSU(2)})=B+w_2(TM)
}
\\[1mm]
\hline
\hline\\[-2mm]
\text{Kramers doublet  ($T^2=-1$) fermionic $W$}
\end{matrix}$ 
&
$\begin{matrix}
{\text{\eqn{w1PB-Link}}}
\\[1.5mm]
\hline
\hline\\[0mm]
\text{ \small{\(  \#(V^4_X\cap V^3_{U_{\bf (i)}}\cap V^3_{U_{\bf (ii)}})\)}}  \\
{\equiv\text{Tlk}^{(5)}_{{w_1 B B} }(\Sigma^3_X,\Sigma^2_{U_{\bf (i)}},\Sigma^2_{U_{\bf (ii)}})}
\\[1mm]
\end{matrix}$ \\
%
%
\hline
\hline
\hline\\[-1mm]
%
$\begin{matrix}
{\text{\bf 4th system } (K_1=1,K_2=1): \quad\quad\quad\quad\quad}\\[1mm]
{\frac{1}{2} \tilde w_1(TM) \mathcal P( B) +w_1(TM)^3  B+\Sq^1(w_2(TM) B)}\\[1mm] 
{\sim w_1  B B + (w_1)^{3} B+    \frac{1}{2} \delta( w_2 B) }
\\[1mm]
\text{or in a closed 5-manifold}:\\[1mm]
{B\Sq^1B+w_2(TM) \Sq^1 B}\\[1mm]
\hline
\hline\\[-2mm]
\text{iTQFT: $\bZ_{\text{SPT}_{(1,1)}}^{5\text{d}}[M^5]$ of \eqn{eq:5dSPT-all}}\\[1mm]  
\hline
\hline\\[-2mm]
\text{TQFT: 
$\bZ_{\text{SET}_{(1,1)}}^{5\text{d}}[M^5]$ of \eqn{eq:5dSET-all}
}\\[1mm]
\end{matrix}$  & 
$\begin{matrix}
{\text{\eqn{Eq.refinedGBC-YM-Z}}}
\\[1.5mm]
\hline
\hline\\[-2mm]
G'={\rm Pin}^-(d) \times_{\Z_2} \SU(2)  \text{ in \eqn{eq:bordism4Pin-}}\\[1mm]  
\hline
\hline\\[0mm]
{w_2(V_{\PSU(2)})=\big(B+}\\[1mm]
{w_1(TM)^2 +w_2(TM)\big)
}
\\[1mm]
\hline
\hline\\[-2mm]
\text{Kramers singlet ($T^2=+1$) fermionic $W$}
\end{matrix}$ 
&
$\begin{matrix}
{\text{\eqn{eq:link-all-K1K2}}}
\\[1.5mm]
\hline
\hline\\[0mm]
{ \frac{1}{2}\#(V^3_{U_h}\cap \Sigma^2_{U_b})}\\[1mm]
{\equiv \frac{1}{2}\text{Lk}^{(5)}_{w_2\dd B}(\Sigma^2_{U_h},\Sigma^2_{U_b})},\\[3mm]
{ \#(V^3_{U_b}\cap \Sigma^2_{U_b})}\\[1mm]
{\equiv \text{Lk}^{(5)}_{B\dd B}(\Sigma^2_{U_b},\Sigma^2_{U_b})}
\\[1mm]
\end{matrix}$ \\
\hline
\hline
\cline{1-3}\\[-2mm]

\end{tabular}
} \hspace*{35mm}
\caption{
A short summary of some results obtained in our work for the Four Siblings of 4d pure non-supersymmetric SU(2)$_{\theta=\pi}$ YM theories
or SO(3) YM theories, and for the 4d-5d-SPT coupled systems or 4d-5d-higher-SET coupled systems. 
}
\label{table:TQFTlink}
}
\end{table}
\restoregeometry

\newpage
\newgeometry{left=0.7cm, right=0.7cm, top=0.2cm, bottom=0.5cm}
\begin{table}[!h]
\centering
\finline[\fontsize{10}{10}]{Alpine}{
\noindent
\makebox[\textwidth][c] 
{
\begin{tabular}{cc} 
%
\hline
\hline
\multicolumn{2}{c}{
Section and Link Invariant}\\ 
 \hline\\[-2mm]
Link Configuration & Intersecting Number Configuration \\
%
\Xhline{2\arrayrulewidth}
\Xhline{2\arrayrulewidth}
\hline\\[-1mm]
 \multicolumn{2}{c}{{\Sec{sec:link-inv-wPB}} and
{\Sec{sec:triple-link-w1BB}}: 
${ \#(V^4_X\cap V^3_{U_{\bf (i)}}\cap V^3_{U_{\bf (ii)}})
\equiv \text{Tlk}^{(5)}_{w_1 BB}(\Sigma^3_X,\Sigma^2_{U_{\bf (i)}},\Sigma^2_{U_{\bf (ii)}})}$
}\\
\hline\\
\includegraphics[width=6.cm]{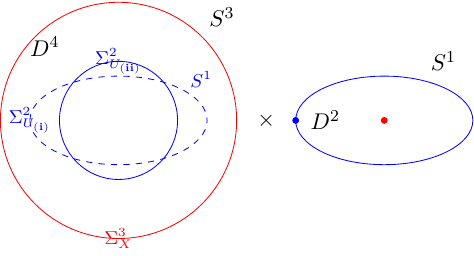} & \includegraphics[width=6.cm]{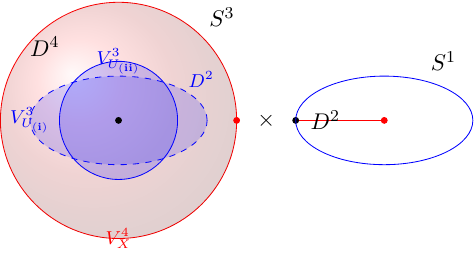}
\\[.25mm]
\hline\\[-1mm]
 \multicolumn{2}{c}{{\Sec{sec:link-inv-w13B-version-2}, \Sec{sec:link-inv-all}} and
{\Sec{sec:triple-link-w1w1dB}}: 
${{  \#(V^4_{X_{\bf (i)}}\cap V^4_{X_{\bf (ii)}}\cap \Sigma^2_U)}
\equiv \text{Tlk}^{(5)}_{w_1 w_1\dd B}(\Sigma^3_{X_{\bf (i)}},\Sigma^3_{X_{\bf (ii)}},\Sigma^2_U)}$
}\\
\hline\\
\includegraphics[width=6.cm]{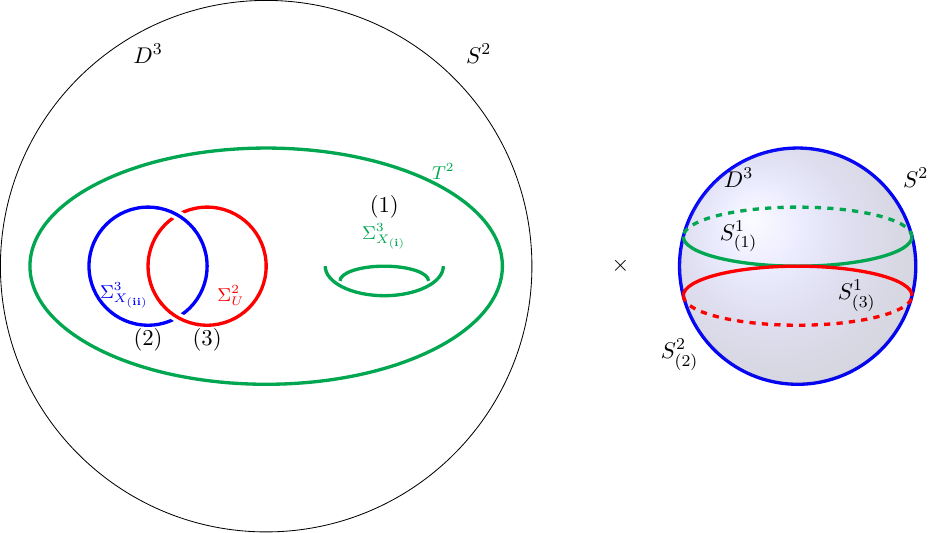} & \includegraphics[width=6.cm]{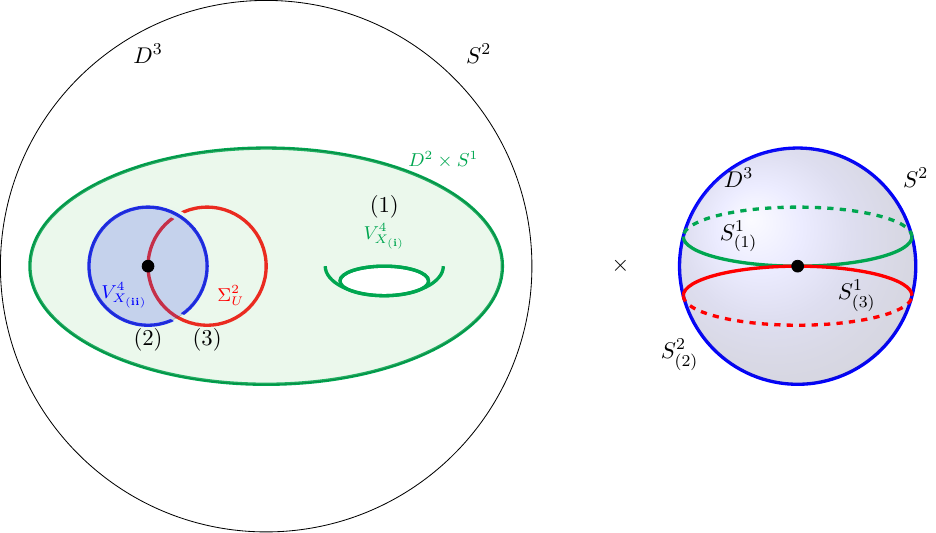}
\\[.25mm]
\hline\\[-1mm]
 \multicolumn{2}{c}{{\Sec{sec:link-inv-w13B-version-1}} and
{\Sec{sec:quadruple-link-w1w1w1B}}: 
${ \#(V^4_{X_{\bf (i)}}\cap V^4_{X_{\bf (ii)}}\cap V^4_{X_{\bf (iii)}}\cap V^3_U)
\equiv {\text{Qlk}^{(5)}_{w_1w_1w_1B}(\Sigma^3_{X_{\bf{(i)}}},\Sigma^3_{X_{\bf{(ii)}}},\Sigma^3_{X_{\bf{(iii)}}},\Sigma^2_U)}}$
}\\
\hline\\
\includegraphics[width=6.5cm]{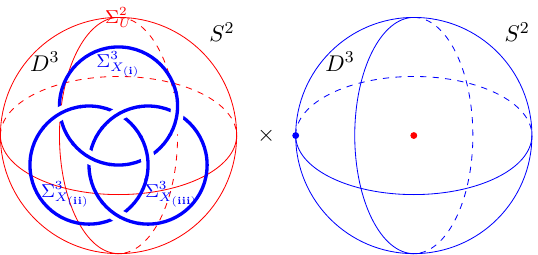} & \includegraphics[width=6.5cm]{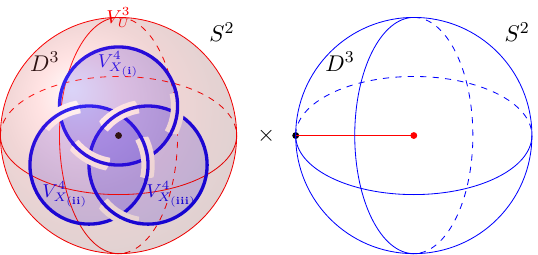}
\\[.25mm]
\hline\\[-1mm]
 \multicolumn{2}{c}{{\Sec{sec:link-inv-all}} and
{\Sec{sec:quadratic-link-BdB}}: 
${{  \#(V^3_{U_{\bf (i)}}\cap \Sigma^2_{U_{\bf (ii)}})}
\equiv \text{Lk}^{(5)}_{B\dd B}(\Sigma^2_{U_{\bf (i)}},\Sigma^2_{U_{\bf (ii)}})   }$
}\\
\hline\\
\includegraphics[width=6.5cm]{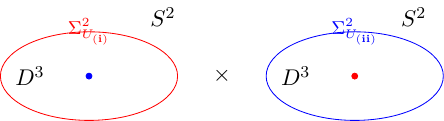} & \includegraphics[width=6.5cm]{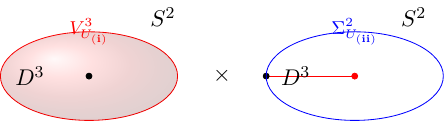}
\\[.25mm]
\hline\\[-1mm]
 \multicolumn{2}{c}{{\Sec{sec:link-inv-w2dB}, \Sec{sec:link-inv-all}} and
{\Sec{sec:quadratic-link-w2dB}}: 
${ \#(V^3_{U'}\cap \Sigma^2_{U})
\equiv \text{Lk}^{(5)}_{w_2 \dd B}(\Sigma^2_{U_{}},\Sigma^2_{U_{}'})  }$
}\\
\hline\\
\includegraphics[width=6.5cm]{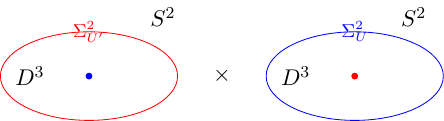} & \includegraphics[width=6.5cm]{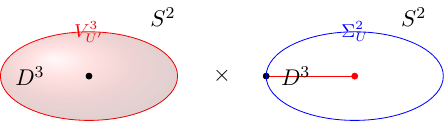}
\\[.25mm]
\hline\\[-1mm]
 \multicolumn{2}{c}{ 
{\Sec{sec:triple-link-AdAB}}: 
${ \#(V^4_{X_{\bf (i)}}\cap \Sigma^3_{X_{\bf (ii)}}\cap V^3_U)
\equiv \text{Tlk}^{(5)}_{(A \dd A) B}(\Sigma^3_{X_{\bf (i)}},\Sigma^3_{X_{\bf (ii)}},\Sigma^2_U)}$
}\\
\hline\\
\includegraphics[width=6.5cm]{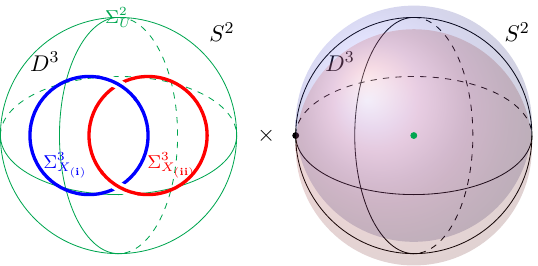} & \includegraphics[width=6.5cm]{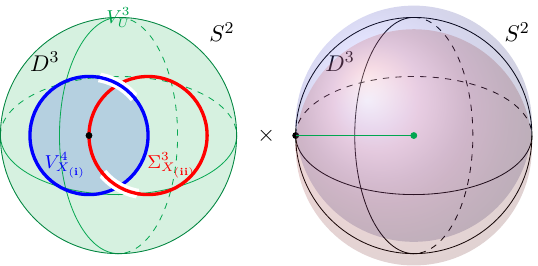}
\\[.25mm]
\hline
\hline
\end{tabular}
} \hspace*{35mm}
\caption{Link invariants and link configurations
{of 2d worldsheet and 3d worldvolume from the \emph{anyonic}-1D-Strings/2D-Branes' spacetime braiding processes in 5d higher-gauge  time-reversal SETs
in \Sec{sec:link-inv} and \ref{sec:link-conf}. Readers can find other related link invariants
in 3d, 4d and others in Tables of \cite{1602.05951, 
Putrov2016qdo1612.09298}.}}
\label{table:link-cong}
}
\end{table}
\restoregeometry

\section{4d SU(2)$_{\theta=\pi}$ Yang-Mills Gauge Theories coupled to 5d Short Range Entangled SPTs}
\label{sec:SPT}

\subsection{Ordinary and Higher Global Symmetries of Yang-Mills Theory}

We discuss the global symmetries of SU(N)$_{\theta}$ YM theory {in Minkowski/Lorentz signature} version of \Eq{eq:YM-pi}:
$\int [{\cal D} {a}] \exp   \big(\ii \int\limits_{M^4} \frac{1}{g^2}\text{Tr}\,(F_a\wedge \star F_a)
-  \ii \int\limits_{M^4} \frac{  \theta}{8 \pi^2}  \text{Tr}\,(F_a\wedge F_a) \big)$
$=
\int [{\cal D} {a}] \exp   \big(\ii \int\limits_{M^4} 
(-\frac{1}{4g^2}\, F_{\mu\nu}^\al F^{\al,\mu\nu}{}  -  \frac{\theta}{ 64 \pi^2 } \epsilon^{{\mu\nu} {\mu'\nu'}} \, F_{\mu\nu}^\al F_{{\mu'\nu'}}^\al ) 
\dd^4 x \big)$, 
for the convenience of studying the anti-unitary time-reversal symmetry
{$\cT$ that sends an imaginary number $\ii \to - \ii$.}
 
\begin{enumerate}
\item
We first focus on the discrete time-reversal symmetry $\Z_2^T$ and its symmetry transformation $\cT$ acting on the gauge field $a_\mu \equiv a_\mu^\alpha T^\alpha$
with the hermitian Lie algebra generator $T^\alpha$ (namely ${T^{\alpha \dagger}}={T^{\alpha}}$)
and the real-valued $a_\mu^\alpha \in \R$.  
The temporal component is $a_0$ and the spatial component is $a_i$. The anti-unitary $\cT$ acts on $a_\mu(t, x_i)$ as an \emph{active} transformation as:
\bea \label{eq:Ta}
\cT: &&\;  a_0^\alpha(t, x_i) \to \cred{\pm} a_0^\alpha(-t, x_i),\;\;\;\; a_i^\alpha(t, x_i) \to \cred{\mp} a_i^\alpha(-t, x_i), \;\;\;\; T^\alpha  \to  \cred{T^{\alpha *}}, \;\;\;\; \ii \to - \ii.  \\
&&\;  a_0(t, x_i) \to \cred{\pm  a_0^*}(-t, x_i) = a_0(-t, x_i),\;\;\;\; a_i(t, x_i) \to \cred{\mp a_i^*}(-t, x_i) = -a_i(-t, x_i). \nn
\eea
\cred{Here and below, the gauge field associated with a real symmetric Lie algebra generator (namely ${T^{\alpha *}}={T^{\alpha}}$) has the upper version of the sign choices.
The gauge field associated with an imaginary antisymmetric Lie algebra generator (namely ${T^{\alpha *}}=-{T^{\alpha}}$) has the lower version of the sign choices.}
The components of the field strength are $F_{ij}= F^\alpha_{ij} T^\alpha$ and $F_{0i}= F^\alpha_{0i} T^\alpha$. 
Under $\cT$, \cred{only one of the $F^\alpha_{ij}$ and $F^{\alpha}_{0i}$ flips its overall sign:} 
\begin{eqnarray}\hspace{-10mm}
\label{eq.T}
\begin{split}
\cT: ~~~&F^\alpha_{ij}(t, x_i) = \partial_i a_j^\alpha - \partial_j a_i^\alpha  + f^{\alpha \beta \gamma} a^\beta_i a^\gamma_j \to 
\cred{\mp} (\partial_i a_j^\alpha - \partial_j a_i^\alpha  + f^{\alpha \beta \gamma} a^\beta_i a^\gamma_j) = \cred{\mp}  F^\alpha_{ij} (-t, x_i), \\
&F^\alpha_{0i}(t, x_i) = \partial_t a_i^\alpha - \partial_i a_0^\alpha  + f^{\alpha \beta \gamma} a^\beta_0 a^\gamma_i \to 
\cred{\pm} ((\partial_{\tilde t} a_i^\alpha) - \partial_i a_0^\alpha  + f^{\alpha \beta \gamma} a^\beta_0 a^\gamma_i)  \cred{\big\vert_{({\tilde t}= -t,x_i)}}= 
\cred{\pm}  F^\alpha_{0i}(-t, x_i). \\
&  
F_{0i} = F^\alpha_{0i} T^\alpha \to \cred{\pm}  F^\alpha_{0i}(-t, x_i)  {T^{\alpha *}} =  F_{0i}(-t, x_i), \quad
F_{ij}  = F^\alpha_{ij} T^\alpha \to {\mp}  F^\alpha_{ij} (-t, x_i) {T^{\alpha *}} = -F_{ij}(-t, x_i).
\end{split}
\end{eqnarray}
Here $f^{\alpha \beta \gamma}$ is the structure constant of the SU(N) Lie algebra which is real. 
The reason that this $\cT$ is a good symmetry choice in contrast to the familiar $\cT$-symmetry of U(1) gauge theory case
is explained in the footnote.\footnote{ \label{CT:Auto}
The familiar U(1) gauge theory's $\cT$ transformation sends $a_0 \to a_0$ and $a_i \to -a_i$.
If we choose instead $a_0^\alpha \to a_0^\alpha$ and $a_i^\alpha \to -a_i^\alpha$ {for $\SU(\rN)$ gauge field}, 
then $F^{\alpha}_{ij}$ and $F^{\alpha}_{0i}$ are not mapped back to themselves (not even up to a sign); thus this does not define any symmetry of SU(N) YM. 
\cred{However, by including the $\cT$'s anti-unitary transformation on the Lie algebra generator $T^\alpha  \to  \cred{T^{\alpha *}}$,
our $\cT$ transformation on a non-abelian gauge field overall still sends $a_0 \to a_0$ and $a_i \to -a_i$.}\\
Given a gauge group $G$, the above discussion 
is related to the center Z$(G)$, the 
automorphism group Aut($G$), the outer automorphism Out($G$) and the inner automorphism Inn($G$). They form short exact sequences:
$$
1 \to \text{Z}(G) \to G \to \text{Inn}(G) \to 1, \quad \text{and    } \quad
1 \to \text{Inn}(G) \to \text{Aut}(G) \to \text{Out}(G) \to 1,
$$
and a combined exact sequence
$$
1 \to \text{Z}(G) \to G  \to \text{Aut}(G) \to \text{Out}(G) \to 1.
$$
%
{If $G$ is a simply-connected compact Lie group and ${\bf{g}}$ is its Lie algebra (which would necessarily be semi-simple),
then $\text{Inn}(G)=\text{Inn}({\bf g})=\mathrm{P}G$,
$\text{Aut}(G)=\text{Aut}({\bf g})$,
and
$ \text{Out}(G)=\text{Out}({\bf g})= \text{Aut}(D_{{\bf{g}} })$ is isomorphic to the automorphism group of the Dynkin diagram $D_{{\bf{g}} }$ of the Lie algebra ${\bf{g}} $.}\\
$\bullet$ For $G=\U(1)$, we have
$\text{Z}(G) =\U(1)$,
$\text{Inn}(G) = 1$, 
$\text{Aut}(G)=\text{Out}(G) = \Z_2$. \\
$\bullet$ For $G=\SU(2)$, 
we have
$\text{Z}(G) =\Z_2$,
$\text{Inn}(G) = \SO(3)$, 
$\text{Aut}(G)=\PSU(2)=\SO(3)$, and
$\text{Out}(G) = 1$. \\
$\bullet$ For $G=\SU(\rN)$ with N $\geq 3$, 
we have
$\text{Z}(G) =\Z_{\rN}$,
$\text{Inn}(G) = \PSU(\rN)$, 
 and
$\text{Out}(G) = \Z_2$. 
We also have {$\text{Aut}(G)=\PSU(\rN) \rtimes \Z_2$}
where $\Z_2$ acts on $\PSU(\rN)$ by $T^\al \to - {T^\al}^*$ on the Lie algebra alone, and by $a \to - a^*$ in \Eq{eq:CT}, 
with a minus sign and a complex conjugation $*$.\\
The validity of the charge conjugation symmetry $\Z_2^C$, with a $\cC$ global symmetry transformation, is based on 
 the validity of the outer automorphism $\text{Out}(G)$ that includes a $\Z_2$ as a $\Z_2^C$.
} 
It is obvious that the kinetic term $ \int_{M^4} \Tr (F \wedge \star F)$ is invariant under  $\cT$.
The $\theta$ term flips the sign under  
$\cT$:\footnote{More explicitly, under $\cT$: (Using \eqref{eq.T})
$$
\cT: \epsilon^{\cred{0} ijk}F^\alpha_{0i}(t,x)F^{\alpha}_{jk}(t,x) \to - \epsilon^{\cred{0} ijk}F^\alpha_{0i}(-t,x)F^{\alpha}_{jk}(-t,x).
$$
The time reversal $\cT$ here is chosen as an \emph{active} transformation that changes the sign \cred{of the time coordinate of the field here $f(t) \to  f(-t)$ up to some phase.}
When we wrote \Eq{eq:Ta}'s $(t, x_i) \to (-t, x_i)$, we mean to say the coordinate assignment to the fields are flipped
so to define an \emph{active} transformation.
The chosen $\cT$ does not do a \emph{passive} transformation here so
that the coordinate $t \to t$
and integration measure maintains $\dd t \to \dd t$. Also for example,
$a_i^\al(t) \to a'^\al_i (t) \equiv {\mp} a_i^\alpha(-t)$ in \eq{eq:Ta},
its time derivative in \eq{eq.T} says
$\partial_t a_i^\alpha \equiv \underset{{\Delta t \to 0}}{\lim} \frac{a_i^\alpha(t+{\Delta t} ) -a_i^\alpha(t ) }{\Delta t}$
maps under $\cT$ to
$\underset{{\Delta t \to 0}}{\lim} \frac{a'^\alpha_i(t+{\Delta t} ) -a'^\alpha_i(t ) }{\Delta t}
=\mp \underset{{\Delta t \to 0}}{\lim} \frac{a_i^\alpha( -t-{\Delta t} ) -a_i^\alpha(-t ) }{\Delta t}
=\mp \underset{{\Delta t \to 0}}{\lim} \frac{ -(a_i^\alpha(-t )- a_i^\alpha( -t-{\Delta t} )) }{\Delta t}
=\pm \underset{{\Delta t \to 0}}{\lim} \frac{ (a_i^\alpha(\tilde t )- a_i^\alpha( \tilde t -{\Delta t} )) }{\Delta t}{\big\vert_{({\tilde t}= -t)}}
=\pm \partial_{\tilde t} a_i^\alpha {\big\vert_{({\tilde t}= -t)}}$ with the time derivative evaluated at $-t$.
}
$$
\cT: \frac{ \theta}{8\pi^2}\int_{M^4} \Tr (F\wedge F) \to -\frac{ \theta}{8\pi^2}\int_{M^4} \Tr (F\wedge F).
$$
The $\theta \in [0, 2 \pi)$ has a $2 \pi$ periodicity, 
thus the theories at $\theta=0$ and $\theta= \pi$ are time-reversal invariant.

\item
We can define the $\Z_2^{CT}$ symmetry associated with the  $\cC\cT$ transformation for an SU(N) gauge theory: 
\bea \label{eq:CT}
\cC\cT: &&\; a_0^\al(t, x_i) \to - a_0^\al (-t, x_i),\;\;\;\; a_i^\al(t, x_i) \to +a_i^\al(-t, x_i), \;\;\;\; T^\al  \to T^{\al*}, \;\;\;\; \ii \to - \ii.\\
&& \:   a_0(t, x_i) \to  \cred{-}a_0^*(-t, x_i),\;\;\;\; a_i(t, x_i) \to \cred{+} a_i^*(-t, x_i), \nn\\
&& \:  F^\alpha_{0i}(t, x_i) \to -F^\alpha_{0i}(-t, x_i), \quad
F_{0i} = F^\alpha_{0i} T^\alpha \to (-F^\alpha_{0i}(-t, x_i) )(T^{\alpha *}) =\cred{-}F_{0i}^*(-t, x_i), \nn\\
&& \:  F^\alpha_{ij}(t, x_i) \to F^\alpha_{ij}(-t, x_i), \quad
F_{ij}  = F^\alpha_{ij} T^\alpha \to F^\alpha_{ij}(-t, x_i) (T^{\alpha *})=\cred{+}F_{ij}^*(-t, x_i).\nn
\eea 
Here $*$ is the complex conjugation. 
{The $\cC\cT$ is anti-unitary so $\cC\cT$ also sends an imaginary number $\ii \to - \ii$.}
We also define the charge conjugation $\Z_2^{C}$ symmetry associated with the $\cC$ transformation for an SU(N) gauge theory:\footnote{The active
$\cC$ transformation acts on the $a_\mu^{\al}$ fields only, 
not on the Lie algebra generator $T^{\al}  \to T^{{\al}}$. But effectively on the
full $a_\mu = a_\mu^\alpha T^\alpha$, the $\cC$ transformation is also equivalent to 
the Lie algebra outer automorphism transformation $T^{\al}  \to -T^{{\al}*}$, $a_\mu^\alpha \to a_\mu^\alpha$, 
and overall $a_\mu^{\al} \to - a_\mu^{\al *}$.} 
\bea \label{eq:C}
\cC: &&\; a_0^{\al} \to \mp a_0^{\al},\;\;\;\; a_i^{\al} \to \mp a_i^{\al}, \;\;\;\; T^{\al}  \to T^{{\al}}, \;\;\;\;  \text{with the same $(t, x_i)$ coordinates}.\;\\
&& \:    a_0 \to \mp a_0^{\al} T^{\al } = -a_0^{\al} T^{\al *}=  -a_0^*,\;\;\;\; a_i \to \mp a_i^{\al} T^{\al } = -a_i^{\al} T^{\al *}= - a_i^*. \nn\\
&& \:  
F_{0i} = F^\alpha_{0i} T^\alpha \to -F_{0i}^*, \nn\\
&& \:  
F_{ij}  = F^\alpha_{ij} T^\alpha \to -F_{ij}^*.\nn
\eea 
However for N = 2, the SU(2) 
YM does not have $\Z_2^{C}$ global symmetry because  SU(2) does not have 
a $\Z_2$  outer automorphism. The $\cC$ transformation is part of the SU(2) \emph{gauge} transformation.
Let $\cC_{SU(2)}=e^{\im\frac{\pi }{2}\sigma_2}\in \SU(2)$ be
the matrix that provides an isomorphism between fundamental representation of SU(2) and its conjugate, and $U_{\SU(2)}=\exp(\ii \frac{\theta}{2} \sigma_{\al})$ be the unitary SU(2) transformation on the SU(2)-fundamentals, where $\sigma_{\al}, {\al}=1,2,3$ are Pauli matrices.  Then 
$
\cC_{\SU(2)} U_{\SU(2)} \cC_{\SU(2)}^{-1}=\exp(-\ii \frac{\theta}{2} \sigma_{\al}^T)=\exp(-\ii \frac{\theta}{2} \sigma_{\al}^*)=U_{\SU(2)}^*
$
In other words,  $\Z_2^T$ and $\Z_2^{CT}$ are the same symmetry for the SU(2) 
YM. See more discussions in the footnote \ref{CT:Auto},
Sec.~2.2 of \cite{2017arXiv171111587GPW} and Sec.~2 of \cite{Wan2018zql1812.11968}.

\item Parity symmetry $\Z_2^{P}$ is another discrete symmetry associated with the transformation $\cP$:
\bea \label{eq:P}
\cP: &&\;  a_0^\alpha(t, x_i) \to a_0^\alpha(t, -x_i),\;\;\;\; a_i^\alpha(t, x_i) \to -a_i^\alpha(t, -x_i). \\
&&\; T^\alpha  \to T^{\alpha}, \;\;\;\; a_0(t, x_i) \to a_0(t, -x_i),\;\;\;\; a_i(t, x_i) \to -a_i(t, -x_i). \nn\\
&&F^\alpha_{ij}(t, x_i) = \partial_i a_j^\alpha - \partial_j a_i^\alpha  + f^{\alpha \beta \gamma} a^\beta_i a^\gamma_j \to \partial_{-i} (-a_j^\alpha) - \partial_{-j} (-a_i^\alpha)  + f^{\alpha \beta \gamma} 
(-a^\beta_i)  (-a^\gamma_j) = F^\alpha_{ij} (t, {-}x_i), \nn\\
&&F^\alpha_{0i}(t, x_i) = \partial_t a_i^\alpha - \partial_i a_0^\alpha  + f^{\alpha \beta \gamma} a^\beta_0 a^\gamma_i \to 
\partial_{t} (-a_i^\alpha) + \partial_{-i} a_0^\alpha  + f^{\alpha \beta \gamma} a^\beta_0 (-a^\gamma_i ) = -F^\alpha_{0i}(t, -x_i). \cr
&&  
F_{0i}  \to   F_{0i}(t, -x_i), \quad
F_{ij}   \to -F_{ij}(t, -x_i).\nn
\eea
$\cP$ is related to $\cC\cT$ via the $\cC\cP\cT$ symmetry:
\bea \label{eq:CPT}
\cC\cP\cT: &&\;  a_0^\alpha(t, x_i) \to -a_0^\alpha(-t, -x_i),\;\;\;\; a_i^\alpha(t, x_i) \to -a_i^\alpha(-t, -x_i). \\
&&\; T^\alpha  \to \cred{+}T^{\alpha*}, \;\;\;\; a_{\mu}(t, x_i) \to  \cred{-}a_{\mu}^*(-t, -x_i). \nn\\
&& \:  F^\alpha_{0i} (t, x_i)\to F^\alpha_{0i}(-t, -x_i), \quad
F_{0i} = F^\alpha_{0i} T^\alpha \to F^\alpha_{0i}(-t, -x_i) (T^{\alpha *}) =\cred{+}F_{0i}^*(-t, -x_i), \nn\\
&& \:  F^\alpha_{ij} (t, x_i) \to F^\alpha_{ij}(-t, -x_i), \quad
F_{ij}  = F^\alpha_{ij} T^\alpha \to F^\alpha_{ij}(-t, -x_i) (T^{\alpha *})=\cred{+}F_{ij}^*(-t, -x_i).\nn
\eea

\item The 1-form electric $\Z_{\rN,[1]}^e$ center global symmetry:
The \emph{charged object} of the 1-form $\Z_{\rN,[1]}^e$-symmetry is a gauge-invariant Wilson line 
\be
W_e^{\text{R}}=\Tr_{\text{R}}( \text{P} \exp(\ii \oint a)).
\ee
%
The gauge field $a$ is Lie algebra su(N) valued. 
The $\text{P} \exp(\ii \oint a)$ specifies a SU(N) group element where P is the path ordering.
Tr is the trace in the  representation R of SU(N). 
For the SU(N) gauge theory, R can be any possible representation.
If R is  an irreducible representation and let $l$ be the number of boxes in the Young diagram of R, then $W_e$ transforms under $\Z_{\rN,[1]}^e$ as  
\begin{eqnarray}
\Z_{\rN,[1]}^e: W_e^{\text{R}} \to e^{2\pi i l/\rN } W_e^{\text{R}}. 
\end{eqnarray}
For the fundamental representation, there is only one box in the Young diagram, hence the Wilson line $W^{\text{fund}}_e$ transforms under  $\Z_{\rN,[1]}^e$  as $W^{\text{fund}}_e \to e^{2\pi i /\rN } W^{\text{fund}}_e$. For N$=$2, the Wilson line in the fundamental representation transforms under $\Z_{2,[1]}^e$ by a sign $W^{\text{fund}}_e \to - W^{\text{fund}}_e$.

The \emph{charge operator} (i.e., \emph{symmetry generator}) of the $\Z_{\rN,[1]}^e$-symmetry
is a co-dimension 2 (thus a 2D operator in 4d spacetime) electric surface operator $U_e$.
For SU(2) gauge theory, we will see that 
\be\label{Eq.UeLambda}
U_e= \exp(\ii \pi \oint \Lambda),
\ee
where $\Lambda \in \H^2(M^4,\Z_2)$ as a cohomology class.

One can couple the SU(2) theory to $\Z_{\rN,[1]}^e$ background gauge field $B$. Following \cite{Kapustin2014gua1401.0740, Gaiotto2014kfa1412.5148, Gaiotto2017yupZoharTTT1703.00501}, we first promote the SU(2) gauge field $a$ to a U(2) gauge field $\widehat{a}$,  
\be
\widehat{a}=a + \frac{1}{2} \widehat{A}\mathbb{I}_2.
\ee
where $\mathbb{I}_2$ is a two dimensional identity matrix. 
The first Chern class of the U(2) bundle is $c_1\equiv c_1(V_{\U(2)})\equiv \frac{\Tr \widehat{F}}{2\pi}\equiv  \frac{\dd \widehat{A}}{ 2 \pi}$ where $\widehat{F}= \dd\widehat{a}- \ii \widehat{a} \wedge \widehat{a}$ is a U(2) field strength.  Then we couple to $B$ by requiring $c_1= B\mod 2$, which can be done via introducing a Lagrangian multiplier $\Lambda$ (see \Eq{eq:SU2YMZ}). This amounts to introducing the following term in the path integral, 
\be 
\label{eq:gauge-bundle-sim}
\int [\cD \Lambda] \dots  \exp \big( \ii \pi \int_{M^4} \Lambda \cup (c_1-B 
)\big).
\ee
The minimal coupling $\exp(\ii \pi \int \Lambda \cup B)$ implies that the symmetry generator (i.e., charge operator) of $\Z_{\rN,[1]}^e$ is precisely $\exp(\ii \pi \int \Lambda)$. This explains \eqref{Eq.UeLambda}. Notice that  integrating out the Lagrangian multiplier $\Lambda$ removes the U(1) degree of freedom, hence the  gauge group is SO(3)$=$PSU(2) (rather than SU(2)), 
\be \label{eq:group-relation}
\frac{\U(2)}{\U(1)}=\frac{\frac{\SU(2)\times \U(1)}{\Z_2}}{\U(1)}=
\frac{\SU(2)}{\Z_2}=\PSU(2)=\SO(3).
\ee
with the gauge bundle constraint $c_1=w_2(V_{\SO(3)})=B$. Here 
the second Stiefel-Whitney class  $w_2(V_{\SO(3)})\in \H^2(M, \Z_2)$  is the obstruction of promoting the SO(3) bundle to SU(2) bundle, which we explain in detail below. 
The nontrivial SU(2) gauge bundle on a manifold $M$ can be constructed by finding an open cover of $M$ and then gluing together trivial bundles from adjacent open patches via the SU(2) transition functions. Suppose $g_{ij} \in \SU(2)$ is the transition function (which plays the role of gauge transformation)
defined on the intersections of two open covers
indexed by $i$ and $j$.  There is a consistency condition  
$$g_{ij} g_{jk}  g_{ki}= 1 \in \SU(2)$$
on the triple overlapping intersections of three open patches indexed by $i$, $j$ and $k$.
However,  the consistency condition of SO(3)-bundle is weaker.
Let $h_{ij}$ be the transition function in the SO(3)-bundle, and $\widehat{h}_{ij}$ is the lift of $h_{ij}$  in the SU(2)-bundle. Then 
\begin{eqnarray}
h_{ij} h_{jk}  h_{ki}= 1 \in \SO(3),
\end{eqnarray}
while 
\begin{eqnarray}
\widehat{h}_{ij} \widehat{h}_{jk} \widehat{h}_{kl} = \exp(\ii \pi w_{ijk}(V_{\SO(3)}) ) \in \{\pm 1\}   \subset \SU(2).
\end{eqnarray}
The $w_{ijk}(V_{\SO(3)}) \in \Z_2$ is related to  $w_2(V_{\SO(3)})$ evaluated on the simplex $(ijk)$.\footnote{The patch $i$ is dual to a 0-simplex $i$ in the dual cell 
decomposition of spacetime. The intersection of two patches $i$ and $j$ is dual to a 1-simplex $(ij)$ in the dual cell decomposition of $M$. 
The intersection of the patches $i,j$ and $k$ is dual to a 2-simplex $(ijk)$ in the dual cell decomposition of $M$. }  Thus SO(3) bundle can be lifted to an SU(2) bundle only when $w_2(V_{SO(3)})$ is trivial, i.e., the $\Z_{\rN,[1]}^e$ background field $B\in \H^2(M,\Z_2)$ is trivial. 
Namely, activating $B$ allows us to study the SU(2) gauge theory with nontrivial SO(3)-gauge bundle.
In short, 
\be
\label{Eq.GBC}
{
\frac{\dd \widehat{A}}{ 2 \pi} = c_1(V_{\U(2)})= B = w_2(V_{\PSU(2)})=w_2(V_{\SO(3)}) =w_2(E)\mod 2},
\ee
and we learn that the SU(2) gauge theory coupled to a background $B$ field 
can be regarded as a path integral summing over SO(3) gauge bundle $E$ 
subject to the gauge bundle constraint $B = w_2(V_{\PSU(2)})=w_2(V_{\SO(3)}) =w_2(E)\mod 2$.
We will soon propose a new generalization of gauge bundle constraint of \Eq{Eq.GBC} on unorientable or non-spin manifolds. See \Eq{Eq.refinedGBC} in \Sec{Sec.thetapiYM}.



Coupling to $\Z_{2,[1]}^e$ background field $B$ allows one to say more on various line and surface operators. First, one can use $B$ to construct a magnetic 2-surface $U_m=\exp(\ii  {{ \pi \int_{\Sigma}}} w_2(V_{\PSU(2)}))=\exp(\ii  {{ \pi \int_{\Sigma}}} B)$. When $\Sigma$ is a surface with boundary, a Wilson line $W_e^{\text{fund}}$ in the fundamental representation (below, we will simply use $W_e$ for simplicity) can be supported on the boundary so that $W_eU_m$ is invariant under the background gauge transformation $B\to B+ \delta \lambda$. 
Second, when the surface $\Sigma$ of the electric 2-surface operator $U_e$, \eqref{Eq.UeLambda}, has a boundary $\partial \Sigma$, a 't Hooft line $T_m$ can be supported on $\partial \Sigma$. Since $U_e$ is dynamical in the SU(2) gauge theory, the 't Hooft line $T_m$ is not a genuine line operator, and $T_m$ has to live on the boundary of $U_e$.  Thus 't Hooft line $T_m$ as the worldline of probe background magnetic monopole must be attached with the
  dynamical and detectable
 open Dirac string, which is visible by $W_e$.  The closed 2d worldsheet of detectable  Dirac string forms the $U_e$ operator. This can be seen from the correlation function
 \be \label{eq:link-We-Ue}
\langle W_e({\gamma^1}) \;  U_e({\Sigma^2}) \rangle =\langle \Tr_{\text{R}}( \text{P} \exp(\ii \oint_{\gamma^1} a))\;  \exp(\ii \pi \oint_{\Sigma^2} \Lambda) \rangle = 
\exp\left(\frac{\ii 2 \pi}{\rN}{\text{Lk}({\gamma^1},{\Sigma^2})}\right)
\cred{\langle W_e({\gamma^1}) \rangle},
 \ee
 where ${\text{R}}$ stands for the fundamental representation. $\text{Lk}({\gamma^1},{\Sigma^2})$ is the linking number between $\gamma^1$ and $\Sigma^2$. 
\cred{The $\langle W_e \rangle$ is on the right-hand side of \eq{eq:link-We-Ue}, because its expectation value depends on a small perturbation
and thus $W_e$ is not a topological operator.} 
 
 
From the Hamiltonian point of view, the spatial Wilson line operator $\hat{W_e}$ and the spatial 't Hooft operator $\hat{T}_m$ (as two canonically quantized line operators) in the SU(N) gauge theory satisfy the commutation relation \cite{tHooft1977nqbQuark}:  
 \be \label{eq:commute-W-T}
 \hat{W}_e(\gamma^1) \hat{T}_m(\gamma^{1'}) = \exp\left(\frac{\ii 2 \pi}{\rN}{\text{Lk}({\gamma^1},{{\gamma^1}'})}\right)\hat{T}_m (\gamma^{1'}) \hat{W}_e (\gamma^1),
 \ee
where ${\text{Lk}({\gamma^1},{{\gamma^1}'})}$ is the linking number between ${\gamma^1}$ and ${\gamma^1}'$ in the 3d space.
For the SU(2) YM, \Eq{eq:commute-W-T} reduces to
 $$
  \hat{W}_e(\gamma^1) \hat{T}_m(\gamma^{1'}) = (-1)^{\text{Lk}({\gamma^1},{{\gamma^1}'})}\hat{T}_m (\gamma^{1'}) \hat{W}_e (\gamma^1).
 $$
The non-commutative nature of \Eq{eq:commute-W-T} implies that the ${W}_e$ and ${T}_m$
are not mutually local, which is consistent with the fact that 
${W}_e$ is a genuine line operator while ${T}_m$ is not a genuine line operator as discussed in the last paragraph.  
  
 

\item The full symmetry $\Z_2^{T} \times \Z_{2,[1]}^e$: 
The full symmetry of SU(2) YM theory relevant in our study is
$\Z_2^{T} \times \Z_{2,[1]}^e$. \footnote{Since $\Z_2^{CT} \times \Z_{2,[1]}^e$ and $\Z_2^{T} \times \Z_{2,[1]}^e$ differ by a SU(2) gauge transformation, we only discuss $\Z_2^{T} \times \Z_{2,[1]}^e$. }
The $\Z_2^T$ symmetry implies the spacetime symmetry has an orthogonal group O($d$) via a short exact sequence extension $1 \to \SO(d) \to \tO(d) \to \Z_2^T \to 1$ where  SO($d$) is the spacetime rotation symmetry.
Knowing the full relevant global symmetry, $\Z_2^{T} \times \Z_{2,[1]}^e$,
we can classify the 't Hooft anomalies based on Thom-Madsen-Tillmann-Freed-Hopkins bordism spectra and cobordism theory\cite{thom1954quelques,MadsenTillmann4, Freed2016.1604.06527}.
In terms of a bordism group 
$\Omega_d^{\mathbb{G}}$ (more precisely, we focus on the torsion part $\Omega_{d,\text{tor}}^{\mathbb{G}}$), the classification of 4d 't Hooft anomalies for 4d SU(2) YM can be written as linear combinations of bordism invariants $\Omega_d^{\mathbb{G}}=
 \Omega_d^{\tO}(\B^2\Z_2)
$ for $d=5$ \cite{Wan2018zql1812.11968, Wan2018bns1812.11967}.  (We leave the details of bordism invariants later in \Eq{eq:bordism5OB2Z2} and in \Sec{Sec.AnotherSU(2)AnomalyInterpretation}.)


\end{enumerate}

\subsection{Derivation of New Higher-Anomalies of SU(2) Yang-Mills Theory at $\theta=\pi$\\ 
on Unorientable Manifolds}
\label{Sec.thetapiYM}

We start with the SU(2) Yang-Mills theory (YM) with $\theta=\pi$, denoted SU(2)$_{\theta=\pi}$. The Euclidean action ${\bf S}_E$ from \eq{eq:YM-pi} is
\begin{eqnarray} \label{eq:SE-YM-theta}
{\bf S}_E[M^4]= \frac{1}{g^2} \int_{M^4} \Tr (F \wedge \star F) -  \frac{\ii \theta}{8\pi^2}\int_{M^4} \Tr (F\wedge F).
\end{eqnarray} 
Since the anomaly is a renormalization group flow invariant, in the following discussion, the kinetic term which is proportional to the running coupling constant $1/g^2$ will not play a role.  Hence we only consider the second term in \eqref{eq:SE-YM-theta}, which we call the theta term. 
To probe the anomaly, we turn on the background gauge field $\CB$ for the $\Z_{2,[1]}^e$ 1-form symmetry. Here $\CB$ is a 
$\Z_2$-valued 2-form 
gauge field with $\oint_\Sigma \CB=\pi \Z$ for any closed surface $\Sigma$. The 2-form gauge field $\CB$ is related to the 2-cochain $B$ via $\CB \sim \pi B$, and 
we also convert the wedge product {$\wedge$}
to the cup product {$\cup$} {when the action is written in terms of cochains}. To couple the SU(2) YM theory to the background gauge field $\CB$, we promote the SU(2) gauge field $a$ to a U(2) gauge field $\widehat{a}$, and the theta term
at $\theta=\pi$ reads\footnote{\label{ft:Eucl} The topological term for the Euclidean action $S_{{\bf E},\text{topological}}$ in the Euclidean partition function ${\bZ}= \exp(-S_{{\bf E},\text{topological}})$ contains a factor of imaginary $\ii$, namely ${\bf S}_E = - \ii (\dots)$ in \eqn{eq:SE-YM-theta}. 
However, by converting $\exp(- {\bf S}_E)=\exp(\ii {\bf S})$,
we have the following {\emph{Minkowski}} ${\bf S}$ in \eqn{Eq.SU21-form}.
}
\begin{eqnarray}\label{Eq.SU21-form}
\frac{\theta}{8\pi^2} \int_{M^4} \Tr \big((\widehat{F}-\CB\mathbb{I}_2)\wedge (\widehat{F}-\CB\mathbb{I}_2)\big)
\end{eqnarray}
where $\widehat{F}= \dd\widehat{a}- \ii \widehat{a} \wedge \widehat{a}$ is the U(2) field strength, and $\mathbb{I}_2$ is the two dimensional identity matrix. To restore the SU(2) gauge field, the U(2) field strength should satisfy the gauge bundle constraint
\begin{eqnarray}\label{Eq.gaugebundleconstraintSU2}
 \frac{\Tr \widehat{F}}{2\pi} = \frac{2\CB}{2\pi}= B = 
 w_2(V_{\PSU(2)})=w_2(V_{\SO(3)})=w_2(E)\mod 2.
\end{eqnarray}
Here 
$w_2(V_{\PSU(2)})=w_2(V_{\SO(3)})$ is the Stiefel-Whitney class of the associated vector bundle 
of the ${\PSU(2)}={\SO(3)}$ (the principal gauge bundle $E$ of ${\PSU(2)}={\SO(3)}$).

To activate the background field for the time-reversal symmetry, we formulate \eqref{Eq.SU21-form} on an unorientable manifold $M^4$. On an unorientable manifold, the top differential form is not well-defined, due to the lack of the volume form whose definition needs an orientation. To make sense of \eqref{Eq.SU21-form} on an unorientable manifold, we reformulate it in terms of the Chern characteristic classes.  We denote the $j$th Chern class of the U(N) gauge bundle as  $c_j(V_{\U(\rN)})$. For $j=1, 2$, we have 
\begin{eqnarray}\label{Eq.defc1c2}
\begin{split}
c_1(V_{\U(\rN)})&= \frac{\Tr \widehat{F}}{2\pi},\\
c_2(V_{\U(\rN)})&= -\frac{1}{8\pi^2} \Tr(\widehat{F}\wedge \widehat{F}) + \frac{1}{8\pi^2} (\Tr \widehat{F}) \wedge(\Tr \widehat{F}).
\end{split}
\end{eqnarray}
Replacing $\frac{1}{8\pi^2}\Tr(\widehat{F}\wedge \widehat{F})$ by ${\frac{c_1\cup c_1 }{2} - c_2}$,
we rewrite \eqref{Eq.SU21-form} as\footnote{ \label{ft:Yau-Eells-1401}
Some of mathematical-oriented readers may wonder
how to rigorously define 
\eqn{Eq.SU21-form}'s
$\frac{\pi}{8\pi^2} \int_{M^4} \Tr \big((\widehat{F}-\CB\mathbb{I}_2)\wedge (\widehat{F}-\CB\mathbb{I}_2)\big)$
to 
a term 
$\frac{\pi}{8\pi^2} \int_{M^4} \Tr \big((\widehat{F}-\pi B\mathbb{I}_2)\wedge (\widehat{F}-\pi B\mathbb{I}_2)\big)$
with the \emph{continuous} differential form $\widehat{F}$ coupling to a \emph{discrete} cohomology class $B$
$\in \H^2(M,\Z_2)$.
In fact, the physics way to interpret this coupling is related to {the coupling between QFT to TQFT}
explained in \cite{Kapustin2014gua1401.0740}. 
More formally, we can also implement mathematical methods \cite{AllendorferEells1957-58-Yau}
to formulate such couplings. JW thanks Shing-Tung Yau for insightful conversations on 
this method \cite{AllendorferEells1957-58-Yau}.
}
\begin{equation}\label{Eq.19}
    \frac{\pi}{8\pi^2} \int_{M^4} \Bigg(\Tr \big(\widehat{F}\wedge \widehat{F}\big)
 - 2 \Tr (\widehat{F}) \wedge  \CB 
+\Tr (\mathbb{I}_2) \CB\wedge \CB  \Bigg)
=
{\pi} \int_{M^4} \Bigg(\frac{1}{8\pi^2} \Tr \big(\widehat{F}\wedge \widehat{F}\big)
 - \frac{1}{2} \Tr (\frac{\widehat{F}}{2\pi}) \wedge \frac{ \CB }{\pi}
+\frac{1}{4} \frac{\CB}{\pi}\wedge \frac{\CB}{\pi} \Bigg), 
\end{equation}
Using \eqref{Eq.defc1c2}, \eqref{Eq.19} can be re-interpreted as  
\bea \label{Eq.SU21-form-2}
\pi \int_{M^4} \Bigg(-c_2(V_{\U(2)}) +\frac{c_1(V_{\U(2)})\cup c_1(V_{\U(2)})}{2}
- \frac{1}{2} c_1(V_{\U(2)}) \cup B+  \frac{\cP(B)}{4} \Bigg)
\eea
where $\cP(B)$ is the Pontryagin square\footnote{Notice it is crucial to treat
$
\frac{\pi}{8\pi^2} \int_{M^4} \Tr \left(\CB\mathbb{I}_2 \wedge \CB\mathbb{I}_2\right)
=
\frac{\pi}{8} \int_{M^4}  B \wedge B \left(\Tr (\mathbb{I}_2)\right)
=
\frac{2\pi}{8} \int_{M^4}  B \wedge B 
=\frac{\pi}{4} \int_{M^4}  B \wedge B 
\simeq\frac{\pi}{4} \int_{M^4} \cP(B)
$
as the more precise re-writing
for the later purposes. The 
$\cP(B): = B\cup B+B\hcup{1}\delta B 
=B\cup B+B\hcup{1} 2 \Sq^1 B$ denotes the Pontryagin square, e.g. see \Refe{Wan2018bns1812.11967,Wan2018zql1812.11968}.} of $B$.

Note that \eqref{Eq.SU21-form-2} is not well-defined even on an orientable manifold. 
In Sec.~\ref{Sec.lift}, we  resolve this problem for the torsion-free  oriented manifolds $M$.   Yet, \eqref{Eq.SU21-form-2} is also not well-defined on an unorientable manifold. 
In general, if $M$ is a $d$-dimensional unorientable manifold and $\omega$ is a $d$-cocycle, $\pi \int_{M} \omega\mod 2\pi $ is well-defined only when $\omega$ is valued in $\Z_2$.\footnote{Using the definition of the fundamental class of an unorientable manifold $M$, i.e., $[M]$, one has $\int_M \omega = \langle \omega, [M]\rangle$ where $\langle \omega, [M]\rangle$ is the $\Z_2$ valued pairing between $\omega$ and $[M]\in \H_d(M, \Z_2)$. } Since $c_2(V_{\U(2)})\in H^4(M^4, \Z)$ is integer valued, the first term in \eqref{Eq.SU21-form-2} makes sense when $M^4$ is unorientable. However, the other terms are fractional, hence the integral of those terms does not make sense if $M^4$ is unorientable. To make sense of \eqref{Eq.SU21-form-2}, we actually need to define
it on both the unorientable $M^4$ and an unorientable $M^5$ such that $\partial M^5 = M^4$.\footnote{
Note that if $M^5$ is orientable, then $M^4$ must be orientable.
Conversely, if $M^4$ is unorientable, $M^5$ must be unorientable.
However, if $M^4$ is orientable, $M^5$ can be orientable or unorientable.}
{To proceed, we extend the integer valued cohomology class $c_1(V_{\U(2)})$ on $M^4$ to an integer valued \emph{cochain} $\widetilde{c}_1(V_{\U(2)})$ on  $M^5$. Note that $\widetilde{c}_1(V_{\U(2)})$ on $M^5$ does not have to be an element in $\H^2(M^5, \Z)$, i.e, $\delta \widetilde{c}_1(V_{\U(2)})=0$ does not have to hold on $M^5$. The requirement of $\widetilde{c}_1(V_{\U(2)})$ will be imposed later by the gauge bundle constraint. The extension means, in particular, that when restricting $\widetilde{c}_1(V_{\U(2)})$ to $M^4$, it reduces to a $\Z$-valued cohomology class ${c}_1(V_{\U(2)})$. We further extend the $\Z_2$-valued cohomology class $B$ on $M^4$ to a $\Z_2$-valued cohomology class on $M^5$, and for simplicity, we use the same notation $B$ on $M^5$ as well. Thus we define \eqref{Eq.SU21-form-2} as follows:}
%
%
%
\begin{eqnarray}\label{Eq.SU(2)thetaterm}
&&-\pi \int_{M^4} c_2(V_{\U(2)}) +\pi\int_{M^5}\delta\Bigg(\frac{c_1(V_{\U(2)})\cup c_1(V_{\U(2)})}{2}- \frac{1}{2}  c_1(V_{\U(2)}) \cup B +  \frac{\cP(B)}{4} \Bigg)\\
&&=-\pi \int_{M^4} c_2(V_{\U(2)}) +\pi\int_{M^5}\Bigg(\frac{\delta(\widetilde{c}_1(V_{\U(2)})\cup \widetilde{c}_1(V_{\U(2)}))}{2}- \frac{1}{2}\delta(\widetilde{c}_1(V_{\U(2)})\cup B)+  \frac{\delta \cP(B)}{4} \Bigg). \nn
\end{eqnarray}
with the background field $B$ properly extended to $M^5$. 
Here $\delta$ is a coboundary operator, such that we apply $\int_{M^4} (\dots) =\int_{M^5} \delta(\dots)$ from \eq{Eq.SU21-form-2} to \eq{Eq.SU(2)thetaterm}.
%
%
{To make sure that the integral on an unorientable $M^5$ is well-defined, and also independent of the dynamical gauge field, we need to utilize the gauge bundle constraint, which relates $\widetilde{c}_1(V_{\U(2)})$ with the background gauge fields $B, w_1(TM)$ and $w_2(TM)$. }
 Below, we will see that the 5-dimensional integral does not depend on the dynamical gauge fields due to the gauge bundle constraints. Hence the 5d integral is an invertible TQFT whose partition function is a \emph{local} function of the background fields. In summary, we find that in order to make sense of the theta term of the SU(2) YM theory with the background fields on an unorientable manifold, one needs to treat the $\SU(2)_{\theta=\pi}$ YM theory as a 4d-5d coupled system. This is a manifestation of the mixed 't Hooft anomaly between the 1-form global symmetry $\Z_{2,[1]}^{e}$ and the time-reversal symmetry $\Z_2^T$.

On an unorientable manifold $M=M^4$, the $w_1(TM)$ is non-trivial and one can treat it as the background gauge field for the time-reversal symmetry. This allows us to 
modify the gauge bundle constraint \eqref{Eq.gaugebundleconstraintSU2} by an additional term $K_1 w_1(TM)^2$, with $K_1=0,1 \in \Z_2$. 
Furthermore, we are also allowed to consider the manifold $M$ with non-trivial $w_2(TM)$ since the underlying manifold does not necessarily allow a Spin/Pin structure, hence we activate the term $K_2 w_2(TM)$ with $K_2=0,1  \in \Z_2$. In summary, there are four choices of gauge bundle constraints labeled by the pair $(K_1, K_2)  \in (\Z_2,\Z_2)$ as
\be
\label{Eq.refinedGBC}
\boxed{
c_1(V_{\U(2)})= B+ K_1 w_1(TM)^2 + K_2 w_2(TM)
= w_2(V_{\PSU(2)})=w_2(V_{\SO(3)})=w_2(E)
 \mod 2, ~~~ K_{1,2}\in \Z_2}. 
\ee
%
{This is a nontrivial constraint between the gauge bundle 
$E$, the spacetime tangent bundle $TM$ and the background field $B$.}
The value of $K_{1,2}$ has physical consequences: when $K_1=0, 1$, the SU(2) gauge charge (in the fundamental representation of SU(2)) 
is a Kramers singlet ($T^2=+1$) or a Kramers doublet ($T^2=-1$)
under time-reversal transformation;\footnote{For an SU(2) gauge theory, 
one can either use $\cT$ or $\cC\cT$ as the time-reversal transformation because the charge conjugation $\cC$ 
of SU(2) is an inner automorphism. The Kramers doublet ($T^2=-1$) 
of Wilson line (in the SU(2) fundamental representation) means that there is a doublet (two-fold) degeneracy associated with the Wilson line. 
The two states of the Wilson line, say $| 1\rangle$ and $| 2\rangle$ forming a 2-dimensional Hilbert space, transforms as 
$| 1\rangle\to | 2 \rangle$ and $| 2 \rangle \to - | 1\rangle$
under the time-reversal transformation.
} when $K_2=0,1$, the SU(2) gauge charge is a boson (spin-statistics as an integer spin) or a fermion (spin-statistics as a half-integer spin). 
More details about the Wilson line properties are derived in \Sec{sec:SU2}.

{The gauge bundle constraint \eqref{Eq.refinedGBC} is defined on $M^4$. We would like to promote it to $M^5$ as follows, 
\begin{eqnarray}\label{Eq.5dGBC}
\widetilde{c}_1(V_{\U(2)})= B+ K_1 w_1(TM)^2 + K_2 w_2(TM)
\mod 2, ~~~ K_{1,2}\in \Z_2.
\end{eqnarray}
\eqref{Eq.5dGBC} imposes additional constraints on $\widetilde{c}_1(V_{\U(2)})$. Since $B, w_1(TM)^2$ and $w_2(TM)$ are $\Z_2$ cohomology on $M^5$, 
the $\widetilde{c}_1(V_{\U(2)})$ is equivalent to a $\Z_2$-valued cohomology $\H^2(M^5, \Z_2)$ mod 2 (although it is not a $\Z$-valued cohomology), i.e., $\delta \widetilde{c}_1(V_{\U(2)})=0\mod 2$.  }

We further apply the gauge bundle constraint \eqref{Eq.refinedGBC} to the 5-dimensional integral \eqref{Eq.SU(2)thetaterm}. 
We should be aware that the 5-manifold ${M^5}$ has a boundary ${M^4}$.
Here we summarize some helpful formulas and mathematical 
definitions in a footnote\footnote{\label{ft:Info}
We clarify the definitions of various fields we used in terms of cochain ($C^n$), cocycle ($Z^n$), coboundary ($B^n$), or cohomology ($\H^n$):
\bea
\left\{\begin{array}{l} 
c_1(V_{\U(2)}) \in \H^2(M,\Z), \\ 
\delta c_1(V_{\U(2)}) =0 \in B^3(M,\Z), \\ 
{\widetilde{c}_1 \in C^2(M,\Z), \quad (\widetilde{c}_1 \mod 2) \in \H^2(M,\Z_2), } \\
c_2(V_{\U(2)}) \in \H^4(M,\Z),\\ 
B \in \H^2(M,\Z_2), \quad \cP(B) \in \H^4(M,\Z_4),\\
w_1(TM) \in \H^1(M,\Z_2),\\
w_2(TM) \in \H^2(M,\Z_2)\\
\lambda \in C^1(M,\Z_2),
\quad \delta \lambda \in B^2(M,\Z_2), \quad \delta^2 \lambda=0 \mod 2,  \quad\Sq^1 \delta \lambda \in Z^3(M,\Z_2).
\end{array}\right.
\eea
Here $C^n$ stands for the $n$-th cochain,
$\H^n$ for the $n$-th cohomology, $Z^n$ for the $n$-th cocycle,
and $B^n$ for the $n$-th coboundary. 
When discussing the cup products, there are subtle distinctions between  (1) cohomology classes in $\H^n$, (2) cocycles in $Z^n$ and (3) cochains in $C^n$, which we enumerate below:
\begin{enumerate}[label=(\arabic*)] 
\item
The cup product between two cohomology classes $u\in \H^p(M, \Z_2), v\in \H^q(M, \Z_2)$ are super-commutative, i.e.,  
	\begin{eqnarray}
	u\cup v= (-1)^{pq} v\cup u.
	\end{eqnarray}
	
\item The cup products between two cocycles are not super-commutative. 
If $u\in Z^p$ and $v\in Z^q$ are general $p$-th and $q$-th cocycles,
their commutation relation is governed by the Steenrod's relation \cite{Steenrod1947}
\begin{eqnarray}
u\cup v - (-1)^{pq} v\cup u = (-1)^{p+q-1} (\delta(u\hcup{1} v)- \delta u\hcup{1} v - (-1)^p u \hcup{1} \delta v)=  (-1)^{p+q-1} \delta(u\hcup{1} v) 
\end{eqnarray}
where we have used the cocycle condition $\delta u=0\mod 2$, $\delta v=0\mod 2$.   
\vspace{\maxdimen}
\item The cup products between two cochains satisfy
Steenrod's relation \cite{Steenrod1947}
\bea \label{eq:Use1}
\delta(u\hcup{i}v)=(-1)^{p+q-i}u\hcup{i-1}v+(-1)^{pq+p+q}v\hcup{i-1}u+\delta u\hcup{i}v+(-1)^pu\hcup{i}\delta v,
\eea
\bea  \label{eq:Use2}
\delta(u\cup v)=\delta u\cup v+(-1)^p u\cup \delta v,
\eea
where $u\in C^p$ and $v\in C^q$ are general $p$-th and $q$-th cochains.
\end{enumerate} 
In this section, all the calculation still 
go through if we regard the $B$ field as a $\Z_2$ 2-cocycle, because we did not use the super-commutativity. 
}.
Since $\widetilde{c}_1(V_{\U(2)})\mod 2$ is in $\H^2(M, \Z_2)$, it makes sense to define its Steenrod square $\Sq^1 \widetilde{c}_1(V_{\U(2)})$. Then the 5d integral in \eqref{Eq.SU(2)thetaterm} can be written as 
{
	\begin{equation}\label{Eq.5dAP}
	\begin{split}
	{\bf S}_{\mathrm{anom}} \equiv&\pi\int_{M^5}\delta\Bigg( \frac{\cP(B)}{4}- \frac{1}{2}   {\widetilde{c}_1(V_{\U(2)})} \cup B +\frac{ {\widetilde{c}_1(V_{\U(2)})}\cup  {\widetilde{c}_1(V_{\U(2)})}}{2}\Bigg)\\
	=& \pi\int_{M^5} \frac{\delta \cP(B)}{4}- \frac{1}{2}   {\widetilde{c}_1(V_{\U(2)})} \cup \delta B- \frac{1}{2}   \delta  {\widetilde{c}_1(V_{\U(2)})}\cup B  +\frac{\delta  {\widetilde{c}_1(V_{\U(2)})}\cup  {\widetilde{c}_1(V_{\U(2)})}}{2}+ \frac{ {\widetilde{c}_1(V_{\U(2)})}\cup \delta  {\widetilde{c}_1(V_{\U(2)})}}{2}\\
	=& \pi\int_{M^5} B \Sq^1 B + \Sq^2 \Sq^1 B-   {\widetilde{c}_1(V_{\U(2)})}\cup \Sq^1 B -   \Sq^1  {\widetilde{c}_1(V_{\U(2)})} \cup B  +\Sq^1  {\widetilde{c}_1(V_{\U(2)})}\cup  {\widetilde{c}_1(V_{\U(2)})}\\&~~~+  {\widetilde{c}_1(V_{\U(2)})}\cup \Sq^1  {\widetilde{c}_1(V_{\U(2)})}\\
	=& \pi \int_{M^5} B \Sq^1 B + \Sq^2 \Sq^1 B-  (B+K_1w_1(TM)^2 + K_2 w_2(TM) ) \cup \Sq^1 B -  (\Sq^1 B + K_2\Sq^1 w_2(TM))\cup B\\&~~~+ (\Sq^1 B+ K_2 \Sq^1 w_2(TM))\cup (B+K_1w_1(TM)^2 + K_2 w_2(TM))\\&~~~+  (B+K_1w_1(TM)^2 + K_2 w_2(TM))\cup  (\Sq^1 B + K_2 \Sq^1 w_2(TM))\\
	=& \pi \int_{M^5} B \Sq^1 B + \Sq^2 \Sq^1 B+K_1 \Sq^1 B \cup w_1(TM)^2 + K_2\Sq^1(  B \cup w_2(TM)  ) \\
	&+K_2 \big( (K_1 w_1^2+ K_2 w_2)\cup  \Sq^1 w_2+\Sq^1 w_2\cup (K_1 w_1^2+ K_2 w_2) \big)
	\\
	=& \boxed{\pi \int_{M^5} B \Sq^1 B + \Sq^2 \Sq^1 B+K_1 \Sq^1 B \cup w_1(TM)^2 + K_2\Sq^1(  B \cup w_2(TM)  ) }.
	\end{split}
	\end{equation}
	In the first equality, we simply stated the initial definition. 
	In the second equality,  we plugged in the coboundary operator $\delta$.
	In the third equality, we used \eqref{eq:Use2} and 
	 replaced $\delta/2$ by $\Sq^1$ which is valid for $\Z_2$-valued cocycles. We also used the identity $\frac{\delta \mathcal{P}(B)}{4}= B \Sq^1 B + \Sq^2 \Sq^1 B$ since $B$ is a $\Z_2$-valued 2-cocycle  \cite{Wan2018zql1812.11968}.\footnote{For a 2-cocycle $B$,  the following equality holds:
$$
\frac{1}{4} \delta \cP(B)
=\frac{1}{4}\delta(B\cup B+B\hcup{1}\delta B)
=(\frac{1}{2}\delta B)\cup B+(\frac{1}{2}\delta B)\hcup{1}(\frac{1}{2}\delta B)
=B\Sq^1B+\Sq^1B\hcup{1}\Sq^1B
=B\Sq^1B+\Sq^2\Sq^1B.
$$ 
See Eq.~(124) in \cite{Wan2018zql1812.11968} for further details. 
} 
In the fourth equality, we plug in the gauge bundle constraint \eqn{Eq.refinedGBC}.
\eqn{Eq.refinedGBC} also implies $\Sq^1 c_1(V_{\U(2)})=\Sq^1 B + K_2\Sq^1 w_2(TM)$.
In the fifth equality, we used $
\Sq^1 (B \cup w_1(TM)^2)
=(\Sq^1 B) \cup w_1(TM)^2+ B \cup \Sq^1(w_1(TM)^2).
$
In the last equality, we used $\big( (K_1 w_1^2+ K_2 w_2)\cup  \Sq^1 w_2+\Sq^1 w_2\cup (K_1 w_1^2+ K_2 w_2) \big)=0\mod 2$ since the Stiefel-Whitney classes are 
\cred{super-commutative}.

Several comments are in order: 
\begin{enumerate}
	\item As mentioned below \eqref{Eq.SU(2)thetaterm}, ${\bf S}_{\mathrm{anom}}$ is a properly quantized integral of the background field $B$ and the Stiefel-Whitney class $w_i(TM)$, which is independent of the dynamical $\U(2)$ gauge field.  Hence ${\bf S}_{\mathrm{anom}}$ is an invertible TQFT. 
	\item In \eqref{Eq.SU(2)thetaterm} and \eqref{Eq.5dAP}, the 5d unorientable manifold $M^5$ has a boundary $M^4$.\\ 
\begin{itemize}
	\item	If $M^5$ does not have a boundary, {the term $K_2\Sq^1(  B \cup w_2(TM)  )$ vanishes, due to }
	\bea \label{eq:K2-vanish}
	&&K_2(  w_2(TM)\cup  \Sq^1 B+    \Sq^1 w_2(TM) \cup B)
	=  K_2  \Sq^1\left( w_2(TM)\cup B\right)\nn\\
	&&= K_2  w_1(TM) w_2(TM)\cup  B= K_2  \Sq^3 B= K_2  u_3 B=0\mod 2\pi. \label{eq:K2Sq3B=0}
	\eea
	In the last step, we have used the Wu-formula $u_3 \equiv u_3(TM)=w_1(TM) w_2(TM)=0\mod 2$,  
	{on a closed 5-manifold}. 
	Hence \eqref{Eq.5dAP} simplifies to  
	\begin{eqnarray}
	\pi \int_{M^5} B \Sq^1 B + \Sq^2 \Sq^1 B + K_1w_1(TM)^2\cup \Sq^1 B.
	\end{eqnarray}
	\item If $M^5$ has a boundary,  $ K_2 \pi \Sq^1\left( w_2(TM)\cup B\right)$ transforms non-trivially under the 
	background gauge transformation $B\to B+ \delta \lambda$, 
	\be \label{eq:K2-variation}
	K_2 \pi \int_{M^5} \Sq^1\left( w_2(TM)\cup B\right)\to  K_2 \pi \int_{M^5} \Sq^1\left( w_2(TM)\cup B\right)+ K_2 \pi \int_{M^5} \Sq^1(w_2(TM)\cup \delta \lambda), \quad
	\ee
	This compensates the non-invariance of  the 4d theory under $B\to B+ \delta \lambda$. 
	Thus although the $K_2$ terms vanish when $M^5$ is a closed manifold,  when $M^5$ has a boundary, it is crucial to keep track of this term. 

\item 
We can show that the term $K_2\int_{M^5} \Sq^1\left( w_2(TM)\cup B\right)$ is \emph{well-defined} in 4d by showing that this terms only depends on the 4d boundary $\partial{M^5}$.
{The triviality of $\int_{M^5}  K_2  \Sq^1\left( w_2(TM)\cup B\right)=0$ on a closed $M^5$ implies that when the 5d manifold has a boundary, such a term does not depend on the choice of  extension, i.e., given two 5d extensions $M^5$ and $\widetilde{M}^5$, 
we know $K_2\int_{M^5 \cup {\widetilde{M}^5} }$ $\Sq^1\left( w_2(TM)\cup B\right)=0$ because $M^5 \cup \widetilde{M}^5$ is closed,
thus we derive $K_2\int_{M^5} \Sq^1\left( w_2(TM)\cup B\right)= K_2 \int_{\widetilde{M}^5} \Sq^1\left( w_2(TM)\cup B\right)$.} Note that when $M^5$ has a boundary, 
$K_2\int_{M^5} \Sq^1\left( w_2(TM)\cup B\right)$ can be nonzero. This is analogues to the WZW term. See Sec.\ref{Sec.Consequences} for further discussions. 

	\end{itemize}
	\item The 4d-5d integral \eqref{Eq.SU(2)thetaterm} is invariant under a 1-form gauge transformation $B\to B+ \delta \lambda$.  We will show this explicitly in Sec.~\ref{Sec.AnotherSU(2)AnomalyInterpretation}. 
	\item Although  ${\bf S}_{\mathrm{anom}}$ only depends on $K_1$ when $M^5$ is closed,  we still label it as the {5d anomaly polynomial parameterized by $(K_1, K_2)$,}
	 due to the subtlety that the 5d integral still depends on $K_2$ when $M^5$ has boundary. 
\end{enumerate}
To summarize, the partition function of the combined 4d-5d coupled system
\begin{eqnarray}\label{Eq.combinedpartitionfunction}
\bZ^{\text{4d}}_{\SU(2)_{\theta=\pi}\text{YM}}[M^4;B,w_j(TM)]\cdot \bZ^{\text{5d}}[M^5;B,w_j(TM)]
\end{eqnarray}
is fully gauge invariant under the transformation of the background gauge field $B$ and time-reversal symmetry,
the full partition function also makes sense when $M^4$ and $M^5$ are unorientable, 
where 
\begin{equation} \label{eq:SU2YMZ}
\begin{split}
&\bZ^{\text{4d}}_{\SU(2)_{\theta=\pi}\text{YM}} [M^4;B,w_j(TM)]= \int [\CD \widehat{a}][\CD \Lambda] 
\exp \Bigg( - \frac{1}{g^2}\int_{M^4} \Tr \big((\widehat{F}-\pi B\mathbb{I}_2)\wedge \star (\widehat{F}-\pi B\mathbb{I}_2)\big) \Bigg)
\\&\quad\quad\quad \quad\quad\quad 
\cdot \exp \Bigg( -\ii \pi \int_{M^4} c_2(V_{\U(2)})   \Bigg)\cdot \exp \Bigg( \ii \pi \int_{M^4} \Lambda \cup (c_1-B- K_1w_1(TM)^2 - K_2 w_2(TM) )\Bigg),
\end{split}
\end{equation}
and 
\begin{equation}  \label{Eq.5danomalypolynomial}
\begin{split}
\bZ^{\text{5d}} [M^5;B,w_j(TM)]= \exp \Bigg(\ii \pi \int_{M^5} & B \Sq^1 B + \Sq^2 \Sq^1 B+K_1w_1(TM)^2\cup \Sq^1 B 
+ K_2  \Sq^1(w_2(TM)\cup  B)   \Bigg).
\end{split}
\end{equation}
The combined 4d-5d system is anomaly free. Equivalently, to couple the background fields of both time-reversal symmetry and the 1-form global symmetry $\Z_{2, [1]}^e$, 
the  $\SU(2)_{\theta=\pi}$ YM theory cannot be background gauge invariant by being placed on an unorientable $M^4$ only. 
Instead, one needs to place $\bZ^{\text{4d}}_{\SU(2)_{\theta=\pi}\text{YM}}$ 
on the boundary of 
an unorientable $M^5$ which supports a 5d invertible TQFT $\bZ^{\text{5d}}$. 
%
{This is the manifestation of the the $\SU(2)_{\theta=\pi}$ YM's mixed 't Hooft anomaly 
between the 1-form global symmetry $\Z_{2,[1]}^{e}$ and the time-reversal symmetry $\Z_2^T$. }

\subsection{Proof of 
Anomaly Matching of 5d-4d Inflow and 5d Cobordism Group Data} 
\label{Sec.AnotherSU(2)AnomalyInterpretation}

In this subsection, we identify the 5d topological terms \eqref{Eq.5dAP} with the mathematically well-defined 
5d bordism invariants, and further explicitly check the invariance of the 4d-5d system \eqref{Eq.combinedpartitionfunction} under $B\to B+\delta \lambda$.  

\subsubsection{Identifying the 4d anomaly with 5d Cobordism Group Data}

We compare the 4d anomaly ${\bf S}_{\mathrm{anom}}$ in \eqref{Eq.5dAP} with the bordism group data given in  \cite{Wan2018zql1812.11968} and \cite{Wan2018bns1812.11967}.  
Since the global symmetries of 4d SU(2)$_{\theta=\pi}$ YM theory is $\Z_2^T\times \Z_{2,[1]}^e$, we compute 
the 5d bordism 
group\footnote{In addition to  \cite{Wan2018zql1812.11968} and \cite{Wan2018bns1812.11967},
we notice that the oriented version of the bordism group $\Omega_5^{\SO}(\B^2\Z_2)$  has been studied recently in \cite{Kapustin2017jrc1701.08264} for different purposes.
Here we study instead the unoriented version of the bordism group $\Omega_5^{\tO}(\B^2\Z_2)$, new to the literature.  See details in Appendix \ref{appendix-bordism}.}
\footnote{
For an ordinary (0-form) global symmetry, we denote $G$ as the $0$-form global symmetry group. 
When gauging a $0$-form symmetry, we introduce a $1$-form flat gauge field with a \emph{gauge group} $G$,
whose classifying space is $BG$.
For an abelian group and for a higher symmetry: We denote $G_{[1]}$ as the $1$-form global symmetry group. 
When gauging a $1$-form symmetry, we introduce a $2$-form flat gauge field with a \emph{higher gauge group},
whose classifying space is associated with $B(BG)=B^2 G$.
Similarly, for an abelian $n$-form global symmetry group $G_{[n]}$, we have the associated classifying space $B^{n+1} G$.
See \cite{Cordova2018cvg2group1802.04790, Delcamp2018wlb1802.10104, Benini2018reh1803.09336}.
} 
\begin{eqnarray}
\Omega_5^{\tO}(\B^2\Z_2)=\Z_2^4.
\end{eqnarray}
Hence there are four independent generators of the bordism group $\Omega_5^{\tO}(\B^2\Z_2)$, 
\begin{equation}\label{eq:bordism5OB2Z2}
\left\{\begin{array}{l} 
B \Sq^1B,\\ 
\Sq^2\Sq^1 B =(w_2(TM)+w_1(TM)^2) \Sq^1 B =(w_3(TM)+w_1(TM)^3) B ,\\ 
w_1(TM)^2\Sq^1 B = w_1(TM)^3 B,\\ 
w_2(TM)w_3(TM). 
\end{array}\right.
\end{equation}
where the equalities
hold only on closed 5-manifolds. 
Clearly, ${\bf S}_{\mathrm{anom}}$ in \Eq{Eq.5dAP} is a bordism invariant except the term proportional to $K_2$. 
Setting $K_2=0$, ${\bf S}_{\mathrm{anom}}$ is identified with the sum of first three bordism invariants in \Eq{eq:bordism5OB2Z2},
\begin{equation}\label{eq:5d-SPT-bord}
\begin{split}
\exp\Bigg(\ii \pi \int_{M^5} (    B\Sq^1B+ \Sq^2\Sq^1B +  K_1 w_1(TM)^3  B  )\Bigg).
\end{split}
\end{equation}
As explained in Sec.~\ref{Sec.thetapiYM}, 
the fourth term in ${\bf S}_{\mathrm{anom}}$ 
is a trivial when $M^5$ does not have a boundary. This is consistent with the fact that there isn't any bordism invariant of $\Omega_5^{\tO}(\B^2\Z_2)$) of the form $\Sq^1(w_2(TM)\cup  B)$.

{Notice that the last invariant in \Eq{eq:bordism5OB2Z2} \footnote{The $w_2(TM)w_3(TM)$ 
is  a
bordism invariant in ${\Omega_5^{\tO}(\B^2\Z_2)}$, ${\Omega_5^{\tO}(pt)}$,  ${\Omega_5^{\SO}(pt)}$ and ${\Omega_5^{\frac{\Spin \times \SU(2)}{\Z_2}}}$, see \cite{Wan2018bns1812.11967}.
Namely, this $w_2(TM)w_3(TM)$ is not only a topological term respecting a spacetime $\tO(d)$ symmetry and 1-form $\Z_{2,[1]}^e$-symmetry,
but also  a topological term respecting a spacetime $\tO(d)$ or $\SO(d)$ symmetry alone, or 
respecting an enhanced spacetime-internal locked symmetry ${\frac{\Spin \times \SU(2)}{\Z_2}}$.
Thus the 4d anomaly from ${\Omega_5^{\frac{\Spin \times \SU(2)}{\Z_2}}}$ is a signature for 
\emph{the 
new SU(2) anomaly} \cite{Wang:2018qoyWWW}.
In fact, the $w_2(TM)w_3(TM)$ topological term plays an important role as the only possible anomaly
of an interacting Spin(10) chiral fermion theory 
--- which is responsible for the anomaly-free of the SO(10) Grand Unification \cite{Wang2018cai1809.11171, Wang:2018qoyWWW}. 
 }
 , i.e. $ w_2(TM)w_3(TM) $,  does not participate in the anomaly of SU(2)$_{\theta=\pi}$ YM.
However
it is responsible for the new SU(2) anomaly \cite{Wang:2018qoyWWW}: 4d SU(2) gauge theory
with an odd
number of fermion multiplets in representations of isospin $4r + 3/2$ of the gauge group is inconsistent, for a non-negative integer $r$. 
The theory is nevertheless consistent on certain manifolds with Spin or Spin$^c$ structure.
The new SU(2) anomaly \cite{Wang:2018qoyWWW} is in contrast of the old SU(2) anomaly \cite{Witten:1982fp}.
The familiar SU(2) anomaly \cite{Witten:1982fp}
states that a 4d SU(2) gauge theory with an odd number of fermion multiplets in the isospin $2r + 1/2$ representation is inconsistent. 
}

In this subsection, we proved that the 4d YM anomaly (derived in \Sec{Sec.thetapiYM}) given in \Eq{Eq.5danomalypolynomial} matches a mathematically well-defined 
 5d bordism invariant from a bordism group data  (given in Appendix \ref{appendix-bordism}).

\subsubsection{Anomaly Matching of 4d-5d Inflow}

We first highlight the distinctions between the derivation of anomalies in \cite{Gaiotto2017yupZoharTTT1703.00501} and in our Sec.~\ref{Sec.thetapiYM}.


\begin{itemize}

\item \Refe{Gaiotto2017yupZoharTTT1703.00501} places the SU(2)$_{\theta=\pi}$ YM on an \emph{orientable} manifold, and  
turns on the 2-form background  field $\cB$ of the 1-form symmetry $\Z_{2,[1]}^e$ (or a 2-cochain $B$). 
By performing a time-reversal $\Z_2^T$ transformation, \Refe{Gaiotto2017yupZoharTTT1703.00501} detects
the $\cT BB$ anomaly, which is linear in $\Z_2^T$ transformation $\cT$ and quadratic to the 2-cochain $B$.

\item
In Sec.~\ref{Sec.thetapiYM}, we have derived the anomaly by first turning on the 2-form gauge field $B$, and further place the theory on an \emph{unorientable manifold}. We find that to make sense of the 4d theta term on an unorientable manifold, we need to promote the original 4d YM theory to a combined 4d-5d system. The 5d theory is an invertible TQFT. 
%
In the following, we reverse the logic: 
\begin{itemize}
\item[(Step 1)] We first formulate the $\SU(2)$ YM on an unorientable manifold before activating $B$.   
\item[(Step 2)] We further match the non-invariance of the 4d SU(2)$_{\theta=\pi}$ YM theory \eqref{eq:SU2YMZ} under $B\to B+\delta \lambda$  with the non-invariance of $\mathbf{S}_{\text{anom}}$ in   \Eq{Eq.5dAP}.  
\end{itemize}
\end{itemize}

\begin{enumerate}[label=(Step \arabic*)]

\item
We first place the $\SU(2)$ Yang-Mills theory on an unorientable manifold without activating the background field $B$. 
If we limit to case that the gauge bundle constraint \Eq{Eq.refinedGBC} as $c_1(V_{\U(2)})=0\mod 2$,
then the theta term is simplified to 
\begin{eqnarray}\label{Eq.boundarytheory}
-\pi \int_{M^4} c_2(V_{\U(2)}).
\end{eqnarray} 
which is a well-defined 4d term. 
If we further change the time-reversal property (i.e., Kramers singlet/doublet)  and the statistics (i.e., bosonic/fermionic)
of the SU(2) gauge charge by modifying the gauge bundle constraint to $c_1(V_{\U(2)})=K_1 w_1(TM)^2+ K_2 w_2(TM)\mod 2$,  the theta term is 
\begin{eqnarray} \label{Eq.noB}
\pi \int_{M^4} \Bigg(- c_2(V_{\U(2)})+\frac{1}{2} c_1(V_{\U(2)}) \cup c_1(V_{\U(2)})\Bigg).
\end{eqnarray}
The second term does not make sense for $M^4$ unorientable, and one needs to define it by promoting the integral to a 5d unorientable manifold $M^5$. 
{Following the discussion around \eqref{Eq.5dGBC}, the $\Z$-valued cohomology class $c_1(V_{\U(2)})$ is extended to a $\Z_2$ cohomology class $\widetilde{c}_1(V_{\U(2)})$, along with the gauge bundle constraint, $\widetilde{c}_1(V_{\U(2)})= K_1 w_1(TM)^2+ K_2 w_2(TM)\mod 2$. Then, \eqref{Eq.noB} shall be re-interpreted as
\begin{equation}\label{Eq.4d5dNoB}
 -\pi \int_{M^4} c_2(V_{\U(2)}) + \pi \int_{M^5} \frac{1}{2}\delta (\widetilde{c}_1(V_{\U(2)}) \cup  \widetilde{c}_1(V_{\U(2)})).
\end{equation}
}
%
%
%
{When $M^5$ does not have a boundary, $\pi \int_{M^5} \frac{1}{2}\delta (\widetilde{c}_1(V_{\U(2)}) \cup  \widetilde{c}_1(V_{\U(2)}))$ vanishes. This means that, for a fixed $M^4$,  the second term in \eqref{Eq.4d5dNoB} does not depend on the choice of $M^5$.} 
Hence, when $B$ is turned off, 
there is no anomaly for generic $(K_1, K_2)$. To summarize, there is {no} pure time-reversal anomaly of $\SU(2)$ Yang-Mills with $\theta=\pi$. 

\item
We further turn on the background field $B$. Under the gauge transformation $B\to B+ \delta \lambda$ where $\lambda$ is a $\Z_2$-valued 1-cochain,  the $\U(2)$ field strength $\widehat{F}$ transforms as 
$$\widehat{F}\to \widehat{F}+ \pi \delta \lambda \mathbb{I}.$$ 
Using \eqref{Eq.defc1c2}, 
we 
  determine that%
\begin{eqnarray}
\label{Eq.gaugetransformation}
\begin{split}
c_1(V_{\U(2)}) &\to c_1(V_{\U(2)})+ \delta \lambda,\\ 
c_2(V_{\U(2)}) &\to c_2(V_{\U(2)}) + \frac{1}{2} c_1(V_{\U(2)}) \cup \delta \lambda + \frac{1}{4} \cP( \delta \lambda).
\end{split}
\end{eqnarray}
The only 4d term in \eqref{Eq.SU(2)thetaterm} is the first term proportional to $c_2(V_{\U(2)})$. Under $B\to B+ \delta \lambda$, 
\begin{equation}\label{Eq.4danomaly}
\begin{split}
-\pi \int_{M^4} c_2 &\to 
-\pi \int_{M^4} \left(c_2 + \frac{1}{2} c_1(V_{\U(2)}) \cup \delta \lambda+ \frac{1}{4} \cP( \delta \lambda)\right)\\
&=-\pi \int_{M^4} c_2 -\pi \int_{M^5} \delta 
\left(\frac{1}{2} \widetilde{c}_1(V_{\U(2)}) \cup \delta \lambda+  \frac{1}{4}\cP( \delta \lambda)\right)\\
&=-\pi \int_{M^4} c_2 -\pi \int_{M^5}\Sq^1 \widetilde{c}_1\cup\delta\lambda+ \widetilde{c}_1\cup\Sq^1\delta\lambda+\delta\lambda\cup \Sq^1\delta\lambda+\Sq^2\Sq^1\delta\lambda\\
&=-\pi \int_{M^4} c_2 -\pi \int_{M^5}\bigg[(\Sq^1B+K_2\Sq^1w_2(TM))\delta\lambda+(B+K_1w_1(TM)^2+K_2w_2(TM))\Sq^1\delta\lambda\\
&~~~+\delta\lambda\Sq^1\delta\lambda+\Sq^2\Sq^1\delta\lambda\bigg].
\end{split}
\end{equation}
In the second equality, we replaced $\delta/2$ by $\Sq^1$ which is valid for $\Z_2$-valued cocycles, and used the identity $\frac{\delta \mathcal{P}(\delta\lambda)}{4}= \delta\lambda \Sq^1 \delta\lambda + \Sq^2 \Sq^1 \delta\lambda$ since $\delta\lambda$ is a cocycle  \cite{Wan2018zql1812.11968}.
On the other hand, the variation of the bulk invertible TQFT $\mathbf{S}_{\mathrm{anom}}$, i.e. the 5d integral in \eqref{Eq.SU(2)thetaterm}, is 
\begin{equation}\label{Eq.5danomalyinflow}
\begin{split}
{\bf S}_{\mathrm{anom}}\equiv& \pi \int_{M^5}  B \Sq^1 B + \Sq^2 \Sq^1 B + K_1w_1(TM)^2\cup \Sq^1 B + K_2   \Sq^1 w_2(TM) \cup B + K_2w_2(TM)\cup  \Sq^1 B    \\
=& \pi \int_{M^5}  \frac{1}{4}\delta\cP(B) + K_1w_1(TM)^2\cup \Sq^1 B + K_2   \Sq^1 w_2(TM) \cup B + K_2w_2(TM)\cup  \Sq^1 B    \\
\longrightarrow~~~ & {\bf S}_{\mathrm{anom}}+ \pi \int_{M^5} \frac{1}{4}\delta\cP(\delta\lambda)+\frac{1}{2}\delta(B\delta\lambda)+K_1w_1(TM)^2\Sq^1\delta\lambda
+ K_2  \Sq^1 w_2(TM)   \delta \lambda + K_2w_2(TM)\Sq^1\delta\lambda  \\
=&{\bf S}_{\mathrm{anom}}+ \pi \int_{M^5}\delta\lambda\Sq^1\delta\lambda+\Sq^2\Sq^1\delta\lambda+\Sq^1B\delta\lambda+B\Sq^1\delta\lambda\\
&+K_1w_1(TM)^2\Sq^1\delta\lambda+ K_2  \Sq^1 w_2(TM)   \delta \lambda + K_2w_2(TM)\Sq^1\delta\lambda.  
\end{split}
\end{equation} 
In the second equality, we used the identity $\frac{\delta \mathcal{P}(B)}{4}= B \Sq^1 B + \Sq^2 \Sq^1 B$ since $B$ is a cocycle  \cite{Wan2018zql1812.11968}, and the formula
$\cP(B+\delta\lambda)=\cP(B)+\cP(\delta\lambda)+2B\delta\lambda$
since $B$ and $\delta\lambda$ are both cocycles.
In the third equality, we replaced $\delta/2$ by $\Sq^1$ which is valid for $\Z_2$-valued cocycles, and used the identity $\frac{\delta \mathcal{P}(\delta\lambda)}{4}= \delta\lambda \Sq^1 \delta\lambda + \Sq^2 \Sq^1 \delta\lambda$ since $\delta\lambda$ is a cocycle  \cite{Wan2018zql1812.11968}.

Comparing \eqref{Eq.4danomaly} and \eqref{Eq.5danomalyinflow}, we find that the non-invariance of the 4d terms \eqref{Eq.4danomaly} precisely cancels the non-invariance of the 5d terms \eqref{Eq.5danomalyinflow}.
Thus the combined 4d-5d coupled system  $-\pi\int_{M^4} c_2(V_{\U(2)})+ {\bf S}_{\mathrm{anom}}$ is symmetric under the background gauge transformation of $B$, thus is  anomaly free under the
1-form background gauge transformation.\footnote{On an unorientable manifold, the mixed time-reversal and 1-form anomaly reduces to 1-form anomaly, since time-reversal symmetry is ``gauged" on an unorientable manifold and it is too late to break $\Z_2^T$. } Furthermore, since both the boundary theory \eqref{Eq.boundarytheory} and the bulk invertible TQFT ${\bf S}_{\mathrm{anom}}$ are well-defined on an unorientable manifold $M^4$ and $M^5$ respectively, the full system $\pi\int_{M^4} c_2(V_{\U(2)})+ {\bf S}_{\mathrm{anom}}$ also respects the time-reversal symmetry. Thus we again arrive at the conclusion that the combined partition function \eqref{Eq.combinedpartitionfunction} is well-defined and free of the 't Hooft anomalies of both 1-form symmetry,  time-reversal symmetry and their mixed anomaly. 

\end{enumerate}

\subsection{
Topological Term On {Torsion-Free} Orientable Manifolds}
\label{Sec.lift}

{In the previous sections \ref{Sec.thetapiYM} and \ref{Sec.AnotherSU(2)AnomalyInterpretation}, we derived the mixed anomaly by first reformulating the theta term in terms of characteristic classes, and then make sense of it on unorientable manifolds by promoting the ill-defined terms on 5-manifolds. However, there is a loop-hole: \eqref{Eq.SU21-form-2} is not well-defined even on an oriented manifold, because $c_1(V_{\U(2)})\cup B$ and $\frac{1}{2}\cP(B)$, as a $\Z_2$ and $\Z_4$ valued cohomology respectively, are ill-defined when the coefficients are fractional. In this subsection, we resolve this issue, for certain manifolds,  by lifting the $\Z_2$ class $B$ to a $\Z$ class $\widetilde{B}$, i.e, 
\begin{eqnarray}
B=\widetilde{B} \mod 2
\end{eqnarray}
Here we restrict to the orientable manifolds $M^4$ with torsion-free cohomology class $\H_1(M^4, \Z)$\cite{kapustin2014topological} where the lifting makes sense. Hence \eqref{Eq.SU21-form-2} becomes 
\bea\label{Eq.orient}
\pi \int_{M^4} -c_2(V_{\U(2)}) +\frac{c_1(V_{\U(2)})\cup c_1(V_{\U(2)})}{2}
- \frac{1}{2} c_1(V_{\U(2)}) \cup \widetilde{B}+  \frac{\widetilde{B}\cup \widetilde{B}}{4}.
\eea

To further formulate \eqref{Eq.orient} on an unorientable manifold, we note that every unorientable manifold $M$ contain nontrivial torsion in $\H_1(M, \Z)$, and thus the lifting does not exist. This implies that on an unorientable manifold $M^4$ and $M^5$, it is not possible to promote a $\Z_2$ cohomology class to a $\Z$ cohomology class. However, the derivation of the 5d anomaly polynomial \eqref{Eq.5dAP} still goes through.

\subsection{
Consequences and Interpretations
of 
Four Siblings of ``Anomalies''}
\label{Sec.Consequences}

In this section, we discuss the two siblings of anomalies labeled by $(K_1=0, K_2)$ and $(K_1=1,K_2)$. We also compare our results with the known mixed $\Z_{2,[1]}^e$-$\Z_2^T$ anomaly discussed in \cite{Gaiotto2017yupZoharTTT1703.00501}. 

\noindent
$\diamond$ When ${(K_1,K_2)}{=(0,0)}$, the bulk anomaly polynomial is
	\begin{eqnarray}\label{Eq.TBBanomaly}
	\pi \int_{M^5}B \Sq^1 B+ \Sq^2 \Sq^1 B= \frac{\pi}{2} \int_{M^5} \tilde{w}_1(TM)\cup \mathcal{P}(B)
	\end{eqnarray}
	which is non-vanishing only on an unorientable $M^5$. 
	This equality has been explored in \Refe{Wan2018zql1812.11968} in relating to the 4d YM theory's anomaly.
	Furthermore, we find that this equality is also 
	explained in a remarkable mathematical note \Refe{Arun}.\\ %
	
	Below let us gain a 
	better understanding based on \Refe{Arun}:
Let $\Z_{w_1}$ be the orientation local system, then $\H^1(\B\tO(1),\Z_{w_1}) = \Z_2$.
Indeed, this is the group cohomology $\H^1(\Z_2,\Z_{\sigma})$, where $\Z_{\sigma}$ denotes $\Z$ with the sign action. The pullback of the nonzero element of $\H^1(\B\tO(1),\Z_{w_1})$ 
under the map $M\to\B\Z_2$ determined by $w_1(TM)\in\H^1(M,\Z_2)$
 is called the twisted first Stiefel-Whitney class 
 $\tilde w_1\in\H^1(M,\Z_{w_1})$.
Its mod 2 reduction is the usual first Stiefel-Whitney class in an untwisted $\Z_2$-cohomology. We consider its reduction $\tilde w_1\in \H^1(M,(\Z_4)_{w_1})$ in a twisted mod 4 cohomology.
Here $\cP$ denotes the Pontryagin square $\cP: \H^2(M,\Z_2)\to \H^4(M,\Z_4)$.
In \eqref{Eq.TBBanomaly}, we use cup and cap products in twisted $\Z_4$-cohomology: if $[M]$ denotes the fundamental class in the twisted $\Z_4$-cohomology, this means that
\bea
 \H^1(M, (\Z_4)_{w_1} )\otimes \H^4(M, \Z_4)\xrightarrow{\cup}  \H^5(M, (\Z_4)_{w_1} ) \xrightarrow{\cap[M]} \Z_4.
 \eea
However, since $2\tilde w_1$ is a twisted coboundary, $2\<\tilde w_1 \cup \cP(B),[M]\>=0\mod4$, $\<\tilde w_1 \cup \cP(B),[M]\>$ is even, hence it makes sense to divide by 2 and obtain an element of $\Z_2$.
This defines $\frac{1}{2}\tilde{w}_1(TM)\cup \mathcal{P}(B)$ as a mod 2 class in
the 5th cohomology group $\H^5(\B \tO \times \B^2 \Z_2, \U(1))$
which is also a bordism invariant of 5th bordism group $\Omega_5^{\tO}(\B^2\Z_2)$.

	 There are two options for the boundary $M^4$: orientable or unorientable. 
	\begin{enumerate}
		\item When $M^4$ is orientable, the time reversal of the $\SU(2)_{\theta=\pi}$ theory is {not gauged.}
		However, there is still a way to probe the mixed $\Z_{2,[1]}^e$-$\Z_2^T$ anomaly, following the approach of \cite{Gaiotto2017yupZoharTTT1703.00501}.  We first couple the $\SU(2)_{\theta=\pi}$ Yang-Mills to background gauge field $B$, and then perform a \emph{global} time-reversal transformation. To determine how the theta term changes under timer reversal, we make use of the fact that shifting $\theta$ by $2\pi$ amounts to change the parameter $p$ of the counter term by $1$, where the counter term is $2\pi \frac{p}{4}\int_{M^4} \mathcal{P}(B)$ and $p\in \Z_4$, i.e.,   
		\begin{eqnarray}\label{Eq.equivalence}
		(\theta+2\pi, p) \longleftrightarrow (\theta, p+1).
		\end{eqnarray}
		Under time reversal, both the theta term \eqref{Eq.SU21-form} and the counter term change sign, i.e., $\Z_2^T: (\pi, p)\to (-\pi, -p)$. Using the identification \eqref{Eq.equivalence}, $(-\pi, -p) \longleftrightarrow (\pi, -p-1)$
		\begin{eqnarray}
		\Z_2^T: (\pi, p)\to (\pi, -p-1).
		\end{eqnarray}
		Equivalently, under time reversal, the theta term is unchanged, but there is a shift of the counter term 
		\begin{eqnarray}\label{Eq.noninvariance}
		\delta {\bf S}_E[M^4] =-  \frac{\pi(2p+1) }{2}\int_{M^4} \mathcal{P}(B).
		\end{eqnarray}
		The non-invariance in \eqref{Eq.noninvariance} cannot be zero  by properly choosing $p\in \Z$, which represents  an anomaly. The anomaly \eqref{Eq.noninvariance} can be canceled by the 't Hooft anomaly inflow \eqref{Eq.TBBanomaly}.

		So it is important to emphasize that the 4d anomaly from $\cT BB$ detected by \cite{Gaiotto2017yupZoharTTT1703.00501} 
		(and Sec.~2 of \Refe{Wan2018zql1812.11968}
		) is precisely captured by the bordism invariant $\frac{1}{2}\tilde{w}_1(TM)\cup \mathcal{P}(B)$ in \Eq{Eq.TBBanomaly} noticed in \Refe{Wan2018zql1812.11968}.
		
		\item When $M^4$ is unorientable, the anomaly can be detected as well, as discussed in Sec.\ref{Sec.thetapiYM} and \ref{Sec.AnotherSU(2)AnomalyInterpretation}. 
	\end{enumerate}
\noindent
$\diamond$ When ${(K_1,K_2)}{=(1,0)}$,
	the bulk action is 
	\begin{eqnarray}\label{Eq.TBBanomalyKramer}
	\pi \int_{M^5}B \Sq^1 B+ \Sq^2 \Sq^1 B+ w_1(TM)^2  \Sq^1 B
	\end{eqnarray}
	which is non-vanishing only when $M^5$ is unorientable. 
	\begin{enumerate}
		\item When $M^4$ is orientable, one cannot probe $K_1$. This is because for  $\int_{M^5} w_1(TM)^2 \Sq^1 B$ to be non-vanishing mod 2 on $M^5$,  there should be at least two or more orientation reversing cycles in $M^5$, hence there should be at least one orientation cycle in $M^4$. 
		{Thus if $M^4$ is orientable, even if $M^5$ is unorientable, we still cannot detect a particular 4d anomaly associated to the 5d term $K_1 w_1(TM)^2  \Sq^1 B$.}
				
		\item When $M^4$ is unorientable, the anomaly can be detected, as discussed in Sec.\ref{Sec.thetapiYM} and \ref{Sec.AnotherSU(2)AnomalyInterpretation}. 
	\end{enumerate}
\noindent
$\diamond$ When ${(K_1,K_2)}{=(0,1)}$,

	$\bullet$ If $M^5$ is a closed 5d manifold (regardless orientable or unorientable), we cannot detect the term $\int_{M^5} K_2 \Sq^1 (w_2(TM)\cup   B)$.
	                  
	                   $\bullet$ If $M^5$ is a 5d manifold with a 4d boundary $M^4$ (regardless orientable or unorientable in 5d or in 4d) and $w_2(TM)$ is nontrivial on both $M^4$ and $M^5$
	                   (e.g., non-Pin$^{+}$ manifolds), 
	                   we can detect the term $\int_{M^5} K_2 \Sq^1 (w_2(TM)\cup           
	                   B)$ on the 4d boundary via the \emph{1-form background gauge transformation}.                
On an $M^5$ with a boundary $M^4$, we regard 
 $\int_{M^5} K_2 \Sq^1 (w_2(TM)\cup   B)$  schematically as a 4d fractional SPTs, which is characterized by a 4d ill-defined term  with a fractional coefficient
$\int_{M^4} K_2 \frac{1}{2}(w_2(TM)\cup   B)$.
Two copies of such 4d fractional SPTs become a well-defined time-reversal $\Z_2^T$ and 1-form $\Z_{2,[1]}^{e}$ symmetric
4d SPTs/bordism invariant $\int_{M^5} (w_2(TM)\cup   B)$, with respect to
a nontrivial $\Z_2$-generator in $\Omega_4^{\tO}(\B^2\Z_2)=\Z_2^4$, see 
\Refe{Wan2018zql1812.11968} and Appendix \ref{appendix-bordism}.
Thus, four layers of such 4d fractional SPTs become a trivial SPTs with respect to $\Omega_4^{\tO}(\B^2\Z_2)$.

{
The $\int_{M^5} \Sq^1 (w_2(TM)   B)$ is similar to \emph{{Wess-Zumino-Witten (WZW)}} term  \cite{Wess1971yuWZ, Witten1983twGlobalACA} in some way but with its own exoticness:\\
(1) The familiar WZW term is an integer $\Z$ class \cite{Wess1971yuWZ, Witten1983twGlobalACA},
here this $\int_{M^5} \Sq^1 (w_2(TM)   B)$ has a fractional discrete class.  
(In some sense, $\int_{M^5} \Sq^1 (w_2(TM)   B)$ seems to be a unit generator in $\Z_4$ respect to a 4d trivial SPTs.)\\
(2) The familiar WZW term is written as a path integral of dynamical fields, but here $\int_{M^5} \Sq^1 (w_2(TM)   B)$ depends on the background probe fields 
$w_2(TM)$ and $B$.\\
(3) Both WZW and $\int_{M^5} \Sq^1 (w_2(TM)   B)$ govern the 4d physics, but they need to be written in one extra higher dimension.
It is tempting to speculate that $\int_{M^5} \Sq^1 (w_2(TM)   B)$ may be a \emph{non-local counter term on $M^4$}, 
which is 4d in nature but cannot be written in 4d alone. The  $\int_{M^5} \Sq^1 (w_2(TM)   B)$ can access the 5d extra bulk,
but it does not depend on how ${M^5}$ is chosen as long as $\partial {M^5}={M^4}$.
}

Related interpretations and facts about $\int_{M^5} K_2 \Sq^1 (w_2(TM)\cup   B)$ are also summarized in \Sec{sec:table}.
	                  
\noindent
$\diamond$ 
When ${(K_1,K_2)}{=(1,1)}$, the interpretation is simply the linear combination of ${(K_1,K_2)}{=(1,0)}$ and ${(K_1,K_2)}{=(0,1)}$ interpretations above.

We will further comment about the fate of dynamics of Four Siblings of SU(2) YM based on their ``anomalies'' 
\Eq{Eq.5danomalypolynomial}, in \Sec{sec:lattice} and in \Sec{sec:conclude}.


\subsection{5d SPTs/Bordism Invariants Whose Boundary Allows 4d SU(2)$_{\theta=\pi}$ YM}

\subsubsection{On a closed manifold}
\label{sec:5dSPT-Z}

We now give various equivalent formulas of the
5d SPTs/bordism invariant in \Eq{Eq.5danomalypolynomial} on a closed 5-manifold $M^5_{\text{closed}}$:
\bea
&&\bZ_{\text{SPT}_{(K_1,K_2)}}^{5\text{d}}[M^5_{\text{closed}}]
=\exp\Bigg({\ii\pi\int_{M^5} B\Sq^1B+\Sq^2\Sq^1B + K_1 w_1(TM)^2 \Sq^1 B +K_2  \Sq^1(w_2(TM) B)}\Bigg) \nn\\
&&\; =
\exp\Bigg({\ii\pi\int_{M^5} B\Sq^1B+\Sq^2\Sq^1B + K_1 w_1(TM)^2 \Sq^1 B }\Bigg)
\nn\\
&&\; =\label{eq.linkuse}
\exp\Bigg({\ii\pi\int_{M^5} B\Sq^1B+
(w_2(TM) + w_1(TM)^2)\Sq^1B
 + K_1 w_1(TM)^2 \Sq^1 B }\Bigg)
\\
&&\; =
\exp\Bigg({\ii\pi\int_{M^5} B\Sq^1B+w_3(TM) B + (1+ K_1) w_1(TM)^3  B }\Bigg)
\\
&&\; = \label{eq:5dSPT-all-w1PB}
\exp\Bigg({\ii\pi\int_{M^5} \frac{1}{2} \tilde w_1(TM)\cup \mathcal P( B) 
 + K_1 w_1(TM)^3  B }\Bigg)
\\
&&\; = \label{eq:5dSPT-all-dPB}
\exp\Bigg({\ii\pi\int_{M^5} \frac{1}{4} \delta(\cP_2(B_2))
 + K_1 w_1(TM)^3  B }\Bigg).
\eea
In the second line, we knew already from the derivation of \Eq{eq:K2Sq3B=0} that 
$\exp({\ii\pi\int_{M^5} K_2  \Sq^1(w_2(TM) B)})=1$ on a closed manifold.\\
In the fourth line, we used
$w_1(TM)^2 \Sq^1 B = \Sq^1(w1(TM)^2 B)= w_1(TM) (w_1(TM)^2 B)= w_1(TM)^3 B$ 
where the second equality uses Wu formula on a closed manifold.
We also used
$$u_3 B=w_1(TM) w_2(TM)  B =\Sq^1 (w_2(TM)  B)= (\Sq^1w_2(TM) ) B + w_2(TM) \Sq^1 B$$
$$=(w_1(TM) w_2(TM)+ w_3(TM)) B + w_2(TM) \Sq^1 B,$$
$$
\Rightarrow   w_2(TM) \Sq^1 B = w_3(TM) B \mod 2,
$$
by the Wu formula on a closed 5-manifold.\footnote{If we consider instead a different 5d SPTs/bordism invariant as $K_3  w_2(TM) \Sq^1 B$, we have the following equalities on a closed 5-manifold:
\bea
\bZ_{\text{SPT}_{(K_3)}}^{5\text{d}}[M^5_{\text{closed}}]
=\exp\Bigg({\ii\pi\int_{M^5} K_3  w_2(TM) \Sq^1 B}\Bigg) =\exp\Bigg({\ii\pi\int_{M^5} K_3  w_3(TM) B}\Bigg).
\eea}
\\
In the fifth line, \Eq{eq:5dSPT-all-w1PB} is based on \Eq{Eq.TBBanomaly} and \Refe{{Wan2018zql1812.11968},{Arun}}.\\
In the sixth line, \Eq{eq:5dSPT-all-dPB} is based on Eq.~(124) in \cite{Wan2018zql1812.11968}.

\subsubsection{On a manifold with a boundary}

We also give various equivalent formulas of the
5d SPTs/bordism invariant in \Eq{Eq.5danomalypolynomial} on a 5-manifold $M^5$ with a non-empty 4d boundary $M^4$:
\bea
&&\bZ_{\text{SPT}_{(K_1,K_2)}}^{5\text{d}}[M^5_{}]
=\exp\Bigg({\ii\pi\int_{M^5} B\Sq^1B+\Sq^2\Sq^1B + K_1 w_1(TM)^2 \Sq^1 B +K_2  \Sq^1(w_2(TM) B)  }\Bigg) \nn\\
&&= \label{eq:5dSPT-all-w1PB-open} \label{eq:5dSPT-all}
\exp\Bigg({\ii\pi\int_{M^5} \frac{1}{2} \tilde w_1(TM)\cup \mathcal P( B) 
 + K_1 w_1(TM)^2 \Sq^1 B +K_2  \Sq^1(w_2(TM) B) }\Bigg)
\\
&&= 
\exp\Bigg({\ii\pi\int_{M^5}  \beta_{(2,4)}  \mathcal P( B)  
 + K_1 w_1(TM)^2 \Sq^1 B +K_2  \Sq^1(w_2(TM) B) }\Bigg)\\
&&= \label{eq:5dSPT-all-dPB-open}
\exp\Bigg({\ii\pi\int_{M^5} \frac{1}{4} \delta(\cP(B))
 + K_1 w_1(TM)^2 \Sq^1 B +K_2  \Sq^1(w_2(TM) B) }\Bigg).
\eea
In the third line,
we followed \Refe{Wan2018zql1812.11968} to
define $\beta_{(n,m)}\equiv \H^*(-,\Z_{m})\to\H^{*+1}(-,\Z_{n})$ as 
the Bockstein homomorphism associated to the extension $\Z_n\stackrel{\cdot m}{\to}\Z_{nm}\to\Z_m$, 
where $\cdot m$ is the group homomorphism given by multiplication by $m$. 
We can show that $\beta_{(2,2^n)}=\frac{1}{2^n}\delta\mod2$ \cite{Wan2018zql1812.11968}.
Using the bordism group data 
and the identities given in  \Refe{Wan2018zql1812.11968} and \cite{Wan2018bns1812.11967}, 
we rewrite the 4d higher-anomalies and 5d higher-SPTs/bordism invariants/anomaly polynomials.

\section{Classification of 4d SU(2)$_{\theta=\pi}$ Yang-Mills Theories 
{and Classification of 4d Time-Reversal Symmetric Bosonic/Fermionic SU(2)-SPTs}}
\label{sec:SU2}

In this section, we explore the physical meaning of the gauge bundle constraint in \Eq{Eq.refinedGBC},  i.e., 
$$
{w_2(V_{\SO(3)})= B+ K_1 w_1(TM)^2 + K_2 w_2(TM) \mod 2, ~~~ K_{1,2}\in \Z_2.}
$$
and discuss their physical consequences. 

\subsection{Kramers Singlet/Doublet under Time-Reversal and Bosonic/Fermionic Wilson line}

\label{sec:W-line}
Below we provide some physical interpretations of the Four Siblings of  4d SU(2) YM theories in terms of the Wilson line properties.

We introduce the standard 4d SU(2) Yang-Mills path integral $\bZ^{\text{$4$d}}_{{\SU(2)} {\text{YM}}}[B]$ coupled to the background 2-form  gauge field $B$.
 $\bZ^{\text{$4$d}}_{{\SU(2)} {\text{YM}}}[B]$ is obtained by replacing $F$ with $\hat F - B$ in   $\bZ^{\text{$4$d}}_{{\text{YM}}}$ in \eqn{eq:YM-pi}. We also need to impose the gauge bundle constraint $w_2(E)\equiv w_2(V_{\SO(3)})= B$, which can be imposed by introducing a Lagrangian multiplier\footnote{We can 
	also introduce an additional Pontryagin square $B$ term $\exp\big(\ii \frac{\pi}{2} p \mathcal{P}(B))$ with $p \in \Z_4$ into the path integral,
	as the pioneer works \Refe{AharonyASY2013hda1305.0318} and \Refe{Gaiotto2014kfa1412.5148} do.
	However, this weight factor term only will result in shifting (thus relabeling) of the classification of 4d SU(2)$_{\theta=\pi}$ theories that we are going to reveal. We use the notations in  \cite{Wan2018zql1812.11968}}, 
%
%
%
$$
\int [D \Lambda] \; \bZ^{\text{$4$d}}_{{\SU(2)} {\text{YM}}}[B]  \exp\big( \ii \pi {\int} \Lambda \cup (w_2(E) - B) \big).
$$
$\bullet$ Electric 2-surface $U_e$:
Mathematically, integrating out the Lagrange multiplier $\Lambda$ sets $(w_2(E) - B)=0 \mod 2$.
Physically, $\exp(\ii \pi \int \Lambda)$ plays the role of
an electric 2-surface $U_e=\exp(\ii \pi \int \Lambda)$,
which measures 1-form $e$-symmetry $\Z_{2,[1]}^e$.
The magnetic 't Hooft line lives on the boundary of an electric 2-surface $U_e=\exp(\ii  \pi {\int} \Lambda)$.
Since $U_e$ is dynamical,
't Hooft line is not genuine thus not in the line spectrum for the SU(2) gauge theory  \cite{Gaiotto2014kfa1412.5148}.

\noindent
$\bullet$ Magnetic 2-surface $U_m$ is given by $\exp( \ii \pi {\int} w_2(E) )$.
%
The boundary of $U_m$ supports the Wilson loop $W_e=\Tr ( \text{P} \exp(\ii \oint a))$.  
Unlinking a 2-surface $U_e$ and a 
Wilson loop $W_e$ yields a nontrivial statistical $\pi$-phase $e^{\ii \pi}=-1$.

Following \Sec{sec:SPT}, we enrich the gauge bundle constraint as \Eq{Eq.refinedGBC} by introducing two couplings labeled by $(K_1, K_2)$, and the partition function is 
\bea \label{Eq.refinedGBC-YM-Z}
\bZ^{\text{$4$d}}_{{\SU(2)} {\text{YM}}_{(K_1,K_2)}}[B]
\equiv \int [D \Lambda] \; \bZ^{\text{$4$d}}_{{\SU(2)} {\text{YM}}}[B]  \exp\big( \ii \pi \int \Lambda \cup (w_2(E) - \Big(B + K_1 w_1(TM)^2 + K_2 w_2(TM)\Big)) \big).\;\;
\eea
As we just deduced, 
the magnetic 2-surface $U_m \sim \exp( \ii \pi {\int} w_2(E) )$
has its boundary as Wilson loop $W_e=\Tr ( \text{P} \exp(\ii \oint a))$. We will apply this relation to the Four Siblings 
 with
the YM partition function \eqn{Eq.refinedGBC-YM-Z} and its constraint \Eq{Eq.refinedGBC} and discuss the properties of the Wilson lines.  
\begin{enumerate}
	
	\item $(K_1,K_2)=(0,0)$:
	The gauge bundle constraint is
	${w_2(E)= B \mod 2}$. The magnetic 2-surface $U_m  \sim \exp( \ii \pi {\int} w_2(E) )$ has no decoration other than the 2-form background $B$ field.
	Thus the 1-Wilson line $W_e$ (which can live on the magnetic 2-surface $U_m$'s boundary) is 
	Kramer singlet ($T^2=+1$) and bosonic.
	
	\item $(K_1,K_2)=(1,0)$:
	The gauge bundle constraint becomes
	${w_2(E)= B+  w_1(TM)^2  \mod 2}$.
	The magnetic 2-surface $U_m  \sim \exp( \ii \pi {\int} w_2(E) )$ has a decoration $\int w_1(TM)^2$  
	other than the 2-form $B$ field.
	But $\int w_1(TM)^2$ is a topological term in a cohomology group $\H^2(\Z_2^T,\U(1))$ also in bordism group $\Omega^\tO_2(pt)$, 
	which is effectively a $1+1$D Haldane's anti-ferromagnetic quantum spin-1 chain (Haldane-chain) protected by time-reversal symmetry.
	It is well-known that there exists two-fold degeneracy due to Kramer doublet ($T^2=-1$) on the boundary of Haldane-chain.
	Thus due to $\int w_1(TM)^2$ decoration,
	the Wilson line $W_e$ is 
	Kramer doublet ($T^2=-1$) and bosonic.
	
	\item $(K_1,K_2)=(0,1)$:
	The gauge bundle constraint becomes
	${w_2(E)= B+  w_2(TM) \mod 2}$.
	The magnetic 2-surface $U_m  \sim \exp( \ii \pi {\int} w_2(E) )$ has a decoration $\int w_2(TM)$  
	other than the 2-form $B$ field.
	But $\int w_2(TM)$ is associated with a spin structure.
	The 1d boundary of the 2d $\int w_2(TM)$ theory supports a worldline of  particle  with fermionic statistics.
	Thus due to $\int w_2(TM)$ decoration, the Wilson line $W_e$ living on the boundary of 
	the magnetic 2-surface $U_m$ is fermionic.
	%
	%
	{Since $w_2(TM)$ specifies the extension of $O(d)$ by the fermionic-parity $\Z_2^F $ via the short exact sequence
		$1 \to \Z_2^F \to \Pin^+(d) \to 
		\tO(d) \to 1$ or equivalently 
		the induced fiber sequence 
		$ \B\Z_2^F \to \B\Pin^+(d) \to \B\tO(d)   \xrightarrow{w_2(TM)} \B^2 \Z_2^F$,
		$w_2(TM)$ specifies a projective representation $\Pin^+(d)$ of the spacetime symmetry $\tO(d)$ \cite{2017arXiv171111587GPW}.
		The $\Pin^+(d)$ demands the Euclidean reflection $R^2=+1$, thus
		the Wick rotated time-reversal transformation $\cT^2=-1$ in Lorentz signature \cite{Kapustin1406.7329}.
		Another way to see $\cT^2=-1$ is to use the methods 
		of \emph{symmetry extension} and the \emph{pullback trivialization} \cite{Wang2017locWWW1705.06728, Wang:2018qoyWWW}. 
		Defining the Wilson line operator on the boundary of the magnetic 2-surface $U_m$  requires a \emph{trivialization}
		of ${w_2(TM)}=0$, which amounts to requiring a $\Pin^+(d)$ structure.  The $\Pin^+(d)$ structure imposes $\cT^2=-1$ and fermionic statistics on the line. 
		In summary, due to the $\int w_2(TM)$ decoration,
		the Wilson line $W_e$ is both Kramer singlet ($T^2=+1$) and fermionic.
	}
	
	\item $(K_1,K_2)=(1,1)$:
	The gauge bundle constraint is
	${w_2(E)= B+  w_1(TM)^2 +  w_2(TM) \mod 2}$.
	{Since $w_2(TM)$ specifies the extension of $O(d)$ by the fermionic-parity $\Z_2^F $ via the short exact sequence
		$1 \to \Z_2^F \to \Pin^-(d) \to 
		\tO(d) \to 1$
		or equivalently  the induced fiber sequence $ \B\Z_2^F \to \B\Pin^-(d) \to \B\tO(d)   \xrightarrow{w_1(TM)^2 +w_2(TM)} \B^2 \Z_2^F$,
		so ${w_1(TM)^2 +w_2(TM)}$ specifies a projective representation $\Pin^-(d)$ of the spacetime symmetry $\tO(d)$ \cite{2017arXiv171111587GPW}.
		The $\Pin^-(d)$ demands the Euclidean reflection $R^2=-1$, thus
	the Wick rotated time-reversal transformation $\cT^2=+1$ in Lorentz signature \cite{Kapustin1406.7329}.
	Another way to see $\cT^2=-1$ is to use the methods 
		of \emph{symmetry extension} and the \emph{pullback trivialization} \cite{Wang2017locWWW1705.06728, Wang:2018qoyWWW}. 
		Defining Wilson line operator on the boundary of the magnetic 2-surface $U_m$ requires the \emph{trivialization}
		of ${w_1(TM)^2 +w_2(TM)}=0$, which amounts to requiring the $\Pin^-(d)$ structure.  The $\Pin^-(d)$ structure imposes $\cT^2=+1$ and fermionic statistics on the line. 
	}
	The combined effect of $\int  w_1(TM)^2 + w_2(TM)$ decoration means that
	the 1-Wilson line $W_e$ is 
	Kramer singlet ($T^2=+1$) and fermionic.
	
\end{enumerate}
In fact, our above discussions are universally applicable to more general SU(N) YM theories!\footnote{\label{unpublish}
	Related studies along this line of analysis have also appeared in \cite{2017arXiv171111587GPW}, \cite{SWW2018}
	and \cite{GPW2018}.} This way of enumerating gauge theories (based on new gauge bundle constraints)
guides us to obtain new classes of gauge theories beyond the frame work of \Refe{AharonyASY2013hda1305.0318}. 
The implications are not restricted to 
merely 4d SU(2)$_{\theta=\pi}$ YM.
{This phenomenon (also in \cite{2017arXiv171111587GPW}) can be poetically phrased as \emph{Lorentz symmetry fractionalization} \cite{Hsin2019gvb}.}

\subsection{Enumeration of Gauge Theories from Dynamically Gauging 4d SPTs:\\
View from 4d Cobordism Group Data}

We have discussed the 
Four Siblings of SU(2)$_{\theta=\pi}$ YM theories given by $\bZ^{\text{$4$d}}_{{\SU(2)} {\text{YM}}_{(K_1,K_2)}}[B]$ in \eqn{Eq.refinedGBC-YM-Z},
with four distinct sets of new anomalies derived in \Sec{sec:SPT},
and with Kramer singlet/doublet ($T^2=+1/-1$) or bosonic/fermionic Wilson lines in \Sec{sec:W-line}.
With these properties shown, we are confident that they are really four distinct classes of SU(2)$_{\theta=\pi}$ YM theories (at least at the UV high energy). The two distinct 't Hooft anomalies of $(K_1,K_2)$ also shows that SU(2)$_{\theta=\pi}$ YM theories with distinct $K_1$ are distinct. 

In this subsection, we would like to construct and enumerate these  Four Siblings of SU(2)$_{\theta=\pi}$ YM theories
by dynamically gauging the SU(2) symmetry from 4d time-reversal symmetric SU(2)-SPTs.
To this end, we follow Freed-Hopkins  \cite{Freed2016.1604.06527} to consider a suitable group extension from the time-reversal symmetry (where the spacetime $d$-manifold requires the orthogonal group O($d$)-structure) via a SU(2) extension:
\bea \label{eq:SU2-G'}
{1 \to \SU(2) \to  G' \to \tO(d) \to 1.}
\eea
These 4d SPTs can be regarded as 4d co/bordism invariants of 
\bea
\Omega^{G'}_{4,\text{tor}},
\eea
which is the torsion subgroup $\Omega^{G'}_{4,\text{tor}}$ of  $\Omega^{G'}_{4}$ for all the possible $G'$ under the above group extension.
The extension is classified by $\H^2(\B\tO(d),\Z_2) = \Z_2 \times \Z_2$ for $d>1$, generated by $w_1^2(TM)$ and $w_2(TM)$.

The solution $G'$ of this extension problem
$
1\to\SU(2)\to G'\to\tO\to1,
$
is given in \cite{Freed2016.1604.06527}
with indeed four choices of $G'=\tO\times \SU(2)$, or $\tE\times_{\Z_2} \SU(2)$, or $\Pin^+\times_{\Z_2}\SU(2)$, 
or $\Pin^-\times_{\Z_2}\SU(2)$.\footnote{The notation $G_1 \times_{N} G_2 := \frac{G_1 \times G_2}{{N}}$
	is defined as the product group ${G_1 \times G_2}$ mod out their ($G_1$'s and $G_2$'s) common normal subgroup $N$ \cite{Freed2016.1604.06527}.}


Following the similar study in Ref.~\cite{2017arXiv171111587GPW},
there is a correspondence between the element 
${\rm b}= K_1 w_1(TM)^2 + K_2 w_2(TM)$ and $\H^2(\B\tO(d),\Z_2)=(\Z_2)^2$.
It will soon become clear that ${\rm b}$ is related to $w_2(V_{\SO(3)} )- B$
(i.e., the difference of the gauge bundle $E=V_{\SO(3)}$ connection and the background gauge connection $B$).
Then the 4 central extension choices labeled by b are:

\begin{enumerate}
	
	\item
	${\rm b} = 0$ $\Rightarrow$                            $G' = {\mathrm{O}(d) \times \SU(2)}$ $\Rightarrow$   After
	gauging SU(2), we gain the gauge bundle constraint with $K_1=K_2=0$, 
	$$
	w_2(V_{\SO(3)}) -B = 0.
	$$
	We compute the co/bordism group in Table \ref{Table:O} (details given in Appendix \ref{appendix-bordism}).
	For $d=4$, we obtain
	\bea
	&&{\boxed{\Omega_{4,\text{tor}}^{\mathrm{O}(d) \times \SU(2)}=\Z_2^3}}, \\
	&&\text{whose bordism invariants are generated by three generators of mod 2 classes:}\nn\\
	&&\left\{\begin{array}{l} 
		w_1(TM)^4,\\ 
		w_2(TM)^2,\\ 
		{c_2\mod2}.
	\end{array}\right.
	\label{eq:bordism4O}
	\eea
	The $c_2$ is the second Chern class of the SU(2) gauge bundle.
	
	\item
	${\rm b} = w_1(TM)^2$               $\Rightarrow$   $G'={\rm{E}(d) \times_{\Z_2} SU(2)}$ %
	\footnote{Here \label{ft:Ed}
		E$(d)$ satisfies the following two short exact sequences:
		$$\left\{\begin{array}{l}
		1\to\Z_2\to  \rE(d)   \to\rO(d) \to1\\
		{1\to\SO(d) \to\tE(d)\to \Z_4\to1},
		\end{array} 
		\right.$$
		given that we also accept the well-known fact $1\to\SO(d)\to\rO(d)\to{\Z_2^T}\to1$.
		Here the above finite groups have physical interpretations: 
		$\Z_2=\Z_2^b$ is a bosonic group, $\Z_4= \Z_4^{Tb}$ is the extended group under $1\to\Z_2^b\to \Z_4^{Tb}\to\Z_2^T \to1$.
		Thus $\rE(d)=\Z_2^b \rtimes \rO(d)=\SO(d) \rtimes  \Z_4^{Tb} =(\SO(d) \times \Z_2^b)\rtimes  \Z_2^T $.
		Another way to define $\rE(d)$ is a specific subgroup of $\tO(d) \times \Z_4$ given in \cite{Freed2016.1604.06527}.
	}    
	$\Rightarrow$  After gauging
	SU(2), we gain the gauge bundle constraint with  $K_1=1$ and $K_2=0$, 
	$$
	w_2(V_{\SO(3)}) -B = w_1(TM)^2.
	$$
	We compute the co/bordism group in Table \ref{Table:E} (details given in Appendix \ref{appendix-bordism}). For $d=4$, we obtain
	\bea
	&&{\boxed{
			\Omega_{4,\text{tor}}^{\rm{E}(d) \times_{\Z_2} SU(2)}=\Z_2^3
	}}, \\
	&&\text{whose bordism invariants are generated by three generators of mod 2 class:}\nn\\
	&&\left\{\begin{array}{l} 
		w_1(TM)^4,\\ 
		w_2(TM)^2,\\
		c_2\mod2.
	\end{array}\right.
	\label{eq:bordism4E}
	\eea
	E$(d)$ is defined in  \cite{Freed2016.1604.06527}
	which
	is a subgroup of $\tO(d) \times \Z_4$, described by two data $(M, j ) \in (\tO(d), \Z_4)$ 
	such that the $\det M = j^2$.
	
	By a different but more physical understanding (see footnote \ref{ft:Ed}), we can further obtain 
	that 
	\bea
	\rE(d)=\Z_2^b \rtimes \rO(d)=\SO(d) \rtimes  \Z_4^{Tb} =(\SO(d) \times \Z_2^b)\rtimes  \Z_2^T
	\eea
	where the bosonic internal symmetry $\Z_2^b$ and the time reversal $\Z_2^T$ form the 
	extended group $\Z_4^{Tb}$ under $1\to\Z_2^b\to \Z_4^{Tb}\to\Z_2^T \to1$.
	
	Here the $c_2$ is the second Chern class of the U(2) gauge bundle.\footnote{
		\label{ft:c2U(2)}
		{Since the constraint $w_1(TM)^2=w_2(V_{\SO(3)})$ is satisfied, let $\beta_2$ denote the Bockstein homomorphism associated to the extension $\Z\to\Z\to\Z_2$, then $W_3(V_{\SO(3)})=\beta_2w_2(V_{\SO(3)})=\beta_2w_1(TM)^2=\beta_2\Sq^1w_1(TM)=0$ where $W_3(V_{\SO(3)})$ is the third integral Stiefel-Whitney class of $V_{\SO(3)}$ and we have used the fact that $\beta_2\Sq^1=0$, hence $V_{\SO(3)}$ lifts to a $\Spin^c(3)=\U(2)$ bundle $V_{\U(2)}$, here $c_2=c_2(V_{\U(2)})$ is the second Chern class of $V_{\U(2)}$.
		}
	}

	\item
	${\rm b} = w_2(TM)$                 $\Rightarrow$   $G' = {\mathrm{Pin}^{+}\times_{\Z_2} \SU(2)}$   $\Rightarrow$  
	After gauging
	SU(2),  we gain the gauge bundle constraint with $K_1=0$ and $K_2=1$,  
	$$
	w_2(V_{\SO(3)}) -B = w_2(TM) .
	$$
	The co/bordism group is computed in \cite{Freed2016.1604.06527, 2017arXiv171111587GPW} and in Table \ref{Table:Pin+} (see also Appendix \ref{appendix-bordism}). For $d=4$, we obtain
	\bea
	&&{\boxed{\Omega^{\mathrm{Pin}^{+}\times_{\Z_2} \SU(2)}_{4,\text{tor}}= \Z_4\times \Z_2}
	}, \\
	&&\text{whose bordism invariants are generated by generators of mod 4 and mod 2 classes:}\nn\\
	&&\left\{\begin{array}{l} 
		\nu  \eta_{\SU(2)},
		\text{with a $\nu \in \Z_4$ class} \\
		{w_2(TM)^2}.
	\end{array}\right.
	\label{eq:bordism4Pin+}
	\eea
	This is related to the interacting version of CI class topological superconductor in condensed matter physics (\cite{CWang1401.1142}, \cite{Freed2016.1604.06527}, and \cite{2017arXiv171111587GPW}).
	Details of these topological terms are discussed in \cite{2017arXiv171111587GPW}.
	%
	\item
	${\rm b} = w_2(TM) +w_1(TM)^2$  $\Rightarrow$  $G' = {\mathrm{Pin}^{-}\times_{\Z_2} \SU(2)}$    $\Rightarrow$  
	After gauging
	SU(2),  we gain the gauge bundle constraint with $K_1=K_2=1$,
	$$
	w_2(V_{\SO(3)}) -B =w_2(TM) + w_1(TM)^2.
	$$
	The co/bordism group is computed in \cite{Freed2016.1604.06527, 2017arXiv171111587GPW} and in Table \ref{Table:Pin-} (see also Appendix \ref{appendix-bordism}). For $d=4$, we obtain
	\bea
	&&{\boxed{
			\Omega^{\mathrm{Pin}^{-}\times_{\Z_2} \SU(2)}_{4,\text{tor}}= (\Z_2)^3}}, \\
	&&\text{whose bordism invariants are generated by three generators of mod 2 classes:}\nn\\
	&&\left\{\begin{array}{l} 
		{ N_0'\mod2},
		\\
		{w_1(TM)^4},
		\\
		{w_2(TM)^2}.
	\end{array}\right.
	\label{eq:bordism4Pin-}
	\eea
	This is related to the interacting version of CII class topological insulator in condensed matter physics (\cite{CWang1401.1142}, \cite{Freed2016.1604.06527}, and \cite{2017arXiv171111587GPW}).
	Details of these topological terms are discussed in \cite{2017arXiv171111587GPW}.
	
	%
\end{enumerate}

More information about these (co)bordism group calculations can be read from \cite{Freed2016.1604.06527, 2017arXiv171111587GPW}.
See Appendix of \cite{2017arXiv171111587GPW} for a quick background review.
In particular, since the computation involves no odd torsion, we can use Adams spectral sequence to compute $\Omega_n^{G'}=\pi_n(MTG')$:
\bea
\Ext_{\A_2}^{s,t}(\H^*(MTG',\Z_2),\Z_2)\Rightarrow\pi_{t-s}(MTG')_2^{\wedge}.
\eea
Here $\pi_{t-s}(MTG')_2^{\wedge}$ is the 2-completion of the group $\pi_{t-s}(MTG')$.
For example, 
\bea
\left\{\begin{array}{l} 
	MT(\tO\times\SU(2))=M\tO\wedge\B\SU(2)_+,\\ 
	MT(\tE \times_{\Z_2}\SU(2))=M\SO\wedge\Sigma^{-3}M\Pin^+(3)=M\SO\wedge\Sigma^{-3}MT\Spin(3)\wedge MT\Z_2,\\
	MT(\Pin^+\times_{\Z_2}\SU(2))=M\Spin\wedge\Sigma^{-3}M\tO(3),\\ 
	MT(\Pin^-\times_{\Z_2}\SU(2))=M\Spin\wedge \Sigma^3MT\tO(3).
\end{array}\right. 
\eea
The $\B\SU(2)_+$ is the disjoint union of $\B\SU(2)$ and a point, while $\Sigma$ is the suspension.

Let $M$ be an $n$-manifold, and $V_{\SO(3)}$ be the associated vector bundle of the SO(3) gauge bundle.
Below we compute the Stiefel-Whitney classes of $(TM-n)\otimes V_{\SO(3)}$. They are used to express the cobordism invariants of $\Omega_d^{\Pin^{\pm}\times_{\Z_2}\SU(2)}$.
Below $w_i$ means the $i$-th Stiefel-Whitney class, $w$ means the total Stiefel-Whitney class, namely, we have $w=1+w_1+w_2+w_3+\cdots$.
We denote $w_i'=w_i(V_{\SO(3)})$ and $\tilde w_i=w_i((TM-n)\otimes V_{\SO(3)})$.
In addition, the $w_i(TM)$ means specifically the $i$-th Stiefel-Whitney class of spacetime tangent bundle $TM$.
\bea\label{Stiefel-Whitney}
&&w((TM-n)\otimes V_{\SO(3)})\nn\\
&=&\frac{w(TM\otimes V_{\SO(3)})}{w(V_{\SO(3)})^n}\nn\\
&=&\frac{1+w_1(TM)+w_1(TM)^2+w_2(TM)+nw_2'+w_1(TM)^3+nw_1(TM)w_2'+w_3(TM)+nw_3'+\cdots}{(1+w_2'+w_3'+\cdots)^n}\nn\\
&=&1+w_1(TM)+w_1(TM)^2+w_2(TM)+w_1(TM)^3+w_3(TM)+\cdots
\eea
So $\tilde w_1=w_1(TM)$, $\tilde w_2=w_1(TM)^2+w_2(TM)$, $\tilde w_3=w_1(TM)^3+w_3(TM)$, etc.

We also use the notation {TP} for the classification of topological phases defined in \cite{Freed2016.1604.06527}, such that
\bea
\TP_{d,\text{tor}}(G')=\Omega_{d,\text{tor}}^{G'}.
\eea
Here are the list of tables summarizing the results in 4d and in 5d: Table \ref{Table:O},  \ref{Table:E},  \ref{Table:Pin+} and 
\ref{Table:Pin-}.

\begin{table}[!h]
	\centering
	\begin{tabular}{c c c c}
		\hline
		$d$ & $\TP_{d,\text{tor}}({\tO(d)\times \SU(2)})$ &co/bordism invariants&manifold generators {$(M,V_{\SO(3)})$} \\
		\hline
		4 & $\Z_2^3$ & $w_1(TM)^4,w_2(TM)^2, {c_2\mod2}$&$\RP^4,\CP^2,(S^4,H)$\\ 
		5 & $\Z_2$ & $w_2(TM)w_3(TM)$&$\SU(3)/\SO(3)$\\ 
		\hline
	\end{tabular}
	\caption{Cobordism groups ${\rm TP}_d({\rm O}(d)\times {\rm SU}(2))$ and co/bordism invariants. Here $w_i(TM)$ is the $i$-th Stiefel-Whitney class of the spacetime tangent bundle, $c_2$ is the second Chern class of the $\rm SU(2)$ gauge bundle. 
		{Here we set $H$ as the Hopf fibration, the $\rm SU(2)$ bundles on $\mathbb{RP}^4$, $\mathbb{CP}^2$ and $\rm SU(3)/\rm SO(3)$ are trivial.}
		{See also Appendix \ref{appendix-bordism}.} 
	}
	\label{Table:O}
\end{table}

\begin{table}[!h]
	\centering
	\begin{tabular}{c c c c}
		\hline
		$d$ & $\TP_{d,\text{tor}}({\tE(d)\times_{\Z_2}\SU(2)})$ &cobordism invariants&manifold generators {$(M,V_{\SO(3)})$} \\
		\hline
		4 &$\Z_2^3$ & $w_1(TM)^4,w_2(TM)^2,c_2\mod2 {}^{\ref{ft:c2U(2)}}$ &$(\RP^4,2L_{\R}+1),(\CP^2,3), (S^4,H)$\\
		5 &$\Z_2$ & $w_2(TM)w_3(TM)$ &$\SU(3)/\SO(3)$ \\
		\hline
	\end{tabular}
	\caption{Cobordism groups ${\rm TP}_d({\rm E}(d)\times_{\mathbb{Z}_2}{\rm SU}(2))$ and cobordism invariants. 
		Here $w_i(TM)$ is the $i$-th Stiefel-Whitney class of the spacetime tangent bundle, $c_2$ is the second Chern class of the U(2) gauge bundle (See the footnote \ref{ft:c2U(2)}).
		The second component in manifold generators  {$(M,V_{\mathrm{SO}(3)})$} is the $\mathrm{SO}(3)$ gauge bundle over the first component. $L_{\mathbb{R}}$ is the real tautological line bundle. $H$ is induced from the Hopf fibration by $\rm SU(2)\to\rm SO(3)$. 
		{The $\rm SO(3)$ bundle on $\rm SU(3)/SO(3)$ is trivial.}
		{See also Appendix \ref{appendix-bordism}.}
	}
	\label{Table:E}
\end{table}


\begin{table}[!h]
	\centering
	\begin{tabular}{c c c c}
		\hline
		$d$ & $\TP_{d,\text{tor}}({\Pin^+(d)\times_{\Z_2}\SU(2)})$ &cobordism invariants&manifold generators {$(M,V_{\SO(3)})$} \\
		\hline
		4 & $\Z_2\times\Z_4$  & $ w_2(TM)^2, \eta_{\SU(2)}$&$(\CP^2,L_{\C}+1),(\RP^4,3)$\\ 
		5 & $\Z_2$ & $ w_2(TM) w_3(TM)$ &$\SU(3)/\SO(3)$\\
		\hline
	\end{tabular}
	\caption{Cobordism groups ${\rm TP}_d({\rm Pin}^+(d)\times_{\mathbb{Z}_2}{\rm SU}(2))$ and cobordism invariants. 
	The $\eta_{{\rm SU}(2)}$ is an eta invariant of Dirac operator defined in \cite{2017arXiv171111587GPW}. More details of computation can be read from \cite{Freed2016.1604.06527, 2017arXiv171111587GPW}. 
		{The second component in manifold generators  {$(M,V_{{\rm SO}(3)})$} is the ${\rm SO}(3)$ gauge bundle over the first component.} 
		$L_{\mathbb{C}}$ is the complex tautological line bundle. {The ${\rm SO}(3)$ bundle on ${\rm SU}(3)/{\rm SO}(3)$ is given by the fibration
			${\rm SO}(3)\to{\rm SU}(3)\to{\rm SU}(3)/{\rm SO}(3)$.}
		{See also Appendix \ref{appendix-bordism}.}
		Note that \cite{2017arXiv171111587GPW} actually derives that the 4d cobordism invariants are $\tilde w_2^2=w_2(TM)^2+w_1(TM)^4$ and $\eta_{\rm SU(2)}$ with $2\eta_{{\rm SU}(2)}=\tilde w_1\tilde w_3=w_1(TM)w_3(TM)+w_1(TM)^4$. 
Here $\tilde w_i$ is the $i$-th Stiefel-Whitney class of $(TM-n)\otimes V_{\rm{SO}(3)}$ 
computed in \eqref{Stiefel-Whitney}, where $V_{\rm{SO}(3)}$ is the associated vector bundle of the $\rm{SO}(3)$ gauge bundle. 
But since the third Wu class $u_3=w_1(TM)w_2(TM)=0$ on any 4-manifold, we have $w_1(TM)w_3(TM)={\rm Sq}^1(w_1(TM)w_2(TM))=0$. 
So by a base change $\tilde w_2^2=w_2(TM)^2+2\eta_{\rm SU(2)}$, 
we can choose the 4d cobordism invariants to be $w_2(TM)^2$ and $\eta_{{\rm SU}(2)}$. 
Also note that the 5d cobordism invariant is actually $\tilde w_2\tilde w_3=(w_2(TM)+w_1(TM)^2)(w_3(TM)+w_1(TM)^3)$. 
But since the third Wu class $u_3=w_1(TM)w_2(TM)=0$ on any 5-manifold, we have ${\rm Sq}^2(w_1(TM)w_2(TM))=w_1(TM)w_2(TM)^2+w_1(TM)^3w_2(TM)+w_1(TM)^2w_3(TM)=0$. 
Also by Wu formula, we have $w_1(TM)w_2(TM)^2={\rm Sq}^1(w_2(TM)^2)=0$ and $w_1(TM)^5={\rm Sq}^1(w_1(TM)^4)=0$ on any 5-manifold, 
so $\tilde w_2\tilde w_3=w_2(TM)w_3(TM)$.
	}
	\label{Table:Pin+}
\end{table}

\begin{table}[!h]
\hspace{-3.45em}
	\centering
	\begin{tabular}{c c c c}
		\hline
		$d$ & $\TP_{d,\text{tor}}({\Pin^-(d)\times_{\Z_2}\SU(2)})$ &cobordism invariants&manifold generators {$(M,V_{\SO(3)})$} \\
		\hline
		4 & $\Z_2^3$  & $ w_2(TM)^2, w_1(TM)^4$, $(N_0'^{(4)}\mod 2)$&$(\CP^2,L_{\C}+1),(\RP^4,2L_{\R}+1),(S^4,H)$\\
		5 & $\Z_2^2$ & $ w_2(TM) w_3(TM)$, $(N_0'^{(5)} \mod 2)$ & $\SU(3)/\SO(3)$, $S^1\times S^4$\\
		\hline
	\end{tabular}
	\caption{Cobordism groups $\rm{TP}_d(\rm{Pin}^-(d)\times_{\mathbb{Z}_2}\rm{SU}(2))$ and cobordism invariants. 
The $N_0'^{(4)}$ 
		is the number of the zero modes of the Dirac operator in 4d. Its value mod 2 is a spin-topological
		invariant known as the mod 2 index defined as 
		$N_0'\mod 2$
		in \cite{2017arXiv171111587GPW}. More details of computation can be read from \cite{Freed2016.1604.06527, 2017arXiv171111587GPW}.
		We find that the bordism invariant of 
		{$N_0'^{(4)} \mod2$} read from Adams chart has the similar form related to $\tilde w_3 \tilde{\eta}$, where $\tilde{\eta}$ is the eta invariant for 1d Dirac operator, given by the
		generator of the 1d spin bordism group $\Omega_{1,\text{tor}}^{\text{Spin}}(pt)=\Z_2$.
		The $N_0'^{(5)}$ 
		is the number of the zero modes of the Dirac operator in 5d. Its value mod 2 is a spin-topological
		invariant known as the mod 2 index defined in \cite{Witten:1982fp, Wang:2018qoyWWW}.
		We find that the bordism invariant of 
		{$N_0'^{(5)} \mod 2$} read from Adams chart has the similar form  related to $\tilde w_3 \text{Arf}$, where Arf is an Arf invariant.
		The second component in manifold generators  {$(M,V_{\rm{SO}(3)})$} is the $\rm{SO}(3)$ gauge bundle over the first component. $L_{\C}$ is the complex tautological line bundle. $L_{\mathbb{R}}$ is the real tautological line bundle.
		$H$ is induced from the Hopf fibration by $\rm{SU}(2)\to\rm{SO}(3)$. 
		{The $\rm{SO}(3)$ bundle on $\rm{SU}(3)/\rm{SO}(3)$ is given by the fibration
			$\rm{SO}(3)\to\rm{SU}(3)\to\rm{SU}(3)/\rm{SO}(3)$. The $\rm{SO}(3)$ bundle on $S^1\times S^4$ is induced from the fibration $S^3\to S^1\times S^7\to S^1\times S^4$ by $\rm{SU}(2)\to\rm{SO}(3)$.}
		{See also Appendix \ref{appendix-bordism}.}
		Note that \cite{2017arXiv171111587GPW} actually derives that the 4d cobordism invariants are $\tilde w_2^2=w_2(TM)^2+w_1(TM)^4$, $\tilde w_1^4=w_1(TM)^4$ and $(N_0'^{(4)}\mod 2)$. 
Here $\tilde w_i$ is the $i$-th Stiefel-Whitney class of $(TM-n)\otimes V_{\rm{SO}(3)}$ 
computed in \eqref{Stiefel-Whitney}, where $V_{\rm{SO}(3)}$ is the associated vector bundle of the $\rm{SO}(3)$ gauge bundle. 
By a base change, we can choose the 4d cobordism invariants to be $w_2(TM)^2$, $w_1(TM)^4$ and $(N_0'^{(4)}\mod 2)$. Also note that the 5d cobordism invariants are actually $\tilde w_2\tilde w_3=(w_2(TM)+w_1(TM)^2)(w_3(TM)+w_1(TM)^3)$ and $(N_0'^{(5)} \mod 2)$. 
But since the third Wu class $u_3=0$ on any 5-manifold and also by Wu formula shown in Table \ref{Table:Pin+}, 
so $\tilde w_2\tilde w_3=w_2(TM)w_3(TM)$.
	}
	\label{Table:Pin-}
\end{table}

We conclude this section with a summary.
The  Four Siblings of  4d SU(2)$_{\theta=\pi}$ YM theories are obtained, specifically, from
summing over the SU(2) gauge connections of the following four topological terms (i.e., gauging the SU(2) global symmetry of the following four distinct SPTs):
\begin{enumerate}
	\item $(-1)^{ c_2 }$ in \eqn{eq:bordism4O}. 
	\item $(-1)^{c_2}$ in \eqn{eq:bordism4E}. (See the footnote \ref{ft:c2U(2)}.)
	\item  ${\exp(2\pi \im \nu  \eta_{\SU(2)})}$
	with an odd class of $\nu =1, 3\in \Z_4$ in \eqn{eq:bordism4Pin+}.
	\item ${(-1)^{N_0'}}$ in \eqn{eq:bordism4Pin-}.
\end{enumerate}
These four theories exactly map to the enumeration of four gauge theories in
\Sec{sec:W-line}.
Adding other SPTs/bordism invariants such as 
$(-1)^{w_1(TM)^4}$ and $(-1)^{w_2(TM)^2}$ (and then dynamically gauging them), 
do not alter or gain new classes of gauge theories. 
They only tensor product the gauge theory with 4d SPTs,
namely (4d SU(2)$_{\theta=\pi}$ YM) $\otimes$ 
(4d SPTs).\footnote{
	Here for the classification of gauge theories, we identify the following phases
	$$
	\text{(gauge theory) $\otimes$ (SPTs)} \simeq \text{(gauge theory)}. 
	$$
	For the classification of 4d SU(2)$_{\theta=\pi}$ YM, we identify the following phases
	$$
	\text{(4d SU(2)$_{\theta=\pi}$ YM) $\otimes$ (4d SPTs)} \simeq \text{(4d SU(2)$_{\theta=\pi}$ YM)}. 
	$$
	See more physically motivated discussions in \cite{2017arXiv171111587GPW}
	and References therein.}

\section{Time-Reversal Symmetry-Enriched 5d Higher-Gauge TQFTs}

\label{sec:5dTRTQFT}

\subsection{Partition Function of 5d Higher-Gauge TQFTs} 

Following the discussions of four classes of 5d time-reversal and 1-form center symmetry  $\Z_2^T \times \Z_{2,[1]}^e$ higher-SPTs $\bZ_{\text{SPT}_{(K_1,K_2)}}^{5\text{d}}[M^5]$ in \Sec{sec:5dSPT-Z} with their partition functions in \eqn{eq:5dSPT-all}, 
 we proceed to dynamically gauge the 1-form symmetry $\Z_{2,[1]}^e$.
Then we obtain the
5d time-reversal symmetric enriched topologically ordered state (SETs) with 2-form $\Z_2$-valued dynamical $B$ gauge fields.
{We expect a precise mathematical formulation requires a certain version of \emph{higher category theory}. 
Below we instead approach from a \emph{higher-gauge TQFT} perspective.}

We can define the four classes of 5d partition functions $\bZ_{\text{SET}_{(K_1,K_2)}}^{5\text{d}}[M^5]$ as:
\bea
\bZ_{\text{SET}_{(K_1,K_2)}}^{5\text{d}}[M^5]
&\equiv&\frac{|\H^0(M,\Z_2)|}{|\H^1(M,\Z_2)|} \sum_{B\in\H^2(M^5,\Z_2)} {\rm e}^{\ii\pi\int_{M^5} \frac{1}{2} \tilde w_1(TM)\cup \mathcal P( B) + K_1 w_1(TM)^2\Sq^1B  }
\label{eq:5dSET-all}\\
&=&\frac{|\H^0(M,\Z_2)|}{|\H^1(M,\Z_2)|} \sum_{B\in\H^2(M^5,\Z_2)}{\rm e}^{\ii\pi\int_{M^5} B\Sq^1B + (1+K_1) w_1(TM)^2 \Sq^1 B+w_2(TM)\Sq^1B } 
\label{eq:5dSET-2}\\
&=&\frac{|\H^0(M,\Z_2)|}{|\H^1(M,\Z_2)|} \sum_{{{}\atop{B, b, h\in \rm{C}^2( M^5, \Z_2)}}\atop{{c \in \rm{C}^3(  M^5, \Z_2)}}}
{\exp}\Big(\ii 
\pi \int_{M^5}    \delta   w_1(TM)\cup c
+   \delta   w_2(TM)\cup h \nn\\
&&+b\cup \delta B  {+B\Sq^1B + (1+K_1) w_1(TM)^2 \Sq^1 B + w_2(TM) \Sq^1 B}\Big)
\label{eq:5dSET-3}\\
&&\cong
\int [\cD B][\cD b][\cD h][\cD c]
{\exp}\Big(\ii 
\pi \int_{M^5}    (\mathrm{d} w_1(TM)) c
+   (\mathrm{d}   w_2(TM)) h \nn\\
&&+b \mathrm{d} B  {+B\frac{1}{2} \mathrm{d} B + (1+K_1) w_1(TM)^2 \frac{1}{2} \mathrm{d}  B + w_2(TM)\frac{1}{2} \mathrm{d} B}\Big).
\label{eq:5dSET-4}
\eea
In the last step (under the symbol $\cong$), we have {converted} the 5d higher-cochain TQFT to 5d higher-form gauge field continuum TQFT for $\bZ_{\text{SET}_{(K_1,K_2)}}^{5\text{d}}[M^5]$. 
Moreover, we can insert extended operators (say $U, X, Y, \dots$) into the  path integral:
\bea
&&\bZ_{\text{SET}_{(K_1,K_2)}}^{5\text{d}}[M^5; U, X, Y, \dots]
\equiv
\int [\cD B][\cD b][\cD h][\cD c]
\; U\cdot X \cdot Y \dots\nn \\
&&\quad\quad \quad\quad\quad\quad \quad\quad{\exp}\Big({\ii 
	\pi \int_{M^5}    (\mathrm{d} w_1(TM)) c
	+   (\mathrm{d}   w_2(TM)) h +b \mathrm{d} B} \nn\\ 
&&\quad \quad\quad\quad \quad\quad {+B\frac{1}{2} \mathrm{d} B + (1+K_1) w_1(TM)^2 \frac{1}{2} \mathrm{d}  B + w_2(TM)\frac{1}{2} \mathrm{d} B}\Big).\label{eq:5dSET-cont}
\eea
{Note that since $K_2 \Sq^1(w_2(TM) B)$ is trivial for closed 5-manifolds, the partition function $\bZ_{\text{SET}_{(K_1,K_2)}}^{5\text{d}}[M^5] $ and the correlation function computed from the path integral $\bZ_{\text{SET}_{(K_1,K_2)}}^{5\text{d}}[M^5; U, X, Y, \dots]$ do not depend on $K_2$. }

\subsection{Partition Function  and Topological Degeneracy }

Below we compute the {partition function $\bZ(M^5)$ on closed manifolds $M^5$.}
When $M^5=M^4 \times S^1$, we can interpret it as topological ground state degeneracy (GSD) of TQFT.
Our computations follow the strategy in  \cite{Wang2018edf1801.05416, Guo2018vij1812.11959}.  We directly summarize the results in Tables
\ref{table:Z-1}, \ref{table:ZM}, and \ref{table:Z-3}.
 
\subsubsection{5d SPTs as Short-Range Entangled Invertible TQFTs}

We evaluate the partition function of various 5d iTQFTs on various manifolds, and enumerate the results in Table \ref{table:Z-1}.  Below we denote the 5-dimensional Wu manifold as $\W \equiv \SU(3)/\SO(3)$.

\begin{center}
\begin{table}[!h]
\begin{tabular}{c c c c c c c c c}
\hline
$\bZ(M^5)$ with $M^5$: & $(\W,0)$ &  $(S^1 \times  \RP^2 \times  \RP^2, \gamma \alpha_1) $  &  $(S^1 \times \RP^4, \gamma \zeta)$ & $(\RP^2\times\RP^3, \alpha \beta )$ &  \\
\hline
\hline\\[-2mm]
$\bZ_{\text{SPT}}^{\text{trivial}}(M^5)$& 1 & 1 & 1 &1\\[2mm]
\hline\\[-2mm]
$\bZ_{\text{SPT}_{B\Sq^1B}}(M^5)$&1 &1&1&$-1$\\[2mm]
\hline\\[-2mm]
$\bZ_{\text{SPT}_{\Sq^2\Sq^1B}}(M^5)$&1 &1&$-1$&1\\[2mm]
\hline\\[-2mm]
$\bZ_{\text{SPT}_{w_1(TM)^2\Sq^1B}}(M^5)$&1 &$-1$&$-1$&1 \\[2mm]
\hline\\[-2mm]
$\bZ_{\text{SPT}_{w_2(TM)\Sq^1B}}(M^5)$& 1&$-1$& 1 &1\\[2mm]
\hline
\end{tabular}
\caption{{Partition Function $\bZ(M^5)$ and Topological Degeneracy (GSD) of 5d higher-SPTs, for example, $\bZ_{\text{SPT}_{B\Sq^1B}}(M^5):=(-1)^{\int_{M^5}B\Sq^1B}$. The notations $\alpha,\beta,\gamma,\zeta$ are explained in the computation below in \Sec{sec:Computation}.
Wu manifold is denoted as $\W \equiv \SU(3)/\SO(3)$.}}
\label{table:Z-1}
\end{table}
\end{center}

\subsubsection{5d SETs, as Long-Range Entangled TQFTs}

We evaluate the partition function of various 5d TQFTs (as 5d SETs) on various manifolds, and enumerate the results in Table \ref{table:ZM}
and Table \ref{table:Z-3}. 

\begin{center}
\begin{table}[!h]
\resizebox{\columnwidth}{!}{
\begin{tabular}{c c c c c c c c c c}
\hline
$\bZ(M^5)$ with $M^5$: & $T^5$ &  $S^1 \times S^4$ & $S^1\times\RP^4$ &  $T^2 \times S^3$  &  $S^1 \times S^2 \times S^2$ &  $S^1\times\RP^2\times\RP^2$ &  $\RP^2 \times\RP^3$ & $S^5$  & W 
\\
\hline
\hline\\[-2mm]
$\bZ_{\text{2-form }B}^{\text{untwist}}(M^5)$ & $
\text{\tiny{\(\frac{2^{10}\cdot 2}{2^5}=64\)}}$ &  
$
\text{\tiny{\(\frac{2^{0}\cdot 2}{2^1}=1\)}}$
& 
$
\text{\tiny{\(\frac{2^{2}\cdot 2}{2^2}=2\)}}$
&  
$
\text{\tiny{\(\frac{2^{1}\cdot 2}{2^2}=1\)}}$
& 
$
\text{\tiny{\(\frac{2^{2}\cdot 2}{2^1}=4\)}}$
& 
$
\text{\tiny{\(\frac{2^{5}\cdot 2}{2^3}=8\)}}$
& 
$
\text{\tiny{\(\frac{2^{3}\cdot 2}{2^2}=4\)}}$
&  $
\text{\tiny{\(\frac{2^{0}\cdot 2}{2^0}=2\)}}$ & 4\\[2mm]
\hline\\[-2mm]
$\bZ_{\text{SET}_{(0,0)}}(M^5)$& 64 & 1   & 1&1 &4 & 2 &   2 & 2 & 4\\[2mm]
\hline\\[-2mm]
$\bZ_{\text{SET}_{(1,0)}}(M^5)$& 64 &  1   & 1& 1& 4& 2 & 2 & 2 & 4\\[2mm]
\hline\\[-2mm]
$\bZ_{\text{SET}_{(0,1)}}(M^5)$& 64 &  1   &1 &1 & 4& 2  &2  & 2 & 4 \\[2mm]
\hline\\[-2mm]
$\bZ_{\text{SET}_{(1,1)}}(M^5)$& 64 &  1  &1 & 1&  4& 2 &2  & 2  & 4\\[2mm]
\hline
\end{tabular}
}
\caption{{Partition Function $\bZ(M^5)$ and Topological Degeneracy (GSD)
of 5d higher-SETs, $\bZ_{\text{SET}_{(K_1,K_2)}}(M^5):=\frac{|\H^0(M^5,\Z_2)|}{|\H^1(M^5,\Z_2)|}\sum_{B\in\H^2(M^5,\Z_2)}(-1)^{\int_{M^5}B\Sq^1B+(1+K_1)w_1(TM)^2\Sq^1B+w_2(TM)\Sq^1B}.$
\protect\footnotemark
}}
\label{table:ZM}
\end{table}
\end{center}

\footnotetext{Interestingly, we notice that the Wu manifold W can assign a closely-related but different partition function  $\bZ'_{\text{SET}}(\text{W})$ with a distinct value: 
 $\bZ'_{\text{SET}}(\text{W}):=\frac{|\H^0(M^5,\Z_2)|}{|\H^1(M^5,\Z_2)|}
 \sum_{B\in\H^2(M^5,\Z_2)}(-1)^{\int_{M^5}B\Sq^1B+(1+K_1)w_1(TM)^2\Sq^1B}=0.$
 }

\begin{center}
\begin{table}[!h]
\centering
\begin{tabular}{c c c c c c c c c}
\hline
$\bZ(M^5)$ with $M^5$: & $  \W$ &  $S^1 \times  \RP^2 \times  \RP^2 $  &  $S^1 \times \RP^4$ & $\RP^2\times\RP^3$ &  \\
\hline
\hline\\[-2mm]
$\bZ_{\text{2-form }B}^{\text{untwist}}(M^5)$&4 &8 & 2 &4\\[2mm]
\hline\\[-2mm]
$\bZ_{\text{SET}_{B\Sq^1B}}(M^5)$&0 &2&1&2\\[2mm]
\hline\\[-2mm]
$\bZ_{\text{SET}_{\Sq^2\Sq^1B}}(M^5)$&0 &8&0&4\\[2mm]
\hline\\[-2mm]
$\bZ_{\text{SET}_{w_1(TM)^2\Sq^1B}}(M^5)$&4 &0&0&4 \\[2mm]
\hline\\[-2mm]
$\bZ_{\text{SET}_{w_2(TM)\Sq^1B}}(M^5)$&0 &0& 2 &4\\[2mm]
\hline
\end{tabular}
\caption{{Partition Function $\bZ(M^5)$ and Topological Degeneracy (GSD)
of 5d higher-SETs, for example, $\bZ_{\text{SET}_{B\Sq^1B}}(M^5):=\frac{|\H^0(M^5,\Z_2)|}{|\H^1(M^5,\Z_2)|}\sum_{B\in\H^2(M^5,\Z_2)}(-1)^{\int_{M^5}B\Sq^1B}.$}}
\label{table:Z-3}
\end{table}
\end{center}

\subsubsection{Computation}
\label{sec:Computation}

Now we illustrate our computation:
\begin{enumerate}
\item
For $M=S^1\times\RP^4$, let $\gamma$ be the generator of $\H^1(S^1,\Z_2)=\Z_2$ and  $\zeta$ be the generator of $\H^1(\RP^4,\Z_2)=\Z_2$. Note that $w_1(TM)=\zeta$. 
The cohomology groups have that
$\H^0(S^1\times\RP^4,\Z_2)=\Z_2$, $\H^1(S^1\times\RP^4,\Z_2)=\Z_2^2$,
and $\H^2(S^1\times\RP^4,\Z_2)=\Z_2^2$ whose two generators are $\gamma\zeta$ and $\zeta^2$.  If $B=\lambda_1\gamma\zeta+\lambda_2\zeta^2$, then $\Sq^1B=\lambda_1\gamma\zeta^2$.  Hence
\bea
\int_{S^1\times\RP^4}B\Sq^1B&=&\lambda_1\lambda_2,\\
\int_{S^1\times\RP^4}B\Sq^1B+w_1(TM)^2\Sq^1B&=&\lambda_1\lambda_2+\lambda_1.
\eea
On the other hand,
since $w_2(TM)=0$ for $S^1\times\RP^4$, 
we have
\bea
\bZ_{\text{SET}_{(0,0)}}(S^1\times\RP^4)&=&\bZ_{\text{SET}_{(0,1)}}(S^1\times\RP^4)=\frac{1}{2}\sum_{\lambda_1,\lambda_2\in\Z_2}(-1)^{\lambda_1(\lambda_2+1)},\\
\bZ_{\text{SET}_{(1,0)}}(S^1\times\RP^4)&=&\bZ_{\text{SET}_{(1,1)}}(S^1\times\RP^4)=\frac{1}{2}\sum_{\lambda_1,\lambda_2\in\Z_2}(-1)^{\lambda_1\lambda_2}.
\eea
Since the number of $(\lambda_1,\lambda_2)$ satisfying the constraint $\lambda_1\lambda_2=1$ is only one:
\bea
\#\{(\lambda_1,\lambda_2)\in\Z_2^2 \mid \lambda_1\lambda_2=1\}=1,
\eea
also note that changing $\lambda_2$ to $\lambda_2+1$ doesn't affect the sum,
so 
\bea
&&\bZ_{\text{SET}_{(0,0)}}(S^1\times\RP^4)=\bZ_{\text{SET}_{(1,0)}}(S^1\times\RP^4)\nn\\
&=&\bZ_{\text{SET}_{(0,1)}}(S^1\times\RP^4)=\bZ_{\text{SET}_{(1,1)}}(S^1\times\RP^4)=\frac{1}{2}(3-1)=1.
\eea
\item
For $M=\RP^2 \times\RP^3$, let $\alpha$ be the generator of $\H^1(\RP^2,\Z_2)=\Z_2$, $\beta$ be the generator of $\H^1(\RP^3,\Z_2)=\Z_2$. Note that $w_1(TM)=\alpha$. 
$\H^0(\RP^2 \times\RP^3,\Z_2)=\Z_2$, $\H^1(\RP^2 \times\RP^3,\Z_2)=\Z_2^2$,
$\H^2(\RP^2 \times\RP^3,\Z_2)=\Z_2^3$ whose three generators are $\alpha^2$, $\beta^2$ and $\alpha\beta$. 
If $B=\lambda_1\alpha^2+\lambda_2\beta^2+\lambda_3\alpha\beta$, then $\Sq^1B=\lambda_3\alpha^2\beta+\lambda_3\alpha\beta^2$.  Hence
\bea
\int_{\RP^2 \times\RP^3}B\Sq^1B&=&\lambda_3^2+\lambda_2\lambda_3,\\
\int_{\RP^2 \times\RP^3}B\Sq^1B+w_1(TM)^2\Sq^1B&=&\lambda_3^2+\lambda_2\lambda_3.
\eea
On the other hand,
since $w_2(TM)+w_1(TM)^2=0$ for $\RP^2 \times\RP^3$, so 
\bea
\bZ_{\text{SET}_{(0,0)}}(\RP^2 \times\RP^3)&=&\frac{1}{2}\sum_{\lambda_1,\lambda_2,\lambda_3\in\Z_2}(-1)^{\lambda_3^2+\lambda_2\lambda_3},\\
\bZ_{\text{SET}_{(1,0)}}(\RP^2 \times\RP^3)&=&\frac{1}{2}\sum_{\lambda_1,\lambda_2,\lambda_3\in\Z_2}(-1)^{\lambda_3^2+\lambda_2\lambda_3}.
\eea
Since
\bea
\#\{(\lambda_1,\lambda_2,\lambda_3)\in\Z_2^3 \mid \lambda_3^2+\lambda_2\lambda_3=1\}=2,
\eea
so 
\bea
&&\bZ_{\text{SET}_{(0,0)}}(\RP^2 \times\RP^3)=\bZ_{\text{SET}_{(1,0)}}(\RP^2 \times\RP^3)\nn\\
&=&\bZ_{\text{SET}_{(0,1)}}(\RP^2 \times\RP^3)=\bZ_{\text{SET}_{(1,1)}}(\RP^2 \times\RP^3)=\frac{1}{2}(6-2)=2.
\eea
\item
For $M=S^1\times\RP^2\times\RP^2$, let $\gamma$ be the generator of $\H^1(S^1,\Z_2)=\Z_2$ and $\alpha_i$ be the generator of $\H^1(\RP^2,\Z_2)=\Z_2$ of the $i$-th factor $\RP^2$ ($i=1,2$). 
Note that $w_1(TM)=\alpha_1+\alpha_2$.  $\H^0(S^1\times\RP^2\times\RP^2,\Z_2)=\Z_2$, $\H^1(S^1\times\RP^2\times\RP^2,\Z_2)=\Z_2^3$,
$\H^2(S^1\times\RP^2\times\RP^2,\Z_2)=\Z_2^5$ whose five generators are  $\alpha_1^2$, $\alpha_2^2$, $\gamma\alpha_1$, $\gamma\alpha_2$ and $\alpha_1\alpha_2$. 
If $B=\lambda_1\alpha_1^2+\lambda_2\alpha_2^2+\lambda_3\gamma\alpha_1+\lambda_4\gamma\alpha_2+\lambda_5\alpha_1\alpha_2$, then 
$\Sq^1B=\lambda_3\gamma\alpha_1^2+\lambda_4\gamma\alpha_2^2+\lambda_5\alpha_1^2\alpha_2+\lambda_5\alpha_1\alpha_2^2$.  Hence
\bea
\int_{S^1\times\RP^2\times\RP^2}B\Sq^1B&=&\lambda_1\lambda_4+\lambda_2\lambda_3+\lambda_3\lambda_5+\lambda_4\lambda_5,\\
\int_{S^1\times\RP^2\times\RP^2}B\Sq^1B+w_1(TM)^2\Sq^1B&=&\lambda_1\lambda_4+\lambda_2\lambda_3+\lambda_3\lambda_5+\lambda_4\lambda_5+\lambda_3+\lambda_4.\nn\\
\eea
On the other hand,
since $w_2(TM)+w_1(TM)^2=0$ for $S^1\times\RP^2\times\RP^2$, so 
\bea
\bZ_{\text{SET}_{(0,0)}}(S^1\times\RP^2\times\RP^2)&=&\frac{1}{4}\sum_{\lambda_1,\lambda_2,\lambda_3,\lambda_4,\lambda_5\in\Z_2}(-1)^{\lambda_1\lambda_4+\lambda_2\lambda_3+\lambda_3\lambda_5+\lambda_4\lambda_5},\\
\bZ_{\text{SET}_{(1,0)}}(S^1\times\RP^2\times\RP^2)&=&\frac{1}{4}\sum_{\lambda_1,\lambda_2,\lambda_3,\lambda_4,\lambda_5\in\Z_2}(-1)^{\lambda_1\lambda_4+\lambda_2\lambda_3+\lambda_3(\lambda_5+1)+\lambda_4(\lambda_5+1)}.\nn\\
\eea
Since
\bea
\#\{(\lambda_1,\lambda_2,\lambda_3,\lambda_4,\lambda_5)\in\Z_2^5 \mid \lambda_1\lambda_4+\lambda_2\lambda_3+\lambda_3\lambda_5+\lambda_4\lambda_5=1\}=12,
\eea
also note that changing $\lambda_5$ to $\lambda_5+1$ doesn't affect the sum,
so 
\bea
&&\bZ_{\text{SET}_{(0,0)}}(S^1\times\RP^2\times\RP^2)=\bZ_{\text{SET}_{(1,0)}}(S^1\times\RP^2\times\RP^2)\nn\\
&=&\bZ_{\text{SET}_{(0,1)}}(S^1\times\RP^2\times\RP^2)=\bZ_{\text{SET}_{(1,1)}}(S^1\times\RP^2\times\RP^2)=\frac{1}{4}(20-12)=2.\nn\\
\eea
\item
For a 5d Wu manifold $\W=\SU(3)/\SO(3)$, with $\H^0(\W,\Z_2)=\Z_2$, $\H^1(\W,\Z_2)=0$, note that $w_1(T\W)=0$, 
$\H^2(\W,\Z_2)=\Z_2$ which is generated by $w_2(T\W)$.  $\Sq^1w_2(T\W)=w_3(T\W)$. 
\bea
\bZ_{\text{SET}_{(0,0)}}(\W)&=&2\sum_{B=0,w_2(T\W)}(-1)^{B\Sq^1B+w_2(T\W)\Sq^1B}=4,
\eea
so
\bea
&&\bZ_{\text{SET}_{(0,0)}}(\W)=\bZ_{\text{SET}_{(1,0)}}(\W) =
\bZ_{\text{SET}_{(0,1)}}(\W)=\bZ_{\text{SET}_{(1,1)}}(\W)=4.
\eea
\end{enumerate}
In the next section, we will use the anyonic string/brane braiding statistics and 
the link invariants of 5d TQFTs to characterize and distinguish these 5d SETs.  

\section{Anyonic String/Brane Braiding Statistics and 
Link Invariants of 5d TQFTs}

\label{sec:link-inv}

Now we compute the path integral \eqn{eq:5dSET-cont} with extended operator insertions. 
To recall the general definitions, we have\\
$\bullet$ Partition or path integral w/out insertion is
$${ \sum_{B\in C^2(M,\Z_2)\atop{\dots}} (e^{\im S}
)}.$$

\noindent
$\bullet$ 
{In physics, the vacuum expectation value (v.e.v)} of a theory $S$ is defined as
\bea
&&\boxed{
\langle {\cal O}\rangle_{\text{(v.e.v)}} =
\frac{\langle \cal{O}\rangle_{\text{(v.e.v)}}}{\langle 1\rangle_{\text{(v.e.v)}}}
 =\frac{
 \sum_{B\in C^2(M,\Z_2)\atop{\dots}} (e^{\im S}
 \cal{O})
 }{ \sum_{B\in C^2(M,\Z_2)\atop{\dots}} (e^{\im S}
)}
  =\frac{
 \sum_{B\in C^2(M,\Z_2)\atop{\dots}} (e^{\im S}
 \cal{O})
 }{\bZ}} \nn\\
&& 
   =\frac{\text{path integral with insertions $ \cal{O}$}}{\text{path integral without insertions}}.
 \quad
\eea
For example, this includes the link invariant that we will focus on in this section:
\bea
\langle \exp(\ii \dots ({\text{Link invariants of }  U, X, Y, \dots }))\rangle_{\text{(v.e.v)}} =
\frac{\bZ_{\text{SET}_{(K_1,K_2)}}^{5\text{d}}[M^5; U, X, Y, \dots]}{
\bZ_{\text{SET}_{(K_1,K_2)}}^{5\text{d}}[M^5]}.
\eea
For conventions of our notations,
we label the 1d Wilson line as $W$,
the 2d surface operator as $U, U'$, etc.
We label the 3d membrane operator as $X$
and the 4d operator as $Y$, etc.
We label the \emph{$d$d-hyper-surface} of general operators that we inserted as $\Sigma^d$,
while we label this $\Sigma^d$'s \emph{$(d+1)$d-Seifert-hyper-volume}  as $V^{d+1}$.


In this section \ref{sec:link-inv}, we focus on deriving the general link invariants for these 5d TQFTs/SETs.\footnote{For more guidance on the physical interpretations of link invariants,
please see \cite{Putrov2016qdo1612.09298} and its Introduction.}
In the next \Sec{sec:link-conf}, we will provide explicit examples of the spacetime braiding process as the link configurations that can be detected by
these link invariants derived here in \Sec{sec:link-inv}.
The techniques for computing all these link invariants below are based on \Refe{Putrov2016qdo1612.09298}.
Below we simply apply the methods and notations introduced in \Refe{Putrov2016qdo1612.09298}.

Caveat: Note that while in the first section \ref{sec:link-inv-wPB}, we explicitly study the discrete cochain version of TQFT,
in the sections below  we implement the continuum formulation of TQFT. 
The reason is related to a fact that the graded non-commutativity of cochain fields is much more complicated to be dealt with than the continuum differential form fields.
The subtle fact will be commented further in footnotes \ref{footnoteSteenrod} and \ref{ft:Deligne}.
We also note that when we deal with the continuum differential form fields later in \Sec{sec:link-inv-w13B} to \Sec{sec:link-inv-all},
we \emph{choose} a normalization of  differential form fields as its enclosed 2-surface integral 
$\oint B \in \Z$ with the periodicity $\oint B\sim \oint B + 2$ (thus more similar to the convention of discrete cochain fields),
\emph{instead of} the more conventional $\oint B \in \pi \Z$ with the periodicity $\oint B\sim \oint B + 2\pi$.

\subsection{$\frac{1}{2} \tilde w_1(TM) \mathcal P( B)$ and a Triple Link Invariant $\text{Tlk}^{(5)}_{w_1BB}(\Sigma^3_X,\Sigma^2_{U_{\bf (i)}},\Sigma^2_{U_{\bf (ii)}})$}
\label{sec:link-inv-wPB}

We start with a 5d TQFT obtained from summing over 2-form field $B$ of $\frac{1}{2} \tilde w_1(TM) \mathcal P( B)$. This amounts to  gauging 1-form $\Z_2$ of this 5d SPTs. The resulting theory is  $\bZ_{\text{SET}_{(K_1=0,K_2=0)}}$ in \eqn{eq:5dSET-cont}. The topological action and the partition function are 
\bea
{\bf S}&=&\pi\int_{M^5}(\frac{1}{2}  \delta  \tilde w_1(TM)\cup \tilde c+b\cup \delta B+\frac{1}{2}  \tilde w_1(TM)\cup \mathcal P( B)). \label{eq:S-w1PB}\\
\bZ&=&\int [\cD B][\cD \tilde c][\cD b]\exp(\ii \bf{S}).\\
\bZ&=& \sum_{{{B, b\in C^2( M^5, \Z_2)}}\atop{\tilde c \in C^3(  M^5, \Z_4)}}
\exp(\ii 
\pi \int_{M^5} \frac{1}{2}  \delta  \tilde w_1(TM)\cup \tilde c+b\cup \delta B+ \frac{1}{2} \tilde w_1(TM)\cup \mathcal P( B)).
\eea
We consider the gauge transformation:\footnote{One may consider
add additional terms on the gauge transformations, such as
$ \tilde w_1(TM)\to \tilde w_1(TM) +\delta\alpha(t,x) +\alpha_1(t,x)$
and
$B\to B+\delta\beta(t,x)+ \alpha_2(t,x)$, etc.
However, terms such as $\alpha_1(t,x)=\alpha_1$ and $\alpha_2(t,x)=\alpha_2$ will need to be constant, which act as the higher-form global symmetry transformation, instead of gauge transformation. 
\label{ft:gauge-tranf}
}
\bea
 \tilde w_1(TM)&\to& \tilde w_1(TM) +\delta\alpha,\nn\\
B&\to&B+\delta\beta,\nn\\
\tilde c&\to&\tilde c+\delta\gamma+\lambda,\nn\\
b&\to&b+\delta\zeta+\mu.
\eea 
under which the action transforms as 
\bea
{\bf S}&\to&\pi\int_{M^5}\frac{1}{2}( \tilde w_1(TM) +\delta\alpha)(B\cup B+B\cup\delta\beta+\delta\beta\cup B+\delta\beta\cup\delta\beta+B\hcup{1}\delta B+\delta\beta\hcup{1}\delta B)\nn\\
&&+\frac{1}{2}\delta  \tilde w_1(TM)(\tilde c+\lambda)+(b+\mu)\delta B.
\eea
The gauge variance of the action is:
\bea
\Delta {\bf S}&=&\pi\int_{M^5}\frac{1}{2} \tilde w_1(TM) (\delta\beta\cup\delta\beta+2\delta\beta\cup B+\delta(\delta\beta\hcup{1}B))\nn\\
&&+\frac{1}{2}\delta\alpha(B\cup B+B\hcup{1}\delta B+\delta\beta\cup\delta\beta+2\delta\beta\cup B+\delta(\delta\beta\hcup{1}B))\nn\\
&&+\frac{1}{2}\delta  \tilde w_1(TM)\lambda+\mu\delta B\label{eq:w1(TM)P(B)1}\\
&=&\pi\int_{M^5}\frac{1}{2}\delta  \tilde w_1(TM) (\beta\delta\beta)+\big( \delta\tilde w_1(TM) (\beta B)+\tilde w_1(TM)\beta\delta B \big)+\frac{1}{2}\delta  \tilde w_1(TM) (\delta\beta\hcup{1}B)\nn\\
&&-\big(\alpha B\delta B+\frac{1}{2}\alpha u_2\delta B\big)-\alpha\delta\beta\delta B+\frac{1}{2}\delta  \tilde w_1(TM)\lambda +\mu\delta B\label{eq:w1(TM)P(B)2}.
\eea
In \eqref{eq:w1(TM)P(B)1}, we have used the formula\footnote{\label{footnoteSteenrod}
This is based on Steenrod's work ``Products of Cocycles and Extensions of Mappings \cite{Steenrod1947},'' which derives 
\bea \label{eq:Steenrod's}
\delta(u\hcup{i}v)=(-1)^{p+q-i}u\hcup{i-1}v+(-1)^{pq+p+q}v\hcup{i-1}u+\delta u\hcup{i}v+(-1)^pu\hcup{i}\delta v
\eea
where $u\in C^p$ and $v\in C^q$.
} 
\bea
B\cup\delta\beta-\delta\beta\cup B+\delta\beta\hcup{1}\delta B
+\delta^2\beta\hcup{1}B=\delta(\delta\beta\hcup{1}B). 
\eea
and $\delta^2\beta=0$. 
In \eqref{eq:w1(TM)P(B)2}, we have used integration by part: for a closed 5-manifold without boundary, after integration by part we can drop the boundary term $\delta(\dots)$.
Since $\delta^2 B = \delta^2 \beta  = \delta^2 \alpha=0$ ,
we drop 
$\delta\alpha(\delta\beta\cup\delta\beta+\delta(\delta\beta\hcup{1}B))$ which has no effect on a closed 5-manifold without boundary.
Denote 
$u_2= w_2(TM) + w_1(TM)^2$ as the second Wu class. We have also used the formula in footnote \ref{footnoteSteenrod} as
\bea
B\cup\delta B-\delta B\cup B +\delta B\hcup{1}\delta B+B\hcup{1}\delta^2 B&=&\delta( B\hcup{1}\delta B),\\
\delta B\hcup{1}\delta B=\Sq^2\delta B&=&u_2\delta B.
\eea
In \eqref{eq:w1(TM)P(B)2}, we used
$\delta (\alpha (  B\cup B+B\hcup{1}\delta B))=\delta\alpha(B\cup B+B\hcup{1}\delta B)+\alpha (\delta B\cup B +  B\cup \delta B+\delta (B\hcup{1}\delta B))
=\delta\alpha(B\cup B+B\hcup{1}\delta B)+\alpha (2  B\cup \delta B+u_2\delta B)$ and we dropped the total derivative term on a closed 5-manifold.
The solution of gauge invariance, i.e., $\Delta{\bf S}=0$, imposes:
\footnote{In general, 
when we study the action \eqn{eq:S-w1PB}, we
have made a convenient choice with a term $\delta  \tilde w_1(TM)\cup \tilde c$ instead of  $\tilde c \cup \delta \tilde w_1(TM)$.
For a generic 3-cochain $x$, $\delta\tilde w_1(TM)x=x\delta\tilde w_1(TM)$ is not true, 
by Steenrod's formula in footnote \ref{footnoteSteenrod} \eqn{eq:Steenrod's}, 
$\delta\tilde w_1(TM)x=x\delta\tilde w_1(TM)+\delta x\hcup{1}\delta\tilde w_1(TM)-\delta(x\hcup{1}\delta\tilde w_1(TM))$, 
we can only drop the total derivative terms (i.e. the coboundary terms). 
{In our present case, we consider
$x=\frac{1}{2}\beta\delta\beta+ \beta B+ \frac{1}{2} \delta\beta\hcup{1}B$.
So if $\delta x\hcup{1}\delta\tilde w_1(TM)$ is a coboundary, then we can also drop it, 
which results in
$$
\lambda=-2 x=-\beta\delta\beta -2\beta B-\delta\beta\hcup{1}B \mod 4.
$$}
{
If $\delta x\hcup{1}\delta\tilde w_1(TM)$ is not a coboundary,
we need the extra term
$$\delta\tilde w_1(TM)x=x\delta\tilde w_1(TM)+\delta x\hcup{1}\delta\tilde w_1(TM) + \text{a total derivative/coboundary term}.$$
When $\delta x\hcup{1}\delta\tilde w_1(TM)$ is not a coboundary, this results in a modified gauge transformation to $\lambda$.
By writing the action as in \eqn{eq:S-w1PB}, we can avoid additional complications, thus we end up
with a simpler gauge transformation \eqn{eq:gauge-transf-final-w1PB}.
The graded non-commutativity of cochain fields is much more complicated than the case for continuum differential form fields.
JW thanks Pierre Deligne for a discussion on the related issues.
 \label{ft:Deligne}
}
}
\bea 
\lambda&=&-\beta\delta\beta -2\beta B-\delta\beta\hcup{1}B \mod 4,\nn\\
\mu&=&-\tilde w_1(TM)\beta+ \alpha B+\frac{1}{2}\alpha u_2 +\alpha\delta\beta\mod 2,
\label{eq:gauge-transf-final-w1PB}
\eea

The 3-submanifold gauge invariant operator is,
\bea
X&=&\exp( \frac{\ii \pi}{2} k (\int_{\Sigma^3}\tilde c+\int_{V^4}\mathcal{P}(B)) )\nn\\
&=&\exp( \frac{\ii \pi}{2} k (\int_{M^5}\delta^{\perp}(\Sigma^3)\tilde c+\delta^{\perp}(V^4) \mathcal{P}(B))) \label{w1PB-X}
\eea
where $k \in \Z_4$.
To verify the gauge invariance, we use $\mathcal{P}(B+\delta \beta)=\mathcal{P}(B) +\delta\beta\cup\delta\beta+2\delta\beta\cup B+\delta(\delta\beta\hcup{1}B)$ and $\delta B=0$ on the 4-submanifold  Seifert volume ${V^4}$. 

The 2-submanifold (2-surface) operator gauge invariant is 
\bea
U&=&\exp( {\ii \pi} \ell ( \int_{\Sigma^2}b-\int_{V^3} \tilde w_1(TM)B-\frac{1}{2}\int_{V^3} \tilde w_1(TM) u_2))\nn\\
&=&\exp( {\ii \pi} \ell (\int_{M^5}b\delta^{\perp}(\Sigma^2)-(\tilde w_1(TM)B+\frac{1}{2} \tilde w_1(TM) u_2)\delta^{\perp}(V^3)))\nn\\
&=&\exp( {\ii \pi} \ell (\int_{M^5}b\delta^{\perp}(\Sigma^2)-(\tilde w_1(TM)B+\frac{1}{2} \tilde w_1(TM) (w_2(TM)+w_1(TM)^2))\delta^{\perp}(V^3)))
 \label{w1PB-V}
\eea
where $\ell \in \Z_2$ is an integer mod 2.
To verify that $U$ is gauge invariant,  we use $\delta B=\delta\tilde w_1(TM)=0$  on the 3-submanifold Seifert volume ${V^3}$.

We insert $X, U_{\bf (i)}$ and $U_{\bf (ii)}$ into the path integral $\bZ$, and 
write the correlation function either in the \emph{continuum} field theory formulation, or in the \emph{discrete} cochain field theory formulation, interchangeably as
\bea
&&\langle XU_{\bf (i)}U_{\bf (ii)} \rangle=\int [\cD B][\cD \tilde c][\cD b] \;XU_{\bf (i)}U_{\bf (ii)} \;\exp(\ii \bf{S}). \\
&&\langle XU_{\bf (i)} U_{\bf (ii)} \rangle= \sum_{{{B, b\in C^2( M^5, \Z_2)}}\atop{\tilde c \in C^3(  M^5, \Z_4)}}
XU_{\bf (i)}U_{\bf (ii)} \;
 \exp(\ii 
\pi \int_{M^5} \frac{1}{2} \delta  \tilde w_1(TM)\cup \tilde c+b\cup \delta B+ \frac{1}{2} \tilde w_1(TM)\cup \mathcal P( B)). \nn
\eea
We compute the correlation functions as follows.
\begin{enumerate}
    \item Integrating out $\tilde c$ yields
\bea
\delta  \tilde w_1(TM) &=&k \delta^{\perp}(\Sigma^3_X),\nn\\
 \tilde w_1(TM) &=&k \delta^{\perp}(V^4_X),
 \eea
hence as a consequence, $\delta^2  \tilde w_1(TM)= \delta (k \delta^{\perp}(\Sigma^3_W))=0$.
 So with the above configuration constraint, we get the double-counting mod 2 cancellation in the exponent of 
 $\exp( \frac{\ii \pi}{2} k (\int_{M^5}\delta^{\perp}(V^4_X)\mathcal{P}(B)))$
$\exp(\ii 
\pi  \int_{M^5}\frac{1}{2} \tilde w_1(TM)$ $\mathcal P( B))$ $=1$.
This boils down to
 \bea
\langle XU_{\bf (i)}U_{\bf (ii)} \rangle&=&\int [\cD B][\cD b] \;
U_{\bf (i)}U_{\bf (ii)}
 \exp(\ii 
\pi \int_{M^5} b\cup \delta B) \vert_{\tilde w_1(TM)=k\delta^{\perp}(V^4_X)}. \label{eq:triple-step1}
\eea

\item Integrating out $b$ yields
 \bea
\delta B&=&\ell_{\bf (i)} \delta^{\perp}(\Sigma^2_{U_{\bf (i)}})+\ell_{\bf (ii)} \delta^{\perp}(\Sigma^2_{U_{\bf (ii)}}),\nn\\
B&=&\ell_{\bf (i)} \delta^{\perp}(V^3_{U_{\bf (i)}})+\ell_{\bf (ii)} \delta^{\perp}(V^3_{U_{\bf (ii)}}).
\eea
\item  We  finally integrate out $B$, from \eqn{eq:triple-step1}:
 \bea  
&&\boxed{\langle XU_{\bf (i)}U_{\bf (ii)} \rangle} \nn\\
&&=\int [\cD B] 
{\rm e}^{( -{\ii \pi}  (\int_{M^5}(\tilde w_1(TM)B+\frac{1}{2} \tilde w_1(TM) (w_2(TM)+w_1(TM)^2))(\ell_{\bf (i)} \delta^{\perp}(V^3_{U_{\bf (i)}})+ \ell_{{\bf (ii)}} \delta^{\perp}(V^3_{U_{\bf (ii)}})  )))}
 \bigg\vert
\text{\tiny{\( {{\tilde w_1(TM)=k\delta^{\perp}(V^4_X), }\atop{{{B=\ell_{\bf (i)} \delta^{\perp}(V^3_{U_{\bf (i)}})}\atop{\;+\ell_{\bf (ii)} \delta^{\perp}(V^3_{U_{\bf (ii)}}).}}
\quad\quad}} \)}}
\nn\\
 &&=\int [\cD B] 
{\rm e}^{( -{\ii \pi}  (\int_{M^5}(\tilde w_1(TM)B+\frac{1}{2} \tilde w_1(TM) (w_2(TM)+w_1(TM)^2))B ))}
 \bigg\vert 
\text{\tiny{\(
 {{{\tilde w_1(TM)=k\delta^{\perp}(V^4_X),\quad\quad\quad\quad\quad\quad\;\;\;\; }\atop{B=\ell_{\bf (i)}\delta^{\perp}(V^3_{U_{\bf (i)}})+\ell_{\bf (ii)}\delta^{\perp}(V^3_{U_{\bf (ii)}}).}}}
\)}} \nn\\
 &&=\int [\cD B] 
{\rm e}^{( -{\ii \pi}  (\int_{M^5}(\tilde w_1(TM)BB+ \Sq^2\big(\frac{1}{2}\tilde w_1(TM) B\big))))}
 \bigg\vert 
\text{\tiny{\(
 {{{\tilde w_1(TM)= k \delta^{\perp}(V^4_X),\quad\quad\quad\quad\quad\quad\;\;\;\; }\atop{B=\ell_{\bf (i)}\delta^{\perp}(V^3_{U_{\bf (i)}})+\ell_{\bf (ii)}\delta^{\perp}(V^3_{U_{\bf (ii)}}).}}}
\)}}
  \label{eq:triple-step3-1}\\
 &&=\int [\cD B] 
{\rm e}^{( -{\ii \pi}  (\int_{M^5}(\tilde w_1(TM)BB+\frac{1}{2}  \tilde w_1(TM)BB+\frac{1}{2} (\frac{1}{2} \delta \tilde w_1(TM)) (\frac{1}{2} \delta B)\big))))}
 \bigg\vert
\text{\tiny{\(
 {{{\tilde w_1(TM)=k\delta^{\perp}(V^4_X),\; 
\delta \tilde w_1(TM) =k\delta^{\perp}(\Sigma^3_X);
 \; }\atop{{{B=\ell_{\bf (i)}\delta^{\perp}(V^3_{U_{\bf (i)}})+\ell_{\bf (ii)}\delta^{\perp}(V^3_{U_{\bf (ii)}}).\;\;}}\atop{
 \delta B=\ell_{\bf (i)} \delta^{\perp}(\Sigma^2_{U_{\bf (i)}})+\ell_{\bf (ii)}\delta^{\perp}(\Sigma^2_{U_{\bf (ii)}}) 
 .}}\quad\quad\quad\quad\quad\;}} \quad\quad\quad 
 \)}}\;
 \label{eq:triple-step3-2}\\
  &&=
  {\rm e}^{( -{\ii \pi \Big(  k \ell_{\bf (i)} \ell_{\bf (ii)}\cdot\#(V^4_X\cap V^3_{U_1}\cap V^3_{U_2}) +\frac{1}{8}
 \delta^{\perp}(\Sigma^3_X)\big(\delta^{\perp}(\Sigma^2_{U_{\bf (i)}})+\delta^{\perp}(\Sigma^2_{U_{\bf (ii)}}) \big)
  \Big)})} \cdot (\text{Self-intersecting $\#$ terms})
  \label{eq:triple-step3-3}\\
  &&\cong
  \boxed{{\rm e}^{( -{\ii \pi \Big(  k \ell_{\bf (i)} \ell_{\bf (ii)}\cdot\text{Tlk}^{(5)}(\Sigma^3_X,\Sigma^2_{U_{\bf (i)}},\Sigma^2_{U_{\bf (ii)}})  
  \Big)})}}.
  \label{eq:triple-step3-4}
\eea
In \eqn{eq:triple-step3-1}, we used  $\tilde w_1(TM) (w_2(TM)+w_1(TM)^2) B=
\tilde w_1(TM) u_2 B= \Sq^2\big(\tilde w_1(TM) B\big)$.
In \eqn{eq:triple-step3-2}, we rewrote  $\frac{1}{2} \tilde w_1(TM)\cP(B)$ via \footnote{We use the Cartan formula of the Steenrod square: 
$\Sq^2( u v)=(\Sq^2u) v +(\Sq^1 u)(\Sq^1 v)+  u\Sq^2( v) $ where $u,v \in \H^*(M,\Z_2)$.
} 
\bea\frac{1}{2} \tilde w_1(TM)u_2B &=&\Sq^2(\frac{1}{2} \tilde w_1(TM)B)=\frac{1}{2}  \tilde w_1(TM)BB+\Sq^1(\frac{1}{2}  \tilde w_1(TM))\Sq^1B\nn\\
&=&\frac{1}{2}  \tilde w_1(TM)BB+\frac{1}{2} (\frac{1}{2} \delta \tilde w_1(TM)) (\frac{1}{2} \delta B).\nn
\eea
We plugged in all the constraints into the path integral 
\eqn{eq:triple-step3-2}
to obtain \eqn{eq:triple-step3-3}.\footnote{\label{footnote:triple}
Here are some more explanations to derive \eqn{eq:triple-step3-3}.\\
$\bullet$ For $
 \int [\cD B] 
{\rm e}^{ -{\ii \pi}  (\int_{M^5}(\tilde w_1(TM)BB))}
 \bigg\vert
\text{\tiny{\(
 {{{\tilde w_1(TM)=k\delta^{\perp}(V^4_X),\quad\quad\quad\quad\quad\quad\;\;\; }\atop{B=\ell_{\bf (i)}\delta^{\perp}(V^3_{U_{\bf (i)}})+\ell_{\bf (ii)}\delta^{\perp}(V^3_{U_{\bf (ii)}}).}}}
\)}}$, we get a mutual-quadratic crossing term $V^3_{U_{\bf (i)}}\cap V^3_{U_{\bf (ii)}}$ with a multiple $2\pi$ exponent in 
$\rm{e}^{\ii 2 \pi \#(V^4_X\cap V^3_{U_{\bf (i)}}\cap V^3_{U_{\bf (ii)}})}$ 
which does not contribute to the expectation value.
There are also two self-quadratic terms $V^3_{U_{\bf (n)}}\cap V^3_{U_{\bf (n)}}$ for ${\bf (n)}={\bf (i)}$ or ${\bf (ii)}$.
These self-quadratic terms contribute, in principle, infinite many intersecting numbers in 
$\#(V^4_X\cap V^3_{U_{\bf (n)}}\cap V^3_{U_{\bf (n)}})$ for ${\bf (n)}={\bf (i)}$ or ${\bf (ii)}$.
Since a multiple $2\pi$ exponent have zero contribution to the expectation value, therefore 
either we can design an even but infinite number of points on each of $\#(V^4_X\cap V^3_{U_{\bf (n)}}\cap V^3_{U_{\bf (n)}})$ for ${\bf (n)}={\bf (i)}$ or ${\bf (ii)}$,
or we can absorb them into the $(\text{Self-intersecting $\#$ terms})$ in \eqn{eq:triple-step3-3}. In either cases, this term does not have any net contribution in the end at \eqn{eq:triple-step3-4}.
 \\
$\bullet$ For $
 \int [\cD B] 
{\rm e}^{ -{\ii \pi}  (\int_{M^5}(\frac{1}{2}\tilde w_1(TM)BB))}
 \bigg\vert
\text{\tiny{\(
 {{{\tilde w_1(TM)=k\delta^{\perp}(V^4_X),\quad\quad\quad\quad\quad\quad\;\; }\atop{B=\ell_{\bf (i)}\delta^{\perp}(V^3_{U_{\bf (i)}})+\ell_{\bf (ii)}\delta^{\perp}(V^3_{U_{\bf (ii)}}).}}}
\)}}$,
 we get a mutual-quadratic crossing term $V^3_{U_{\bf (i)}}\cap V^3_{U_{\bf (ii)}}$ with a multiple $\pi$ exponent in 
$\rm{e}^{\ii \pi \#(V^4_X\cap V^3_{U_{\bf (i)}}\cap V^3_{U_{\bf (ii)}})}$, 
 which \emph{does} contribute to the expectation value when this intersecting number $\#$ is odd, in a 1 mod 2 effect.
 There are also two self-quadratic terms $V^3_{U_{\bf (n)}}\cap V^3_{U_{\bf (n)}}$ for ${\bf (n)}={\bf (i)}$ or ${\bf (ii)}$.
 Again either we can design an quadruple/four-multiplet but infinite number of points for each of $\#(V^4_X\cap V^3_{U_{\bf (n)}}\cap V^3_{U_{\bf (n)}})$, 
 or  we can absorb them into the $(\text{Self-intersecting $\#$ terms})$ in \eqn{eq:triple-step3-3}.
 \\
$\bullet$ For $\int [\cD B] 
{\rm e}^{( -{\ii \pi}  (\int_{M^5}(\frac{1}{8} \delta \tilde w_1(TM) \delta B)))}
 \bigg\vert
\text{\tiny{\(
{\delta \tilde w_1(TM) =k\delta^{\perp}(\Sigma^3_X);
 \; \quad\quad\quad\quad\;\;}\atop{
 \delta B=\ell_{\bf (i)} \delta^{\perp}(\Sigma^2_{U_{\bf (i)}})+\ell_{\bf (ii)}\delta^{\perp}(\Sigma^2_{U_{\bf (ii)}}) } 
 \)}}\;
 =
 {\rm e}^{ -{\ii \pi}  \int_{M^5}(\frac{1}{8} (k\delta^{\perp}(\Sigma^3_X))\Big(\ell_{\bf (i)} \delta^{\perp}(\Sigma^2_{U_{\bf (i)}})+\ell_{\bf (ii)}\delta^{\perp}(\Sigma^2_{U_{\bf (ii)}}) \Big))}
 $, we find the exponent depends on the intersecting number
 $\#(\Sigma^3_X\cap \Sigma^2_{U_{\bf (n)}})$ for ${\bf (n)}={\bf (i)}$ or ${\bf (ii)}$, between 3-surface and 2-surface in a 5 manifold ---
 although generically this number $\#(\Sigma^3_X\cap \Sigma^2_{U_{\bf (n)}})$ is finite but can be nonzero, we design by default that there is no intersection
 between any of our insertions of 3-surface and 2-surface into the path integral. Thus we set $\#(\Sigma^3_X\cap \Sigma^2_{U_{\bf (n)}})=0$ by default.
}
We propose a set-up to remove or renormalize the (\text{Self-intersecting $\#$ terms}) appeared in \eqn{eq:triple-step3-4}, described in the footnote \ref{footnote:triple}.
The second exponent in  \eqn{eq:triple-step3-3}  shows
that 
$\int_{M^5} \delta^{\perp}(\Sigma^3_W)\big(\delta^{\perp}(\Sigma^2_{U_{\bf (i)}})+\delta^{\perp}(\Sigma^2_{U_{\bf (ii)}}))
=
\#(\Sigma^3_X\cap \Sigma^2_{U_{\bf (i)}})
+
\#(\Sigma^3_X\cap \Sigma^2_{U_{\bf (ii)}})
$, which counts the number of intersections between our insertions of 3-surface and 2-surface.
However, we choose \emph{by default} that our insertions of 3-surface and 2-surface have no intersections (to avoid unnecessary singularities).
Namely, we set $\#(\Sigma^3_W\cap \Sigma^2_{U_{\bf (n)}})=0$  for ${\bf (n)}={\bf (i)}$ or ${\bf (ii)}$, 
and $\#(\Sigma^2_{U_{\bf (i)}}\cap \Sigma^2_{U_{\bf (ii)}} )=0$ by default.
Overall, under the default assumption and the clarifications in footnote  \ref{footnote:triple},
we obtain a final relation between \eqn{eq:triple-step3-3} and our final effective answer \eqn{eq:triple-step3-4}.
We use the congruence symbol ($\cong$) to express that other unwanted terms that can be removed by design.

\end{enumerate}

In summary, we have derived the link invariant for the 5d TQFT $\bZ_{\text{SET}_{(K_1=0,K_2=0)}}[M^5]$ in \eqn{eq:triple-step3-4}: 
\bea \label{w1PB-Link}
\boxed{ \#(V^4_X\cap V^3_{U_{\bf (i)}}\cap V^3_{U_{\bf (ii)}})\equiv\text{Tlk}^{(5)}_{w_1BB}(\Sigma^3_X,\Sigma^2_{U_{\bf (i)}},\Sigma^2_{U_{\bf (ii)}})}.
\eea 
The path integral, with appropriate insertions of extended operators, becomes
\eqn{eq:triple-step3-4} which provides the above  link invariant.

\subsection{$w_1(TM)^3 B=w_1(TM)^2\Sq^1B$}

\label{sec:link-inv-w13B}

\subsubsection{Version I: $w_1(TM)^3 B$ and a Quartic Link Invariant {Qlk}$^{(5)}(\Sigma^3_{X_{\bf{(i)}}},\Sigma^3_{X_{\bf{(ii)}}},\Sigma^3_{X_{\bf{(iii)}}},\Sigma^2_U)$}

\label{sec:link-inv-w13B-version-1}

As a test example, we consider a 5d TQFT obtained from summing over 2-form field $B$ with the topological action $w_1(TM)^3 B$ 
(i.e., gauging the 1-form $\Z_2$  symmetry of this 5d higher-SPTs). 
For simplicity,  we convert the cochain TQFT to a differential-form continuum TQFT. 
The partition function and the topological action of the gauged theory (see footnote \ref{ft:Eucl}) are:
\bea
\bZ&=&\int[\cD B][\cD b][\cD c]\exp(\ii {\bf S}).\\
{\bf S}&=&\pi\int_{M^5} c \dd w_1(TM) +b\dd B+w_1(TM)^3B.
\eea
This 5d TQFT is distinct from any of four classes of $\bZ_{\text{SET}_{(K_1,K_2)}}$, but it still serves as a useful toy model.

We first specify the gauge transformations of various fields. Let us assume the gauge transformations take the following form (see footnote \ref{ft:gauge-tranf}):
\bea
w_1(TM)&\to&w_1(TM)+\dd\alpha,\nn\\
B&\to&B+\dd\beta,\nn\\
{c}&\to&{c+\dd\gamma+\lambda},\nn\\
b&\to&b+\dd\zeta+\mu.
\eea
The  variation of action under the gauge transformations is
\bea
{\bf S}&\to&{\bf S}+\pi\int_{M^5}\dd\gamma\dd w_1(TM)+\lambda\dd w_1(TM)+\dd\zeta\dd B+\mu\dd B\nn\\
&&+(\dd\alpha\dd\alpha w_1(TM)+w_1(TM)^2\dd\alpha+\dd\alpha\dd\alpha\dd\alpha)B\nn\\
&&+(w_1(TM)^3+\dd\alpha\dd\alpha w_1(TM)+w_1(TM)^2\dd\alpha+\dd\alpha\dd\alpha\dd\alpha)\dd\beta\\
&=&{\bf S}+\pi\int_{M^5}\lambda\dd w_1(TM)+\mu\dd B+(\alpha\dd\alpha B\dd w_1(TM)-\alpha\dd\alpha w_1(TM)\dd B)\nn\\
&&-\alpha w_1(TM)^2\dd B-\alpha\dd\alpha\dd\alpha\dd B+w_1(TM)^2\beta\dd w_1(TM)+\alpha\dd\alpha\dd\beta\dd w_1(TM)
\eea
where we have used integration by part. 
For a closed 5-manifold without boundary, after integration by part  we drop the total derivative terms $\dd(\dots)$ 
which have no effect on a closed 5-manifold without boundary.
The gauge variance of the action, i.e. 
$\Delta {\bf S}=0$, requires 
\bea
\lambda&=&-\alpha\dd\alpha B-w_1(TM)^2\beta-\alpha\dd\alpha\dd\beta,\nn\\
\mu&=&\alpha\dd\alpha w_1(TM)+\alpha w_1(TM)^2+\alpha\dd\alpha\dd\alpha.
\eea
%
%

The gauge invariant 3-submanifold operator is
\bea
X&=&\exp(\ii\pi k(\int_{\Sigma^3}c+\int_{V^4}w_1(TM)^2 B))\nn\\
&=&\exp(\ii\pi k(\int_{M^5}(\delta^{\perp}(\Sigma^3)c+\delta^{\perp}(V^4)w_1(TM)^2 B)))
\eea
and the gauge invariant 2-surface operator is 
\bea
U&=&\exp(\ii\pi\ell(\int_{\Sigma^2}b-\int_{V^3}w_1(TM)^3))\nn\\
&=&\exp(\ii\pi\ell(\int_{M^5}(\delta^{\perp}(\Sigma^2)b-\delta^{\perp}(V^3)w_1(TM)^3)))
\eea
where $k,\ell\in \Z_2$.
To verify the gauge invariance, we need to use $\dd w_1(TM)=\dd B=0$ on the 2-surfaces and 3-submanifolds.

To compute the link invariants, we insert $X_{\bf{(i)}},X_{\bf{(ii)}},X_{\bf{(iii)}}, U$ into the path integral $\bZ$.  In the \emph{continuum} field theory formulation, the link invariant is 
\bea
\langle X_{\bf{(i)}}X_{\bf{(ii)}}X_{\bf{(iii)}} U \rangle&=&\int [\cD B][\cD c][\cD b] \;X_{\bf{(i)}}X_{\bf{(ii)}}X_{\bf{(iii)}} U \;\exp(\ii \bf{S}). \\
&=&\int [\cD B][\cD c][\cD b]
X_{\bf{(i)}}X_{\bf{(ii)}}X_{\bf{(iii)}} U \;
 \exp(\ii 
\pi \int_{M^5} c \dd w_1(TM)+b\dd B+ w_1(TM)^3 B). \nn
\eea
We compute $\langle X_{\bf{(i)}}X_{\bf{(ii)}}X_{\bf{(iii)}} U \rangle$ as follows. 
\begin{enumerate}
    \item Integrating out $c$, we get
\bea
\dd w_1(TM)&=&k_{\bf (i)}\delta^{\perp}(\Sigma^3_{X_{\bf{(i)}}})+k_{\bf (ii)}\delta^{\perp}(\Sigma^3_{X_{\bf{(ii)}}})+k_{\bf (iii)}\delta^{\perp}(\Sigma^3_{X_{\bf{(iii)}}}),\nn\\
w_1(TM) &=&k_{\bf (i)}\delta^{\perp}(V^4_{X_{\bf{(i)}}})+k_{\bf (ii)}\delta^{\perp}(V^4_{X_{\bf{(ii)}}})+k_{\bf (iii)}\delta^{\perp}(V^4_{X_{\bf{(iii)}}}).
\eea

 With the above configuration constraint, we get the double-counting mod 2 cancellation in the exponent of 
 $\exp( \ii \pi (\int_{M^5}w_1(TM)^2B(k_{\bf (i)}\delta^{\perp}(V^4_{X_{\bf{(i)}}})+k_{\bf (ii)}\delta^{\perp}(V^4_{X_{\bf{(ii)}}})+k_{\bf (iii)}\delta^{\perp}(V^4_{X_{\bf{(iii)}}}))))
 \exp(\ii 
\pi   \int_{M^5}w_1(TM)^3 B) =1$.
Consequently the link invariant boils down to 
 \bea
\langle X_{\bf{(i)}}X_{\bf{(ii)}}X_{\bf{(iii)}} U \rangle&=&\int [\cD B][\cD b] \;
U
 \exp(\ii 
\pi \int_{M^5} b\dd B) \vert_{ w_1(TM)=k_{\bf (i)}\delta^{\perp}(V^4_{X_{\bf{(i)}}})+k_{\bf (ii)}\delta^{\perp}(V^4_{X_{\bf{(ii)}}})+k_{\bf (iii)}\delta^{\perp}(V^4_{X_{\bf{(iii)}}})}. \nn\\\label{eq:quadruple-step1}
\eea

\item Integrating out $b$ further yields the constraint
 \bea
\dd B&=&\ell \delta^{\perp}(\Sigma^2_U),\nn\\
B&=&\ell \delta^{\perp}(V^3_U).
\eea

\item We finally integrate out $B$ as follows
 \bea  
&&\boxed{\langle X_{\bf{(i)}}X_{\bf{(ii)}}X_{\bf{(iii)}} U \rangle} \nn\\
&&=\int [\cD B] 
{\rm e}^{ -{\ii \pi}  (\int_{M^5} w_1(TM)^3\ell \delta^{\perp}(V^3_U))}
 \bigg\vert
\text{\tiny{\( {{w_1(TM)=k_{\bf (i)}\delta^{\perp}(V^4_{X_{\bf{(i)}}})+k_{\bf (ii)}\delta^{\perp}(V^4_{X_{\bf{(ii)}}})+k_{\bf (iii)}\delta^{\perp}(V^4_{X_{\bf{(iii)}}}), }\atop{{{B=\ell \delta^{\perp}(V^3_U)}.}
\quad\quad\quad\quad\quad\quad\quad\quad\quad\quad\quad\quad\quad\quad\quad\quad\quad\quad\quad\quad\quad\quad}} \)}}
\nn\\
 &&=\int [\cD B] 
{\rm e}^{( -{\ii \pi}  (\int_{M^5}w_1(TM)^3B))}
 \bigg\vert 
\text{\tiny{\(
 {{{w_1(TM)=k_{\bf (i)}\delta^{\perp}(V^4_{X_{\bf{(i)}}})+k_{\bf (ii)}\delta^{\perp}(V^4_{X_{\bf{(ii)}}})+k_{\bf (iii)}\delta^{\perp}(V^4_{X_{\bf{(iii)}}}),}\atop{B=\ell\delta^{\perp}(V^3_U).\quad\quad\quad\quad\quad\quad\quad\quad\quad\quad\quad\quad\quad\quad\quad\quad\quad\quad\quad\quad\quad\quad}}}
\)}} \nn\\
  \label{eq:quadruple-step3-1}\\
    &&=
  {\rm e}^{( -{\ii \pi \Big( k_{\bf (i)}k_{\bf (ii)}k_{\bf (iii)} \ell \big(
  \#(V^4_{X_{\bf{(i)}}}\cap V^4_{X_{\bf{(ii)}}}\cap V^4_{X_{\bf{(iii)}}}\cap V^3_U) 
  +
    \#(V^4_{X_{\bf{(ii)}}}\cap V^4_{X_{\bf{(iii)}}}\cap V^4_{X_{\bf{(i)}}}\cap V^3_U) 
      +
    \#(V^4_{X_{\bf{(iii)}}}\cap V^4_{X_{\bf{(i)}}}\cap V^4_{X_{\bf{(ii)}}}\cap V^3_U) }}
  \nn\\
  &&
 {}^{{
  \#(V^4_{X_{\bf{(i)}}}\cap V^4_{X_{\bf{(iii)}}}\cap V^4_{X_{\bf{(ii)}}}\cap V^3_U) 
  +
    \#(V^4_{X_{\bf{(iii)}}}\cap V^4_{X_{\bf{(ii)}}}\cap V^4_{X_{\bf{(i)}}}\cap V^3_U) 
      +
    \#(V^4_{X_{\bf{(ii)}}}\cap V^4_{X_{\bf{(i)}}}\cap V^4_{X_{\bf{(iii)}}}\cap V^3_U)
 \big) \Big)})} \nn\\
  &&\quad\quad \cdot (\cdots) \cdot (\text{Self-intersecting $\#$ terms})
  \label{eq:quadruple-step3-2}
  \\
  &&\cong
  {\rm e}^{( -{\ii \pi \Big( k_{\bf (i)}k_{\bf (ii)}k_{\bf (iii)} \ell\cdot6\#(V^4_{X_{\bf{(i)}}}\cap V^4_{X_{\bf{(ii)}}}\cap V^4_{X_{\bf{(iii)}}}\cap V^3_U) 
  \Big)})} \cdot (\cdots)
  \label{eq:quadruple-step3-3}\\
  &&\cong
  \boxed{{\rm e}^{( -{\ii \pi \Big(  k_{\bf (i)}k_{\bf (ii)}k_{\bf (iii)} \ell\cdot6\text{Qlk}^{(5)}(\Sigma^3_{X_{\bf{(i)}}},\Sigma^3_{X_{\bf{(ii)}}},\Sigma^3_{X_{\bf{(iii)}}},\Sigma^2_U)  
  \Big)})} \cdot (\cdots)}.
  \label{eq:quadruple-step3-4}
 \eea
We propose a set-up to remove or renormalize the (\text{Self-intersecting $\#$ terms}) appeared in \eqn{eq:quadruple-step3-2}, 
following the same strategy in footnote \ref{footnote:triple}.
\end{enumerate}

For ${\bf S}=\pi\int_{M^5} c \dd w_1(TM) +b\dd B+w_1(TM)^3B$,
we derive the link invariant for the 5d TQFT $\bZ_{\text{SET}}[M^5]$ in \eqn{eq:quadruple-step3-2} and \eqn{eq:quadruple-step3-3}: 
\bea
\boxed{ \#(V^4_{X_{\bf{(i)}}}\cap V^4_{X_{\bf{(ii)}}}\cap V^4_{X_{\bf{(iii)}}}\cap V^3_U)\equiv\text{Qlk}^{(5)}(\Sigma^3_{X_{\bf{(i)}}},\Sigma^3_{X_{\bf{(ii)}}},\Sigma^3_{X_{\bf{(iii)}}},\Sigma^2_U)}.
\eea 
The path integral with appropriate extended operator insertions become
\eqn{eq:quadruple-step3-3} which provides the above  link invariant.
However, note that the factorial $3!=6$ trivializes the complex $\rm{e}^{\ii \pi}$ phase to  $\rm{e}^{\ii 6 \pi}$. 
It may be possible to take into account (see footnote \ref{ft:Deligne})
from the subtle graded non-commutativity of cochain field effect. 
Thus one may need to go beyond the continuum differential form TQFT formulation 
by using the cochain TQFT formulation in order to see the subleading effect.

\subsubsection{Version II: $w_1(TM)^2\Sq^1B$ and a Triple Link Invariant {Tlk}$^{(5)}_{w_1w_1 \dd B}(\Sigma^3_{X_{\bf{(i)}}},\Sigma^3_{X_{\bf{(ii)}}},\Sigma^2_U)$}

\label{sec:link-inv-w13B-version-2}

As another test example, we consider a 5d TQFT obtained from summing over 2-form field $B$ with the topological action $w_1(TM)^2\Sq^1B$. 
We again use the  continuum version of the TQFT.\footnote{Even though
$w_1(TM)^2\Sq^1B$ is a rewriting of $w_1(TM)^3 B$ on a closed 5-manifold, it turns out that we still gain new insights about an additional link invariant.} 
Its partition function and the topological action (see footnote \ref{ft:Eucl}) are:
\bea
\bZ&=&\int[\cD B][\cD b][\cD c]\exp(\ii {\bf S}),\\
{\bf S}&=&\pi\int_{M^5} c \dd w_1(TM) +b\dd B+w_1(TM)^2\Sq^1B,\\
{\bf S}&=&\pi\int_{M^5} c \dd w_1(TM) +b\dd B+w_1(TM)^2\frac{1}{2} \dd B.
\eea
We assume the gauge transformations have the following ansatz
\bea
w_1(TM)&\to&w_1(TM)+\dd\alpha,\nn\\
B&\to&B+\dd\beta,\nn\\
{c}&\to&{c+\dd\gamma+\lambda},\nn\\
b&\to&b+\dd\zeta+\mu.
\eea
Under the gauge transformations, the action transforms as 
\bea
{\bf S}&\to&{\bf S}+\pi\int_{M^5}\dd\gamma\dd w_1(TM)+\lambda\dd w_1(TM)+\dd\zeta\dd B+\mu\dd B\nn\\
&&+(w_1(TM)\dd\alpha+\dd\alpha w_1(TM)+\dd\alpha\dd\alpha)\frac{1}{2}\dd B\nn\\
&&+(w_1(TM)^2+w_1(TM)\dd\alpha+\dd\alpha w_1(TM)+\dd\alpha\dd\alpha)\frac{1}{2}\dd^2\beta\\
&=&{\bf S}+\pi\int_{M^5}\lambda\dd w_1(TM)+\mu\dd B+\frac{1}{2}(w_1(TM)\dd\alpha+\dd\alpha w_1(TM)+\dd\alpha\dd\alpha)\dd B
\eea
where we have used integration by part. 
$\Delta {\bf S}=0$ requires 
\bea
\lambda&=&0,\nn\\
\mu&=&-\frac{1}{2}(w_1(TM)\dd\alpha+\dd\alpha w_1(TM)+\dd\alpha\dd\alpha).
\eea
%
%
The gauge invariant 3-submanifold operator is 
\bea
X&=&\exp(\ii\pi k(\int_{\Sigma^3}c))\nn\\
&=&\exp(\ii\pi k(\int_{M^5}(\delta^{\perp}(\Sigma^3)c)))
\eea
and the gauge invariant 2-surface operator is
\bea
U&=&\exp(\ii\pi\ell(\int_{\Sigma^2}(b+\frac{1}{2}w_1(TM)^2)))\nn\\
&=&\exp(\ii\pi\ell(\int_{M^5}(\delta^{\perp}(\Sigma^2)(b+\frac{1}{2}w_1(TM)^2) ))
\eea
where $k,\ell \in \Z_2$. 

We proceed to compute the link invariants by inserting $X_{\bf{(i)}},X_{\bf{(ii)}}, U$ into the path integral $\bZ$,  
\bea
&&\langle X_{\bf{(i)}}X_{\bf{(ii)}} U \rangle=\int [\cD B][\cD c][\cD b] \;X_{\bf{(i)}}X_{\bf{(ii)}} U \;\exp(\ii \bf{S}). \\
&&\langle X_{\bf{(i)}}X_{\bf{(ii)}} U \rangle=\int [\cD B][\cD c][\cD b]
X_{\bf{(i)}}X_{\bf{(ii)}} U \;
 \exp(\ii 
\pi \int_{M^5} c \dd w_1(TM)+b\dd B+ w_1(TM)^2\frac{1}{2}\dd B). \nn
\eea
To evaluate $\langle X_{\bf{(i)}}X_{\bf{(ii)}} U \rangle$, we integrate out various fields step by step. 
\begin{enumerate}
    \item Integrating out $c$, we get 
\bea
\dd w_1(TM)&=&k_{\bf (i)}\delta^{\perp}(\Sigma^3_{X_{\bf{(i)}}})+k_{\bf (ii)}\delta^{\perp}(\Sigma^3_{X_{\bf{(ii)}}}),\nn\\
w_1(TM) &=&k_{\bf (i)}\delta^{\perp}(V^4_{X_{\bf{(i)}}})+k_{\bf (ii)}\delta^{\perp}(V^4_{X_{\bf{(ii)}}}).
\eea
The link invariant thus boils down to 
 \bea
\langle X_{\bf{(i)}}X_{\bf{(ii)}} U \rangle&=&\int [\cD B][\cD b] \;
U
 \exp(\ii 
\pi \int_{M^5} b\dd B+ w_1(TM)^2\frac{1}{2}\dd B) \vert_{ w_1(TM)=k_{\bf (i)}\delta^{\perp}(V^4_{X_{\bf{(i)}}})+k_{\bf (ii)}\delta^{\perp}(V^4_{X_{\bf{(ii)}}})}. \nn\\\label{eq:triple2-step1}
\eea

\item Integrate out $b$, we get the constraint
 \bea
\dd B&=&\ell \delta^{\perp}(\Sigma^2_U),\nn\\
B&=&\ell \delta^{\perp}(V^3_U).
\eea

\item We finally integrate out $B$ in \eqn{eq:triple2-step1}:
 \bea  
&&\boxed{\langle X_{\bf{(i)}}X_{\bf{(ii)}} U \rangle} \nn\\
&&=\int [\cD B] 
{\rm e}^{ -{\ii \pi}  (\int_{M^5} \frac{1}{2} w_1(TM)^2\ell \delta^{\perp}(\Sigma^2_U)+ w_1(TM)^2\frac{1}{2}\dd B)}
 \bigg\vert
\text{\tiny{\( {{w_1(TM)=k_{\bf (i)}\delta^{\perp}(V^4_{X_{\bf{(i)}}})+k_{\bf (ii)}\delta^{\perp}(V^4_{X_{\bf{(ii)}}}), }\atop{{{B=\ell \delta^{\perp}(V^3_U)}}
\quad\quad\quad\quad\quad\quad\quad\quad\quad\quad\quad\quad\quad}} \)}}
\nn\\
 &&=\int [\cD B] 
{\rm e}^{( -{\ii \pi}  (\int_{M^5} \frac{1}{2} w_1(TM)^2 \dd B+ w_1(TM)^2\frac{1}{2}\dd B))}
 \bigg\vert 
\text{\tiny{\(
 {{{w_1(TM)=k_{\bf (i)}\delta^{\perp}(V^4_{X_{\bf{(i)}}})+k_{\bf (ii)}\delta^{\perp}(V^4_{X_{\bf{(ii)}}}), }\atop{B=\ell\delta^{\perp}(V^3_U).\quad\quad\quad\quad\quad\quad\quad\quad\quad\quad\quad\quad\quad}}}
\)}} \nn\\
  \label{eq:triple2-step3-1}\\
    &&=
  {\rm e}^{( -{\ii \pi \Big( k_{\bf (i)}k_{\bf (ii)} \ell \big(
  \#(V^4_{X_{\bf{(i)}}}\cap V^4_{X_{\bf{(ii)}}}\cap \Sigma^2_U)
  +  \#(V^4_{X_{\bf{(ii)}}}\cap V^4_{X_{\bf{(i)}}}\cap \Sigma^2_U) 
 \big) \Big)})} \nn\\
  &&\quad\quad \cdot (\cdots) \cdot (\text{Self-intersecting $\#$ terms})
  \label{eq:triple2-step3-2}
  \\
  &&\cong
  \boxed{{\rm e}^{( -{\ii \pi \Big(  k_{\bf (i)}k_{\bf (ii)} \ell\cdot
  (\text{Tlk}^{(5)}_{w_1w_1 \dd B}(\Sigma^3_{X_{\bf{(i)}}},\Sigma^3_{X_{\bf{(ii)}}},\Sigma^2_U) + \text{Tlk}^{(5)}_{w_1w_1 \dd B}(\Sigma^3_{X_{\bf{(ii)}}},\Sigma^3_{X_{\bf{(i)}}},\Sigma^2_U)   )
  \Big)})} \cdot (\cdots)}.
  \label{eq:triple2-step3-3}
 \eea
 We propose a set-up to remove or renormalize the (\text{Self-intersecting $\#$ terms}) appeared in \eqn{eq:triple2-step3-2}, following the same strategy as footnote \ref{footnote:triple}.
\end{enumerate}
 
For ${\bf S}=\pi\int_{M^5} c \dd w_1(TM) +b\dd B+w_1(TM)^2\frac{1}{2}\dd B$,
we derive the link invariant for the 5d TQFT $\bZ_{\text{SET}}[M^5]$ in \eqn{eq:triple2-step3-2} and \eqn{eq:triple2-step3-3}: 
\bea
\boxed{ \#(V^4_{X_{\bf{(i)}}}\cap V^4_{X_{\bf{(ii)}}}\cap \Sigma^2_U)\equiv\text{Tlk}^{(5)}_{w_1w_1 \dd B}(\Sigma^3_{X_{\bf{(i)}}},\Sigma^3_{X_{\bf{(ii)}}},\Sigma^2_U)}.
\eea 
The path integral with appropriate extended operators insertions become
\eqn{eq:triple2-step3-3} which provides the above  link invariant.
{
However, note that  the two terms on the exponent of \eqn{eq:triple2-step3-3} are the same, which 
trivializes the complex $\rm{e}^{\ii \pi}$ to  $\rm{e}^{\ii 2 \pi}$. 
It may be possible to take into account (see footnote \ref{ft:Deligne})
from the subtle graded non-commutativity of cochain field effect. 
Thus one may need to go beyond the continuum differential form TQFT formulation 
by using the cochain TQFT formulation in order to see the subleading effect.
}

\subsection{$w_3(TM) B=w_2(TM) \Sq^1 B$ and a Quadratic Link Invariant $\text{Lk}^{(5)}_{w_2 \dd B}(\Sigma^2_{U'},\Sigma^2_U)$}

\label{sec:link-inv-w2dB}


We further consider a 5d TQFT obtained from summing over 2-form field $B$ of $w_3(TM) B=w_2(TM) \Sq^1 B$. 
We again adopt the continuum version of TQFT. 
The partition function and action (see footnote \ref{ft:Eucl}) are:
\bea
\bZ&=&\int[\cD B][\cD b][\cD h]\exp(\ii {\bf S}),\\
{\bf S}&=&\pi\int_{M^5} h \dd w_2(TM) +b\dd B+w_2(TM) \Sq^1 B,\\
{\bf S}&=&\pi\int_{M^5} h \dd w_2(TM) +b\dd B+w_2(TM) \frac{1}{2}\dd B.
\eea
We assume the gauge transformations have the following ansatz
\bea
w_2(TM)&\to&w_2(TM)+\dd\alpha,\nn\\
B&\to&B+\dd\beta,\nn\\
h&\to&h+\dd\gamma+\lambda,\nn\\
b&\to&b+\dd\zeta+\mu.
\eea
Under the gauge transformation, the action transforms as 
\bea
{\bf S}&\to&{\bf S}+\pi\int_{M^5}\dd\gamma\dd w_2(TM)+\lambda\dd w_2(TM)+\dd\zeta\dd B+\mu\dd B\nn\\
&&+\dd\alpha \frac{1}{2}\dd  B+w_2(TM) \frac{1}{2}\dd^2\beta+\dd\alpha\frac{1}{2}\dd^2\beta\\
&=&{\bf S}+\pi\int_{M^5}\lambda \dd w_2(TM)+\mu\dd B+ (\frac{1}{2}\dd\alpha) \dd  B+(-\frac{1}{2} \dd \beta)\dd w_2(TM).
\eea
Thus
$\Delta {\bf S}=0$ requires 
\bea
\lambda&=&(\frac{1}{2} \dd \beta),\nn\\
\mu&=&-(\frac{1}{2}\dd\alpha).
\eea
%
%
%
%
There are two types of gauge invariant 2-surface operators, 
\bea
U'&=&\exp(\ii\pi k(\int_{\Sigma^2}h- \int_{V^3} {\frac{1}{2} \dd B}))\nn\\
&=&\exp(\ii\pi k(\int_{M^5}(\delta^{\perp}(\Sigma^2)h-\delta^{\perp}(V^3) \frac{1}{2}\dd B)))\\
&=&\exp(\ii\pi k(\int_{M^5}(\delta^{\perp}(\Sigma^2) (h- \frac{1}{2} B) )))
\eea
and 
\bea
U&=&\exp(\ii\pi\ell(\int_{\Sigma^2}b+\int_{V^3}  \frac{1}{2} \dd w_2(TM)))\nn\\
&=&\exp(\ii\pi\ell(\int_{\Sigma^2}b+\int_{\Sigma^2}  \frac{1}{2} w_2(TM)))\nn\\
&=&\exp(\ii\pi\ell(\int_{M^5}\delta^{\perp}(\Sigma^2)(b+\frac{1}{2} w_2(TM) ) ))
\eea
where $k,\ell \in \Z_2$. 

We define the link invariant by inserting $U', U$ into the path integral $\bZ$
\bea
&&\langle U' U \rangle=\int [\cD B][\cD h][\cD b]
U' U \;
 \exp(\ii 
\pi \int_{M^5} h \dd w_2(TM) +b\dd B+w_2(TM) \frac{1}{2}\dd B). \nn
\eea
Below we evaluate $\langle U' U \rangle$ by integrating out various fields. 
\begin{enumerate}
    \item Integrating out $h$, 
we get
\bea
\dd w_2(TM)&=&k\delta^{\perp}(\Sigma^2_{U'}),\nn\\
w_2(TM) &=&k\delta^{\perp}(V^3_{U'}).
\eea
Plugging the above constraints into the partition function, we find the double-counting mod 2 cancellation in the exponent of 
 $\exp( \ii \pi (\int_{M^5}  \delta^{\perp}(V^3_{U'}) {\frac{k}{2} \dd B}
+w_2(TM) \frac{1}{2}\dd B)) =1$.
Thus the link invariant boils down to
 \bea
\langle U' U \rangle&=&\int [\cD B][\cD b] \;
U
 \exp(\ii 
\pi \int_{M^5} b\dd B) \vert_{ w_2(TM) =k\delta^{\perp}(V^3_{U'})}. \nn\\\label{eq:link2-step1}
\eea

\item Integrating out $b$,
we get the constraint
 \bea
\dd B&=&\ell \delta^{\perp}(\Sigma^2_U),\nn\\
B&=&\ell \delta^{\perp}(V^3_U).
\eea

\item  We finally integrate out $B$  in \eqn{eq:link2-step1}:
 \bea  
&&\boxed{\langle U' U \rangle} \nn\\
&&=\int [\cD B] 
{\rm e}^{ -{\ii \pi}  (\int_{M^5} \frac{1}{2} w_2(TM)\ell \delta^{\perp}(\Sigma^2_U))}
 \bigg\vert
\text{\tiny{\( {{ w_2(TM) =k\delta^{\perp}(V^3_{U'}),}\atop{{{B=\ell \delta^{\perp}(V^3_U)}}
\quad\quad\quad}} \)}}
\nn\\
 &&=\int [\cD B] 
{\rm e}^{( -{\ii \pi}  (\int_{M^5} \frac{1}{2} w_2(TM) \dd B))}
 \bigg\vert
\text{\tiny{\( {{ w_2(TM) =k\delta^{\perp}(V^3_{U'}),}\atop{{{B=\ell \delta^{\perp}(V^3_U)}}
\quad\quad\quad}} \)}}
\nn\\
  \label{}\\
  &&=
  {\rm e}^{( -{\ii \pi \Big( \frac{k\ell}{2}\cdot\#(V^3_{U'}\cap \Sigma^2_U) 
  \Big)})} 
  \label{}\\
  &&\cong
  \boxed{{\rm e}^{( -{\ii \pi \Big(  \frac{k\ell}{2}\cdot\text{Lk}^{(5)}(\Sigma^2_{U'},\Sigma^2_U)  
  \Big)})}}.
  \label{eq:link2-step3}
 \eea
\end{enumerate}
 We derive the link invariant for the 5d TQFT $\bZ_{\text{SET}}[M^5]$ for
${\bf S}=\pi\int_{M^5} h \dd w_2(TM) +b\dd B+w_2(TM) \Sq^1 B$ 
in \eqn{eq:link2-step3}: 
\bea
\boxed{ \#(V^3_{U'}\cap \Sigma^2_U)\equiv\text{Lk}^{(5)}_{w_2 \dd B}(\Sigma^2_{U'},\Sigma^2_U) }.
\eea 
The path integral with appropriate extended operators insertions become
\eqn{eq:link2-step3} which provides the above  link invariant.

\subsection{ 
$B\Sq^1B+(1+K_1)w_1(TM)^2\Sq^1B+w_2(TM)\Sq^1B$
and More Link Invariants:\\ 
$\text{Tlk}^{(5)}_{w_1w_1 \dd B}(\Sigma^3_{X_{\bf{(i)}}},\Sigma^3_{X_{\bf{(ii)}}},\Sigma^2_{U_{\bf}})$, 
$\text{Lk}^{(5)}_{B \dd B}(\Sigma^2_{U_{\bf (i)}},\Sigma^2_{U_{\bf (ii)}})$
and  $  \text{Lk}^{(5)}_{w_2 \dd B}(\Sigma^2_{U'},\Sigma^2_{U_{\bf }})$}

\label{sec:link-inv-all}

 We finally consider the generic form including   the four classes of $\bZ_{\text{SET}_{(K_1,K_2)}}^{5\text{d}}$ in \eqn{eq:5dSET-cont}  by gauging
$\bZ_{\text{SPT}_{(K_1,K_2)}}^{5\text{d}}$ in  \eqref{eq.linkuse}, with $(K_1,K_2) \in (\Z_2, \Z_2)$ labeling the Four Siblings.
Below, we find it convenient to introduce $K_1'$ via  $K_1':=1+K_1$ mod 2.

The partition function and action (see footnote \ref{ft:Eucl}) are:
\bea
\bZ&=&\int[\cD B][\cD b][\cD h][\cD c]\exp(\ii {\bf S}).\\
{\bf S}&=& \pi\int_{M^5} {K_1' } c \dd w_1(TM) +{ }h\dd w_2(TM) +b\dd B+B\Sq^1B+K_1'w_1(TM)^2\Sq^1B 
+w_2(TM)\Sq^1B. \quad\quad \\
{\bf S}&=& \pi\int_{M^5} {K_1'} c \dd w_1(TM) +{} h\dd w_2(TM) +b\dd B+B\frac{1}{2}\dd B+K_1'w_1(TM)^2\frac{1}{2} \dd B 
+w_2(TM)\frac{1}{2} \dd B. \quad\quad
\eea

\subsubsection{Gauge Invariance}
We assume the following ansatz of the gauge transformations 
\bea
w_1(TM)&\to&w_1(TM)+\dd\alpha_1,\nn\\
w_2(TM)&\to&w_2(TM)+\dd\alpha_2,\nn\\
B&\to&B+\dd\beta,\nn\\
{c}&\to&{c+\dd\gamma_1+\lambda_1},\nn\\
h&\to&{h+\dd\gamma_2+\lambda_2},\nn\\
b&\to&b+\dd\zeta+\mu.
\eea
The gauge variation of the action is
\bea
{\bf S}&\to&{\bf S}+\pi\int_{M^5}  {K_1' }  \dd\gamma_1\dd w_1(TM)+ {K_1' }  \lambda_1\dd w_1(TM)+
 { }  \dd\gamma_2\dd w_2(TM)+ { }  \lambda_2\dd w_2(TM)\nn\\
 &&+\dd\zeta\dd B+\mu\dd B+\dd\beta\frac{1}{2}\dd B+B\frac{1}{2}\dd^2\beta+\dd\beta\frac{1}{2}\dd^2\beta\nn\\
&&+K_1'(w_1(TM)\dd\alpha_1+\dd\alpha_1 w_1(TM)+\dd\alpha_1\dd\alpha_1)\frac{1}{2}\dd  B\nn\\
&&+K_1'(w_1(TM)^2+w_1(TM)\dd\alpha_1+\dd\alpha_1 w_1(TM)+\dd\alpha_1\dd\alpha_1)\frac{1}{2}\dd^2\beta\nn\\
&&+\dd\alpha_2 \frac{1}{2}\dd  B+w_2(TM) \frac{1}{2}\dd^2\beta+\dd\alpha_2\frac{1}{2}\dd^2\beta\\
&=&{\bf S}+\pi\int_{M^5} 
{K_1' }  \lambda_1\dd w_1(TM)+ { }  \lambda_2 \dd w_2(TM)+\mu\dd B+ (\frac{1}{2}\dd\alpha_2) \dd  B+(-\frac{1}{2} \dd \beta)\dd w_2(TM)\nn\\
&&+K_1'\frac{1}{2}(w_1(TM)\dd\alpha_1+\dd\alpha_1 w_1(TM)+\dd\alpha_1\dd\alpha_1)\dd B
\eea
where we have used integration by part. Gauge invariance, i.e.
$\Delta {\bf S}=0$, requires 
\bea
 {K_1' }  \lambda_1&=&0,\nn\\
 { }  \lambda_2&=& \frac{1}{2}\dd\beta,\nn\\
\mu&=&-K_1'\frac{1}{2}(w_1(TM)\dd\alpha_1+\dd\alpha_1 w_1(TM)+\dd\alpha_1\dd\alpha_1)-\frac{1}{2}\dd\alpha_2.
\eea

\subsubsection{Extended 2-Surface/3-Brane Operators and Link Invariants}

The gauge invariant 3-manifold operator is 
\bea
X&=&\exp(\ii\pi k K_1'(\int_{\Sigma^3}c))\nn\\
&=&\exp(\ii\pi k (1+K_1)(\int_{M^5}(\delta^{\perp}(\Sigma^3)c))).
\eea
$X$  is trivial when $K_1'=1+K_1=0$ mod 2. 

There are two types of gauge invariant 2-surface operators, 
\bea
U'&=&\exp(\ii\pi k' (\int_{\Sigma^2}h- \int_{V^3} {\frac{1}{2} \dd B}))\nn\\
&=&\exp(\ii\pi k' (\int_{M^5}(\delta^{\perp}(\Sigma^2)h-\delta^{\perp}(V^3) \frac{1}{2}\dd B)))\nn\\
&=&\exp(\ii\pi k' (\int_{M^5}(\delta^{\perp}(\Sigma^2) (h- \frac{1}{2} B) ))) \\
&=&\exp(\ii\pi k' (\int_{M^5}(\delta^{\perp}(\Sigma^2) (h- \frac{1}{2} B) ))).\nn
\eea
and
\bea
U&=&\exp(\ii\pi\ell(\int_{\Sigma^2}(b+K_1'\frac{1}{2}w_1(TM)^2+\frac{1}{2}w_2(TM))))\nn\\
&=&\exp(\ii\pi\ell(\int_{M^5}(\delta^{\perp}(\Sigma^2)(b+K_1'\frac{1}{2}w_1(TM)^2+\frac{1}{2}w_2(TM)) )))\\
&=&\exp(\ii\pi\ell(\int_{M^5}(\delta^{\perp}(\Sigma^2)(b+(1+K_1)\frac{1}{2}w_1(TM)^2+\frac{1}{2}w_2(TM)) ))).\nn
\eea
where $k,k',\ell\in \Z_2$. 

Inserting $X_{\bf{(i)}},X_{\bf{(ii)}},U', U_{\bf{(i)}}, U_{\bf{(ii)}}$ into the path integral $\bZ$, 
we define the link invariant as
\bea
\langle X_{\bf{(i)}}X_{\bf{(ii)}} U'U_{\bf{(i)}} U_{\bf{(ii)}} \rangle&=&\int [\cD B][\cD b][\cD h][\cD c] \;X_{\bf{(i)}}X_{\bf{(ii)}}U' U_{\bf{(i)}} U_{\bf{(ii)}} \;\exp(\ii \bf{S}). \\
&=&\int [\cD B][\cD b][\cD h][\cD c]
X_{\bf{(i)}}X_{\bf{(ii)}} U'U_{\bf{(i)}} U_{\bf{(ii)}} \;
 \exp(\ii 
\pi \int_{M^5} {K_1' }   c \dd w_1(TM)\nn\\
&&+ { }  h\dd w_2(TM)+b\dd B+ B\frac{1}{2}\dd B+ K_1'w_1(TM)^2\frac{1}{2}\dd B+w_2(TM)\frac{1}{2}\dd B). \nn
\eea
We evaluate the path integral below. 
\begin{enumerate}
    \item Integrating out $c$, 
we get
\bea
 {K_1' }  \dd w_1(TM)&=&K_1'\big(k_{\bf (i)}\delta^{\perp}(\Sigma^3_{X_{\bf{(i)}}})+k_{\bf (ii)}\delta^{\perp}(\Sigma^3_{X_{\bf{(ii)}}})\big),\nn\\
{K_1' } w_1(TM) &=&{K_1' } \big(k_{\bf (i)}\delta^{\perp}(V^4_{X_{\bf{(i)}}})+k_{\bf (ii)}\delta^{\perp}(V^4_{X_{\bf{(ii)}}})\big).
\eea
We keep ${K_1' }$ on both sides because when ${K_1' }=1$ mod 2 we have this constraint; while when 
${K_1' }=0$ mod 2 the constraint is trivial.
Using the above constraints, the path integral boils down to 
 \begin{equation}
 \begin{split}
\langle X_{\bf{(i)}}X_{\bf{(ii)}}U' U_{\bf{(i)}} U_{\bf{(ii)}} \rangle&=\int [\cD B][\cD b][\cD h] \;
U'U_{\bf{(i)}} U_{\bf{(ii)}}
 \exp(\ii 
\pi \int_{M^5}  { }  h\dd w_2(TM)+b\dd B+B\frac{1}{2}\dd B\\
&+ K_1'w_1(TM)^2\frac{1}{2}\dd B+ w_2(TM)\frac{1}{2}\dd B) 
 \bigg\vert 
\text{\tiny{\(
{K_1' w_1(TM)=K_1'\big(k_{\bf (i)}\delta^{\perp}(V^4_{X_{\bf{(i)}}})+k_{\bf (ii)}\delta^{\perp}(V^4_{X_{\bf{(ii)}}})\big)}
 \)}}. 
\quad\quad\quad\label{eq:triple-link3-step1}
 \end{split}
 \end{equation}

\item Integrating out $h$, 
we get
\bea
 { }  \dd w_2(TM)&=&\ k'\delta^{\perp}(\Sigma^2_{U'}),\nn\\
{ } w_2(TM) &=&{ } \ k'\delta^{\perp}(V^3_{U'}).
\eea
Substituting these into the path integral, we find
 \bea
\langle X_{\bf{(i)}}X_{\bf{(ii)}}U' U_{\bf{(i)}} U_{\bf{(ii)}} \rangle&=&\int [\cD B][\cD b] \;
U_{\bf{(i)}} U_{\bf{(ii)}}
 \exp(\ii 
\pi \int_{M^5} b\dd B+B\frac{1}{2}\dd B\nn\\
&&+ {K_1' }  w_1(TM)^2\frac{1}{2}\dd B) \bigg\vert 
\text{\tiny{\(
{ {K_1'w_1(TM)=K_1'\big(k_{\bf (i)}\delta^{\perp}(V^4_{X_{\bf{(i)}}})+k_{\bf (ii)}\delta^{\perp}(V^4_{X_{\bf{(ii)}}})\big),}\atop{w_2(TM) = k'\delta^{\perp}(V^3_{U'}).
\quad\quad\quad\quad\quad\quad\quad\quad\quad\;\;\;}}
 \)}}. 
 \nn\\\label{eq:triple-link3-step2}
\eea

\item Integrate out $b$, 
we get the constraint
 \bea
\dd B&=&\ell_{\bf (i)} \delta^{\perp}(\Sigma^2_{U_{\bf (i)}})+\ell_{\bf (ii)} \delta^{\perp}(\Sigma^2_{U_{\bf (ii)}}),\nn\\
B&=&\ell_{\bf (i)} \delta^{\perp}(V^3_{U_{\bf (i)}})+\ell_{\bf (ii)} \delta^{\perp}(V^3_{U_{\bf (ii)}}).
\eea

\item  We finally integrate out $B$ in \eqn{eq:triple-link3-step1}:
 \bea  
&&\boxed{\langle X_{\bf{(i)}}X_{\bf{(ii)}} U'U_{\bf{(i)}} U_{\bf{(ii)}} \rangle} \nn\\
&&=\int [\cD B] 
\exp( -{\ii \pi}  (\int_{M^5} \frac{1}{2} (K_1'w_1(TM)^2+w_2(TM))(\ell_{\bf (i)} \delta^{\perp}(\Sigma^2_{U_{\bf (i)}})+\ell_{\bf (ii)} \delta^{\perp}(\Sigma^2_{U_{\bf (ii)}}))\nn\\
&&+B\frac{1}{2}\dd B+ K_1'w_1(TM)^2\frac{1}{2}\dd B))
 \bigg\vert
\text{\tiny{\( {{{K_1'w_1(TM)=K_1'\big(k_{\bf (i)}\delta^{\perp}(V^4_{X_{\bf{(i)}}})+k_{\bf (ii)}\delta^{\perp}(V^4_{X_{\bf{(ii)}}})\big),\quad}\atop{
w_2(TM) = k'\delta^{\perp}(V^3_{U'}),\quad\quad\quad\quad\quad\quad\quad\quad\quad\quad\;\;\;} }\atop{{{B=\ell_{\bf (i)} \delta^{\perp}(V^3_{U_{\bf (i)}})+\ell_{\bf (ii)} \delta^{\perp}(V^3_{U_{\bf (ii)}}) \quad\quad\quad\;}}
\quad\quad\quad}} \)}}
\nn\\
 &&=\int [\cD B] 
\exp( -{\ii \pi}  (\int_{M^5} \frac{1}{2} (K_1'w_1(TM)^2+w_2(TM)) \dd B+B\frac{1}{2}\dd B\nn\\
&&+ K_1'w_1(TM)^2\frac{1}{2}\dd B))
 \bigg\vert 
\text{\tiny{\( {{{K_1'w_1(TM)=K_1'\big(k_{\bf (i)}\delta^{\perp}(V^4_{X_{\bf{(i)}}})+k_{\bf (ii)}\delta^{\perp}(V^4_{X_{\bf{(ii)}}})\big),\quad}\atop{
w_2(TM) = k'\delta^{\perp}(V^3_{U'}),\quad\quad\quad\quad\quad\quad\quad\quad\quad\quad\;\;\;} }\atop{{{B=\ell_{\bf (i)} \delta^{\perp}(V^3_{U_{\bf (i)}})+\ell_{\bf (ii)} \delta^{\perp}(V^3_{U_{\bf (ii)}}) \quad\quad\quad\;}}
\quad\quad\quad}} \)}} \nn\\
  \label{eq:triple-link3-step3-1}\\
    &&=
  \exp( -\ii \pi \Big( K_1'k_{\bf (i)}k_{\bf (ii)}  \cdot
 2 \#(V^4_{X_{\bf{(i)}}}\cap V^4_{X_{\bf{(ii)}}}\cap(\ell_{\bf (i)} \delta^{\perp}(\Sigma^2_{U_{\bf (i)}})+\ell_{\bf (ii)} \delta^{\perp}(\Sigma^2_{U_{\bf (ii)}})))\nn\\
 && 
 +(\frac{k'\ell_{\bf (i)}}{2}\cdot\#(V^3_{U'}\cap\Sigma^2_{U_{\bf (i)}})+\frac{k'\ell_{\bf (ii)}}{2}\cdot\#(V^3_{U'}\cap\Sigma^2_{U_{\bf (ii)}}))
 +\frac{\ell_{\bf (i)}\ell_{\bf (ii)}}{2} \cdot(\#(V^3_{U_{\bf (i)}}\cap \Sigma^2_{U_{\bf (ii)}})+ \#(V^3_{U_{\bf (ii)}}\cap \Sigma^2_{U_{\bf (i)}}))
   \Big)) \nn\\
  &&\quad\quad \cdot (\cdots) \cdot (\text{Self-intersecting $\#$ terms})
  \label{eq:triple-link3-step3-2}
  \\
  &&\cong
  \exp( -\ii \pi \Big(  K_1'(k_{\bf (i)}k_{\bf (ii)} \ell_{\bf (i)}\cdot
  2\text{Tlk}^{(5)}(\Sigma^3_{X_{\bf{(i)}}},\Sigma^3_{X_{\bf{(ii)}}},\Sigma^2_{U_{\bf (i)}})+k_{\bf (i)}k_{\bf (ii)} \ell_{\bf (ii)}\cdot
  2\text{Tlk}^{(5)}(\Sigma^3_{X_{\bf{(i)}}},\Sigma^3_{X_{\bf{(ii)}}},\Sigma^2_{U_{\bf (ii)}}))\nn\\
  &&
  +(\frac{k'\ell_{\bf (i)}}{2}\cdot\text{Lk}^{(5)}(\Sigma^2_{U'},\Sigma^2_{U_{\bf (i)}})+\frac{k'\ell_{\bf (ii)}}{2}\cdot\text{Lk}^{(5)}(\Sigma^2_{U'},\Sigma^2_{U_{\bf (ii)}}))
  +\ell_{\bf (i)}\ell_{\bf (ii)}\cdot\text{Lk}^{(5)}(\Sigma^2_{U_{\bf (i)}},\Sigma^2_{U_{\bf (ii)}})
    \Big))\cdot (\cdots).
  \label{eq:triple-link3-step3-3}
 \eea
We propose a set-up to remove or renormalize the (\text{Self-intersecting $\#$ terms}) appeared in \eqn{eq:triple-link3-step3-2}, 
following the same strategy as footnote \ref{footnote:triple}.
 
\end{enumerate}

For ${\bf S}=\pi\int_{M^5} {K_1'} c \dd w_1(TM) + {} h \dd w_2(TM)+b\dd B+B\frac{1}{2}\dd B+K_1'w_1(TM)^2\frac{1}{2}\dd B+w_2(TM)\frac{1}{2}\dd B$,
we derive the link invariant for the 5d TQFT $\bZ_{\text{SET}}[M^5]$ in \eqn{eq:triple-link3-step3-2} and \eqn{eq:triple-link3-step3-3}: \\
{
\fbox{\parbox{7in}{\parindent=0pt 
\bea
&&{ K_1'k_{\bf (i)}k_{\bf (ii)}  \cdot
 2 \#(V^4_{X_{\bf{(i)}}}\cap V^4_{X_{\bf{(ii)}}}\cap(\ell_{\bf (i)} \delta^{\perp}(\Sigma^2_{U_{\bf (i)}})+\ell_{\bf (ii)} \delta^{\perp}(\Sigma^2_{U_{\bf (ii)}})))}\nn \\
  &&{
    +(\frac{k'\ell_{\bf (i)}}{2}\cdot\#(V^3_{U'}\cap\Sigma^2_{U_{\bf (i)}})+\frac{k'\ell_{\bf (ii)}}{2}\cdot\#(V^3_{U'}\cap\Sigma^2_{U_{\bf (ii)}}))
  +\frac{\ell_{\bf (i)}\ell_{\bf (ii)}}{2} \cdot(\#(V^3_{U_{\bf (i)}}\cap \Sigma^2_{U_{\bf (ii)}})+ \#(V^3_{U_{\bf (ii)}}\cap \Sigma^2_{U_{\bf (i)}}))
  }\nn\\
  &\equiv&{(1+K_1)(k_{\bf (i)}k_{\bf (ii)} \ell_{\bf (i)}\cdot
  2\text{Tlk}^{(5)}_{w_1w_1 \dd B}(\Sigma^3_{X_{\bf{(i)}}},\Sigma^3_{X_{\bf{(ii)}}},\Sigma^2_{U_{\bf (i)}})+k_{\bf (i)}k_{\bf (ii)} \ell_{\bf (ii)}\cdot
  2\text{Tlk}^{(5)}_{w_1w_1 \dd B}(\Sigma^3_{X_{\bf{(i)}}},\Sigma^3_{X_{\bf{(ii)}}},\Sigma^2_{U_{\bf (ii)}}))}\nn\\
 && {
   +(\frac{k'\ell_{\bf (i)}}{2}\cdot\text{Lk}^{(5)}_{w_2 \dd B}(\Sigma^2_{U'},\Sigma^2_{U_{\bf (i)}})+\frac{k'\ell_{\bf (ii)}}{2}\cdot
  \text{Lk}^{(5)}_{w_2 \dd B}(\Sigma^2_{U'},\Sigma^2_{U_{\bf (ii)}})
  )
 +\ell_{\bf (i)}\ell_{\bf (ii)}\cdot\text{Lk}^{(5)}_{B \dd B}(\Sigma^2_{U_{\bf (i)}},\Sigma^2_{U_{\bf (ii)}})
  }.\nn\\
  \label{eq:link-all-K1K2}
\eea 
}}
}\\
The path integral with appropriate extended operators insertions becomes
\eqn{eq:triple-link3-step3-3} which provides the above  link invariant.

\subsubsection{$(K_1,K_2)=(0,0)$ or $(0,1)$: 1st and 3rd Sibling}
\label{sec:1-SET-sibling}

The $\bZ_{\text{SET}_{(K_1=0,K_2=0)}}^{5\text{d}}$ gives rise to a 5d triple link invariant: 
\begin{itemize}
    \item $\text{Tlk}^{(5)}_{w_1BB}$ in \Sec{sec:link-inv-wPB}'s \Eq{w1PB-Link}. 
We present an exemplary link configuration later in (\Sec{sec:triple-link-w1BB}) that can be detected by this link invariant. 
\end{itemize}

In another equivalent expression, 
$\bZ_{\text{SET}_{(K_1=0,K_2=0)}}^{5\text{d}}$ in \eqn{eq:5dSET-cont} gives rise to other link invariants in \eqn{eq:link-all-K1K2} 
including:
\begin{itemize}
    \item $\text{Tlk}^{(5)}_{w_1w_1 \dd B}(\Sigma^3_{X_{\bf{(i)}}},\Sigma^3_{X_{\bf{(ii)}}},\Sigma^2_{U_{\bf}})$, a second type of triple link in 5d (although seemly undetectable due to an exponent factor $2 \pi$ in the expectation value).
We present an exemplary link configuration later in (\Sec{sec:triple-link-w1w1dB}) that can be detected by this link invariant.

\item $\text{Lk}^{(5)}_{B \dd B}(\Sigma^2_{U_{\bf (i)}},\Sigma^2_{U_{\bf (ii)}})$, a quadratic link of 2-surfaces in 5d. 
We present an exemplary link configuration later in (\Sec{sec:quadratic-link-BdB}) that can be detected by this link invariant.

\item $  \text{Lk}^{(5)}_{w_2 \dd B}(\Sigma^2_{U'},\Sigma^2_{U})$, another quadratic link of 2-surfaces in 5d. 
We present an exemplary link configuration later in (\Sec{sec:quadratic-link-w2dB}) that can be detected by this link invariant. 
\end{itemize}

Physically, these link invariants may be related to each other by re-arranging the spacetime braiding process of strings/branes. 
It will be interesting to find a precise mathematical equality to relate these link invariants.

\subsubsection{$(K_1,K_2)=(1,0)$ or $(1,1)$: 2nd and 4th Sibling}

$\bZ_{\text{SET}_{(K_1=1,K_2=0)}}^{5\text{d}}$ in \eqn{eq:5dSET-cont} gives rise to link invariants in \eqn{eq:link-all-K1K2} 
including:
\begin{itemize}
    \item $\text{Lk}^{(5)}_{B \dd B}(\Sigma^2_{U_{\bf (i)}},\Sigma^2_{U_{\bf (ii)}})$, a quadratic link of 2-surfaces in 5d. 
We present an exemplary link configuration later in (\Sec{sec:quadratic-link-BdB}) that can be detected by this link invariant.

\item $  \text{Lk}^{(5)}_{w_2 \dd B}(\Sigma^2_{U'},\Sigma^2_{U})$, another quadratic link of 2-surfaces in 5d. 
We present an exemplary link configuration later in (\Sec{sec:quadratic-link-w2dB}) that can be detected by this link invariant. 
\end{itemize}

Similar to our comments above in \Sec{sec:1-SET-sibling}, it will be interesting to find a precise mathematical equality to relate these link invariants.

\section{Anyonic String/Brane Spacetime Braiding Processes and Link Configurations of Extended Operators}
\label{sec:link-conf}

We provide the exemplary 
spacetime braiding processes of anyonic strings and branes in general dimensions (with an emphasis on $5$d), and the link configurations of extended operators,
which can be detected by the link invariants that we derived in \Sec{sec:link-inv}.

\subsection{Quadratic Link (Aharanov-Bohm) Configuration in Any Dimension}

To warm up, we first discuss the quadratic link, associated with the Aharanov-Bohm statistics in $d$d due to the linking of  1-worldline of the charged particle and the $(d-2)$d-worldsheet of the fractional flux.
In 3d spacetime, we have the following presentation
\begin{equation*}
\begin{tikzpicture}
    \draw[color=red] (-1.5,0) arc (180:360:1.5cm and 0.75cm);
    \draw[color=red] (-1.5,0) arc (180:0:1.5cm and 0.75cm);
    \draw[fill=blue,color=blue] (0,0) circle(0.05);
    \node at (2.25,0) {$\times$};
    \node at (-1.,0) {$D^2$};
         \node at (1.25,1) {$S^1$};

     \draw[color=blue] (3,0) arc (180:360:1.5cm and 0.75cm);
    \draw[color=blue] (3,0) arc (180:0:1.5cm and 0.75cm);

    \node at (5.5,1) {$S^1$};
      \node at (3.5,0) {$D^2$};
    \draw[fill=red,color=red] (4.5,0) circle(0.05);
\end{tikzpicture}
\end{equation*}
where gluing two solid tori $D^2 \times S^1$ gives rise to a 3-sphere:
$(D^2_{\rm L} \times S^1_{\rm R})\cup  (S^1_{\rm L} \times D^2_{\rm R})=S^3$. 
We represent the two solid tori as a blue solid tori  and a red solid tori, 
$(\ccblue{D^2_{\rm L} \times S^1_{\rm R}})\cup  (\ccred{S^1_{\rm L} \times D^2_{\rm R}})=S^3$.
The quadratic link invariant detecting this Aharanov-Bohm configuration is given by (\Refe{Putrov2016qdo1612.09298} and References therein):
 ${\rm Lk}( {(0_{\rm pt})_{\rm L} \times S^1_{\rm R}} , {S^1_{\rm L} \times (0_{\rm pt})_{\rm R}} )$,
 which we also express as
\bea
  {\rm Lk}( \ccblue{(0_{\rm pt})_{\rm L} \times S^1_{\rm R}} , \ccred{S^1_{\rm L} \times (0_{\rm pt})_{\rm R}} )
\eea
based on the color labeling of the inclusion of two $S^1$ circles belonging to  which of the two solid tori.
This link invariant can be computed from the intersection number,
\begin{equation*}
\begin{tikzpicture}
  \draw[color=red] (-1.5,0) arc (180:360:1.5cm and 0.75cm);
    \draw[color=red] (-1.5,0) arc (180:0:1.5cm and 0.75cm);
\strand [ultra thick, red,fill=red!20!white,opacity=0.20] (-1.5,0) arc (180:360:1.5cm and 0.75cm);
   \strand[ultra thick, red,fill=red!20!white,opacity=0.20] (-1.5,0) arc (180:0:1.5cm and 0.75cm);

    \draw[fill=black,color=black] (0,0) circle(0.05);
    \node at (2.25,0) {$\times$};
    \node at (-1.,0) {$D^2$};
         \node at (1.25,1) {$S^1$};

     \draw[color=blue] (3,0) arc (180:360:1.5cm and 0.75cm);
    \draw[color=blue] (3,0) arc (180:0:1.5cm and 0.75cm);

\draw[color=red] (3,0) -- (4.5,0);      
    \draw[fill=black,color=black] (3,0) circle(0.05);

    \node at (5.5,1) {$S^1$};
      \node at (3.5,0) {$D^2$};
    \draw[fill=red,color=red] (4.5,0) circle(0.05);

\end{tikzpicture}
\end{equation*}
where
\bea
\#(( \ccblue{0_{\rm pt})_{\rm L} \times S^1_{\rm R}}) \cap (\ccred{D^2_{\rm L} \times (0_{\rm pt-})_{\rm R}}))=1.
\eea
$(0_{\rm pt-})$ means the point $(0_{\rm pt})$ now is attached to a line.
The intersection number 
$\#(( \ccblue{0_{\rm pt})_{\rm L} \times S^1_{\rm R}}) \cap (\ccred{D^2_{\rm L} \times (0_{\rm pt-})_{\rm R}}))=1$ 
 precisely corresponds to the black dot $\bullet$.


In $d$d spacetime,  $S^d$ can be obtained by
\bea
(\ccblue{D^{d-1}_{\rm L} \times S^1_{\rm R}})\cup  (\ccred{S^{d-2}_{\rm L} \times D^2_{\rm R}})=S^d.
\eea
which can be graphically represented as 
\begin{equation*}
\begin{tikzpicture}
    \draw[color=red] (-1.5,0) arc (180:360:1.5cm and 0.75cm);
    \draw[color=red] (-1.5,0) arc (180:0:1.5cm and 0.75cm);
    \draw[fill=blue,color=blue] (0,0) circle(0.05);
    \node at (2.25,0) {$\times$};
    \node at (-1.,0) {$D^{d-1}$};
         \node at (1.25,1) {$S^{d-2}$};

     \draw[color=blue] (3,0) arc (180:360:1.5cm and 0.75cm);
    \draw[color=blue] (3,0) arc (180:0:1.5cm and 0.75cm);

    \node at (5.5,1) {$S^1$};
      \node at (3.5,0) {$D^2$};
    \draw[fill=red,color=red] (4.5,0) circle(0.05);
\end{tikzpicture}
\end{equation*}
The associated link invariant is 
\bea  
{\rm Lk}( \ccblue{(0_{\rm pt})_{\rm L} \times S^1_{\rm R}} , \ccred{S^{d-2}_{\rm L} \times (0_{\rm pt})_{\rm R}} )
\eea
with the color prescription explained earlier.
This link invariant can be computed from the intersection number,
\begin{equation*}
\begin{tikzpicture}
    \draw[color=red] (-1.5,0) arc (180:360:1.5cm and 0.75cm);
    \draw[color=red] (-1.5,0) arc (180:0:1.5cm and 0.75cm);
\strand [ultra thick, red,fill=red!20!white,opacity=0.20] (-1.5,0) arc (180:360:1.5cm and 0.75cm);
   \strand[ultra thick, red,fill=red!20!white,opacity=0.20] (-1.5,0) arc (180:0:1.5cm and 0.75cm);
 \shade[ball color=red!80!white, opacity=0.20]  (0,0) ellipse (1.5cm and 0.75cm);
     
    \draw[fill=black,color=black] (0,0) circle(0.05);
    \node at (2.25,0) {$\times$};
    \node at (-1.,0) {$D^{d-1}$};
         \node at (1.25,1) {$S^{d-2}$};

     \draw[color=blue] (3,0) arc (180:360:1.5cm and 0.75cm);
    \draw[color=blue] (3,0) arc (180:0:1.5cm and 0.75cm);

\draw[color=red] (3,0) -- (4.5,0);      
    \draw[fill=black,color=black] (3,0) circle(0.05);

    \node at (5.5,1) {$S^1$};
      \node at (3.5,0) {$D^2$};
    \draw[fill=red,color=red] (4.5,0) circle(0.05);
\end{tikzpicture}
\end{equation*}
\bea
\#(( \ccblue{0_{\rm pt})_{\rm L} \times S^1_{\rm R}}) \cap (\ccred{D^{d-1}_{\rm L} \times (0_{\rm pt-})_{\rm R}}))=1.
\eea 
Here, $(0_{\rm pt-})$ means the point $(0_{\rm pt})$ now is attached with a line.
We see the intersection number 
$\#(( \ccblue{0_{\rm pt})_{\rm L} \times S^1_{\rm R}}) \cap (\ccred{D^{d-1}_{\rm L} \times (0_{\rm pt-})_{\rm R}}))=1
$ 
 precisely corresponds to the black dot $\bullet$.



\subsection{The 1st Triple Link
${ \#(V^4_X\cap V^3_{U_{\bf (i)}}\cap V^3_{U_{\bf (ii)}})
\equiv\text{Tlk}^{(5)}_{w_1 BB}(\Sigma^3_X,\Sigma^2_{U_{\bf (i)}},\Sigma^2_{U_{\bf (ii)}})}$ Configuration in 5d}
\label{sec:triple-link-w1BB}
We proceed to discuss the triple link configuration for $\text{Tlk}^{(5)}_{w_1 BB}(\Sigma^3_X,\Sigma^2_{U_{\bf (i)}},\Sigma^2_{U_{\bf (ii)}})$ 
derived in \Sec{sec:link-inv-wPB}.\footnote{Effectively,
$\text{Tlk}^{(5)}_{w_1 BB}(\Sigma^3_X,\Sigma^2_{U_{\bf (i)}},\Sigma^2_{U_{\bf (ii)}})$ can be also regarded as
$\text{Tlk}^{(5)}_{ABB}(\Sigma^3_X,\Sigma^2_{U_{\bf (i)}},\Sigma^2_{U_{\bf (ii)}})$ where $A$ is other $\Z_n$ 1-form gauge field.
}
We propose that this link invariant  derived in \Sec{sec:link-inv-wPB} can detect the link configuration in \fig{fig:link-inv-wPB-1}.
\begin{figure}[!h]
\centering
\includegraphics[scale=1.05]{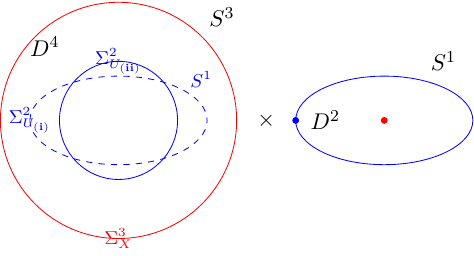}
\caption{$S^5=\partial D^6=\partial(D^4\times D^2)=S^3\times D^2\cup D^4\times S^1=S^3\times D^2\cup D^2\times D^2\times S^1$, the intersection of the two copies of $D^2\times S^1$ in the second piece
($D^2\times 0_{\text{pt}}\times S^1$ and $0_{\text{pt}}\times D^2\times S^1$) is $0_{\text{pt}}\times 0_{\text{pt}}\times S^1=0_{\text{pt}}\times S^1$, this $0_{\text{pt}}\times S^1$ and $S^3\times0_{\text{pt}}$ in the first piece are linked. In this figure, $\Sigma^3_X=S^3\times0_{\text{pt}}$,
$\Sigma^2_{U_{\bf (i)}}=\partial(D^2\times 0_{\text{pt}}\times S^1)$, $\Sigma^2_{U_{\bf (ii)}}=\partial(0_{\text{pt}}\times D^2\times S^1)$.}
\label{fig:link-inv-wPB-1}
\end{figure}
\begin{figure}[!h]
\centering
\includegraphics[scale=1.05]{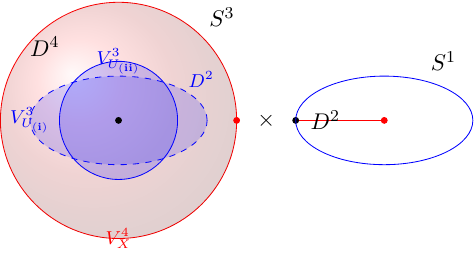}
\caption{Following the last \Fig{fig:link-inv-wPB-1}, $V^4_X=D^4\times0_{\rm pt}$ which bounds $\Sigma^3_X$, $V^3_{U_{\bf (i)}}=D^2\times 0_{\text{pt}}\times S^1$ which bounds $\Sigma^2_{U_{\bf (i)}}$, $V^3_{U_{\bf (ii)}}=0_{\text{pt}}\times D^2\times S^1$ which bounds $\Sigma^2_{U_{\bf (ii)}}$. The intersection of $V^3_{U_{\bf (i)}}$ and $V^3_{U_{\bf (ii)}}$ is $0_{\text{pt}}\times S^1$, the intersection of $V^4_X$ and this $0_{\text{pt}}\times S^1$ is a point which is the point in black in this figure.}
\label{fig:link-inv-wPB-2}
\end{figure}

To explain, we start by constructing the 5-sphere via $S^5=\partial D^6=\partial(D^4\times D^2)=S^3\times D^2\cup D^4\times S^1=S^3\times D^2\cup D^2\times D^2\times S^1$.
More explicitly, we color the different components as
$S^5= (\ccred{S^3_{\rm L}\times D^2_{\rm R}})\cup (\ccblue{D^4_{\rm L}\times S^1_{\rm R}})$
and 
$S^5
= (\ccred{S^3_{\rm L}\times D^2_{\rm R}})\cup (\ccblue{D^2_{\rm L}\times D^2_{\rm L}\times S^1_{\rm R}})$.

Consider the link invariant defined by
${ \#(V^4_X\cap V^3_{U_{\bf (i)}}\cap V^3_{U_{\bf (ii)}})
\equiv\text{Tlk}^{(5)}_{w_1 BB}(\Sigma^3_X,\Sigma^2_{U_{\bf (i)}},\Sigma^2_{U_{\bf (ii)}})}$,
we see that the link configuration in \fig{fig:link-inv-wPB-1} gives the intersection number 1
in  \fig{fig:link-inv-wPB-2}.
Again in  \fig{fig:link-inv-wPB-2} associated with the intersection number  ${ \#(V^4_X\cap V^3_{U_{\bf (i)}}\cap V^3_{U_{\bf (ii)}})}$, 
$(0_{\rm pt-})$ means the point $(0_{\rm pt})$ now is attached to a line.
We see the intersection number 
${ \#(V^4_X\cap V^3_{U_{\bf (i)}}\cap V^3_{U_{\bf (ii)}})}=1$ precisely corresponds to  the black dot $\bullet$.


\subsection{
The 2nd Triple Link  
${ \#(V^4_{X_{\bf (i)}}\cap V^4_{X_{\bf (ii)}}\cap \Sigma^2_U)\equiv
\text{Tlk}^{(5)}_{w_1 w_1\dd B}(\Sigma^3_{X_{\bf (i)}},\Sigma^3_{X_{\bf (ii)}},\Sigma^2_U)
}$
 Configuration in 5d}
\label{sec:triple-link-w1w1dB}

We now discuss 
$
\text{Tlk}^{(5)}_{w_1 w_1\dd B}(\Sigma^3_{X_{\bf (i)}},\Sigma^3_{X_{\bf (ii)}},\Sigma^2_U)
$,
or schematically $\text{Tlk}^{(5)}_{AA\dd B}(\Sigma^3_{X_{\bf (i)}},\Sigma^3_{X_{\bf (ii)}},\Sigma^2_U)
$.
This link invariant is derived in \Sec{sec:link-inv-w13B-version-2}.

\begin{figure}[!h]
\centering
{
\includegraphics[scale=0.85]{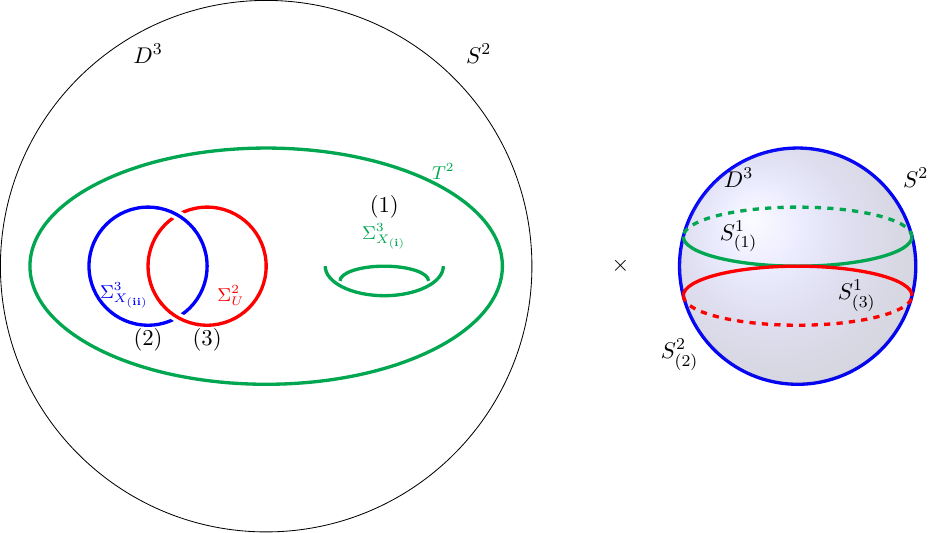}
}
\caption{$S^5=\partial D^6=\partial(D^3\times D^3)=S^2\times D^3\cup D^3\times S^2$. Put a 2-torus (denoted by (1)) in $D^3\times 0_{\text{pt}}$, and put a Hopf link (the two circles are denoted by (2) and (3) respectively) in the solid 2-torus. Put two circles (denoted by $S^1_{(1)}$ and $S^1_{(3)}$ respectively) which intersect in only one point in $0_{\text{pt}}\times S^2$ (denoted by $S^2_{(2)}$). In this figure, $\Sigma^3_{X_{\bf (i)}}$ is the cartesian product of the 2-torus (1) and $S^1_{(1)}$, $\Sigma^3_{X_{\bf (ii)}}$ is the cartesian product of the circle (2) and $S^2_{(2)}$, $\Sigma^2_U$ is the cartesian product of the circle (3) and $S^1_{(3)}$.}
\label{fig:link-inv-3}
\end{figure}

\begin{figure}[!h]
\centering
{
\includegraphics[scale=0.85]{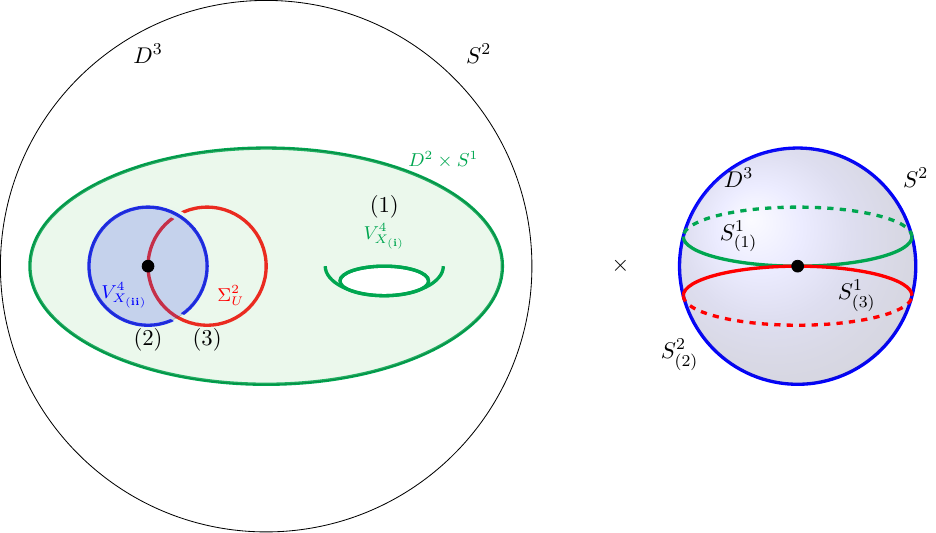}
}
\caption{Following the last \Fig{fig:link-inv-3}, if we fill in $\Sigma^3_{X_{\bf (i)}}$ and $\Sigma^3_{X_{\bf (ii)}}$, we get $V^4_{X_{\bf (i)}}=D^2\times S^1\times S^1$ and $V^4_{X_{\bf (ii)}}=D^2\times S^2$, $V^4_{X_{\bf (i)}}$, $V^4_{X_{\bf (ii)}}$ and $\Sigma^2_U$ will intersect in only one point which is the point in black in this figure.}
\label{fig:link-inv-4}
\end{figure}

Let us consider the link invariant defined by
${{ \#(V^4_{X_{\bf (i)}}\cap V^4_{X_{\bf (ii)}}\cap \Sigma^2_U)\equiv
\text{Tlk}^{(5)}_{w_1 w_1\dd B}(\Sigma^3_{X_{\bf (i)}},\Sigma^3_{X_{\bf (ii)}},\Sigma^2_U)
}}$. 
We see that the link configuration in \fig{fig:link-inv-3} gives the intersection number 1
in  \fig{fig:link-inv-4}.

\subsection{Quadruple Link\\
 ${ \#(V^4_{X_{\bf (i)}}\cap V^4_{X_{\bf (ii)}}\cap V^4_{X_{\bf (iii)}}\cap V^3_U)\equiv\text{Qlk}^{(5)}_{w_1w_1w_1 B}(\Sigma^3_{X_{\bf (i)}},\Sigma^3_{X_{\bf (ii)}},\Sigma^3_{X_{\bf (iii)}} ,\Sigma^2_U)}$ Configuration in 5d}
 \label{sec:quadruple-link-w1w1w1B}
 
We now discuss $\text{Qlk}^{(5)}_{w_1w_1w_1 B}(\Sigma^3_{X_{\bf (i)}},\Sigma^3_{X_{\bf (ii)}},\Sigma^3_{X_{\bf (iii)}})$, or schematically
$\text{Qlk}^{(5)}_{aaa b}(\Sigma^3_{X_{\bf (i)}},\Sigma^3_{X_{\bf (ii)}},\Sigma^3_{X_{\bf (iii)}})$. This link invariant is derived in \Sec{sec:link-inv-w13B-version-1}.

\begin{figure}[!h]
\centering
{
\includegraphics[scale=1.25]{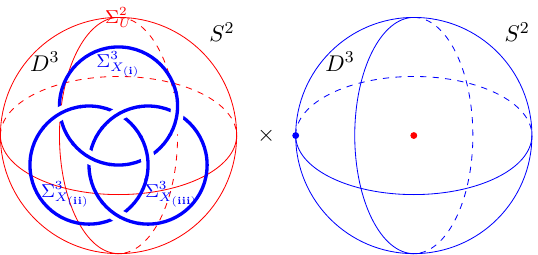}
}
\caption{$S^5=\partial D^6=\partial(D^3\times D^3)=S^2\times D^3\cup D^3\times S^2$. Put Borromean rings in $D^3\times 0_{\text{pt}}$, If we fill in each of the three circles of the Borromean rings, then we get an intersection point, we can think of this point as $0_{\text{pt}}$ in $D^3$, then the cartesian product of each of the three circles and $S^2$ (denoted by $\Sigma^3_{X_{\bf (i)}}$, $\Sigma^3_{X_{\bf (ii)}}$ and $\Sigma^3_{X_{\bf (iii)}}$ respectively) will intersect in $0_{\text{pt}}\times S^2$, this $0_{\text{pt}}\times S^2$ and $S^2\times0_{\text{pt}}$ ($\Sigma^2_U$ in this figure) are linked.}
\label{fig:link-inv-5}
\end{figure}

\begin{figure}[!h]
\centering
{
\includegraphics[scale=1.25]{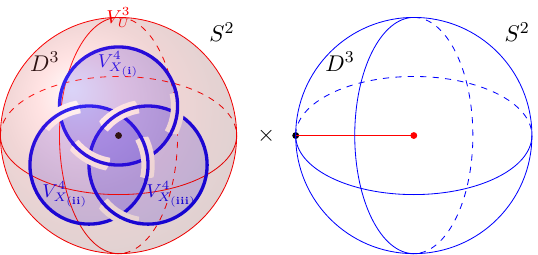}
}
\caption{Following the last \Fig{fig:link-inv-5}, we denote the three $D^2\times S^2$ which bound the cartesian product of the three circles and $S^2$ as $V^4_{X_{\bf (i)}}$, $V^4_{X_{\bf (ii)}}$, $V^4_{X_{\bf (iii)}}$ respectively. The intersection of $V^4_{X_{\bf (i)}}$, $V^4_{X_{\bf (ii)}}$ and $V^4_{X_{\bf (iii)}}$ is $0_{\text{pt}}\times S^2$. The intersection of $V^3_U=D^3\times0_{\text{pt}}$ which bounds $\Sigma^2_U$ and $0_{\text{pt}}\times S^2$ is a point which is the point in black in this figure.}
\label{fig:link-inv-6}
\end{figure}
Let us consider the link invariant defined by
$${ \#(V^4_{X_{\bf (i)}}\cap V^4_{X_{\bf (ii)}}\cap V^4_{X_{\bf (iii)}}\cap V^3_U)\equiv\text{Qlk}^{(5)}_{w_1w_1w_1 B}(\Sigma^3_{X_{\bf (i)}},\Sigma^3_{X_{\bf (ii)}},\Sigma^3_{X_{\bf (iii)}} ,\Sigma^2_U)}$$. 
We see that the link configuration in \fig{fig:link-inv-5} gives the intersection number 1
in  \fig{fig:link-inv-6}.

\subsection{Quadratic Link $ \#(V^3_{U_{\bf (i)}}\cap \Sigma^2_{U_{\bf (ii)}})\equiv\text{Lk}^{(5)}_{B\dd B}(\Sigma^2_{U_{\bf (i)}},\Sigma^2_{U_{\bf (ii)}}) $}
\label{sec:quadratic-link-BdB}

Now we discuss  $\text{Lk}^{(5)}_{B\dd B}(\Sigma^2_{U_{\bf (i)}},\Sigma^2_{U_{\bf (ii)}})$.
This link invariant is derived in \Sec{sec:link-inv-all}.
\begin{figure}[!h]
\centering
{
\includegraphics{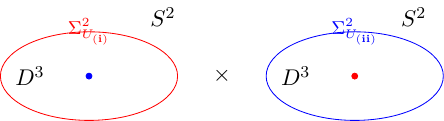}
}
\caption{$S^5=\partial D^6=\partial(D^3\times D^3)=S^2\times D^3\cup D^3\times S^2$. The $S^2\times 0_{\text{pt}}$ in the first piece and the $0_{\text{pt}}\times S^2$ in the second piece are linked. In this figure, $\Sigma^2_{U_{\bf (i)}}=S^2\times 0_{\text{pt}}$, $\Sigma^2_{U_{\bf (ii)}}=0_{\text{pt}}\times S^2$.}
\label{fig:link-inv-7}
\end{figure}
\begin{figure}[!h]
\centering
{
\includegraphics{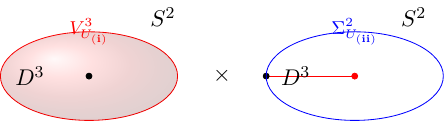}
}
\caption{Following the last \Fig{fig:link-inv-7}, if we fill in $S^2\times0_{\text{pt}}$, we get $V^3_{U_{\bf (i)}}=D^3\times 0_{\text{pt}}$, the intersection of $D^3\times0_{\text{pt}}$ and $0_{\text{pt}}\times S^2$ is a point which is the point in black in this figure.}
\label{fig:link-inv-8}
\end{figure}

Let us consider the link invariant defined by
$${ \#(V^4_{X_{\bf (i)}}\cap V^4_{X_{\bf (ii)}}\cap V^4_{X_{\bf (iii)}}\cap V^3_U)\equiv\text{Qlk}^{(5)}_{w_1w_1w_1 B}(\Sigma^3_{X_{\bf (i)}},\Sigma^3_{X_{\bf (ii)}},\Sigma^3_{X_{\bf (iii)}} ,\Sigma^2_U)}$$.
We see that the link configuration in \fig{fig:link-inv-7} gives the intersection number 1
in  \fig{fig:link-inv-8}.

\subsection{Quadratic Link $ \#(V^3_{U'}\cap \Sigma^2_{U})\equiv\text{Lk}^{(5)}_{w_2 \dd B}(\Sigma^2_{U_{}},\Sigma^2_{U_{}'}) $}
\label{sec:quadratic-link-w2dB}

Now we discuss $\text{Lk}^{(5)}_{w_2\dd B}(\Sigma^2_{U'},\Sigma^2_U)$
or $\text{Lk}^{(5)}_{B'\dd B}(\Sigma^2_{U'},\Sigma^2_U)$. This link invariant is derived in \Sec{sec:link-inv-w2dB}.

\begin{figure}[!h]
\centering
{
\includegraphics{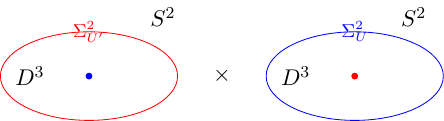}
}
\caption{$S^5=\partial D^6=\partial(D^3\times D^3)=S^2\times D^3\cup D^3\times S^2$. The $S^2\times 0_{\text{pt}}$ in the first piece and the $0_{\text{pt}}\times S^2$ in the second piece are linked. In this figure, $\Sigma^2_{U'}=S^2\times 0_{\text{pt}}$, $\Sigma^2_U=0_{\text{pt}}\times S^2$.}
\label{fig:link-inv-9}
\end{figure}

\begin{figure}[!h]
\centering
{
\includegraphics{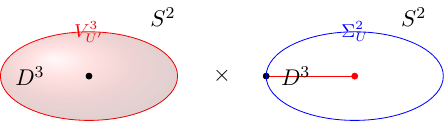}
}
\caption{Following the last \Fig{fig:link-inv-9}, if we fill in $S^2\times0_{\text{pt}}$, we get $V^3_{U'}=D^3\times 0_{\text{pt}}$, the intersection of $D^3\times0_{\text{pt}}$ and $0_{\text{pt}}\times S^2$ is a point which is the point in black in this figure.}
\label{fig:link-inv-10}
\end{figure}
Let us consider the link invariant defined by
$ \#(V^3_{U'}\cap \Sigma^2_{U})\equiv\text{Lk}^{(5)}_{w_2 \dd B}(\Sigma^2_{U_{}},\Sigma^2_{U_{}'}) $. 
We see that the link configuration in \fig{fig:link-inv-9} gives the intersection number 1
in  \fig{fig:link-inv-10}.

\subsection{The 3rd Triple Link  
${ \#(V^4_{X_{\bf (i)}}\cap \Sigma^3_{X_{\bf (ii)}}\cap V^3_U)\equiv
\text{Tlk}^{(5)}_{(A \dd A) B}(\Sigma^3_{X_{\bf (i)}},\Sigma^3_{X_{\bf (ii)}},\Sigma^2_U)
}$
 Configuration in 5d}
\label{sec:triple-link-AdAB}

Finally, we discuss a third triple link invariant 
${ \#(V^4_{X_{\bf (i)}}\cap \Sigma^3_{X_{\bf (ii)}}\cap V^3_U)\equiv
\text{Tlk}^{(5)}_{(A \dd A) B}(\Sigma^3_{X_{\bf (i)}},\Sigma^3_{X_{\bf (ii)}},\Sigma^2_U)
}$. We have not derived these from 4d YM-5d SET coupled systems.
However,  
 to get this, we need a topological term $(w_1(TM) \dd w_1(TM)) B$. This is possible however from $(A_I \dd A_J) B$ type of TQFTs.
 %
 %
\begin{figure}[!h]
\centering
{
\includegraphics[scale=1.15]{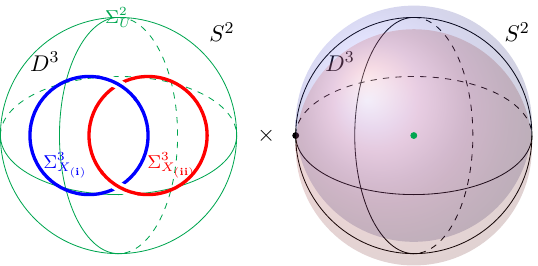}
}
\caption{$S^5=\partial D^6=\partial(D^3\times D^3)=S^2\times D^3\cup D^3\times S^2$, put a Hopf link in $D^3\times0_{\rm pt}$. In this figure, $\Sigma^3_{X_{\bf (i)}}$ and $\Sigma^3_{X_{\bf (i)}}$ are the cartesian product of the two circles in the Hopf link and $S^2$ respectively, namely, they are both $S^1\times S^2$, $\Sigma^2_U=S^2\times0_{\rm pt}$.}
\label{fig:link-inv-11}
\end{figure}

\begin{figure}[!h]
\centering
{
\includegraphics[scale=1.15]{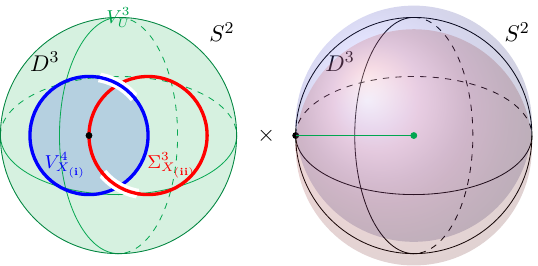}
}
\caption{Following the last \Fig{fig:link-inv-11}, if we fill in $\Sigma^3_{X_{\bf (i)}}$, we get $V^4_{X_{\bf (i)}}=D^2\times S^2$, the intersection of $V^4_{X_{\bf (i)}}$ and $\Sigma^3_{X_{\bf (ii)}}$ is the cartesian product of a point (we can think of the point as $0_{\rm pt}$) and $S^2$. If we fill in $\Sigma^2_U$ further, we get $V^3_U=D^3\times0_{\rm pt}$, the intersection of $D^3\times0_{\rm pt}$ and $0_{\rm pt}\times S^2$ is a point which is the point in black in this figure.}
\label{fig:link-inv-12}
\end{figure}

Let us consider the link invariant defined by
${ \#(V^4_{X_{\bf (i)}}\cap \Sigma^3_{X_{\bf (ii)}}\cap V^3_U)\equiv
\text{Tlk}^{(5)}_{(A \dd A) B}(\Sigma^3_{X_{\bf (i)}},\Sigma^3_{X_{\bf (ii)}},\Sigma^2_U)
}$. 
We see that the link configuration in \fig{fig:link-inv-11} gives the intersection number 1
in  \fig{fig:link-inv-12}.

\section{4d SO(3)$_{\theta=\pi}$ Yang-Mills Gauge Theories coupled to the Boundary of 5d SETs/Long-Range Entangled TQFTs}
\label{sec:SO3-SET}

In \Sec{sec:SPT}, we have shown that that the $\SU(2)$ Yang-Mills theory with $\theta=\pi$, with the gauge bundle constraint $w_2(V_{\PSU(2)} )= B+ K_1 w_1(TM)^2+ K_2 w_2(TM)$, has  {two} distinct 't Hooft anomalies as shown in \Eq{Eq.5danomalypolynomial}. In this section, we 
further comment on gauging the 1-form $\Z_{2,[1]}^e$ center symmetry of {the Four Siblings of} $\SU(2)_{\theta=\pi}$ YM
to obtain $\SO(3)_{\theta=\pi}$ YM theories. 
Since the  't Hooft anomalies involve the 1-form center symmetry and the spacetime symmetries (whose background fields are the Stiefel-Whitney classes $w_i(TM)$), depending on which manifold we formulate the $\SU(2)$  Yang-Mills, one obtains different theories. 
 
\subsection{From SU(2) to SO(3) Gauge Theory}

To illustrate, we start by gauging the 1-form symmetry \cite{Gaiotto2014kfa1412.5148, Hsin2018vcg1812.04716}  of
the  $\SU(2)_{\theta=0}$ YM theories   {which is time-reversal symmetric and anomaly free}. There are still four choices of gauge bundle constraints labeled by $(K_1, K_2)$, i.e. \Eq{Eq.refinedGBC-YM-Z}. 
Let  $\bZ^{\text{$4$d}}_{{\SU(2)} {\text{YM}}}[B] $ be the path integral without specifying the gauge bundle constraint, the partition function with the gauge bundle constraint $w_2(E) = (B + K_1 w_1(TM)^2 + K_2 w_2(TM))\mod 2$ is 
$$ 
\bZ^{\text{$4$d}}_{{\SU(2)} {\text{YM}}_{(K_1,K_2)}}[B]
\equiv \int [D \Lambda] \; \bZ^{\text{$4$d}}_{{\SU(2)} {\text{YM}}}[B]  \exp( \ii \pi \Lambda \cup (w_2(E) - (B + K_1 w_1(TM)^2 + K_2 w_2(TM))) ),
$$  
More generally, we can add the counter term $\frac{p \pi}{2}\mathcal{P}(B)$ labeled by an integer $p$, which modifies the partition function as
\be
\bZ^{\text{$4$d}}_{{\SU(2)} {\text{YM}}_{(K_1,K_2)}}[B]
\equiv \int [D \Lambda] \; \bZ^{\text{$4$d}}_{{\SU(2)} {\text{YM}}}[B]  \exp\big( \ii \pi (
\Lambda \cup (w_2(E) - (B + K_1 w_1(TM)^2 + K_2 w_2(TM))) +\frac{p}{2} \mathcal{P}(B) )  \big),
\ee
Below we would like to obtain $\SO(3)$ YM by gauging 1-form $\Z_{2,[1]}^e$ center symmetry. The theta angle of the resulting theory is $2\pi p$. If $w_2(TM)$ is nontrivial, the resulting $\SO(3)$ theory is time-reversal symmetric only when $p\in 2\Z$ and $p \sim p+4$. When $w_2(TM)$ is trivial, the resulting $\SO(3)$ theory is time-reversal symmetric for $p\in \Z$ and $p \sim p+2$. In the following, we always restrict to the time-reversal symmetric case. 
Gauging 1-form center symmetry amounts to  summing over the background gauge field $B$ (promoting $B$ to a dynamical gauge field),
\begin{multline} \label{eq:ZSO(3)-4d}
\bZ^{\text{$4$d}}_{{\SO(3)} {\text{YM}}_{(K_1,K_2)}}=\\ 
\int [D \Lambda] [D B] \; \bZ^{\text{$4$d}}_{{\SU(2)} {\text{YM}}}[B]  \exp\big( \ii \pi (
\Lambda \cup (w_2(E) - (B + K_1 w_1(TM)^2 + K_2 w_2(TM))) +\frac{p}{2} \mathcal{P}(B))  \big).
\end{multline}
{Integrating} out $\Lambda$ enforces the relation between SO(3)-gauge bundles and 2-form dynamical gauge field $B$. 
This outputs the SO(3)-gauge theory $\bZ^{\text{$4$d}}_{{\SO(3)} {\text{YM}}_{(K_1,K_2)}}$ with $\theta=2\pi p$.

\subsection{
Gauging 1-form $\Z_{2,[1]}^e$-symmetry of $\SU(2)$ Gauge Theory with $\theta=\pi$}

We proceed to discuss gauging the 1-form symmetry of $\SU(2)$ Yang-Mills with $\theta=\pi$. 

If one formulates the {$\SU(2)_{\theta=\pi}$}  Yang-Mills on an orientable and spin manifold, i.e., $w_1=w_2=0$ (hence $w_3=0$ as well), for spacetime tangent bundle $TM$,
 there is the freedom to ignore the time reversal as a symmetry of the theory. The only symmetry of interest is the 1-form symmetry, which does not have  anomaly with itself. Hence one can gauge the 1-form symmetry and the resulting theory is $\PSU(2)=\SO(3)$ Yang-Mills with $\theta=\pi$. Indeed, $\SO(3)$ Yang-Mills with $\theta=\pi$ does not respect time reversal, which maps $\theta=\pi$ to $\theta=3\pi$ due to the identification $\theta\sim \theta+4\pi$ on a spin manifold. 

If one formulates the   {$\SU(2)_{\theta=\pi}$}  Yang-Mills on an orientable and non-spin manifold, one still has the freedom to ignore the time reversal as a symmetry of the theory. However, in this case,   {there is a counter term}
{\begin{eqnarray}
\int_{M^5} K_2 \pi  \Sq^1 (w_2(TM)\cup B)
\end{eqnarray}}
which {is a WZW-like term of background fields (i.e., probe fields in condensed matter language)}. Denoting the partition function of the $\SU(2)_{\theta=\pi}$ Yang-Mills coupled to $B$ as $\bZ_{\SU(2)\mathrm{YM}_{(0, K_2)}}[M^4, B]$, after promoting $B$ to a dynamical field, the partition function of the entire 4d-5d system is
\begin{eqnarray}
{\int [D B]  \bZ_{\SU(2)\mathrm{YM}_{(0, K_2)}}[M^4, B] \exp \Bigg(  \ii \pi \int_{M^5} K_2 \pi  \Sq^1 (w_2(TM)\cup B)  \Bigg).}
\end{eqnarray}
If $K_2=0$, the 4d-5d system reduces to {an }intrinsic 4d system. Physically, this corresponds to the case where the gauge charge is a boson.  It makes sense to gauge the 1-form symmetry which again gives raise to the time-reversal broken $\SO(3)$ Yang-Mills theory. If $K_2=1$, {the theory is still an intrinsic 4d system.} Physically, this corresponds to the case where the gauge charge is a fermion.

If one formulates the $\SU(2)$  Yang-Mills on an unorientable manifold, the time-reversal symmetry is built in, so time-reversal symmetry is too late to be abandoned. 
Promoting $B$ to a dynamical gauge field, the partition function for the entire 4d-5d system is 
\begin{equation}
{\int [D B]   \bZ_{\SU(2)\mathrm{YM}_{(K_1, K_2)}}[M^4, B] \exp\Bigg[ \ii \pi \int_{M^5} \Bigg(    B\Sq^1B+ \Sq^2\Sq^1B +  K_1 w_1(TM)^2 \Sq^1 B +  K_2 \Sq^1 (w_2(TM)\cup B) \Bigg)\Bigg].}
\end{equation}
Since $M^5$ is unorientable, for all four choices of $(K_1, K_2)$, the 5d terms do not vanish (because $  B\Sq^1B+ \Sq^2\Sq^1B $ is always non-vanishing on unorientable manifold). Hence one can only discuss the 4d-5d system rather than discussing the 4d system alone. 
We summarize all the above cases in Table \ref{Table:Gauge1-form}.

\begin{table}[h!]
	\begin{equation}
	\begin{tabular}{ |c|c| c| c| c|} 
	\hline
	$(w_1, w_2)\backslash (K_1, K_2)$ & $(0,0)$ & $(1,0)$ & $(0,1)$ & $(1,1)$ \\
	\hline
	$(0,0)$ & $\checkmark$ & $\checkmark$ & $\checkmark$& $\checkmark$   \\
	\hline
	$(1,0)$ & $\times$  & $\times$ &$\times$ & $\times$\\
	\hline
	$(0,1)$ & $\checkmark$ &  $\checkmark$ & WZW&WZW\\
	\hline
	$(1,1)$ & $\times$ & $\times$ &$\times$ & $\times$ \\
	\hline
	\end{tabular}
	\nonumber
	\end{equation}
	\caption{Possibilities of gauging the $\SU(2)_{\theta=\pi}$ Yang-Mills theory with gauge bundle constraint $(K_1, K_2)$ on a manifold with Stiefel-Whitney (SW) class {$(w_1, w_2)=(w_1(TM), w_2(TM))$, where
	0 and 1 mean trivial or nontrivial SW classes respectively.
	} The $\checkmark$ means that there is a way to make sense of the resulting gauged theory as a purely 4d theory. The theories labeled by $\times$ means that it only makes sense to discuss the combined 4d-5d systems. {The WZW (Wess-Zumino-Witten) means the theory is intrinsically 4d, however, there is a Wess-Zumino-Witten-like term of background fields, which involves a 5d integral (but does not depend on the choice of 5d manifold $M^5$).} }
	\label{Table:Gauge1-form}
\end{table}

\newpage

\section{Lattice Regularization, UV completion and Symmetric Anomalous TQFT}
\label{sec:lattice}

In this section, we formulate the   partition function of the 5d higher-SPT 
$\bZ_{\text{SPT}_{(K_1=0,K_2=0)}}^{5\text{d}}[M^5; B]$ on a simplicial complex spacetime. This provides a lattice regularization of the 5d SPT. We also provide lattice realization of  (i) 4d higher-symmetry-extended boundary theory or
(ii) 4d higher-symmetry-enriched anomalous topologically ordered boundary theory. We will generalize the approach in \cite{Wang2017locWWW1705.06728}
and follow the Section IX of \cite{Wan2018djlW2.1812.11955}.
In condensed matter physics, this (ii) phenomenon is known as the \emph{anomalous surface topological order}
(firstly noticed in \cite{VishwanathSenthil2012tq1209.3058}) 
typically for the 2+1D boundary of 3+1D SPTs, see a review \cite{Senthil1405.4015}.

\subsection{Lattice Realization of 4d Higher-SPTs and Higher-Gauge TQFT: 4d Simplicial Complex and 3+1D Condensed Matter Realization}

We warm up by considering a lattice realization of 4d Higher-SPTs given by a probe-field partition function
\bea
\bZ_{\text{SPT}}^{4\text{d}}[M^4; B]=\exp( \ii \frac{\pi}{2} \int_{M^4}   \cP(B)  )
= \exp( \ii \frac{\pi}{2} \int_{M^4}   B\cup B+ B\hcup{1} \delta B ).
\eea
The path integral can be regularized on a triangulated 4-manifold $M^4$.   The building blocks of $M^4$ are 4-simplices. Without loss of generality, we consider an 
arbitrary 4-simplex which we denote as $(01234)$ where each number labels one vertex. See Fig.~\ref{Fig.4simplex} for a graphical representation of a 4-simplex. We denote $B_{ijk}$ as restricting the 2-cochain $B$ on the 2-simplex $(ijk)$.   We label the path integral amplitude on $(01234)$ as $\omega_4(01234)$, i.e., 
\begin{equation}
	\begin{split}
		\omega_4(01234) &=  \exp\Bigg[ \ii  \frac{\pi}{2}  \Bigg( B\cup  B+ B\hcup{1}\delta B\Bigg)_{01234} \Bigg]\\
		&= \exp\Bigg[ \ii \frac{\pi}{2}  \Bigg( B_{012}B_{234}+ B_{034}(B_{123}-B_{023}+B_{013}-B_{012})+ B_{014}(B_{234}-B_{134}+B_{124}-B_{123}) \Bigg) \Bigg].\\
	\end{split}
\end{equation}
It is straightforward to verify that  $\omega_4(01234)$ satisfies the cocycle condition:
\begin{eqnarray}
	(\delta \omega_4)(012345)= \frac{\omega_4(12345) \cdot \omega_4(01345)\cdot \omega_4(01235) }{ \omega_4(02345)\cdot \omega_4(01245)\cdot \omega_4(01234)}= 1.
\end{eqnarray}

\begin{figure}[!h]
	\centering
	\includegraphics[width=8.cm]{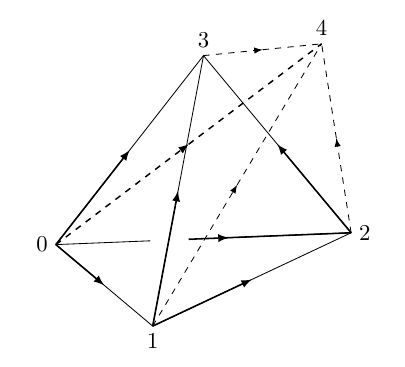} 
	\caption{Graphical representation of a 4-simplex $(01234)$.}
	\label{Fig.4simplex}
\end{figure}

\subsection{Lattice Realization of 5d Higher-SPTs and Higher-Gauge SETs: 5d Simplicial Complex and 4+1D Condensed Matter Realization}

The 5d partition function with $(K_1=0,K_2=0)$ is
\begin{eqnarray}\label{Eq.cochain}
	\bZ_{\text{SPT}_{(K_1=0,K_2=0)}}^{5\text{d}}[M^5]= \exp\Bigg( \ii \pi \int_{M^5}   B\Sq^1B+\Sq^2\Sq^1B \Bigg).
\end{eqnarray}
We start by triangulating the 5d closed spacetime manifold (without boundary) into 5-simplicial complex.  There are some useful identities:
\begin{eqnarray}\label{Eq.cupiden}
	\begin{split}
		\Sq^1 B&=B\hcup{1} B= \frac{1}{2} \delta B,\\
		\Sq^2 \Sq^1 B &= (\Sq^1 B)\hcup{1} (\Sq^1 B)= \frac{1}{4} (\delta B) \hcup{1}  (\delta B).
	\end{split}
\end{eqnarray}
{Note that  in the second equality of the first line, we have used the cocycle condition that $\delta B=0\mod 2$.}  One can express the SPT action \eqref{Eq.cochain} in terms of the sum of  cup-products of $B$ cochains over 5-simplices
\begin{eqnarray}
	\bZ_{\text{SPT}_{(K_1=0,K_2=0)}}^{5\text{d}}[M^5]= \exp\Bigg( \ii \frac{\pi}{2} \sum_{M^5}   B\cup \delta B+ \ii \frac{\pi}{4} \sum_{M^5}\delta B\hcup{1}\delta B \Bigg).
\end{eqnarray}


\begin{figure}[!h]
	\centering
	\includegraphics[width=8.cm]{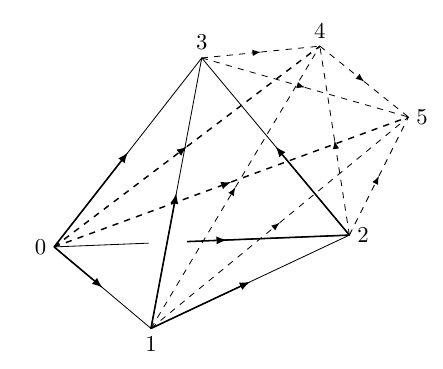} 
	\caption{Graphical representation of a 5-simplex $(012345)$.}
	\label{Fig.5simplex}
\end{figure}



Without loss of generality, we consider an arbitrary 5-simplex which we denote as $(012345)$ where each number labels one vertex. See Fig.~\ref{Fig.5simplex} for a graphical representation of a 5-simplex.   We will label the path integral amplitude on the simplex $(012345)$ as $\omega_5(012345)$, i.e., 
\begin{eqnarray}
	\omega_5(012345)=  \exp\Bigg[ \ii \pi  \Bigg(\frac{1}{2} B\cup \delta B+ \frac{1}{4}\delta B\hcup{1}\delta B\Bigg)_{012345} \Bigg]
\end{eqnarray}
so that the partition function can be simplified as $\bZ_{\text{SPT}_{(K_1=0,K_2=0)}}^{5\text{d}}[M^5]= \prod_{(ijklmn)\in M^5}\omega(ijklmn)$. 
Using the definition of the cup products on simplices and the identities \eqref{Eq.cupiden}, we have
\begin{equation}
	\begin{split}
		\left(\Sq^1 B\right)_{0123}&= \frac{1}{2}(B_{123}-B_{023}+B_{013}-B_{012})\\
		\left(\Sq^2 \Sq^1 B \right)_{012345}&=\frac{1}{4} ((\delta B)_{0345}(\delta B)_{0123}+ (\delta B)_{0145}(\delta B)_{1234}+(\delta B)_{0125}(\delta B)_{2345}),\\
		&=\frac{1}{4}(\left(-B_{045}-B_{034}+B_{035}+B_{345}\right)\left(-B_{023}-B_{012}+B_{013}+B_{123}\right)\\&~~~+ \left(-B_{045}-B_{014}+B_{015}+B_{145}\right)\left(-B_{134}-B_{123}+B_{124}+B_{234}\right)\\&~~~+ \left(-B_{025}-B_{012}+B_{015}+B_{125}\right)\left(-B_{245}-B_{234}+B_{235}+B_{345}\right)).
	\end{split}
\end{equation}
Hence the path integral amplitude  on the simplex $(012345)$ is 
\begin{equation}
	\begin{split}
		\omega_5(012345)= &\exp \Bigg[ \frac{\ii \pi}{2} B_{012}(-B_{245}-B_{234}+B_{235}+B_{345}) \\&+ \frac{\ii \pi}{4} \left(-B_{045}-B_{034}+B_{035}+B_{345}\right)\left(-B_{023}-B_{012}+B_{013}+B_{123}\right)\\&+\frac{\ii \pi}{4}\left(-B_{045}-B_{014}+B_{015}+B_{145}\right)\left(-B_{134}-B_{123}+B_{124}+B_{234}\right) \\&+ \frac{\ii \pi}{4}\left(-B_{025}-B_{012}+B_{015}+B_{125}\right)\left(-B_{245}-B_{234}+B_{235}+B_{345}\right)  \Bigg].
	\end{split}
\end{equation}
It is straightforward to verify that  $\omega_5(012345)$ satisfies the cocycle condition:
\begin{eqnarray}
	(\delta \omega_5)(0123456)= \frac{\omega_5(123456) \cdot \omega_5(013456)\cdot \omega_5(012356) \cdot\omega_5(012345)}{ \omega_5(023456)\cdot \omega_5(012456)\cdot \omega_5(012346)}= 1.
\end{eqnarray}
We emphasize that $\omega(012345)$ is a cocycle only when $B$ is a cocycle, i.e., $\delta B=0$. If $B$ is a cochain rather than a cocycle, \eqref{Eq.cochain} is not a cocycle, hence cannot be a partition function of a topological field theory. \footnote{The cocycle condition is crucial in proving the partition function to be invariant under re-triangulating the spacetime manifold $M^5$. }

We further comment on the lattice regularization of theory with various choices of $(K_1, K_2)$. 
\begin{enumerate}
	\item When $(K_1, K_2)=(0,0)$, as we derived above, there is a lattice regularization of the 5d SPT partition function. 
	\item When $(K_1, K_2)=(1,0)$, the path integral amplitude depends on the first Stiefel-Whitney class $w_1(TM)$. Using the method of \cite{Chen2011pg1106.4772}, one can write down the simplicial form of $ w_1(TM)^2$ using the twisted cocycle, with the coefficient in $U(1)_{T}$ due to anti-unitary  nature of time-reversal symmetry
	(in the Hamiltonian formalism of \cite{Chen2011pg1106.4772}). 
	We will not write down the explicit expression for the cocycle. 
	\item $(K_1, K_2)=(0,1), (1,1)$ has the same anomaly polynomial as  $(K_1, K_2)=(0,0), (1,0)$ respectively. 
\end{enumerate}
Other than the  treating $B$ as the background gauge field, we can also sum over $B$ to get the
the topologically ordered 5d SET $\bZ_{\text{SET}_{(K_1,K_2)}}^{5\text{d}}[M^5]$.

Given that the 5d SPT and 5d SET path integral can be regularized on a lattice, following \cite{Chen2011pg1106.4772}, one can write down the quantum wavefunction via the spacetime path integral. It is also possible to construct a lattice quantum Hamiltonian on the 4D space (on a constant time slice), for both SPTs and SETs,
similar to the formulations of \cite{KitaevToric, Chen2011pg1106.4772, Hu2012wx1211.3695, Wan:2014woa1409.3216}.
For the topologically ordered 5d SET, we implement the method of \cite{Hu2012wx1211.3695, Wan:2014woa1409.3216}: 
\bea
\hat {\bf H}= - \sum_{\text{1-link } \ell} \hat {\bf A}_{\ell} - \sum_{\text{3-simplex}} \hat {\bf B}_{\text{3-simplex}}
\eea
where $\hat {\bf A}_{\ell}$ is an operator acting on the plaquettes (2-simplex) adjacent to the 1-link $\ell$, and $\hat {\bf B}_{\text{3-simplex}}$ is an operator acting on the boundary of 
a given {\text{3-simplex}} which again are plaquettes (2-simplex). 
The $\hat {\bf A}_{\ell}$  has its effect on imposing the time evolution constraint as the same as the path integral formulation: 
$\hat {\bf A}_{\ell}$ lifting the state vector to a next time slice locally around the 1-link $\ell$.
The $\hat {\bf B}_{\text{3-simplex}}$ imposes the zero flux condition enclosed by the {3-simplex} (which is a 2-sphere $S^2$ in topology).
We will not give the explicit expression of the quantum Hamiltonian $\hat {\bf H}$ in this work. 

\subsection{Higher-Symmetry-Extended 3+1D Gapped Boundaries}
\label{sec:higher-sym-extended-3+1D}

One option to saturate the anomaly inflow from the bulk 5d (4+1D) SPT is to extend the global symmetry on the 4d (3+1D) boundary, based on the
\emph{symmetry-extension method} of \cite{Wang2017locWWW1705.06728}.  
{Note that $(K_1, K_2=0)$ and $(K_1,K_2=1)$ theories have the same 4d anomaly, they are differed by a 4d WZW-like counter term written on a 5d $M^5$, 
$$
\bZ_{\text{SPT}_{(K_1,K_2)}}^{5\text{d}}[M^5]= \bZ_{\text{SPT}_{(K_1,0)}}^{5\text{d}}[M^5]
 \exp \Big( {\ii\pi\int_{M^5}   K_2 \Sq^1 (w_1(TM)^2  B)} \Big).
$$
The 4d counter term is shown to be vanished on a closed $M^5$ in 5d in \Eq{eq:K2-vanish}. 
Due to \Eq{eq:K2-vanish}, this $K_2$-dependent term has no consequences via any anomaly consideration on 4d dynamics.
Thus we only discuss the  $3+1$D  gapped boundary for the two siblings $(K_1,K_2=0)$.  The 5d partition function is 
\begin{eqnarray}\label{Eq.5dPartitionfunction}
	\begin{split}
	\bZ_{\text{SPT}_{(K_1,0)}}^{5\text{d}}[M^5]&= \exp \Big( {\ii\pi\int_{M^5}   B\Sq^1 B + \Sq^2 \Sq^1 B + K_1 w_1(TM)^2 \Sq^1 B} \Big) \\&
	=\exp \Big( {\ii\pi\int_{M^5}  ( B + (1+K_1) w_1(TM)^2 +w_2(TM) ) \cup \Sq^1 B}\Big) \cdot \exp\Big( \ii \pi\int_{M^5} \delta (...) \Big).
	\end{split}
\end{eqnarray}
In the second line, we have used $\int_{M^5}\Sq^2 \Sq^1 B= \int_{M^5}(w_1(TM)^2 + w_2(TM))\Sq^1 B+\int_{M^5} \delta(...)$. Note that when $M^5$ is closed, the total derivative vanishes and we have the standard Wu formula $\int_{M^5}\Sq^2 \Sq^1 B=\int_{M^5}(w_1(TM)^2 + w_2(TM))\Sq^1 B$. When $M^5$ has a boundary, the standard Wu formula may no longer hold, and they differ at most by a co-boundary term.\footnote{{By the Wu formula for $n$-manifolds with boundary, 
$\Sq^j x=u_j x$ where $u_j$ is the relative Wu class, and $x \in \H^{n-j}(M,\partial M;\Z_2)$.  
The total relative Wu class $u$ is related to the total Stiefel-Whitney class  $w$ of M as $\Sq(u)=w$. So $u_1=w_1(TM)$, $u_2=w_1(TM)^2 + w_2(TM)$, etc., still hold but $\Sq^1 B$ may not be in  $\H^3(M,\partial M;\Z_2)$ for a 5-manifold $M$ with boundary. Therefore, $\Sq^2 \Sq^1 B= (w_1(TM)^2 + w_2(TM))\Sq^1 B$ may not hold for a 5-manifold $M$ with boundary. See for instance \Refe{Kervaire10.2307}.
}} We denote the co-boundary term as $\delta(...)$ in the second line. Since $(...)$ is a well-defined term of background gauge fields, it is a 4d invertible TQFT, which does not contribute to the 4d dynamics. Hence in the construction of 4d symmetric boundary TQFT below, we only focus on the first part in the second line of \eqref{Eq.5dPartitionfunction}. }
Using the systematic way in  \cite{Wang2017locWWW1705.06728} and its generalized higher-symmetry extension \cite{Wan2018djlW2.1812.11955}, 
we find that the boundary of 5d SPT can support a 4d TQFT via the higher-symmetry extension from a 1-form $\Z_2$ to a 1-form $\Z_4$ symmetry.  
Schematically, let {$\omega^{(K_1,0)}_5$} be the 5-cocycle whose product over the 5d manifold $M^5$ gives the 5d SPT partition function \eqref{Eq.5dPartitionfunction}. Let {$\beta^{(K_1,0)}_4$} be a 4-cochain which trivializes the 5d cocycle, i.e., 
{\begin{eqnarray}\label{Eq.reductionExtension}
	\omega^{(K_1,0)}_5=\delta \beta^{(K_1,0)}_4.
\end{eqnarray}}
We find that the following {$\beta^{(K_1,0)}_4$} satisfies \eqref{Eq.reductionExtension}:
\begin{eqnarray} \label{eq:beta4}
	\beta^{(K_1,0)}_4=\exp \Bigg[ \ii \pi \int_{M^4}  (B + (1+K_1) w_1(TM)^2 + w_2(TM)) \cup \gamma(C)   \Bigg],
\end{eqnarray}
where $C$ is a $\Z_4$ valued 2-cochain satisfying $B=C\mod 2$, and $\gamma: \Z_4\to \Z_2$ is a function which maps the $\Z_4$ 2-cochain to a $\Z_2$ 2-cochain: 
\begin{eqnarray}
	\left(\gamma(C)\right)_{ijk}= \frac{(C_{ijk})^2- C_{ijk}}{2}.
\end{eqnarray}

In summary, the $\beta^{(K_1,0)}_4$ in \eq{eq:beta4} is the partition function of the \emph{higher-symmetry-extended 3+1D 
gapped boundary}, while the original 1-form anomalous global symmetry $\Z_{2,[1]}$ is extended to 
the anomaly-free 1-form global symmetry $\Z_{4,[1]}$.
We can also rephrase that the higher-anomaly associated to the bordism group $\Omega_5^{\tO}(\B^2\Z_2)$ with a higher classifying space $\B^2 \Z_{2}$
can be \emph{pulled back and trivialized} as 
fully anomaly-free in the bordism group $\Omega_5^{\tO}(\B^2\Z_4)$ with an extended higher classifying space $\B^2 \Z_{4}$:
{\begin{equation}
	\begin{array}{c}
	\B^2 \Z_{2} \\ 
	\text{}\\
	\text{}\\
	\text{}
	\end{array}
	\begin{array}{c}
	\longrightarrow\\
	\text{}\\
	\text{}\\
	\text{}
	\end{array}
	\begin{array}{c}
	\B^2 \Z_{4}  \\ 
	\text{(Extended global symmetry }\\
	\text{$\Z_4$ 1-form symmetry probe}\\
	\text{by $\Z_4$ 2-cochain $C$ background field)}
	\end{array}
	 \begin{array}{c}
	 \longrightarrow\\
	\text{}\\
	\text{}\\
	\text{}
	\end{array}
	\begin{array}{c}
	 \B^2 \Z_{2}\\ 
	 \text{(Global Symmetry} \\
	 \text{$\Z_2$ 1-form symmetry probe}\\
	\text{by $\Z_2$ 2-cochain $B$ background field)}
		\end{array}.
	\end{equation}
	 }	 
This \emph{higher-symmetry-extended 3+1D 
gapped boundary}, described by \eq{eq:beta4}, has no long-range entanglement and no intrinsic topological order.
This \emph{higher-symmetry-extended 3+1D 
gapped boundary} is known as the System (i) in Section 7 of \Refe{Wang2018edf1801.05416} as a \emph{short-range entangled} state, both 
in the bulk and on the boundary (denoted as SRE/SRE in Section 7 of \Refe{Wang2018edf1801.05416}). 
In fact, this whole SRE/SRE bulk-boundary theory is still an \emph{invertible TQFT} with
a partition function $|\bZ|=1$ on an $M^5$ with a 4d gapped boundary $M^4$.

\subsection{Higher-Symmetry Anomalous 3+1D Topologically Ordered Gapped Boundaries: Spontaneous Higher-Symmetry Breaking}
\label{sec:higher-sym-pre-break-3+1D}

There also exists another boundary theory of the 5d SPTs in \Eq{Eq.5dPartitionfunction}, with $\Z_{2}$ 2-cochain $b$ summed over as dynamical fields on the boundary. The boundary theory is a dynamical $\Z_2$ gauge theory, which can be obtained from gauging the normal $\Z_{2,[1]}$ subgroup of $\Z_{4,[1]}$ in the symmetry extended gapped boundary from the previous section. Schematically, we promote the 2-form (or 2-cochain) gauge field $b$ coupling to 1-form $\Z_{2,[1]}$-symmetry to dynamical, in a normal subgroup {of $\Z_{4,[1]}$}. The resulting boundary theory has \emph{long range entanglement} in contrast to the short range entanglement of the symmetric extended boundary theory in the previous section
\Sec{sec:higher-sym-extended-3+1D}. This can be
summarized as an induced fiber sequence of their higher classifying space
$\B^2 \Z_{2} \to \B^2 \Z_{4} \to \B^2 \Z_{2}$:
	{\begin{equation} \label{eq:sequence-H-ext}
	\begin{array}{c}
	\B^2 \Z_{2} \\ 
	\text{(Dynamical/emergent}\\
	\text{gaugeable}\\
	\text{$\Z_2$ 2-cochain $b$ field)}
	\end{array}
	\begin{array}{c}
	\longrightarrow\\
	\text{}\\
	\text{}\\
	\text{}
	\end{array}
	\begin{array}{c}
	\B^2 \Z_{4}  \\ 
	\text{}\\
	\text{}\\
	\text{}
	\end{array}
	 \begin{array}{c}
	 \longrightarrow\\
	\text{}\\
	\text{}\\
	\text{}
	\end{array}
	\begin{array}{c}
	 \B^2 \Z_{2}\\ 
	 \text{(Global Symmetry} \\
	 \text{$\Z_{2,[1]}$ symmetry probed by}\\
	\text{$\Z_2$ 2-cochain background $B$ field)}
		\end{array}.
	\end{equation}
	 Or we may denote the above as $\B^2 \Z_{2}^{\text{gauged}} \to \B^2 \Z_{4}^{\text{total}} \to \B^2 \Z_{2}^{G}$, implementing the notations of
	 \cite{Wang2017locWWW1705.06728,{Wang2018edf1801.05416}}.

{However, as noticed in \Refe{Wang2017locWWW1705.06728} and in 
section 7 of \Refe{Wang2018edf1801.05416},
when the boundary theory is long-range entangled (after gauging a normal subgroup), 
it is possible that the \emph{new fate} of low energy dynamics  may not preserve the global symmetry. 
Specifically,  \Refe{Wang2017locWWW1705.06728} finds that, under the 
exact sequence $K \to H \to G$,
even for a successful \emph{$H$-symmetry extended construction of gapped boundary},
in certain cases, dynamically gauging the normal subgroup $K$ may still result in $G$ spontaneously broken.
In short, we should question:\\ 

\emph{Is it possible that $\Z_{2,[1]}$ global symmetry in the long-range entangled boundary theory happens to be spontaneously broken? }

Namely,  following the notations in Sec.~7 of \Refe{Wang2018edf1801.05416} on the bulk/boundary (denoted bulk/bdry) systems of the SRE/SRE and SRE/LRE types,
the symmetry-extension construction  \cite{Wang2017locWWW1705.06728,{Wang2018edf1801.05416}} under the 
exact sequence $K \to H \to G$, may result in different dynamical fates:
$$
\left\{\begin{array}{l}
	\text{\emph{Case 1. $H$-symmetry-extended gapped boundary: SRE bulk/SRE bdry.}}\\
	\text{e.g. The 1-form $\Z_{4,[1]}$-symmetry extended in \Sec{sec:higher-sym-extended-3+1D}. Many 0-form $G$-symmetry examples given in \Refe{Wang2017locWWW1705.06728}}\\[2mm]
	\text{\emph{Case 2. $G$-symmetry-preserving anomalous $K$-gauge gapped boundary: SRE bulk/LRE bdry.}}\\
	 \text{e.g. Many 0-form $G$-symmetry examples given in \Refe{Wang2017locWWW1705.06728}}.\\[2mm]
        	\text{\emph{Case 3. $G$-symmetry-breaking $K$-gauge gapped boundary: SRE bulk/LRE bdry.}}\\
        \text{e.g. \Refe{Wang2017locWWW1705.06728}'s  Sec.~3.4 and Appendix A.2.4, and \Refe{Wang2018edf1801.05416}'s Sec.~7.1.}
	\end{array}.
\right.
$$
Thus below what we aim to examine is whether a proposal of \emph{Case 2} associated with \Eq{eq:sequence-H-ext} is in fact the \emph{Case 3} in disguise, when $K$ is dynamically gauged,
as  
	{\begin{equation}
	\begin{array}{c}
	\B^2 \Z_{2} \\ 
	\text{(Dynamical/emergent}\\
	\text{gaugeable}\\
	\text{$\Z_2$ 2-cochain $b$ field)}
	\end{array}
	\begin{array}{c}
	\longrightarrow\\
	\text{}\\
	\text{}\\
	\text{}
	\end{array}
	\begin{array}{c}
	\B^2 \Z_{4}  \\ 
	\text{}\\
	\text{}\\
	\text{}
	\end{array}
	 \begin{array}{c}
	 \longrightarrow\\
	\text{}\\
	\text{}\\
	\text{}
	\end{array}
	\begin{array}{c}
	 \B^2 \Z_{2}\\ 
	 \text{(Global Symmetry} \\
	 \text{$\Z_{2,[1]}$ symmetry}\\
	\text{spontaneously broken?)}
		\end{array}.
	\end{equation}
Concretely,  we propose a $\Z_{2}$ gauge theory (endorsed with a $\Z_2$ 2-cochain $b$ field and the dual $\Z_2$ 1-cochain ${\tilde a}$ field,
as a candidate IR theory of the UV SU(2) YM) which has $\Z_{2,[1]}$ global symmetry and saturates the 't Hooft anomaly \Eq{Eq.5dPartitionfunction}. We will find that the $\Z_{2,[1]}$ global symmetry is spontaneously broken. 
}



We consider the  4d $\Z_2$-gauge TQFT  
\begin{eqnarray}\label{Eq.Z2gaugetheory}
\int [D b] [D {\tilde a}] \exp( \ii \pi \int_{M^4}  b \delta {\tilde a} + \frac{1}{4}\cP(\delta {\tilde a} ))
=\int [D b] [D {\tilde a}] \exp( \ii \pi \int_{M^4}  b \delta {\tilde a} + {\tilde a} {\tilde a} {\tilde a} {\tilde a}).
\end{eqnarray}
Here  $\int [D b] [D {\tilde a}]$ means we 
sum over the $\Z_2$ valued 2-cochain $b\in C^2(M, \Z_2)$ and $\Z_2$ valued 1-cochain  ${\tilde a} \in C^1(M, \Z_2)$. 
The first term is the standard BF term of discrete gauge theory, while the second term is the 4d analogue of the Dijkgraaf-Witten type action\cite{Putrov2016qdo1612.09298}. We also refer it as the twisting term. Integrating out $b$ enforces ${\tilde a}$ to be a $\Z_2$ valued cocycle, hence the twisting term  
$\frac{1}{4}\cP(\delta {\tilde a} )=\frac{1}{4}\delta {\tilde a} \delta {\tilde a} = {\tilde a} {\tilde a} {\tilde a} {\tilde a}$ mod 2 is time-reversal symmetric. 
%
To match the anomaly from 5d SPTs in \Eq{Eq.5dPartitionfunction}, we couple to background fields $B$ and $w_1(TM)$, as follows, 
\begin{eqnarray}\label{Eq.Z2theoryBgd}
\pi \int_{M^4}  b (\delta {\tilde a}- B-K_1 w_1(TM)^2) + \frac{1}{4}\cP(\delta {\tilde a}-B )
\end{eqnarray}
Summing over the $\Z_2$ valued 2-cochain $b$ enforces the gauge bundle constraint:
\begin{eqnarray}\label{Eq.GBCZ2}
\delta {\tilde a}= B + K_1 w_1(TM)^2\mod 2.
\end{eqnarray}
Notice that under the gauge transformation of the background field $B\to B+ \delta \lambda$, we demand $\tilde{a}\to \tilde{a}+ \lambda$. It is obvioius that the action \eqref{Eq.Z2theoryBgd} is gauge invariant. However, due to the gauge bundle constraint \eqref{Eq.GBCZ2}, the twisting term $\frac{\pi}{4}\cP(\delta \tilde{a}-B)$ is no longer $0$ or $\pi\mod 2\pi$, hence it is not time-reversal invariant, and \eqref{Eq.Z2theoryBgd} is not well defined on an unorientable manifold. To make sense of the theory on an unorientable manifold, we use the same idea in \Sec{sec:SPT} where we promote the twisting term to a 5d integral, 
\begin{eqnarray}
\begin{split}
\pi \int_{M^5}\frac{1}{4}\delta \cP(\delta {\tilde a}-B )&
= \pi \int_{M^5} \frac{1}{2} \delta \tilde{a} \delta (\delta \tilde{a}) + \frac{1}{2} \delta B \delta \tilde{a} + \frac{1}{4} \delta \cP(B)\\&
= \pi \int_{M^5} B\Sq^1 B + \Sq^2 \Sq^1 B + K_1 w_1(TM)^2 \Sq^1 B
\end{split}
\end{eqnarray}
where we have used $\frac{\pi}{4} \delta \cP(B)= \pi (B\Sq^1 B + \Sq^2 \Sq^1 B)$ and the gauge bundle constraint \eqref{Eq.GBCZ2}. Thus when we couple theory to the background field $B$ and formulate it on an unorientable manifold, only the 4d-5d coupled system is 
well defined
\begin{multline}
\int [D b] [D {\tilde a}] \exp\big( \ii \pi 
\int_{M^4}  b (\delta {\tilde a}- B-K_1 w_1(TM)^2) + 
 \ii   \pi \int_{M^5} (B \Sq^1 B+ \Sq^2 \Sq^1 B+ K_1 w_1(TM)^2 \Sq^1 B) \big).
\end{multline}
Indeed, the 5d integral implies that the $\Z_2$ gauge theory saturates the anomaly \Eq{Eq.5dPartitionfunction}. 
Let us comment on the dynamics of the above 4d $\Z_2$ gauge theory.  The fact that $\tilde{a}$ transforms as $\tilde{a}\to \tilde{a}+ \lambda$ under 1-form background gauge transformation $B\to B+ \delta \lambda$ suggests that the Wilson line
\begin{eqnarray}
\exp(  \ii \pi \oint {\tilde a})
\end{eqnarray}
has a charge 1 under $\Z_{2,[1]}$. In a TQFT, any genuine line operators are tensionless, thus the expectation does not satisfy area law.  Since $\exp(  \ii \pi \oint {\tilde a})$ is a genuine line operator, it obeys a perimeter law, and it spontaneously breaks the $\Z_{2,[1]}$ 1-form symmetry. 

\cred{We comment that in Section 7.1 of \Refe{Wang2018edf1801.05416},  similar arguments have been used to show that the 0-form $\Z_2$ global symmetry of a 2d TQFT is spontaneously broken. In its 2d action $\pi \int \alpha (\delta \phi -A) +\dots$, 
there is a scalar $\phi$ with charge 1 under $\Z_2$ 0-form global symmetry whose background field is $A$. 
(The $\phi$ and $\alpha$ are dynamical 0-cochain and 1-cochain fields.)
In the 2d TQFT, $\phi$ has a nontrivial expectational value $\< \phi \> \neq 0$. 
Thus the $\Z_2$ 0-form symmetry is spontaneously broken. See  \Refe{Wang2018edf1801.05416} for more details.}

The fact that the dynamical $\Z_2$ gauge theory we constructed spontaneously breaks the 1-form $\Z_{2,[1]}$-symmetry is consistent with the conclusion pointed out by Cordova and Ohmori (see \cite{KO-Strings-2019-talk} and \cite{CordovaCO2019}) where the authors showed the impossibility of having any $\Z_{2}^T\times \Z_{2,[1]}$ symmetric anomalous gapped 4d TQFT saturating the 4d higher-anomaly of 5d SPTs \Eq{Eq.5dPartitionfunction}.



\newpage

\section{Conclusions, Discussions and Dynamics}

\label{sec:conclude}

\begin{enumerate}

\item \emph{Summary}: In this work, we  {proved} (physically via the quantum field theory method) that 
4d time-reversal symmetric pure Yang-Mills of an SU(2) gauge group with a second-Chern-class topological term at $\theta=\pi$ (i.e., SU(2)$_{\theta=\pi}$ YM)
have new higher 't Hooft anomalies in 4d, given by a 5d topological term
\Eq{Eq.5danomalypolynomial} and \Eq{eq:5dSPT-all-w1PB-open}: 
$$
{\pi \int_{M^5} \Bigg(    B \cup \Sq^1B+ \Sq^2\Sq^1B +  K_1 w_1(TM)^2 \cup \Sq^1 B +  K_2 \Sq^1 (w_2(TM) \cup  B) \Bigg)}.
$$
The 5d term $B  \Sq^1B+ \Sq^2\Sq^1B +  K_1 w_1(TM)^2 \Sq^1 B$
is a 5d bordism invariant (or equivalently 5d iTQFT/SPTs/counter term) specifies the 4d 't Hooft anomaly.
However, the 5d term $K_2 \Sq^1 (w_2(TM)   B)$
is \emph{not} a 5d bordism invariant but only a 4d WZW-like counterterm, 
thus strictly speaking it does not indicate any 4d 't Hooft anomaly.
We found that there are at least Four Siblings of SU(2)$_{\theta=\pi}$ YM with bosonic UV completion, labeled by $(K_1,K_2) \in (\Z_2, \Z_2)$.  
Their higher 't Hooft anomalies of generalized global symmetries indicate that 4d SU(2)$_{\theta=\pi}$ YM,  in order to realize all global symmetries locally, 
necessarily couple to 5d higher symmetry-protected topological states (SPTs, as
 invertible TQFTs [iTQFTs], as 5d 1-form-center-symmetry-protected 
interacting ``topological superconductors'' in condensed matter). 

We explored various 4d YM gauge theories living 
as boundary conditions of 5d gapped short/long-range entangled (SRE/LRE) topological states. 
We revisited 4d SU(2)$_{\theta=\pi}$ YM-5d SRE-higher-SPTs coupled systems \cite{Gaiotto2017yupZoharTTT1703.00501, Wan2018zql1812.11968}. 
Follow Weyl's gauge principle, by dynamically gauging the 1-form center symmetry, we transformed a 5d bulk SRE SPTs into an LRE symmetry-enriched topologically ordered state (SETs); thus we obtained the 4d SO(3)$_{\theta=\pi}$ YM-5d LRE-higher-SETs coupled system with dynamical higher-form gauge fields. 
We illustrate such 4d-5d systems schematically in \Fig{Fig-SU(2)-SPT} and \Fig{Fig:curve-SU(2)-SO(3)}.


\begin{figure}[!h]
	{\centering
(a) \includegraphics[width=8.cm]{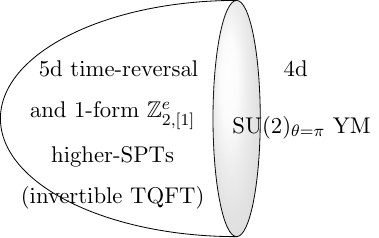} (b) \includegraphics[width=8.cm]{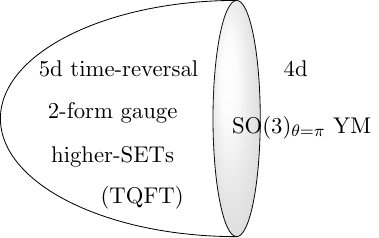} 
	\caption[caption]{
An alternative illustration of \Fig{Fig-SU(2)-SPT}: 
Various 4d Yang-Mills gauge theories (YM) live as the boundary conditions of certain 5d invertible TQFT (iTQFT) or 5d TQFT,
in order to realize YM's (higher) global symmetries locally.
	}
	\label{Fig:curve-SU(2)-SO(3)}}
\end{figure}


{The 4d SO(3) YM has a $\theta$ periodicity $\theta \sim \theta + 4 \pi$ on a spin manifold \cred{(due to that
the instanton number becomes fractional as half integer $\int_{M^4} \frac{1}{8 \pi^2}  \text{Tr}\,(F_a\wedge F_a)  \in \frac{1}{2}\Z$)}, and 
$\theta \sim \theta + 8 \pi$ on a non-spin manifold \cred{(due to that
the instanton number becomes quarter integer as $\int_{M^4} \frac{1}{8 \pi^2}  \text{Tr}\,(F_a\wedge F_a) \in \frac{1}{4}\Z$)}. Since time-reversal symmetry is preserved if and only if  $\theta \to -\theta$ is identified,
thus SO(3)$_{\theta=\pi}$ YM  explicitly breaks the time-reversal symmetry at ${\theta=\pi}$.
In the right-hand side (b) of \Fig{Fig-SU(2)-SPT} and \Fig{Fig:curve-SU(2)-SO(3)}, we actually have a 5d SETs whose 4d boundary
 explicitly breaks time-reversal symmetry.}

Apply the tool introduced in \cite{Putrov2016qdo1612.09298}, we derive new 
exotic anyonic statistics of extended objects such as 2d worldsheet of strings and 3d worldvolume of branes, which physically characterize the 5d SETs. We 
discover new triple and quadruple link invariants associated with the underlying 5d higher-gauge TQFT, hinting a new intrinsic relation between 
 non-supersymmetric 4d pure YM and topological links in 5d.

\item
\emph{Appearances of mod 2 anomalies}:
We note that the anomaly associated to the 5d term $\exp(\ii \pi \int w_3(TM) B) $ has also appeared in the context of an adjoint QCD$_4$ theory (\cite{AnberPoppitz2018tcj1805.12290, Cordova2018acb1806.09592, Bi:2018xvr} and \cite{Wan2018djlW2.1812.11955}). 
The $\exp(\ii \pi \int w_2(TM) w_3(TM)) $  has also appeared as a new SU(2) anomaly in the SU(2) gauge theory \cite{Wang:2018qoyWWW}.
All these anomalies and all our anomalies in \Eq{eq:5dSPT-all} are mod 2 non-perturbative global anomalies, like the SU(2) anomalies \cite{Witten:1982fp, Wang:2018qoyWWW}.

\item \emph{Mathematical relation between 5d and 4d bordism groups}: Mathematically there seems to be an amusing relation between (1) gauging the SU(2) gauge bundle/connection
under the coupling of 4d YM with 4d SPTs (4d bordism invariants of $\Omega^{G'}_4$) with $G'$ derived from a group extension
\Eq{eq:SU2-G'}:
$$
1 \to \SU(2) \to  G' \to \tO(d) \to 1;
$$
and (2) some of the 5d bordism invariants given by
${\Omega_5^{\tO}(\B^2\Z_2)=\Z_2^4}$.
We provide the computations of bordism groups and topological invariants in Appendix \ref{appendix-bordism}.
It will be illuminating to explore this relation further in the future.

\begin{figure}[!h] %
	{\centering
\includegraphics[width=10.cm]{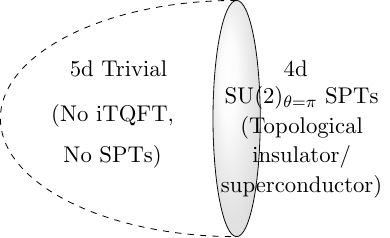} 
	\caption[caption]{4d time-reversal and SU(2) symmetric-protected topological states (SPTs)
	can be defined as 4d iTQFTs/bordism invariants. Their symmetries can be realized locally in 4d, without the need to an extra dimensional 5d system. 
Gauging SU(2) symmetry of this 4d SPTs renders the 4d-5d coupled system in \Fig{Fig:curve-SU(2)-SO(3)}(a).
Further gauging $\mathbb{Z}_{2,[1]}^e$-symmetry of the whole system in \Fig{Fig:curve-SU(2)-SO(3)}(a) renders the 4d-5d coupled system in 
\Fig{Fig:curve-SU(2)-SO(3)}(b). 
	See the remark 4 of \Sec{sec:conclude} for details. 	
	}
	\label{Fig-5d-4d-curved-SU(2)-SPT}}
\end{figure}

\item \emph{Classes of 4d SU(2)$_{\theta=\pi}$ YM}: In \Refe{2017arXiv171111587GPW}, it was noted that the Pin$^+$ and Pin$^-$ version of the above group extensions
$G' = {\mathrm{Pin}^{+}\times_{\Z_2} \SU(2)}$ and $G' = {\mathrm{Pin}^{-}\times_{\Z_2} \SU(2)}$ provide two different SPTs vacua after dynamically gauging the SU(2) symmetry give rise to two distinct 4d SU(2)$_{\theta=\pi}$ YM theories.
Although  \Refe{2017arXiv171111587GPW} suggested that the Pin$^+$ and Pin$^-$ of 4d SU(2)$_{\theta=\pi}$ YM are secretly indistinguishable by correlators of local operators on orientable spacetimes nor by gapped SPT states, can be distinguished on non-orientable spacetimes or potentially by correlators of extended operators.

In this work,  we  {have} shown that Pin$^+$ and Pin$^-$ of 4d SU(2)$_{\theta=\pi}$ YM indeed have distinct
new higher 't Hooft anomalies, given by
\Eq{Eq.5danomalypolynomial} and \Eq{eq:5dSPT-all}, with $(K_1,K_2)=(0,1)$ and $(K_1,K_2)=(1,1)$ respectively. 
Thus we confirm that Pin$^+$ and Pin$^-$ of 4d SU(2)$_{\theta=\pi}$ YM live in distinct Hilbert spaces, thus
they are indeed distinct vacua.

{More generally, in this work, we propose a classification 4d SU(2)$_{\theta=\pi}$ YM with a bosonic UV completion (e.g., on a lattice with bosonic degrees of freedom) and without fermionic parity symmetry $\Z_2^F$.
We propose that a classification can be all obtained from dynamically gauging the SU(2) normal subgroup of 4d $G'$-SPTs
where $G'$ is given by \Eq{eq:SU2-G'}:
$
1 \to \SU(2) \to  G' \to \tO(d) \to 1
,$
i.e., follow \cite{2017arXiv171111587GPW} by coupling the 4d $G'$-SPTS to a pure 4d SU(2) YM theory and dynamically gauge 
their $\SU(2)$, see \Sec{sec:SU2}.\footnote{Although 
we mainly focus on 4d SU(2)$_{\theta=\pi}$ YM here,
this gauge principle works for more general 4d SU(2) YM, or equivalently 4d SU(2)-gauge quantum spin liquids in condensed matter}
Other than the Pin$^+$ and Pin$^-$  cases, this extension \Eq{eq:SU2-G'} includes two more cases: $G'={\rm{O}(d) \times SU(2)}$ and 
${\rm{E}(d) \times_{\Z_2} SU(2)}$. In summary, in terms of the un-gauged 4d SPTs, we have the symmetry group $G'$
\cite{Freed2016.1604.06527, 2017arXiv171111587GPW}:
\be \label{eq:4-G'}
G'=\left\{\begin{array}{l} 
{\rm{O}(d) \times SU(2)} \text{ (bosonic, relates to $(K_1,K_2)=(0,0)$)},\\
{\rm{E}(d) \times_{\Z_2} SU(2)} \text{ (bosonic, relates to $(K_1,K_2)=(1,0)$)},\\
{\mathrm{Pin}^{+}\times_{\Z_2^F} \SU(2)} \text{ (fermionic, relates to $(K_1,K_2)=(0,1)$)},\\ 
{\mathrm{Pin}^{-}\times_{\Z_2^F} \SU(2)} \text{ (fermionic, relates to $(K_1,K_2)=(1,1)$)}.
\end{array}\right.
\ee
The global symmetries of SPTs for the first two cases are purely bosonic since they do not contain $\Z_2^F$;
the later two cases are fermionic since they do contain $\Z_2^F$ (the $\Z_2^F$ is shared by Pin and by the center of SU(2)).
\begin{itemize}
\item
Before gauging, see \Fig{Fig-5d-4d-curved-SU(2)-SPT}, the 4d SPTs are well-defined 4d topological terms/bordism invariants that can live on 4d with fully 
local onsite symmetry
without the need of an extra 5d  {bulk}.
\item
After gauging SU(2) normal subgroup of $G'$ symmetry of these SPTs, 
see \Fig{Fig:curve-SU(2)-SO(3)}(a), 
there is an emergent 1-form center $\Z_{2,[1]}^e$-symmetry.
After gauging SU(2), all four systems become \emph{bosonic} without $\Z_2^F$ symmetry. 
We obtain new theories:  4d SU(2)$_{\theta=\pi}$ YM as boundary conditions of 5d iTQFT
 in order to realize all global symmetries locally.
 \item
Furthermore, after gauging this $\Z_{2,[1]}^e$-symmetry of the 4d SU(2)$_{\theta=\pi}$ YM and 5d iTQFT coupled system, 
see \Fig{Fig:curve-SU(2)-SO(3)}(b), we obtain new theories: 4d SO(3)$_{\theta=\pi}$ YM as boundary conditions of 5d TQFT.
\end{itemize}
}

\item \emph{Quantum spin liquids in condensed matter}:
Strong coupled gauge theories have condensed matter implications as quantum spin liquids.
Time-reversal symmetric U(1) gauge theories as quantum spin liquids \cite{Senthil1405.4015} have been explored and classified based on
the quantum numbers of gapped electric and magnetic excitations (Wilson and 't Hooft line operators) in \Refe{Wang2016cto1505.03520, Zou2017ppq1710.00743}, see also  {recent works} \cite{Hsin2019fhf1904.11550, NingZouMeng2019}. 
Time-reversal symmetric SU(N) gauge theories as quantum spin liquids have been explored in \cite{2017arXiv171111587GPW}.
We will leave additional interpretations of our results of non-abelian SU(2) gauge theories in the context of quantum spin liquids 
for a future work.

\item \emph{Relations of link invariants and braiding statistics in various dimensions}:
{We have applied the tools developed in \cite{Putrov2016qdo1612.09298} to compute link invariants of 5d higher-gauge TQFTs in \Sec{sec:5dTRTQFT}.
We anticipate a precise mathematical formulation of this 5d \emph{higher-gauge TQFT} 
requires a certain \emph{higher category theory}.}
We remark that several link invariants that we find here in 5d have dimensionally reduction analogy to 4d and 3d,
such that the dimensional reduced link configurations in 4d and 3d are related to what had been studied in \cite{1602.05951}, 
\cite{Putrov2016qdo1612.09298} and References therein.

\item \emph{Fate of IR dynamics of gauge theories, UV completion and lattice regularizations at $\theta=\pi$}:
For the 4d-5d systems that we explore (schematically in \Fig{Fig-SU(2)-SPT} and \Fig{Fig:curve-SU(2)-SO(3)}),
we mainly focus on their Four Siblings as the UV theories. 
We do not yet know the IR fate of their dynamics of these strongly coupled gauge theories.
However, given the potentially complete 't Hooft anomalies in \Eq{Eq.5danomalypolynomial} and \Eq{eq:5dSPT-all} (at zero temperature),
we can constrain the IR dynamics by UV-IR anomaly matching.\newpage
The consequence of anomaly matching implies that the IR theories must be at least one of the following scenarios: 
\begin{enumerate}[label= (\roman*)]
\item 
Time-reversal $\Z_2^T$ symmetry broken (spontaneously or explicitly): The conventional standard lore suggests the two fold $\Z_2^T$-breaking degenerate ground states \cite{Gaiotto2017yupZoharTTT1703.00501}.
\item 
1-form center $\Z_{2,[1]}^e$ symmetry broken (spontaneously or explicitly) as \emph{deconfinement}:
\begin{enumerate}[label=(2-\roman*)]
    \item 1-form $\Z_{2,[1]}^e$-breaking and deconfined TQFTs, i.e., \emph{topological order} in condensed matter. 
    A $\Z_2$ gauge theory with the 1-form symmetry spontaneously broken has been proposed in \Sec{sec:higher-sym-pre-break-3+1D}.
    \item 1-form $\Z_{2,[1]}^e$-breaking and deconfined gapless theories or deconfined CFTs.
\end{enumerate}
\item 
1-form symmetry unbroken as \emph{confinement}:
\begin{enumerate}[label=(3-\roman*)]
    \item 1-form symmetry-extended invertible TQFT: This exotic scenario is discussed in Sec. \ref{sec:higher-sym-extended-3+1D}. 
     In Sec.~\ref{sec:higher-sym-extended-3+1D}, 1-form $\Z_{2,[1]}$-symmetry is extended to $\Z_{4,[1]}$ in order to trivialize, thus saturate, the anomaly.
    \item 1-form symmetry-preserving TQFT: 
    Cordova and Ohmori \cite{KO-Strings-2019-talk, CordovaCO2019} have proved the nonexistence of TQFTs preserving both the 1-form symmetry and time-reversal symmetry while
    saturating the 4d SU(N)$_{\theta=\pi}$ YM's anomaly. 
    This is consistent with Sec.~\ref{sec:higher-sym-pre-break-3+1D}'s analysis, which results in the previous phase of (2-i).
\end{enumerate}
\item 
Full symmetry-preserving gapless theory (CFT): This is a fairly exotic case which seems to be less likely to happen.
\end{enumerate}

Let us comment more on the recent Cordova-Ohmori's result 
\cite{KO-Strings-2019-talk, CordovaCO2019} which rules out the 1-form and time-reversal symmetry-preserving gapped phases for 4d SU(N)$_{\theta=\pi}$ YM,
namely the phase of (3-ii).
To recall, although in Sec.~\ref{sec:higher-sym-extended-3+1D}, we show the phase of (3-i) 
1-form symmetry-extended invertible TQFT can be constructed, but in reality, such phases are unlikely to happen \cite{Wang2017locWWW1705.06728} and should be only regarded
as an intermediate step to construct the phase of (3-ii) 1-form symmetry-preserving TQFT.
Furthermore, we show in Sec.~\ref{sec:higher-sym-pre-break-3+1D}
that dynamically gauging the extended symmetry of symmetry-extended invertible TQFT (of Sec.~\ref{sec:higher-sym-extended-3+1D})
results in the spontaneous symmetry breaking phases, instead of the symmetry-preserving gapped phases.
Thus our Sec.~\ref{sec:lattice} is consistent with \cite{KO-Strings-2019-talk, CordovaCO2019}.

The fate of any of the four scenarios of IR phases
above is meant to match the anomaly (or match Lieb-Schultz-Mattis type of theorem in condensed matter physics).
Since we have the Four Siblings of 4d SU(2)$_{\theta=\pi}$ YM at UV, labeled by $(K_1,K_2) \in (\Z_2, \Z_2)$,
we can discuss each of their IR dynamics.  {We leave more systematic discussions of the IR dynamics for a future work.}    

\underline{$\spadesuit$ 4d SU(2)$_{\theta=\pi}$ YM with $(K_1,K_2)=(0,0)$ and $(1,0)$}:\\
For $(K_1,K_2) = (0,0)$ or $(1,0)$ with $K_2=0$, we see that 
$\int_{M^5} B \Sq^1B+ \Sq^2\Sq^1B +  K_1 w_1(TM)^2  \Sq^1 B =\int_{M^5} \frac{1}{2} \tilde w_1(TM) \mathcal P( B) +   K_1 w_1(TM)^2  \Sq^1 B$
(or schematically $\sim \int_{M^5} \cT BB + K_1 \cT^3 B$)
vanishes on 5d orientable manifolds. In other words,
 if $\Z_2^T$-symmetry is spontaneously or explicitly broken,  we can match the anomaly at IR.
This means that when $K_2=0$, 
the
${\rm{O}(d) \times SU(2)}$
and ${\rm{E}(d) \times_{\Z_2} SU(2)}$ versions of 4d SU(2)$_{\theta=\pi}$ YM in \Eq{eq:4-G'} can indeed flow to the $\Z_2^T$-breaking Scenario (1) at IR.

\underline{$\spadesuit$ 4d SU(2)$_{\theta=\pi}$ YM with $(K_1,K_2)=(0,1)$ and $(1,1)$}:\\
For $(K_1,K_2) = (0, 1)$ or $(1,1)$  with $K_2=1$, we see that $K_2 \int_{M^5} \Sq^1 (w_2(TM)   B)$ is nonzero if $M^5$ has a boundary, and this 
$\int_{M^5} \Sq^1 (w_2(TM)  B)$ term does not vanish even if we restrict to orientable manifolds. In other words,
even if we break $\Z_2^T$-symmetry (spontaneously or explicitly),
%
the $\int \Sq^1 (w_2(TM)   B)$ still can suffer from
background gauge variance.
In this case, we should define the physical 4d theory to include not only $\bZ^{\text{4d}}_{\SU(2)_{\theta=\pi}\text{YM}}[M^4;B,w_j(TM)]$ given in
\Eq{eq:SU2YMZ}  but also the 4d counter term $\int \Sq^1 (w_2(TM)   B)$, which combine to
\begin{equation} \label{eq:SU2YMZ-Conc-4d}
\begin{split}
\bZ^{\text{4d}}_{\underset{\text{+ counterterms}}{\SU(2)_{\theta=\pi}\text{YM}}}[M^4;B,w_j(TM)]
\equiv
\bZ^{\text{4d}}_{\SU(2)_{\theta=\pi}\text{YM}}[M^4;B,w_j(TM)]
\cdot  \exp \Bigg(\ii \pi  \int K_2 \Sq^1 (w_2(TM)   B)\Bigg).
\end{split}
\end{equation}
The intrinsic 5d theory thus contains only the 5d bordism invariant/SPTs/iTQFT:
\footnote{Alternatively, \label{ft:theta=pi-dynamics}
if we instead interpret 
the background gauge variance of $\int \Sq^1 (w_2(TM)   B)$ as a 4d higher 't Hooft \emph{anomaly} (rather than just a 4d counter term), 
then it has a consequence on 4d SU(2)$_{\theta=\pi}$ YM  dynamics.
This means that when $K_2=1$, the
${\rm{Pin}^+(d) \times SU(2)}$
and ${\rm{Pin}^-(d) \times_{\Z_2} SU(2)}$ versions of 4d SU(2)$_{\theta=\pi}$ YM in \Eq{eq:4-G'} \emph{cannot} flow merely to the 
$\Z_2^T$-breaking Scenario (1) at IR.
However, Scenario (2), (3), and (4) are still possible IR fates.
It seems that the 
Scenario (2) with $\Z_{2,[1]}^e$-breaking with deconfinement (due to the perimeter law of Wilson loop) 
can be the most likely outcome. Scenario (2-ii) 
for ${\rm{Pin}^+(d) \times SU(2)}$
and ${\rm{Pin}^-(d) \times_{\Z_2} SU(2)}$ versions of 4d SU(2)$_{\theta=\pi}$ YM,
are proposed in 
\cite{2017arXiv171111587GPW} as two distinct versions of deconfined gapless theories or deconfined CFTs:
In this work, we show that these two siblings are indeed distinct theories with different Hilbert spaces at UV, 
due to their distinct 4d anomalies differed by a 5d invariant $\int_{M^5}  K_1 w_1(TM)^2  \Sq^1 B \sim \int_{M^5}  K_1 \cT^3 B$.\\
However, we will see that interpreting $\int_{M^5} \Sq^1 (w_2(TM)   B)$ as 4d higher 't Hooft anomaly will lead to a rather bizarre and strong constraint
on the dynamics of 4d SU(2)$_{\theta=0}$ YM, see the footnote \ref{ft:theta=0-dynamics}
}
\begin{equation}   \label{Eq.5danomalypolynomial-Conc}
\begin{split}
\bZ^{\text{5d}}_{\text{SPTs}}[M^5;B,w_j(TM)]= \exp \Bigg(\ii \pi \int_{M^5} & B \Sq^1 B + \Sq^2 \Sq^1 B+K_1w_1(TM)^2 \Sq^1 B    \Bigg),
\end{split}
\end{equation}

In fact,  in \Sec{sec:lattice}, we construct the 4d boundary based of the Scenario (3) above as a boundary TQFT with a lattice spacetime path integral or a lattice Hamiltonian regularization;
in this case, the full spacetime partition function $\bZ[M]$ of 4d-5d system can be explicitly computed as a number (by following Sec.~9 of \cite{Wang2017locWWW1705.06728}).  

We will revisit other issues of dynamics in the future.


\item \emph{Refinement and modification of 
Yang-Mills existence and mass gap problem} at $\theta=0$: 
The original statement of the Clay Math Millennium Prize Problem \cite{YMMP-Jaffe-Witten}
is ``Prove that for any compact simple gauge group $G$, a non-trivial quantum Yang-Mills theory exists on $\mathbb {R}^{4}$ and has a mass gap $\Delta  > 0$.''
It seems that in addition to the specification of gauge group $G$ and topological term $\frac{ \theta}{8\pi^2}\int_{M^4} \Tr (F\wedge F)$ with $\theta=0$,
we may also need to specify the data $(K_1,K_2) \in (\Z_2, \Z_2)$ for Kramers single/doublet and bosonic/fermionic statistics for quantum number of Wilson lines $W_e$,
as we did in \Eq{Eq.refinedGBC}
and \Eq{eq:SU2YMZ},
say for 4d SU(2)$_{\theta=0}$ YM.
The data $(K_1,K_2) \in (\Z_2, \Z_2)$ may have been ignored in the past.

Here are possible outcomes for Four Siblings $(K_1,K_2) \in (\Z_2, \Z_2)$ of 4d SU(2)$_{\theta=0}$ YM.
Notice that the transition from the vacua of 4d SU(2)$_{\theta=\pi}$ YM to 4d SU(2)$_{\theta=0}$ YM
must break $\Z_2^T$ in between $0<\theta< \pi$.

\underline{$\spadesuit$ 4d SU(2)$_{\theta=0}$ YM $(K_1,K_2)=(0,0)$ and $(1,0)$}:\\
Since the anomalies associated to 4d SU(2)$_{\theta=\pi}$ YM with $(K_1,K_2)=(0,0)$ and $(1,0)$ can be removed by $\Z_2^T$-breaking,
therefore 4d SU(2)$_{\theta=0}$ YM $(K_1,K_2)=(0,0)$ and $(1,0)$ can have no 't Hooft anomaly, thus it can be trivially gapped as a trivial vacuum.

\underline{$\spadesuit$ 4d SU(2)$_{\theta=0}$ YM with $(K_1,K_2)=(0,1)$ and $(1,1)$}:\\
These two siblings only differ from the  trivially gapped vacuum of the previous two siblings, $(K_1,K_2)=(0,0)$ and $(1,0)$,
by the 4d counter term 
$\int K_2 \Sq^1 (w_2(TM)   B)$.\footnote{\label{ft:theta=0-dynamics}
Follow the earlier footnote \ref{ft:theta=pi-dynamics},
if we instead interpret 
the background gauge variance of $\int \Sq^1 (w_2(TM)   B)$ as a 4d higher 't Hooft \emph{anomaly} (rather than just a 4d counter term), then it has a consequence on 4d SU(2)$_{\theta=0}$ YM dynamics also. Notice that $\int \Sq^1 (w_2(TM)   B)$ survive without $\Z_2^T$-protection, 
therefore if there is an ``anomaly'' $\int \Sq^1 (w_2(TM)   B)$ at $\theta= \pi$, then it remains for all $0 \leq \theta \leq \pi$ including at $\theta= 0$. 
The only way to saturate the if-anomaly of $\int K_2 \Sq^1 (w_2(TM)   B)$ for SU(2)$_{\theta=0}$ YM is breaking the 1-form symmetry.
If so, this means that  SU(2)$_{\theta=0}$ YM with fermionic Wilson line (i.e., $K_2=1$)
has 1-form symmetry spontaneously broken thus \emph{deconfined},
which cannot be trivially gapped nor a trivial vacuum! 
In this case, if 4d SU(2)$_{\theta=0}$ YM with $K_2=1$  is still gapped as the conventional wisdom goes,
they belong to the scenarios:\\
$\bullet$ {(2-i) 1-form $\Z_{2,[1]}^e$-breaking and deconfined TQFTs, i.e., \emph{topological order}.}\\
This deconfined scenario seems to be too exotic for SU(2)$_{\theta=0}$ YM merely with fermionic Wilson line.\\
Therefore, our canonical interpretation with $\int \Sq^1 (w_2(TM)   B)$ being a 4d counter term in 4d YM (see the main text around \Eq{eq:SU2YMZ-Conc-4d})
avoids leading to this bizarre deconfinement scenario for SU(2)$_{\theta=0}$ YM.
}

It will be enlightening to contemplate more consequences of their IR dynamics for these Four Siblings $(K_1,K_2) \in (\Z_2, \Z_2)$ of 4d SU(2) YM. 
\end{enumerate}

\section{Acknowledgements} 

The authors are listed in the alphabetical order by the standard convention.
JW thanks Pierre Deligne, Juan Maldacena, Max Metlitski, Kantaro Ohmori, Nathan Seiberg, Edward Witten and Shing-Tung Yau for encouraging conversations,
 thanks Robert Gompf and Pavel Putrov 
 for research collaborations in Ref.~\cite{JWGompf}, and in Ref.~\cite{Putrov2016qdo1612.09298, 2017arXiv171111587GPW}  respectively.
We also thank Clay Cordova, Zohar Komargodski and Ho Tat Lam for conversations.
JW thanks Harvard CMSA for the seminar invitation (March 11,  April 10 and September 10, 2019) on presenting this specific work \cite{CMSA}   
and thanks the audience for the feedback at the IAS Princeton seminar \cite{IAS}.
ZW  acknowledges support from NSFC grants 11431010 and 11571329. 
ZW is supported by the Shuimu Tsinghua Scholar Program.
JW is supported by
NSF Grant PHY-1606531. 
YZ thanks the support from Physics Department of Princeton University.
This work is also supported by 
NSF Grant DMS-1607871 ``Analysis, Geometry and Mathematical Physics'' 
and Center for Mathematical Sciences and Applications at Harvard University.

\appendix

\section{Computation of bordism groups}
\label{appendix-bordism}

In this Appendix, we compute the bordism groups $\Omega_d^{G'}$ where
$G'$ is a solution of all possible extensions of
$$
1 \to \SU(2) \to  G' \to \tO(d) \to 1;
$$
given by \Eq{eq:4-G'}
$$ 
G'=\left\{\begin{array}{l} 
{\rm{O}(d) \times SU(2)} \text{ (bosonic, relates to $(K_1,K_2)=(0,0)$) in Appendix \ref{app:1}},\\
{\rm{E}(d) \times_{\Z_2} SU(2)} \text{ (bosonic, relates to $(K_1,K_2)=(1,0)$) in Appendix \ref{app:2}},\\
{\mathrm{Pin}^{+}\times_{\Z_2^F} \SU(2)} \text{ (fermionic, relates to $(K_1,K_2)=(0,1)$)  in Appendix \ref{app:3}},\\ 
{\mathrm{Pin}^{-}\times_{\Z_2^F} \SU(2)} \text{ (fermionic, relates to $(K_1,K_2)=(1,1)$)  in Appendix \ref{app:4}}.
\end{array}\right.
$$
The bordism groups and their bordism invariants (topological invariants and SPTs) are used in the main text, for example, 
$\Omega_{d=4}^{G'}$ in \Sec{sec:SU2}.
We also compute $\Omega_{d=5}^{\tO}(\B^2 \Z_2)$, used in \Sec{sec:SPT}, in Appendix \ref{app:5}.

In the two subsections Appendix \ref{app:3}-\ref{app:4}, we will encounter the $\A_2(1)$ module structure due to the appearance of $M\Spin$ in the decomposition of $MT(\Pin^{\pm}\times_{\Z_2}\SU(2))$.

Readers can find the introduction to this computation in \Refe{Freed2016.1604.06527,{2017arXiv171111587GPW},{Wan2018bns1812.11967}}.  
For a short summary of the used concepts and terminology here, the readers may consult a succinct summary in the Appendix B of \cite{2017arXiv171111587GPW}.
For readers who are not familiar with the details of mathematical calculations, 
we will help by stating the results explicitly.

\subsection{Bordism group of  $\mathrm{O} \times \SU(2)$: $\Omega_{d}^{\mathrm{O}\times\SU(2)}$}

\label{app:1}

We first notice that $MT(\tO\times\SU(2))=M\tO\wedge\B\SU(2)_+$, where $\wedge$ is the smash product and $\SU(2)_+$ is the disjoint union of the topological space $\SU(2)$ and a point. 
$MTH$ is the Madsen-Tillmann spectrum of the group $H$,
$MH$ is the Thom spectrum of the group $H$.

By the Adams spectral sequence, we have
\bea
\Ext_{\A_2}^{s,t}(\H^*(M\tO\wedge\B\SU(2)_+,\Z_2),\Z_2)\Rightarrow\Omega_{t-s}^{\tO\times\SU(2)}.
\eea

The Thom spectrum $M\rO$ is the wedge sum of suspensions of the mod 2 Eilenberg-MacLane spectrum $H\Z_2$, $\H^*(M\tO,\Z_2)$ is the direct sum of suspensions of the mod 2 Steenrod algebra $\A_2$, actually Thom proved
that
\bea
\pi_*(M\rO)=\Omega_*^{\rO}=\Z_2[y_2,y_4,y_5,y_6,y_8,\dots]
\eea
where the generators are in each degree other than $2^n-1$.

\begin{table}[!h]
	\centering
	\begin{tabular}{c c c c}
		\hline
		$i$ & $\Omega^{\tO}_i$  & manifold generators & cobordism invariants\\
		\hline
		0& $\Z_2$\\
		1& $0$\\
		2& $\Z_2$ &  $\RP^2$  & $w_1^2$ \\
		3 & $0$\\
		4 & $\Z_2^2$ &  $\RP^4,\RP^2\times\RP^2$ & $w_1^4,w_2^2$ \\ 
		5 & $\Z_2$&  Wu manifold or Dold manifold & $w_2w_3$\\
		\hline
	\end{tabular}
\end{table}

So $M\tO=H\Z_2\vee\Sigma^2H\Z_2\vee2\Sigma^4H\Z_2\vee\Sigma^5H\Z_2\vee\cdots$ and $\H^*(M\tO,\Z_2)=\A_2\oplus\Sigma^2\A_2\oplus2\Sigma^4\A_2\oplus\Sigma^5\A_2\oplus\cdots$.

Since
\bea
\H^*(\B\SU(2),\Z_2)=\Z_2[c_2],
\eea
by the K\"unneth formula, we get
\bea
\H^*(M\tO\wedge\B\SU(2)_+,\Z_2)&=&\H^*(M\tO,\Z_2)\otimes\H^*(\B\SU(2),\Z_2)\nn\\
&=&(\A_2\oplus\Sigma^2\A_2\oplus2\Sigma^4\A_2\oplus\Sigma^5\A_2\oplus\cdots)\otimes\Z_2[c_2] \nonumber\\
&=&\A_2\oplus\Sigma^2\A_2\oplus3\Sigma^4\A_2\oplus\Sigma^5\A_2\oplus\cdots.
\eea
Here $\Sigma^n\A_2$ is the $n$-th iterated shift of the graded algebra $\A_2$.

Since 
\bea
\Ext_{\A_2}^{s,t}(\Sigma^r\A_2,\Z_2)=\left\{\begin{array}{ll}\Hom_{\A_2}^t(\Sigma^r\A_2,\Z_2)=\Z_2&\text{ if }t=r, s=0\\ 0 &\text{ else}\end{array}\right.,
\eea 

we have
$\Omega_4^{\tO\times\SU(2)}=\Z_2^3$,
$\Omega_5^{\tO\times\SU(2)}=\Z_2$.

 The bordism invariants of $\Omega_4^{\tO\times\SU(2)}=\Z_2^3$ are $w_1^4$, $w_2^2$, and $c_2\mod2$.
Namely, in physics terms,
the topological invariants/SPTs from $\Omega_4^{\tO\times\SU(2)}$ 
are $w_1(TM)^4$, $w_2(TM)^2$, and $c_2\mod2$.

 The bordism invariant of $\Omega^{\tO\times\SU(2)}_5=\Z_2$ is $w_2w_3.$
Namely, in physics terms,
the topological invariants/SPTs from  $\Omega^{\tO\times\SU(2)}_5$
 is $w_2(TM)w_3(TM).$

\subsection{Bordism group of  $\mathrm{E}\times_{\Z_2}\SU(2)$: $\Omega_{d}^{\mathrm{E}\times_{\Z_2}\SU(2)}$}

\label{app:2}

Recall that $\rE$ is defined to be the subgroup of $\rO\times\Z_4$ consisting of the pairs $(A,j)$ such that $\det A=j^2$, there is a fibration $\B\rE\to\B\rO\xrightarrow{w_1^2}\B^2\Z_2$. 

We can also think of the space $\B\rE$ as the fiber of $w_1 + x: \B\rO \times \B\Z_4 \to \B\Z_2$, where $x$ is the generator of $\H^1(\B\Z_4, \Z_2)$.

Note that $\SU(2)\times_{\Z_2}\Z_4=\Pin^+(3)$, we can think of the space $\B(\rE\times_{\Z_2}\SU(2))$ as the fiber of $w_1+w_1':\B\rO\times\B\Pin^+(3)\to\B\Z_2$, where $w_1'$ is the generator of $\H^1(\B\Pin^+(3),\Z_2)$. Take $W$ to be the rank 3 vector bundle on $\B\Pin^+(3)$ determined by $\B\Pin^+(3)\to\B\rO(3)$. 

Define a map $f: \B\rO \times \B\Pin^+(3) \to \B\rO \times \B\Pin^+(3)$ by $(V, V') \to (V + W - 3, V')$, with inverse $(V, V') \to (V - W + 3, V')$.  
Observe $f^*(w_1) = w_1 + w_1'$, so that $\cred{\B(\rE\times_{\Z_2}\SU(2))}$ is homotopy equivalent to $\B\SO \times \B\Pin^+(3)$.  
The canonical bundle $\cred{\B(\rE\times_{\Z_2}\SU(2))} \to \B\rO$ corresponds to $V - W + 3$ on $\B\SO \times \B\Pin^+(3)$.  
So $MT(\rE\times_{\Z_2}\SU(2)) = MT\SO \wedge \text{Thom}(\B\Pin^+(3), \cred{W-3}) = M\SO\wedge\Sigma^{-3}M\Pin^+(3)$.  

Note that $M\Pin^+(3)=MT\Pin^-(3)=MT(\Spin(3)\times\Z_2)=MT\Spin(3)\wedge MT\Z_2$.

So $MT(\rE\times_{\Z_2}\SU(2)) = M\SO\wedge\Sigma^{-4}M\SU(2)\wedge\Sigma^1MT\rO(1)\simeq M\rO\wedge\Sigma^{-4}M\SU(2)$, here $\wedge$ is the smash product, $\Sigma$ is the suspension, $MTH$ is the Madsen-Tillmann spectrum of the group $H$ and
$MH$ is the Thom spectrum of the group $H$.

 By the Adams spectral sequence,
\bea
\Ext_{\A_2}^{s,t}(\H^*(M\tO\wedge\Sigma^{-4}M\SU(2),\Z_2),\Z_2)\Rightarrow\Omega_{t-s}^{\tE\times_{\Z_2}\SU(2)}.
\eea
Since
\bea
\H^*(\Sigma^{-4}M\SU(2),\Z_2)=\Z_2[c_2]U
\eea
where $c_2$ is the Chern class of the $\SU(2)$ bundle and $U$ is the Thom class.

By the K\"unneth formula,
\bea
\H^*(M\tO\wedge\Sigma^{-4}M\SU(2),\Z_2)&=&\H^*(M\tO,\Z_2)\otimes\H^*(\Sigma^{-4}M\SU(2),\Z_2)\nn\\
&=&(\A_2\oplus\Sigma^2\A_2\oplus2\Sigma^4\A_2\oplus\Sigma^5\A_2\oplus\cdots)\otimes\Z_2[c_2]U\nn\\
&=&\A_2\oplus\Sigma^2\A_2\oplus3\Sigma^4\A_2\oplus\Sigma^5\A_2\oplus\cdots.
\eea

Hence we have
$\Omega_4^{\tE\times_{\Z_2}\SU(2)}=\Z_2^3$,
$\Omega_5^{\tE\times_{\Z_2}\SU(2)}=\Z_2$.

The bordism invariants of $\Omega_4^{\tE\times_{\Z_2}\SU(2)}=\Z_2^3$ are $w_1^4$, $w_2^2$, and $c_2\mod2$.
Namely, in physics terms,
the topological invariants/SPTs from $\Omega_4^{\tE\times_{\Z_2}\SU(2)}$ 
are $w_1(TM)^4$, $w_2(TM)^2$, and $c_2\mod2$. 
{Since the constraint $w_1(TM)^2=w_2(V_{\SO(3)})$ is satisfied, let $\beta_2$ denote the Bockstein homomorphism associated to the extension $\Z\to\Z\to\Z_2$, then $W_3(V_{\SO(3)})=\beta_2w_2(V_{\SO(3)})=\beta_2w_1(TM)^2=\beta_2\Sq^1w_1(TM)=0$ where $W_3(V_{\SO(3)})$ is the third integral Stiefel-Whitney class of $V_{\SO(3)}$ and we have used the fact that $\beta_2\Sq^1=0$, hence $V_{\SO(3)}$ lifts to a $\Spin^c(3)=\U(2)$ bundle $V_{\U(2)}$, here $c_2=c_2(V_{\U(2)})$ is the second Chern class of $V_{\U(2)}$.
}

The bordism invariants of $\Omega^{\tE\times_{\Z_2}\SU(2)}_5$  is $w_2w_3.$
Namely, in physics terms,
the topological invariants/SPTs from  $\Omega^{\tE\times_{\Z_2}\SU(2)}_5$
 is $w_2(TM)w_3(TM).$

\subsection{Bordism group of  $\Pin^+\times_{\Z_2}\SU(2)$: $\Omega_{d}^{\Pin^+\times_{\Z_2}\SU(2)}$}

\label{app:3}

Since there is a homotopy pullback square
\begin{center}
	\begin{tikzcd}
		\B H \ar[r,"\sim"] \ar[d] & \B\Pin^+\times \B\SO(3) \ar[d]\\
		\B\tO\times \B\SO(3) \ar[r,"f"] \ar[d,"{pr_1,V}"] \ar[rr,bend right=10,"w_2+w'_2"' ] & \B\tO\times \B\SO(3) \ar[r,"{w_2+0}"] \ar[dl,"V+W-3"]& \B^2\Z_2 \\
		\B\tO
	\end{tikzcd}
\end{center}
where $f$ maps $(V,W)$ to $(V-W+3,W)$,
we have $MTH=MT\Pin^+\wedge\Sigma^{-3}M\SO(3)=M\Spin\wedge\Sigma^{-3}M\tO(3)$. 
By the Adams spectral sequence,
\bea
\Ext_{\A_2}^{s,t}(\H^*(M\Spin\wedge\Sigma^{-3}M\tO(3),\Z_2),\Z_2)\Rightarrow\Omega_{t-s}^{\Pin^+\times_{\Z_2}\SU(2)}
\eea
The mod 2 cohomology of Thom spectrum $M\Spin$ is
\bea
\H^*(M\Spin,\Z_2)=\A_2\otimes_{\A_2(1)}\{\Z_2\oplus M\}
\eea
where $M$ is a graded $\A_2(1)$-module with the degree $i$ homogeneous part $M_i=0$ for $i<8$. Here $\A_2(1)$ stands for the sub-algebra of $\A_2$ generated by $\Sq^1$
and $\Sq^2$.

For $t-s<8$, we can identify the $E_2$ page with
\bea
\Ext_{\A_2(1)}^{s,t}(\H^{*+3}(M\tO(3),\Z_2),\Z_2).
\eea

For other details and the computation of ${\A_2(1)}$ module structure and Adams chart, please consult \Refe{Freed2016.1604.06527, 2017arXiv171111587GPW}.
We can extract the bordism group and their bordism invariants from \cite{2017arXiv171111587GPW}.

\subsection{Bordism group of  $\Pin^-\times_{\Z_2}\SU(2)$: $\Omega_{d}^{\Pin^-\times_{\Z_2}\SU(2)}$}

\label{app:4}

Since there is a homotopy pullback square
\begin{center}
	\begin{tikzcd}
		\B H \ar[r,"\sim"] \ar[d] & \B\Pin^-\times \B\SO(3) \ar[d]\\
		\B\tO\times \B\SO(3) \ar[r,"f"] \ar[d,"{pr_1,V}"] \ar[rr,bend right=10,"w_2+w_1^2+w'_2"' ] & \B\tO\times \B\SO(3) \ar[r,"{w_2+w_1^2+0}"] \ar[dl,"V+W-3"]& \B^2\Z_2 \\
		\B\tO
	\end{tikzcd}
\end{center}
where $f$ maps $(V,W)$ to $(V-W+3,W)$,
we have $MTH=MT\Pin^-\wedge\Sigma^{-3}M\SO(3)=M\Spin\wedge\Sigma^{3}MT\tO(3)$. 
By the Adams spectral sequence,
\bea
\Ext_{\A_2}^{s,t}(\H^*(M\Spin\wedge\Sigma^{3}MT\tO(3),\Z_2),\Z_2)\Rightarrow\Omega_{t-s}^{\Pin^-\times_{\Z_2}\SU(2)}
\eea
For $t-s<8$, we can identify the $E_2$ page with
\bea
\Ext_{\A_2(1)}^{s,t}(\H^{*-3}(MT\tO(3),\Z_2),\Z_2).
\eea

For other details and the computation of ${\A_2(1)}$ module structure and Adams chart, please consult \Refe{Freed2016.1604.06527, 2017arXiv171111587GPW}.
We can extract the bordism group and their bordism invariants from \cite{2017arXiv171111587GPW}.

\subsection{Bordism group of $\mathrm{O} \times \Z_{2,[1]}^e$: $\Omega_{d}^{\mathrm{O}}(\B^2 \Z_{2})$}

\label{app:5}

By the Adams spectral sequence 
\bea
\Ext_{\A_2}^{s,t}(\H^*(M\tO\wedge (\B ^2\Z_2)_+,\Z_2),\Z_2)\Rightarrow\Omega^{\tO}_{t-s}(\B ^2\Z_2).
\eea

Since $\H^*(\B ^2\Z_2,\Z_2)=\Z_2[x_2,x_3,x_5,x_9,\dots] $
where $x_2$ is the generator of $\H^2(\B^2\Z_2,\Z_2)$, $x_3=\Sq^1x_2$, $x_5=\Sq^2x_3$, $x_9=\Sq^4x_5$, etc, so by the K\"unneth formula,
\bea
\H^*(M\tO\wedge (\B ^2\Z_2)_+,\Z_2)&=&\H^*(M\tO,\Z_2)\otimes\H^*(\B ^2\Z_2,\Z_2)\nn\\
&=&(\A_2\oplus\Sigma^2\A_2\oplus2\Sigma^4\A_2\oplus\Sigma^5\A_2\oplus\cdots)\otimes\Z_2[x_2,x_3,x_5,x_9,\dots] \nonumber\\
&=&\A_2\oplus2\Sigma^2\A_2\oplus\Sigma^3\A_2\oplus4\Sigma^4\A_2\oplus4\Sigma^5\A_2\oplus\cdots
\eea

Hence we have 
$\Omega^{\tO}_4(\B ^2\Z_2)=\Z_2^4$, $\Omega^{\tO}_5(\B ^2\Z_2)=\Z_2^4$.

The bordism invariants of $\Omega_4^{\tO}(\B ^2\Z_2)=\Z_2^4$ are $x_2^2$, $w_1^4$, $w_1^2x_2$, and $w_2^2$.
Namely, in physics terms,
the topological invariants/SPTs from  $\Omega_4^{\tO}(\B ^2\Z_2)$
 are $B^2$, $w_1(TM)^4$, $w_1(TM)^2 B$, and $w_2(TM)^2$.

The bordism invariants of $\Omega^{\tO}_5(\B ^2\Z_2)=\Z_2^4$ are $x_2x_3$, $x_5$, $w_1^2x_3$, and $w_2w_3.$
Namely, in physics terms,
the topological invariants/SPTs from  $\Omega_5^{\tO}(\B ^2\Z_2)$
 are $B\Sq^1 B$, $\Sq^2 \Sq^1 B$, $w_1(TM)^2 \Sq^1 B$, and $w_2(TM)w_3(TM)$. 
Readers can find more detailed discussions and calculations of the cobordism theory of higher symmetries in \Refe{Wan2018bns1812.11967}.



\bibliographystyle{Yang-Mills.bst}
\bibliography{Yang-Mills.bib,Yang-Mills-JW-SET.bib, Ref_new.bib}

\end{document}